\documentclass[12pt, letterpaper]{report}   % Single-sided printing for the library

\usepackage[bf]{caption} % Make nice captions with bold "Figure..." and "Table..."
\usepackage{bu_math_thesis}

\usepackage[titletoc]{appendix}  % <-- all this does is display "Appendix" in the Table of Contents Entry for each appendix.  This is a UWM requirement.

\usepackage{orcidlink}  % In order for this to work, must include the oridlink.sty file in the project folder!

\graphicspath{{figures/}{tables/}}

\hypersetup{urlcolor=cyan}

\begin{document}
%%%%%%%%%%%%%%%%%%%%%%%%%%%%%%%%%%%%%%%%%%%%%%%%%%%%%%%%%%%%%%%%%%%%%%%%%%
% Setup commands for the bu thesis style file

\title{Gravitational Wave Timing Residual Models for Pulsar Timing Experiments}
\author{Casey McGrath}

% Type of document prepared for this degree:
%   2 = Doctor of Philosophy dissertation.
%   4 = Doctoral Dissertation Prospectus
\degree=2

% List your bachelors degree before your masters - HG 2018
\prevdegrees{BS in Physics, University of Redlands, 2014\\ 
BS in Mathematics, University of Redlands, 2014}
\department{Department of Physics}
\university{The University of Wisconsin-Milwaukee}
\faculty{Graduate School of Arts and Sciences}

% Degree year is the year the diploma is expected, and defense year is the year the dissertation is written up and defended. Often, these will be the same, except for January graduation, when your defense will be in the fall of year X, and your graduation will be in January of year X+1
\defenseyear{2021}
\degreeyear{2021}
\degreemonth{August}

% For each reader, specify appropriate label {First, second, third}, then name, then title. Warning: If you have more than five readers you are out of luck, because it will overflow to a new page. Sometimes you may wish to put part of the title in with the name
% Do NOT put the chair on your approval page - HG 2018
\reader{First}{Jolien Creighton, PhD}{Professor, Physics}
\reader{Second}{Patrick Brady, PhD}{Professor, Physics}
\reader{Third}{David Kaplan, PhD}{Associate Professor,  Physics}
\reader{Fourth}{Dawn Erb, PhD}{Associate Professor,  Physics}
\reader{Fifth}{Dan Agterberg, PhD}{Professor,  Physics}

% The Major Professor is the same as the first reader, but must be specified again for the abstract page
% Just copy and paste the same information
\majorprof{Jolien Creighton}{PhD}

%%%%%%%%%%%%%%%%%%%%%%%%%%%%%%%%%%%%%%%%%%%%%%%%%%%%%%%%%%%%%%%%%%%%%
% other set up commands which are a good idea

%the bottom margins should be "as close as possible" to 1 inch, so allowdisplaybreaks is a good idea for theses with a lot of equations
\allowdisplaybreaks

%%%%%%%%%%%%%%%%%%%%%%%%%%%%%%%%%%%%%%%%%%%%%%%%%%%%%%%%%%%%%%%%%%%%%
%                       PRELIMINARY PAGES
% According to the BU guide the preliminary pages consist of: title, copyright (optional), approval,  acknowledgments (opt.), abstract, preface (opt.), Table of contents, List of tables (if any), List of illustrations (if any). The \tableofcontents, \listoffigures, and \listoftables commands can be used in the appropriate places. For other things like preface, do it manually with something like \newpage\section*{Preface}.

%%%%%%%%%%%%%%%%%%%%%%%%%%%%%%%%%%%%%%%%%%%%%%%%%%%%%%%%%%%%%%%%%%%%
% This is an additional page (do not hand it in at the library) to print boxed-in title, author and degree statement so that they are visible through the opening in BU covers used for reports. This makes a nicely bound copy.

%\buecethesistitleboxpage

%%%%%%%%%%%%%%%%%%%%%%%%%%%%%%%%%%%%%%%%%%%%%%%%%%%%%%%%%%%%%%%%%%%%
%%%% TITLE PAGE 
% Make the titlepage based on the above information.  If you need something special and can't use the standard form, you can specify the exact text of the titlepage yourself.  Put it in a titlepage environment and leave blank lines where you want vertical space. The spaces will be adjusted to fill the entire page.
\maketitle

%%%%%%%%%%%%%%%%%%%%%%%%%%%%%%%%%%%%%%%%%%%%%%%%%%%%%%%%%%%%%%%%%%%%
%%%% ABSTRACT		 
\newpage
\cleardoublepage
\phantomsection
\addcontentsline{toc}{chapter}{Abstract}
% The abstract page environment sets up everything on the page except the text itself.  The title and other header material are put at the top of the page, and the supervisors are listed at the bottom.  A new page is begun both before and after.  Of course, an abstract may be more than one page itself.  If you need more control over the format of the page, you can use the abstract environment, which puts the word "Abstract" at the beginning and single spaces its text.

% 1.5.1 Purpose in writing the abstract

% The abstract should contain a clear and brief statement of the problem, the procedure(s) and/or method(s) followed, the result(s), and the conclusion(s). The purpose of an abstract is to help a reader decide if they want to consult the complete work. 
% As with the title, the abstract is searchable in many databases, including ProQuest Dissertations & Theses Full Text. Include relevant place names, full personal names, and other proper nouns, which can be very useful keywords for scholars locating resources.

% 1.5.2 Written in English and limited in length

% The abstract must be written in English. A dissertation abstract is limited to 350 words or approximately 2,450 characters. A thesis abstract is limited to 250 words or approximately 1,750 characters.

\begin{abstractpage}

The ability to detect gravitational waves now gives scientists and astronomers a new way in which they can study the universe.  So far, 
%the scientific collaboration LIGO has been successful in detecting binary black hole and binary neutron star mergers. 
%
the LIGO and Virgo scientific collaborations have been successful in detecting binary black hole and binary neutron star mergers~\citep{gw150914, gw170817_h0}.
These types of sources produce gravitational waves with frequencies of the order hertz to millihertz.  But there do exist other theoretical sources which would produce gravitational waves in different parts of the frequency spectrum.  Of these are the theoretical mergers of supermassive black hole binaries (SMBHBs), which could occur upon the merging of two galaxies with supermassive black holes at their cores.  Sources like these would produce gravitational waves generally around the nanohertz regime, and the current main effort for detecting and measuring these waves comes from pulsar timing experiments.  Detection of gravitational waves in these experiments would come as small fluctuations in the otherwise extremely regular period of pulsars over a long period of time (months to decades).

There are numerous goals for this dissertation.  The first is to re-present much of the fundamental physics and mathematics concepts behind the calculations in this paper.  While there exist many reference sources in the literature, we simply try to offer a fully self-contained explanation of the fundamentals of this research which we hope the reader will find helpful.  It is often a challenge when jumping into a new field of study to quickly learn and understand the fundamentals (like the derivations of various formulae and the assumptions behind the models), so if this dissertation can help future readers to connect the dots between the blanks not filled in by other literature sources, then this goal will be accomplished.  The pedantic approach to this dissertation is also helpful since much of the initial work for this dissertation was theoretical development of the mathematical models used in pulsar timing.

The next goal broadly speaking has been to combine the efforts of two previous studies by \citet{DF_main_paper} and \citet{CC_main_paper} to further develop the mathematics behind the currently used pulsar timing models for detecting gravitational waves with pulsar timing experiments.  Previous timing residual models have first been derived assuming that the pulsar timing array receives plane-waves coming from distant sources (with the notable exception of~\citeauthor{DF_main_paper}).  Then these models can either treat the SMBHB as a monochromatic gravitational wave source, or model the frequency evolution of the gravitational waves over the thousands to tens of thousands of years it takes light to travel from the pulsar to the Earth.  Our research began by first generalizing these models by removing the plane-wave assumption.  In Chapter~\ref{ch: The Continuous Wave Timing Residual} we classify four regimes of interest (Figure~\ref{fig: 4 regimes}), governed by the main assumptions made when deriving each regime.  Of these four regimes the plane-wave models are well established in previous literature.  We add a new regime which we label ``Fresnel,'' as we will show it becomes important for significant Fresnel numbers describing the curvature of wavefronts.

With these mathematical models developed, in Chapter~\ref{ch:results - first investigation} we present the first main study investigated which was to forecast the ability of future pulsar timing experiments to probe and measure these Fresnel effects.  Here we show the constraints needed on the pulsar timing experiments themselves (largely explained by the discussion in Chapter~\ref{ch:L-wrapping problem}), and the types of precision measurements which could theoretically be achieved.

Then we generalize our models to a cosmologically expanding universe in Chapter~\ref{ch:results - H0 measurement}.  We show that in the fully general Fresnel frequency evolution regime, the Hubble constant enters the model and can now be measured directly.  In this chapter we investigate what we will need of future experiments in order to obtain a measurement of this parameter.  This offers future pulsar timing experiments the unique possibility of being able to procure a purely gravitational wave-based measurement of the Hubble constant. 

Finally, Chapter~\ref{ch: General doppler tracking} shows the initial steps taken to extend this work in the future, specifically for Doppler tracking experiments.  The main goal of the inclusion of this final section, which was not the primary focus of this dissertation, is to point out how the mathematics and models derived in Chapters~\ref{ch: GWs from a Binary Source} and~\ref{ch: The Continuous Wave Timing Residual} can be applied and extended more generally.

\end{abstractpage}

% Now you can include a preface. Again, use something like
% \newpage\section*{Preface} followed by your text

%%%%%%%%%%%%%%%%%%%%%%%%%%%%%%%%%%%%%%%%%%%%%%%%%%%%%%%%%%%%%%%%%%%%
%%%% COPYRIGHT PAGE 
% The copyright page is blank except for the notice at the bottom. You must provide your name in capitals.
\copyrightpage

%%%%%%%%%%%%%%%%%%%%%%%%%%%%%%%%%%%%%%%%%%%%%%%%%%%%%%%%%%%%%%%%%%%%
%%%% APPROVAL PAGE 
% Now include the approval page based on the readers information
% \approvalpage

%%%%%%%%%%%%%%%%%%%%%%%%%%%%%%%%%%%%%%%%%%%%%%%%%%%%%%%%%%%%%%%%%%%%
%%%% DEDICATION			 
% \addcontentsline{toc}{chapter}{Dedication}
% \section*{Dedication}
% This dissertation is dedicated to....

%%%%%%%%%%%%%%%%%%%%%%%%%%%%%%%%%%%%%%%%%%%%%%%%%%%%%%%%%%%%%%%%%%%%
%%%%% CODE FOR CLEANING UP PDF HYPERLINKS - HG 2018
%remove the hyperlink colors for table of contents
\begingroup
  \hypersetup{linkbordercolor=white,linkcolor=black,
    filecolor=black, urlcolor=black}

%%%%%%%%%%%%%%%%%%%%%%%%%%%%%%%%%%%%%%%%%%%%%%%%%%%%%%%%%%%%%%%%%%%%
%%%% TABLE OF CONTENTS	 
% Table of contents comes after preface
\tableofcontents

%%%%%%%%%%%%%%%%%%%%%%%%%%%%%%%%%%%%%%%%%%%%%%%%%%%%%%%%%%%%%%%%%%%%
%%%% LIST OF FIGURES	
% If you have figures this goes here - needs to be in appendix HG 2018
\newpage
\cleardoublepage
\phantomsection
\addcontentsline{toc}{chapter}{List of Figures}
\listoffigures

%%%%%%%%%%%%%%%%%%%%%%%%%%%%%%%%%%%%%%%%%%%%%%%%%%%%%%%%%%%%%%%%%%%%
%%%% LIST OF TABLES	 
% If you have tables this goes here - needs to be in appendix HG 2018
\newpage
\cleardoublepage
\phantomsection
\addcontentsline{toc}{chapter}{List of Tables}
\listoftables

\endgroup

%%%%%%%%%%%%%%%%%%%%%%%%%%%%%%%%%%%%%%%%%%%%%%%%%%%%%%%%%%%%%%%%%%%%
%%%% LIST OF SYMBOLS AND ABBREVIATIONS	 
% List of Abbrevs is NOT optional (PUT IN ALPHABETICAL ORDER)
% For mathematics a list of symbols is perhaps more appropriate, but fulfills the same role
\newpage
\cleardoublepage
\phantomsection
\addcontentsline{toc}{chapter}{List of Symbols and Abbreviations}
\chapter*{List of Abbreviations}\label{RefSymbols}
  \begin{longtable}{lp{0.75\textwidth}}
    PTA \dotfill & Pulsar Timing Array \\
    TOA \dotfill & Time-of-arrival \\
    SMBHB \dotfill & Supermassive Black Hole Binary \\
    NANOGrav \dotfill &  North American Nanohertz Observatory for Gravitational Waves (a PTA) \\
    LIGO \dotfill & Laser Interferometer Gravitational-wave Observatory \\
    KAGRA \dotfill & Kamioka Gravitational Wave Detector \\
    CV \dotfill & Coefficient of Variation \\
    SNR \dotfill & Signal-to-Noise Ratio \\
\end{longtable}

%%%%%%%%%%%%%%%%%%%%%%%%%%%%%%%%%%%%%%%%%%%%%%%%%%%%%%%%%%%%%%%%%%%%
%%%% ACKNOWLEDGEMENTS	 
\newpage
\cleardoublepage
\phantomsection
\addcontentsline{toc}{chapter}{Acknowledgements}
\chapter*{Acknowledgments}

I am grateful to have known many wonderful people throughout my entire life, who have stimulated my curiosity and nurtured me along the path which has now culminated in the completion of this PhD work.  This path for me began in 5th grade, when I discovered my love for science and astronomy.  The last nearly two decades have been a remarkable journey of personal growth, and this PhD marks the end of a large phase of my life.  All of the people who I have known and supported me during this time have my love and thanks for everything they have done for me.

My time as a graduate student has been extremely trying, and not what I envisioned.  But even still there have been certain individuals who have given me their support and help during this time, and for that I am very grateful.  I am thankful for the relationships I had with all of the graduate students in my department - many of whom helped and supported me through their friendship.  Especially Alex McEwen, Siddharth Mohite, Deep Chatterjee, Ian Brown, Caitlin Rose, and Kristina Islo who always had my back and would help me when I needed it.

From the encouragement of some of the previously mentioned individuals, during my PhD work I began seeing a therapist for help with my mental health.  This was the single greatest decision I made during this period of my life.  The friendship I found with Katt Cochran has been one of the greatest I have ever known, and what I needed the most in my life at this time.

Of the post-docs in the department I am especially grateful to have known and worked with Angela Van Sistine, Megan DeCesar, Shaon Ghosh, Shasvath Kapadia, Danielle Berg, Annalisa Citro, Laleh Sadeghian, and Sydney Chamberlin.  I am especially grateful to have worked with them in organizing our public outreach group ``CoffeeShop Astrophysics,'' to bring science and a love for astronomy to the greater Milwaukee community.

In the department I am very thankful to Heidi Matera, Robert Wood, Tonia Klein, and Swapnil Tripathi for their immense support.

I am very grateful to have met and collaborated with Dan D'Orazio near the end of my PhD work.  Working with him was what I had imagined my PhD was going to be like when I had started my time here, so I am very glad that we met.  Additionally, Luke Kelley is an incredibly important person to me for the help and support that he gave me.  Being able to open up to him and have honest conversations was very much needed and refreshing.  I am also thankful for Michael Lam, Stephen Taylor, and Joe Swiggum for their help throughout my research work, answering my questions and pointing me to helpful resources when I needed it.

Much of my work required extensive computing capabilities for my numerical simulations, which was accomplished using the University of Wisconsin-Milwaukee's High Performance Computing resources on campus.  I'd especially like to thank Darin Peetz and Daniel Siercks who maintain and run these resources, as they were always extremely willing to help answer any and all of my questions, and moreover teach me more about high performance computing so that I could better improve my skills.

Lastly, my PhD work was supported by the National Science Foundation (NSF) through the NANOGrav collaboration Physics Frontier Center award NSF PHY-1430284, as well as the awards NSF PHY-1607585 and NSF PHY-1912649. This work also received support from the Wisconsin Space Grant Consortium Graduate and Professional Research Fellowship Program under the NASA Training Grant \#NNX15AJ12H, as well as from the University of Wisconsin-Milwaukee, for which I am grateful for access to their computational resources supported by NSF PHY-1626190.

% END OF THE PRELIMINARY PAGES
\newpage
\endofprelim
\cleardoublepage

%%%%%%%%%%%%%%%%%%%%%%%%%%%%%%%%%%%%%%%%%%%%%%%%%%%%%%%%%%%%%%%%%%
% the body of the thesis goes here.
\chapter{Introduction}\label{ch:intro}

The mathematics and physics presented in this chapter are generally true for any binary black hole system.  However, as this work is building towards a pulsar timing specific study, we will specifically focus our discussion towards ``supermassive black hole binaries'' (SMBHBs).

It is theorized that during the collision of galaxies with supermassive black holes at their centers, if the two black holes come close enough then they could form a binary pair (a SMBHB), and their orbital interaction would cause massive distortions in spacetime which would propagate away as gravitational waves.  SMBHB sources would likely be slowly orbiting each other with periods ranging from months to decades, and hence they are thought to produce gravitational waves in and around the nanohertz regime.

Current efforts to detect these types of gravitational waves comes from pulsar timing techniques.  Some pulsars in our Galaxy can be used as extremely regular clocks when their pulses are timed by radio telescopes.  By timing a collection of such pulsars, known as a pulsar timing array (PTA), over long periods of time, scientists can look for small deviations in their otherwise very precise ``clock ticks.''  It is understood that the presence of gravitational waves hitting the timed pulsars would induce a periodic redshifting and blueshifting of the timing signals as measured on Earth (discussed in detail in Chapter~\ref{ch: The Continuous Wave Timing Residual}).  Hence by looking for such deviations in the timing signals of pulsars, scientists hope to find the evidence of the presence of gravitational waves.
%produced within the frequency regime.
%
A discovery such as this would be highly complimentary to 
%LIGO, as not only would it be in a gravitational frequency regime that the LIGO experiment cannot probe, 
%
ground-based detectors such as LIGO~\citep{ligo_ref}, Virgo~\citep{virgo_ref}, and KAGRA~\citep{kagra_ref}, as it would be in a gravitational frequency regime that these experiments cannot probe.
%but it would also be the product of different types of astrophysical sources in the universe.

SMBHBs would produce what are called ``continuous waves,'' as the black holes would likely be sufficiently far apart that as they orbited each other the loss of energy in the system radiated away as gravitational waves would not cause a significant collapse of their orbit.  As such these sources would maintain a nearly constant or ``continuous'' gravitational wave frequency, hence the name of this type of radiation.  It would not be until later times in their binary evolution that the frequency of the orbit would begin to evolve notably over the observation time scale (a process called ``frequency chirping'').  And unlike 
%the LIGO experiment, 
ground-based experiments,
PTAs would not be sensitive to the much later inspiral or merger of such objects where the frequency chirping is very strong as the binary coalesces.  However, to try and head-off any misconceptions early on, we are referring to the frequency of the gravitational waves measured on \textit{observational time scales}.  As we will see later in Chapter~\ref{ch: The Continuous Wave Timing Residual}, the gravitational wave frequency can evolve in the measured timing residual of a pulsar due to the nature of the incredibly large physical scale of a PTA experiment.

% Currently, pulsar timing collaborations such as NANOGrav have been regularly collecting timing data on numerous pulsars for over a decade, which makes it hopeful that a confirmed detection of gravitational waves using this method will be possible~\citep{NG_11yr_cw}.
%
Currently, pulsar timing collaborations such as the North American Nanohertz Observatory for Gravitational Waves (NANOGrav), the Parkes Pulsar Timing Array, and the European Pulsar Timing Array have been regularly collecting timing data on numerous pulsars for over a decade, which makes it hopeful that a confirmed detection of gravitational waves using this method will be possible~\citep{PPTA_cw_2014,EPTA_cw_2016,NG_11yr_cw}.
While the first detection may be the cumulative effect of gravitational waves coming from many sources all across the sky as a stochastic background, in this work we focus on the possibility of detecting individually loud continuous wave signals coming from SMBHBs.

\begin{minipage}{0.9\linewidth}
\fbox {
    \parbox{\linewidth}{ \textbf{Publications Connected to this Dissertation}\newline
    The work presented in \textbf{Chapters~\ref{ch: The Continuous Wave Timing Residual}, \ref{ch: Fisher matrix analysis}, \ref{ch:L-wrapping problem}, and~\ref{ch:results - first investigation}} was published in the following paper:
    
    \begin{displayquote}
    McGrath, C., \& Creighton, J. (2021). \textit{Fresnel Models for Gravitational Wave Effects on Pulsar Timing}. Monthly Notices of the Royal Astronomical Society.\\
    $\rightarrow$ \url{https://doi.org/10.1093/mnras/stab1417} \\
    $\rightarrow$ \url{https://arxiv.org/abs/2011.09561}
    \end{displayquote}
    
    At the time of finishing and submitting this dissertation, the work in \textbf{Chapter~\ref{ch:results - H0 measurement}} was being prepared in multiple manuscripts to submit for publication.

    $\longrightarrow$ \textbf{Update:} The following paper presents the final results of this work:

    \begin{displayquote}
    McGrath, C., D'Orazio, D., \& Creighton, J. (2022). \textit{Measuring the Hubble Constant with Double Gravitational Wave Sources in Pulsar Timing}. Monthly Notices of the Royal Astronomical Society.\\
    $\rightarrow$ \url{https://doi.org/10.1093/mnras/stac2593} \\
    $\rightarrow$ \url{https://arxiv.org/abs/2208.06495}
    \end{displayquote}
    }
}
\end{minipage}

\cleardoublepage
\chapter{Gravitational Waves from a Binary Source}\label{ch: GWs from a Binary Source}

%---------------------------------------------------------------------------------
%---------------------------------------------------------------------------------
    \section{Multipole Expansion}\label{sec: Multipole Expansion}
    In order to build a mathematical base and help gain deeper physical insights into the quadrupole formula (equation~\ref{eqn: quadrupole formula}) and gravitational waves, we begin by looking at the multipole expansion in general.  The mathematics here are extremely important and useful in physics, especially in fields like electromagnetism.  However, understanding the multipole expansion can often be very difficult the first time you learn it, as the fundamental ideas can easily be lost in the sometimes tedious and difficult math.  Here, we aim to cut directly through that math and show that fundamentally the multipole expansion is a rather simple idea.
    
    At its heart all the multipole expansion is, is a Taylor expansion of the function:
    \begin{align}
        f\left(\vec{x}\right) &= \frac{1}{\left| \vec{x} - \vec{x}^{'} \right|}, \label{eqn: multipole expansion function} \\
        &= \frac{1}{\sqrt{\left(x-x^{'}\right)^2 + \left(y-y^{'}\right)^2 + \left(z-z^{'}\right)^2}}, \nonumber \\
        &\equiv \frac{1}{\sqrt{\Delta x^2 + \Delta y^2 + \Delta z^2}}, \nonumber \\
        &= \frac{1}{\sqrt{\delta_{ab}\Delta x^a \Delta x^b}} = \frac{1}{\sqrt{\Delta x_a \Delta x^a}}, \nonumber
    \end{align}
    where $\vec{x}$ is the field point of interest, and $\vec{x}^{'}$ is the ``source'' location.  In the final line, Einstein index notation (and implied summations) are used.  This is a function of three variables (in Cartesian coordinates, $x$, $y$, and $z$), so the Taylor expansion will be multivariable Taylor expansion.  We are going to imagine that the source is far from the origin, and Taylor expand the function about the origin, that is about $\vec{x}=0$.
    
    Looking at equation~\ref{eqn: multipole expansion function}, notice the many different ways in which we can express this function.  We can write it in terms of the vectors $\vec{x}$ and $\vec{x}^{'}$, in terms of the variables $x$, $y$, and $z$, or by using index notation over the variables $x^i$.  Where this derivation typically becomes difficult is choosing the mathematical tool you are going to use to perform the Taylor series expansion.  Arguably the most straight forward approach is to write everything in terms of $x$, $y$, and $z$, and Taylor expand in all of those variables near $0$.  Conceptually this is the easiest, but mathematically this ends up getting really ugly, really fast.  Alternatively we could use the vectorized version of the Taylor series formula and apply vector derivatives, but personally I find this formula conceptually harder to grasp.  In my opinion, the easiest and best way to do this derivation in a manner that keeps in sight of all of the important conceptual and mathematical ideas is to exploit index notation.  We often don't learn index summation notation until after first learning the multipole expansion, which I believe is why this derivation is often not presented using this notation.  However, it helps to retain the most basic expression of the Taylor series expansion, removes some of the ugliness of using vector notation, and requires the least amount of written work.  The trade off is simply first \textit{learning} how to use the notation properly, but after that this derivation is pretty straight forward.
    
    We need to Taylor expand equation~\ref{eqn: multipole expansion function}, which can be written incredibly concisely (with index notation!) as:
    \begin{equation}
        f\left(\vec{x}\right) \quad\approx\quad f\Big\rvert_{\vec{x}=0} \quad+\quad \partial_i f\Big\rvert_{\vec{x}=0} x^i \quad+\quad \frac{1}{2!}\partial_i\partial_j f\Big\rvert_{\vec{x}=0} x^i x^j \quad+\quad \ldots ,
    \end{equation}
    where $\partial_i = \frac{\partial}{\partial x^i}$.  The math for the needed derivatives is given explicitly in the ``Working out the terms'' box below.  When we evaluate the terms at $\vec{x}=0$, note that we define the distance to the source as $R \equiv \sqrt{x^{'}_a x^{'a}} = \left|\vec{x}^{'}\right|$.  Therefore, the multipole (Taylor) expansion of equation~\ref{eqn: multipole expansion function} is:
    \begin{equation}
        f\left(\vec{x}\right) \quad\approx\quad \underbrace{\frac{1}{R}}_{\text{``monopole''}} \quad+\quad \underbrace{\frac{x^{'}_i}{R^3}x^i}_{\text{``dipole''}}  \quad+\quad \underbrace{\frac{3}{2R^5}\left[x^{'}_i x^{'}_j - \frac{1}{3}\delta_{ij}R^2\right] x^i x^j}_{\text{``quadrupole''}} \quad+\quad \ldots ,
    \label{eqn: multipole expansion function - computed}
    \end{equation}
    where we have given the names of each order term as they will be referred to in what follows.\\
    
    \begin{minipage}{0.9\linewidth}
    \fbox {
        \parbox{\linewidth}{ \textbf{Working out the terms}\newline
        Here we work out explicitly the needed derivatives in our Taylor series.  Really it is just careful application of the chain and product rules.  Note that $\partial_i \Delta x^a = \frac{\partial\left(x^a - x^{'a}\right)}{\partial x^i} = \delta_i^a$, and $\partial_i \Delta x_a = \frac{\partial\left(x_a - x_a^{'}\right)}{\partial x^i} = \delta_{ia}$.
        \begin{align*}
            \partial_i f &= \partial_i \frac{1}{\sqrt{\Delta x_a \Delta x^a}} = -\frac{1}{2}\frac{\delta_{ia} \Delta x^a + \Delta x_a \delta^{ia}}{\left[\Delta x_a \Delta x^a\right]^{3/2}} = -\frac{1}{2}\frac{2\Delta x_i}{\left[\Delta x_a \Delta x^a\right]^{3/2}} = -\frac{\Delta x_i}{\left[\Delta x_a \Delta x^a\right]^{3/2}} , \\[10pt]
            \partial_i\partial_j f &= -\partial_i \frac{\Delta x_j}{\left[\Delta x_a \Delta x^a\right]^{3/2}} = \frac{-\delta_{ij}}{\left[\Delta x_a \Delta x^a\right]^{3/2}} -\Delta x_j\cdot-\frac{3}{2}\frac{2\Delta x_i}{\left[\Delta x_a \Delta x^a\right]^{5/2}} , \\
            &= \frac{-\delta_{ij}}{\left[\Delta x_a \Delta x^a\right]^{3/2}} + 3\frac{\Delta x_i \Delta x_j}{\left[\Delta x_a \Delta x^a\right]^{5/2}} , \\
            &= \frac{3}{\left[\Delta x_a \Delta x^a\right]^{5/2}}\left[\Delta x_i \Delta x_j -\frac{1}{3}\delta_{ij}\Delta x_a \Delta x^a\right] .
        \end{align*}
        Now evaluating these derivatives at $\vec{x}=0$, and using $R \equiv \sqrt{x^{'}_a x^{'a}} = \left|\vec{x}^{'}\right|$, we find:
        \begin{align*}
            \partial_i f\Big\rvert_{\vec{x}=0} &= +\frac{x^{'}_i}{R^3} , \\[10pt]
            \partial_i\partial_j f\Big\rvert_{\vec{x}=0} &= \frac{3}{R^5}\left[x^{'}_i x^{'}_j - \frac{1}{3}\delta_{ij}R^2\right] .
        \end{align*}
        }
    }
    \end{minipage}

%---------------------------------------------------------------------------------
%---------------------------------------------------------------------------------
    \section{Multipole Potentials}\label{sec: Multipole Potentials}
    The multipole expansion is often used to write out the electric or gravitational potentials, due to a distribution of charge or mass, respectively \citep{creighton_anderson_2011,harrison_2020}.  For continuous distributions, these two expressions are computed in the same way, simply with different scaling factors out front:
    \begin{equation}
        V\left(\vec{x}\right) = \begin{cases}
        - G \mathlarger{\int} \frac{dm\left(\vec{x}^{'}\right)}{\left|\vec{x}-\vec{x}^{'}\right|} &= - G \mathlarger{\int} \frac{\rho\left(\vec{x}^{'}\right)}{\left|\vec{x}-\vec{x}^{'}\right|}d^3 x^{'} , \qquad \text{(Gravitational Potential)} \\[14pt]
         \frac{1}{4\pi\epsilon_0} \mathlarger{\int} \frac{dm\left(\vec{x}^{'}\right)}{\left|\vec{x}-\vec{x}^{'}\right|} &=  \frac{1}{4\pi\epsilon_0} \mathlarger{\int} \frac{\rho\left(\vec{x}^{'}\right)}{\left|\vec{x}-\vec{x}^{'}\right|}d^3 x^{'} .\qquad \text{(Electric Potential)}
        \end{cases}
    \end{equation}
    Focusing our attention on the gravitational potential (an analogous approach can be taken for the electric potential), we can multipole expand the denominator using our result in equation~\ref{eqn: multipole expansion function - computed} and write:
    \begin{align}
        V\left(\vec{x}\right) &\quad=\quad -\frac{GM}{R} \quad-\quad \frac{Gx^i D_i}{R^3} \quad-\quad \frac{3Gx^i x^j \textit{\sout{I}}_{ij}}{2R^5} \quad-\quad \ldots , \\[10pt]
        &\mathrm{where}\quad\begin{cases}
            \begin{tabular}{l l c}
                $M$ & $\equiv \mathlarger{\int} \rho\left(\vec{x}^{'}\right) d^3x^{'}$ , & (Monopole) \\[6pt]
                $D_i$ & $\equiv \mathlarger{\int} x^{'}_i \rho\left(\vec{x}^{'}\right) d^3x^{'}$ , & (Dipole) \\[6pt]
                $\textit{\sout{I}}_{ij}$ & $\equiv \mathlarger{\int} \left[x^{'}_i x^{'}_j - \frac{1}{3}\delta_{ij}R^2 \right]\rho\left(\vec{x}^{'}\right) d^3x^{'}$ , & (Quadrupole)
            \end{tabular}
        \end{cases} \label{eqn: gravitational moments - m, d, q}
    \end{align}
    The monopole term in our expression is the standard Newtonian result for a point mass, but with more complicated mass distributions we gain higher order corrections.
    
    There is a crucially important difference, however, for a mass distribution in gravity as compared to a charge distribution in electromagnetism.  We can always choose a center-of-mass coordinate system for mass distributions.  The consequence of this is that the dipole term vanishes identically when working in this chosen coordinate system.  Conceptually this is because in a center-of-mass coordinate system, by definition the mass is equally distributed in all directions.  Therefore integrating up all of the mass along each direction in $D_i = \mathlarger{\int} x^{'}_i \rho\left(\vec{x}^{'}\right) d^3x^{'} = \mathlarger{\int} x^{'}_i dm\left(\vec{x}^{'}\right)$ will just cancel out and net $0$.
    
    We cannot \textit{always} guarantee the same though for electromagnetism - that is, we cannot always choose a center-of-charge coordinate system.  As a simple example, just consider two equal and opposite charges lying at $\pm x$.  There is no center-of-charge for this distribution, and thus there will always be some non-zero dipole moment, no matter the chosen set of coordinates.  This helps point out one of the fundamental differences between gravity and electromagnetism.  In gravity there is no ``negative mass particle'' like in electromagnetism with both positive and negative charges.
    
    Another important point to take notice of here is the physical significance of the quadrupole tensor.  The quadrupole tensor in equation~\ref{eqn: gravitational moments - m, d, q} is the \textit{negative traceless} version of the familiarly defined ``moment of inertia tensor.''  The normal moment of inertia tensor describes all of the moments of inertia about an object's chosen orthogonal basis axes, and is defined as:
    \begin{equation}
        \text{I}_{ij} \equiv \mathlarger{\int} \left[\delta_{ij}R^2 - x^{'}_i x^{'}_j\right]\rho\left(\vec{x}^{'}\right) d^3x^{'} .
    \label{eqn: moment of inertia tensor}
    \end{equation}
    The trace of the moment of inertia tensor is:
    \begin{align*}
        \text{Trace}\left(\text{I}_{ij}\right) = \delta^{ij}\text{I}_{ij} &= \mathlarger{\int} \left[\delta^{ij}\delta_{ij}R^2 - \delta^{ij} x^{'}_i x^{'}_j\right]\rho\left(\vec{x}^{'}\right) d^3x^{'} = \mathlarger{\int} \left[\delta^i_i R^2 - x^{'}_i x^{'i}\right]\rho\left(\vec{x}^{'}\right) d^3x^{'} , \\
        &= \mathlarger{\int} \Big[(3)R^2 - R^2 \Big]\rho\left(\vec{x}^{'}\right) d^3x^{'} \equiv \mathlarger{\int} 2R^2\rho\left(\vec{x}^{'}\right) d^3x^{'} .
    \end{align*}
    If we remove the trace of this tensor from itself by subtracting $1/3$ of this value off of the diagonal elements, we end up with the negative of our quadrupole tensor.  Mathematically, this statement is written as:
    \begin{align*}
        \text{I}_{ij} - \frac{1}{3}\delta_{ij}\text{Trace}\left(\text{I}_{ij}\right) &= \mathlarger{\int}\left[ \delta_{ij}R^2 - x^{'}_i x^{'}_j - \frac{1}{3}\delta_{ij}\Big(2R^2\Big)\right]\rho\left(\vec{x}^{'}\right) d^3x^{'} , \\
        &= \mathlarger{\int}\left[ \frac{1}{3}\delta_{ij}R^2 - x^{'}_i x^{'}_j \right]\rho\left(\vec{x}^{'}\right) d^3x^{'} , \\
        &\equiv - \ \textit{\sout{I}}_{ij} .
    \end{align*}
    
    The tensor $\textit{\sout{I}}_{ij}$ is typically referred to as the ``reduced quadrupole tensor.''  A similarly defined ``quadrupole tensor'' will appear later and the two are defined as follows:
    \begin{equation}
        \begin{cases}
            \begin{tabular}{l l c}
                $\textit{I}_{ij}$ & $\equiv \mathlarger{\int} x^{'}_i x^{'}_j \rho\left(\vec{x}^{'}\right) d^3x^{'}$ , & (Quadrupole Tensor) \\[6pt]
                $\textit{\sout{I}}_{ij}$ & $\equiv \mathlarger{\int} \left[x^{'}_i x^{'}_j - \frac{1}{3}\delta_{ij}R^2 \right]\rho\left(\vec{x}^{'}\right) d^3x^{'}$ . & (Reduced Quadrupole Tensor)
            \end{tabular}
        \end{cases} \label{eqn: quadrupole and reduced quadrupole tensors}
    \end{equation}
    The regular quadrupole tensor does not have the traceless property like the reduced quadrupole tensor.

%---------------------------------------------------------------------------------
%---------------------------------------------------------------------------------
    \section{The Gravitational Wave Solution - General Form}

    For the gravitational waves produced by the sources we are interested in (i.e. massive binaries) we solve the Einstein field equations under the following assumptions:\\ 
    
    \begin{minipage}{0.9\linewidth}
    \fbox{
        \parbox{\linewidth}{
        \textbf{Assumptions:} Field Equation Solutions
        \begin{enumerate}
            \item Cosmologically static universe.
            \item Cosmologically flat empty universe (hence the background metric is Minkowski). \label{as:empty}
            \item Weak-field limit (gravitational waves are a metric perturbation on top of Minkowski). \label{as:weak}
            \item Small source compared to the distance to the observer and the wavelength of the wave. \label{as:small}
            \item Slow-moving source (non-relativistic, no post-Newtonian analysis required). \label{as:slow}
            \item Transverse-traceless (``TT'') gauge.  \label{as:TT}      
            \item Far field approximation of the metric perturbation \textit{amplitude} - the binary is sufficiently far from the field point of interest that $|\vec{x}_\mathrm{field} - \vec{x}_\mathrm{source}| \approx R$ in the amplitude of the metric perturbation.  This assumption won't be made when evaluating the retarded time (which won't affect the amplitude of the wave, but rather the phase of the wave). \label{as:far field amplitude}
        \end{enumerate}
        }
    }
    \end{minipage}\\ \\
   
    Assumptions~\ref{as:weak}-\ref{as:slow} together make the ``weak-small-slow'' assumption for our source - or in other words, we are in the ``zeroth order Newtonian'' regime.  We will not be considering sources near coalescence when this assumption might otherwise be broken.  There are numerous helpful resources including~\citet{maggiore_2008,creighton_anderson_2011,moore_2013,carroll_2013} and~\citet{zee_2013} that this discussion is based upon.
    
    Under assumption~\ref{as:weak}, we begin by proposing a metric solution that looks like flat spacetime with a small perturbation $h_{\mu\nu}$ on top of it:
    \begin{equation}
        g_{\mu\nu} = \eta_{\mu\nu} + h_{\mu\nu}, \quad \mathrm{where} \ \left|h_{\mu\nu}\right| \ll 1 .
        \label{eqn:metric with perturbation}
    \end{equation}
    The next step in the solution process is typically to re-express the metric perturbation in a different form, known as the trace-reversed metric perturbation:
    \begin{equation}
        H_{\mu\nu} \equiv h_{\mu\nu} - \frac{1}{2}\eta_{\mu\nu}h \qquad \longleftrightarrow \qquad h_{\mu\nu} = H_{\mu\nu} - \frac{1}{2}\eta_{\mu\nu}H,
        \label{eqn: trace-reversed metric perturbation}
    \end{equation}    
    where $h=\text{Trace}(h_{\mu\nu})=\eta^{\mu\nu}h_{\mu\nu}$.  The reason for this choice is simply mathematical convenience, as it helps further reduce the complexity of the Einstein equations into a form more easily solved.  If we were to find a solution for $H_{\mu\nu}$, then equation~\ref{eqn: trace-reversed metric perturbation} would give us the conversion back to $h_{\mu\nu}$.  Under the Lorenz gauge choice (which is part of assumption~\ref{as:TT}), the resulting Einstein equations can be expressed in this weak-field regime as:
    \begin{equation}
    \begin{cases}
        \Box^2 H_{\mu\nu} &= \frac{-16\pi G}{c^4} T_{\mu\nu} , \\[6pt]
        \partial^\mu H_{\mu\nu} &= 0 ,
    \end{cases}\label{eqn: Einstein equations}
    \end{equation}
    where $T_{\mu\nu}$ is the ``effective stress energy tensor'' describing the source of our gravitational wave solution.  Here the ``effective'' differentiates the tensor from the typically defined ``stress energy tensor'' in that it contains $\mathcal{O}\left(h^2\right)$ terms which results above when substituting equation~\ref{eqn:metric with perturbation} into the Einstein equations.
    
    In general, we know that the sourced (inhomogeneous) wave equation solution can be constructed by integrating the wave equation Green's function.  Specifically, with the wave equation written in the following form for some sourcing function $f\left(\vec{x},t\right)$, the solution can be written as:
    \begin{align}
        \Box^2 \psi = -4\pi f\left(\vec{x},t\right) \qquad \xrightarrow{\text{solution}} \qquad \psi\left(\vec{x},t\right) &= \int G\left(\vec{x},\vec{x}^{'},t,t^{'}\right) f\left(\vec{x}^{'},t^{'}\right) d^3x^{'}dt^{'} , \nonumber \\
        &= \int \frac{\delta\left(t-t_\mathrm{ret}\right)}{\left|\vec{x}-\vec{x}^{'}\right|} f\left(\vec{x}^{'},t^{'}\right) d^3x^{'}dt^{'} , \nonumber \\
        &= \int \frac{f\left(\vec{x}^{'},t_\mathrm{ret}\right)}{\left|\vec{x}-\vec{x}^{'}\right|} d^3x^{'} ,
    \end{align}
    where $t_\mathrm{ret}$ is the retarded time (see equation~\ref{eqn:retarded time}).  Comparing this to equation~\ref{eqn: Einstein equations}, we can see that the solution in terms of the trace-reversed metric perturbation is:
    \begin{equation}
        H_{\mu\nu} = \frac{4G}{c^4}\int \frac{T_{\mu\nu}\left(\vec{x}^{'},t_\mathrm{ret}\right)}{\left|\vec{x}-\vec{x}^{'}\right|} d^3x^{'} .
    \label{eqn: trace-reversed metric solution}
    \end{equation}
    At this point we have a few options.  We could multipole expand this solution using the results and ideas developed in Sections~\ref{sec: Multipole Expansion} and~\ref{sec: Multipole Potentials}.  But as we will see, for our level of precision we actually only need the first term in the multipole expansion, the ``monopole'' or ``far-field'' term (this is our assumption~\ref{as:far field amplitude}).  However, the reason we developed the previous two sections so extensively is to show that the final answer we will find will actually be expressed in terms of the quadrupole tensor!
    
    So with this in mind, we can approximate the solution in equation~\ref{eqn: trace-reversed metric solution} as:
    \begin{equation}
        H_{\mu\nu} \approx \frac{4G}{c^4 R}\int T_{\mu\nu}\left(\vec{x}^{'},t_\mathrm{ret}\right) d^3x^{'} .
    \label{eqn: trace-reversed metric solution approx}
    \end{equation}    
    Now we are in a position to being evaluating all of the individual terms of the metric tensor solution here.  Since this is a wave solution to the wave equation, we are looking for components that are behaving like a ``wave.''  The $tt$- and $it$/$ti$-components of the effective stress energy tensor represent the energy density and the $i$-momentum density solutions.  The integrals of these components over all space which encompasses the source are:
    \begin{align*}
        \int T_{tt}\left(\vec{x}^{'},t_\mathrm{ret}\right) d^3x^{'} &= \int \rho\left(\vec{x}^{'},t_\mathrm{ret}\right) d^3x^{'} \equiv M ,\\
        \int T_{it}\left(\vec{x}^{'},t_\mathrm{ret}\right) d^3x^{'} &= \int T_{ti}\left(\vec{x}^{'},t_\mathrm{ret}\right) d^3x^{'} \equiv P_i \equiv 0 ,
    \end{align*}
    The $i$-momentum $P_i$ vanishes identically here because if we move the coordinate system to the center-of-mass frame, then the source itself is not moving - it has zero momentum.  And integrating up the energy content of the source just gives us its mass $M$, so neither of these terms will behave like waves.  So these two non-waving sets of solutions look like:
    \begin{align}
        H_{tt} &= \frac{4GM}{c^4 R} , \\
        H_{it} = H_{ti} &= 0 .
    \end{align}
    At this point we just need to focus our attention on the $ij$-components of the metric perturbation, which will produce the final wave solution.
    
    Thanks to the conservation of energy, the divergence theorem, and some clever algebra/calculus, the $ij$-components of the metric perturbation solution can actually be re-expressed in terms of the energy density ($tt$-component) alone.  Details of this are given explicitly in the ``A Helpful Identity'' box below.  The result is:
    \begin{align*}
        H_{ij} \approx \frac{4G}{c^4 R}\int T_{ij}\left(\vec{x}^{'},t_\mathrm{ret}\right) d^3x^{'} &= \frac{4G}{c^4 R}\cdot\frac{1}{2}\frac{\partial^2}{\partial t^2} \int x^{'}_i x^{'}_j T_{tt}\left(\vec{x}^{'},t_\mathrm{ret}\right) d^3x^{'} , \\
        &= \frac{2G}{c^4 R}\frac{\partial^2}{\partial t^2} \int x^{'}_i x^{'}_j \rho\left(\vec{x}^{'},t_\mathrm{ret}\right) d^3x^{'} . \\
    \end{align*}
    Now we can make our connection back to the multipole potentials of Section~\ref{sec: Multipole Potentials}, and note that the final answer can be expressed in terms of the quadrupole tensor from equation~\ref{eqn: quadrupole and reduced quadrupole tensors} as:
    \begin{equation}
        H_{ij} \approx \frac{2G}{c^4 R}\frac{\partial^2}{\partial t^2} I_{ij}\left(t_\mathrm{ret}\right) .
    \label{eqn: quadrupole formula}
    \end{equation}
    This is known as the ``quadrupole formula'' - the gravitational wave solution to the weak-field Einstein equations.  So the gravitational waves produced by an object are proportional to the acceleration of its quadrupole tensor, which as we discussed can sort of be thought of as a different version of the moment of inertia tensor of that object.
    
    Fundamentally, this is different from electromagnetism.  An insightful discussion of the significance of equation~\ref{eqn: quadrupole formula} is given in the following quote:
    \begin{displayquote}
    ``In contrast, the leading contribution to electromagnetic radiation comes from the changing \textit{dipole} moment of the charge density.  The difference can be traced back to the universal nature of gravitation.  A changing dipole moment corresponds to motion of the center of density - charge density in the case of electromagnetism, energy density in the case of gravitation.  While there is nothing to stop the center of charge of an object from oscillating, oscillation of the center of mass of an isolated system violates conservation of momentum.  (You can shake a body up and down, but you and the earth shake ever so slightly in the opposite direction to compensate.)  The quadrupole moment, which measures the shape of the system, is generally smaller than the dipole moment, and for this reason, as well as the weak coupling of matter to gravity, gravitational radiation is typically much weaker than electromagnetic radiation.''
    
    \rightline{--- Sean Carroll~\citep{carroll_2013}}
    \end{displayquote}
    With regards to Carroll's conservation of momentum statement, we already explained that the dipole moment in center-of-mass coordinates is necessarily zero, but to further illustrate this consider a simple example of a system of point particles with mass $m$.  The dipole moment in equation~\ref{eqn: gravitational moments - m, d, q} would be $D_i = \sum m x^{'}_i$.  Imagine if the metric perturbation now were proportional to the acceleration of the dipole moment instead of the quadrupole moment.  Then $\frac{d^2}{d t^2}D_i = \frac{d^2}{d t^2} \sum m x^{'}_i = \frac{d}{d t} \sum m v^{'}_i = \frac{d}{d t} 0 = 0$, because of the conservation of momentum of the entire system.  So in general, gravitational radiation is such a small perturbation to spacetime largely because gravity is such a weak force, but also since there is no dipole and the quadrupole correction tends to be a smaller effect. \\
    
    \begin{minipage}{0.9\linewidth}
    \fbox {
        \parbox{\linewidth}{ \textbf{A Helpful Identity}\newline
        First consider the following quantity and carefully apply the product rule to expand it out term by term in the following way:
        \begin{align*}
            \partial_i \partial_j\left(T^{ij}x^m x^n\right) &= \partial_i \Big[ x^m x^n \partial_j T^{ij} + x^n T^{ij}\delta^m_j + x^m T^{ij}\delta^n_j \Big] , \\
            &= \partial_i \Big[ x^m x^n \partial_j T^{ij} + \left(x^n T^{im} + x^m T^{in}\right) \Big] , \\
            &= x^m x^n \partial_i\partial_j T^{ij} + x^n\partial_j T^{mj} + x^m\partial_j T^{nj} + \partial_i\left(x^n T^{im} + x^m T^{in}\right) , \\
            &= x^m x^n \partial_i\partial_j T^{ij} + \partial_j\left(x^n T^{mj} + x^m T^{nj}\right) - 2T^{mn} + \partial_i\left(x^n T^{im} + x^m T^{in}\right) , \\
            &= x^m x^n \partial_i\partial_j T^{ij} + 2\partial_i\left(x^n T^{im} + x^m T^{in}\right) -2T^{mn} .
        \end{align*}
        Going from the 2nd to 3rd line we used the following twice:
        \begin{equation*}
            \partial_i \left(x^n T^{mj}\right) = \delta^n_j T^{mj} + x^n\partial_j T^{mj} = T^{mn} + x^n\partial_j T^{mj} \  \implies \  x^n \partial_j T^{mj} = \partial_i \left(x^n T^{mj}\right) - T^{mn} .
        \end{equation*}
        Then going from the 3rd to the 4th line we recognize that in the second term the $j$-index is summed over, so we can re-label it to anything we want, namely we can relabel it from $j\rightarrow i$.  We can re-arrange the terms in this expression and write:
        }
    }
    \end{minipage}
    
    \begin{minipage}{0.9\linewidth}
    \fbox {
        \parbox{\linewidth}{
        \begin{equation*}
            T^{mn} = \frac{1}{2}\Big[ -\partial_i \partial_j\left(T^{ij}x^m x^n\right) + 2\partial_i\left(x^n T^{im} + x^m T^{in}\right) + x^m x^n \partial_i \partial_j T^{ij} \Big] .
        \end{equation*}
        
        Next let's integrate this expression for $T^{mn}$ over all space.  The first crucial mathematical exploit we are going to make comes from the divergence theorem.  The generalized form of the divergence theorem states:
        \begin{equation*}
            \int \partial_i F^{ij} dV = \oint F^{ij} dA_i .
        \end{equation*}
        The key idea here is that when we apply the divergence theorem, we will be completely enclosing our source as we integrate over all space, and far from our source, the effective stress energy tensor will be zero, which is where the surface integral is evaluated.  So looking at the first two terms in this integral expression we have terms that look like:
        \begin{align*}
            \int \partial_i \partial_j\left(T^{ij}x^m x^n\right) dV &= \oint\oint \left(T^{ij}x^m x^n\right) dA_i dA_j \equiv 0 \\
            \int \partial_i\left(x^n T^{im}\right) dV &= \oint \left(x^n T^{im}\right) dA_i \equiv 0
        \end{align*}
        (It is important here to understand that quantities like $T^{ij}x^m x^n$ can be thought of as a single tensor objects - for example, we could just re-define this quantity as some tensor $T^{ij}x^m x^n \equiv A^{ijmn}$).
        \\ \\
        The final critical step comes from the conservation of energy, which tells us $\partial_\mu T^{\mu\nu} = 0$.  Expanding this statement out slightly means we can write:
        \begin{equation*}
            \partial_t T^{tj} + \partial_i T^{ij} = 0 \quad \implies \quad \partial_t T^{tj} = - \partial_i T^{ij}.
        \end{equation*}
        }
    }
    \end{minipage}%\end{wrapfigure}

    \begin{minipage}{0.9\linewidth}
    \fbox {
        \parbox{\linewidth}{
        Two applications of this conservation statement (along with the symmetry $T^{\mu\nu} = T^{\nu\mu}$), means we can replace the third term in the expression for $T^{mn}$ with the following:
        \begin{equation*}
            x^m x^n \partial_i \partial_j T^{ij} = + x^m x^n \partial_t \partial_t T^{tt} = \partial_t \partial_t \left(x^m x^n T^{tt}\right) .
        \end{equation*}
        With this we are replacing spatial derivatives over the $ij$-components of our effective stress energy tensor with two time derivatives over the energy density component alone.  This gives us our final result, that:
        \begin{equation*}
            \int T^{mn} dV = \frac{1}{2}\int \partial_t \partial_t \left(x^m x^n T^{tt}\right) dV = \frac{1}{2}\partial_t \partial_t\int \left(x^m x^n T^{tt}\right) dV .
        \end{equation*}
        }
    }
    \end{minipage}
    \\ \\
    
    At this point we could specify an objects energy density, and solve equation~\ref{eqn: quadrupole formula} (we could then convert that back to the original metric perturbation using equation~\ref{eqn: trace-reversed metric perturbation}).  However we still need to get the solution into the transverse-traceless gauge (``TT-gauge'') as stated in assumption~\ref{as:TT}.  We know that the reduced quadrupole tensor in equation~\ref{eqn: quadrupole and reduced quadrupole tensors} is the traceless version of the quadrupole tensor, so that would get us half-way there.  But while the reduced quadrupole tensor is traceless, it may not necessarily be transverse to the direction of interest along which the gravitational wave is propagating.
    
    We can cast any matrix into the TT-gauge by use of the special TT-projection operator:
    \begin{align}
        \textbf{M}^{TT} &= \Lambda(\textbf{M}) = \textbf{PMP} - \frac{1}{2}\mathrm{Trace}(\textbf{PM})\textbf{P} , \nonumber \\
        &\longrightarrow  M^{TT}_{ij} = \Lambda_{ijkl} M^{kl} = \left(P_{ik}P_{jl} - \frac{1}{2}P_{ij}P_{kl}\right)M^{kl}, \label{eqn: TT-projection operator} \\
        &\text{where}\quad P_{ij} = \delta_{ij} - \hat{n}_i \hat{n}_j . \nonumber
    \end{align}
    Here $\hat{n}$ is the desired direction of propagation of the gravitational wave that we want to project onto.  The TT-gauge gives us a number of useful conceptual and mathematical results, namely:
    \begin{equation}
    \begin{cases}
        \textit{I}^{TT}_{ij} &= \textit{\sout{I}}^{TT}_{ij} , \\[6pt]
        H^{TT}_{ij} &= h^{TT}_{ij} .
    \end{cases}
    \end{equation}
    
    With regards to the quadrupole tensors, the regular quadrupole tensor is equivalent to the reduced quadrupole tensor after it is made traceless, so if both of these are made traceless then transverse, then necessarily they are equivalent.  In practice we will often work with the reduced quadrupole tensor, so this realization is helpful.  As for the regular and trace-reversed metric perturbations, the proof is given in the ``TT Metric Projection'' box.
    
    With this we finalize our wave solution to the original proposed solution (equation~\ref{eqn:metric with perturbation} to the Einstein equations.  In the transvserse-traceless gauge, the wave-like weak-field solutions are:
    \begin{equation}
        h^{TT}_{ij} = \frac{2G}{c^4 R}\frac{\partial^2}{\partial t^2} \textit{\sout{I}}^{TT}_{ij}\left(t_\mathrm{ret}\right) .
    \label{eqn: metric perturbation solution (TT)}
    \end{equation} \\
    
    \begin{minipage}{0.9\linewidth}
    \fbox {
        \parbox{\linewidth}{ \textbf{TT Metric Projection}\newline
        Conveniently, when both the regular and the trace-reversed metric perturbations are cast into the TT-gauge they are equivalent, meaning once we have the solution to one we have the solution to both.  To see this we simply apply the TT operator to equation~\ref{eqn: trace-reversed metric perturbation} and see that:
        }
    }
    \end{minipage}
    
    \begin{minipage}{0.9\linewidth}
    \fbox {
        \parbox{\linewidth}{
        \begin{align*}
            H^{TT}_{ij} = \Lambda_{ijkl} H^{kl} &= \Lambda_{ijkl} h^{kl} -\frac{1}{2}\Lambda_{ijkl} \eta^{kl} h , \\
            &= h^{TT}_{ij} - \frac{1}{2}{\Lambda_{ijk}}^k h , \\
            &= h^{TT}_{ij} - 0.
        \end{align*}
        The trace of our projection operator along the $kl$-indices is identically zero, which we can get by simply working out all of the terms explicitly.  First noting that $P_k^k = \delta_k^k - n_k n^k = 3 - 1 = 2$, we can write:
        \begin{align*}
            {\Lambda_{ijk}}^k &= P_{ik}P_j^k-\frac{1}{2}P_{ij}P_k^k = P_{ik}P_j^k-P_{ij} , \\
            &=\left(\delta_{ik}-n_i n_k\right)\left(\delta_j^k-n_j n^k\right)-P_{ij} , \\
            &=\delta_{ij} -n_j n_i -n_i n_j + n_i n_j(1) - P_{ij} , \\
            &= P_{ij} - P_{ij} = 0.
        \end{align*}
        }
    }
    \end{minipage}

%---------------------------------------------------------------------------------
%---------------------------------------------------------------------------------
    \section{The Gravitational Wave Solution - Binary System}
    
\begin{figure}
    \centering
      \begin{subfigure}[t]{0.48\linewidth}
      \centering
        \includegraphics[width=\linewidth]{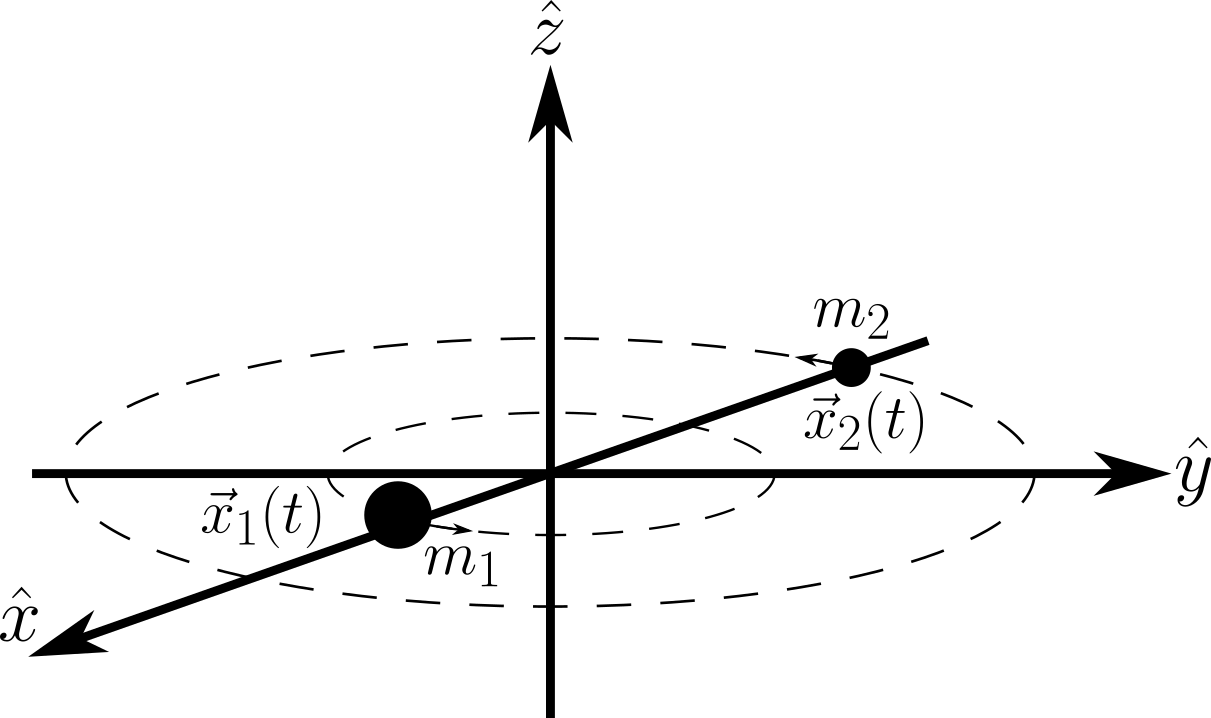}
        \caption{Binary system orbiting in the $x$/$y$-plane with the origin placed at the system's center-of-mass.}
        \label{fig:starting source orientation}
      \end{subfigure}
      \hfill
      \begin{subfigure}[t]{0.48\linewidth}
      \centering
        \includegraphics[width=\linewidth]{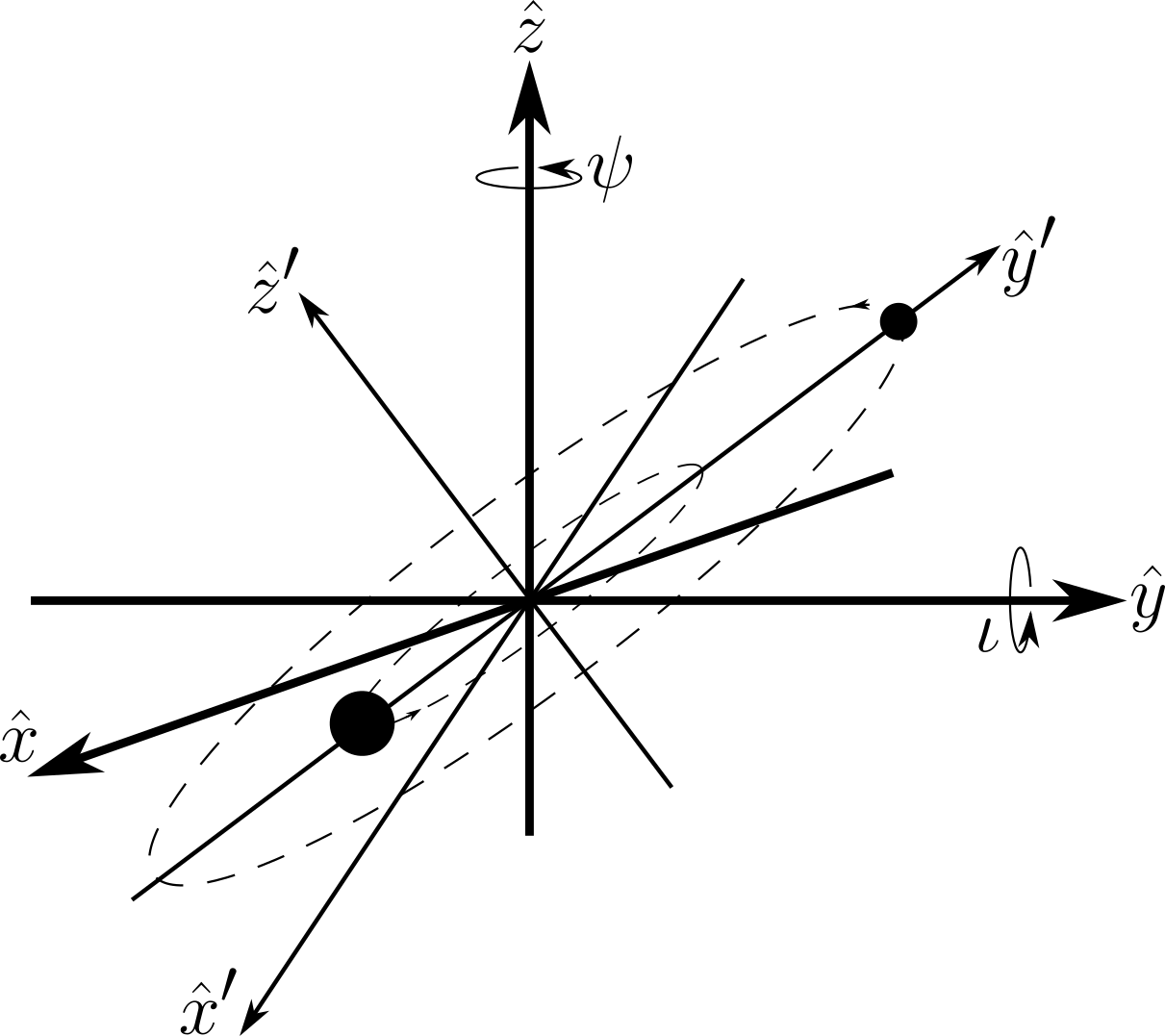}
        \caption{Generalized source orientation.  Two Euler angles $\iota$ (``inclination'') and $\psi$ (``polarization''), and a phase offset $\theta_0$ in the binary system's motion (see equations~\ref{eqn:monochrome phase} and~\ref{eqn:freq evolution phase}), fully generalize the binary's orientation.}
        \label{fig:generalized source orientation}
      \end{subfigure}
    \caption[Binary System: Orientation]{}
    \label{fig: source - starting & generalized orientations}
    \end{figure}

    We begin with a simplified problem which we will then generalize.  First consider an arbitrary binary system aligned in the configuration shown in Figure~\ref{fig:starting source orientation}.  The binary orbits in the $x/y$-plane, ``face-on'' with no inclination to the $\hat{z}$-axis.  Gravitational waves produced by the binary propagate out in all directions, but we begin by focusing our attention to the gravitational waves which propagate along the $\hat{z}$-axis.  Therefore let's start by writing the solution in the TT-gauge for the gravitational waves along that direction.  To remind ourselves of the axis of propagation for our TT-gauge, we will include the unit vector next to the ``TT'' symbol in the notation that follows.  (We start with $TT\hat{z}$, then later generalize to an arbitrary $\hat{r}$ vector, denoted by $TT\hat{r}$).
    
    The following assumptions are made about the binary system:\\
    
    \begin{minipage}{0.9\linewidth}
    \fbox {
        \parbox{\linewidth}{
        \textbf{Assumptions:} Binary Mechanics
        \begin{enumerate}
            \item The binary is circular. \label{as:circular}
            \item The source is located at a fixed angular sky position and distance from the Earth.
        \end{enumerate}
        }
    }
    \end{minipage}\\ \\
    
    Under these assumptions, the metric perturbation solution to equation~\ref{eqn: metric perturbation solution (TT)} can be written compactly in the following form:
    \begin{align}
        h^{TT\hat{z}}_{ij} &= e^{\hat{z}+}_{ij}h_{+} + e^{\hat{z}\times}_{ij}h_{\times} = e^{\hat{z}\textsc{A}}_{ij}h_{\textsc{A}} \qquad \mathrm{for}\quad \textsc{A} \in [+, \times], \\
    \nonumber \\
        &\mathrm{where}\quad \begin{cases}
            h_{+}(t) \equiv -h(t) \cos\big(2\Theta(t)\big), \\
            h_{\times}(t) \equiv -h(t) \sin\big(2\Theta(t)\big), \label{eqn: h plus and cross}
            \end{cases}\\
        &\hspace{1.8cm} h(t) \equiv \frac{4(G\mathcal{M})^{5/3}}{c^4 R}\omega(t)^{2/3}, \label{eqn: amplitude of metric perturbation} \\
        &\qquad \qquad \begin{cases}
            e^{\hat{z}+}_{ij} \equiv \left( \begin{array}{ccc}
            1 & 0 & 0 \\
            0 & -1 & 0 \\
            0 & 0 & 0 \end{array} \right) = \hat{x}_i\hat{x}_j - \hat{y}_i\hat{y}_j, \\
            \\[-25pt]
            e^{\hat{z}\times}_{ij} \equiv \left( \begin{array}{ccc}
            0 & 1 & 0 \\
            1 & 0 & 0 \\
            0 & 0 & 0 \end{array} \right) = \hat{y}_i\hat{x}_j + \hat{x}_i\hat{y}_j,
        \end{cases} \label{eqn:basic polarization tensors}
    \end{align}
    where the metric decomposes into two polarizations, labeled ``plus'' and ``cross,'' $\omega(t)$ and $\Theta(t)$ are the orbital frequency and phase (specified in Section~\ref{sec: two frequency models}), $R$ is the Earth-source coordinate distance, $M$ is the total binary system mass, and $\mathcal{M}$ is the ``chirp mass,'' defined as $\mathcal{M} \equiv \frac{\left(m_1 m_2\right)^{3/5}}{\left(m_1 + m_2\right)^{1/5}}$.  Note here we emphasize in equation~\ref{eqn: amplitude of metric perturbation} that the amplitude or ``strain'' is time-dependent, which will be important to consider later in Chapter~\ref{ch: The Continuous Wave Timing Residual}. The polarization tensors $e^{\hat{z}+}_{ij}$ and $e^{\hat{z}\times}_{ij}$ can be conveniently decomposed into outer products of the basis vectors $\hat{x}$ and $\hat{y}$, as shown in equation~\ref{eqn:basic polarization tensors}.

%---------------------------------------------------------------------------------
%---------------------------------------------------------------------------------    
    \section{Generalized Source Orientation and Location}\label{sec: generalized source orientation & location}
    
    This solution represents a highly simplified and specific geometry - here the orbital rotation axis is along the $\hat{z}$-direction, and the TT-gauge has also been chosen for the gravitational waves propagating along the $\hat{z}$-axis.  However, in general we want a formalism for:
    \begin{enumerate}
        \item a source oriented with it's rotation axis in any arbitrary direction, 
        \item and the TT-gauge to be for a gravitational wave coming from the source and going along any arbitrary axis (not necessarily the same as the rotation axis).  
    \end{enumerate}
    This can be achieved in two steps.
    
    \paragraph{\underline{Step one}} - we want the same TT-gauge as before (along the $\hat{z}$-axis), but now for a binary source in an arbitrary orientation, as shown in Fig~\ref{fig:generalized source orientation}.  Apply two active (``alibi'') rotation transformations to the above result - first about the $\hat{y}$-axis by angle $\iota$ and then about the $\hat{z}$-axis by angle $\psi$ to construct $h^{TT\hat{z}^{'}}_{ij} = R_z(\psi)R_y(\iota)h^{TT}_{ij}R_y^\mathrm{Transpose}(\iota)R_z^\mathrm{Transpose}(\psi)$.
    
    It is important to note that we actually only need two Euler angles to fully generalize our problem, not the usual three.  There is no unique way of defining \textit{how} you rotate the axes, but in our case here let us choose to perform the ``extrinsic'' rotations (about the $\hat{x}$,$\hat{y}$, and $\hat{z}$ axes) in the following order: $\hat{z} \rightarrow \hat{y} \rightarrow \hat{z}$ by the angles $\theta_0$, $\iota$, and $\psi$, respectively.  (This is one of six ``proper Euler angle'' definitions).  An initial rotation about the $\hat{z}$ axis by angle $\theta_0$ can be equivalently achieved by beginning the binary's orbit with some initial orbital phase angle, which is exactly what we will later have the equations~\ref{eqn:monochrome phase} and \ref{eqn:freq evolution phase} which describe the orbital motion.  So in this problem the first Euler angle is already accounted for by interpreting it as a phase factor in the orbit, which is why we don't need to explicitly perform this rotation here, and why we simply perform the other two rotations by $\iota$ and $\psi$, which are the inclination and polarization angles, respectively.
    
    The result $h^{TT\hat{z}^{'}}_{ij}$ will be in the TT-gauge for a gravitational wave traveling in the $\hat{z}^{'} = R_z(\psi)R_y(\iota)\hat{z}$ direction.  To get back the TT-gauge along the $\hat{z}$-direction, we can apply the TT-projection operator equation~\ref{eqn: TT-projection operator} once again.  Setting $\hat{n} = \hat{z}$ we write out $\Lambda_{ijkl}h_{TT\hat{z}^{'}}^{kl} = h^{TT\hat{z}}_{ij}$.  The resulting metric perturbation solution can be written compactly in a couple of different ways:
    \begin{align}
    h^{TT\hat{z}}_{ij} &= e^{\hat{z}+}_{ij}H_{+} + e^{\hat{z}\times}_{ij}H_{\times} = e^{\hat{z}\textsc{A}}_{ij}H_{\textsc{A}}, \nonumber \\
    &= E^{\hat{z}+}_{ij}h_{+} + E^{\hat{z}\times}_{ij}h_{\times} = E^{\hat{z}\textsc{A}}_{ij}h_{\textsc{A}} \qquad \mathrm{for}\quad \textsc{A} \in [+, \times], \label{eqn: primed metric perturbation} \\
        &\mathrm{where}\quad \begin{cases}
            H_{+} \equiv h_{+}\frac{1}{2}\left(1 + \cos^2(\iota)\right)\cos(2\psi) - h_{\times}\cos(\iota)\sin(2\psi), \\
            H_{\times} \equiv h_{+}\frac{1}{2}\left(1 + \cos^2(\iota)\right)\sin(2\psi) + h_{\times}\cos(\iota)\cos(2\psi),
            \end{cases} \label{eqn: h prime}\\
        &\qquad\quad \ \ \begin{cases}
            E^{\hat{z}+}_{ij} \equiv \frac{1}{2}\left(1 + \cos^2(\iota)\right) \left[\cos(2\psi)e^{\hat{z}+}_{ij} + \sin(2\psi)e^{\hat{z}\times}_{ij} \right], \\
            E^{\hat{z}\times}_{ij} \equiv \cos(\iota) \left[-\sin(2\psi)e^{\hat{z}+}_{ij} + \cos(2\psi)e^{\hat{z}\times}_{ij} \right].
            \end{cases} \label{eqn: e plus/cross prime}
    \end{align}
    In words, we can either look at this as a transformation to the original $h_{+}$ and $h_{\times}$ terms, or a transformation to the polarization tensors, which can be useful interpretations.
    
    \paragraph{\underline{Step two}} - we now want the TT-gauge to be along any arbitrary axis.  Considering our problem, we would like to be able to anchor the observer at the origin of the coordinate system, and place the gravitational wave source anywhere in the sky.  The axis along which the gravitational wave travels which connects the observer to the source is then the desired TT-gauge axis.
    
    There is no unique way of doing this, but perhaps an intuitive and straight-forward approach is as follows.  Let $\hat{x}$, $\hat{y}$, and $\hat{z}$ be the basis vectors for the observer, and let the unit vectors in spherical coordinates be the basis vectors of the source:
    \begin{equation}
        \begin{cases}
            \begin{tabular}{c l c r l}
                $\hat{r}$ &$= \big[\sin(\theta)\cos(\phi),$ &$\sin(\theta)\sin(\phi),$ &$\cos(\theta) \big]$ , \ & (Earth to gravitational wave \\
                & & & & source unit vector) \\
                $\hat{\theta}$ &$= \big[\cos(\theta)\cos(\phi),$ &$\cos(\theta)\sin(\phi),$ &$\sin(\theta) \big]$ , \ & (transverse plane basis vector) \\
                $\hat{\phi}$ &$= \big[-\sin(\phi),$ &$\cos(\phi),$ &$0 \big]$ , \ & (transverse plane basis vector)
            \end{tabular}
            \end{cases} \label{eqn:source basis vectors} 
    \end{equation}
    where the angles $\theta$ and $\phi$ indicate the sky position of the source relative to the observer.  The source will be a distance $R$ away from the observer, and the vector $\hat{r}$ points toward the source, so $\vec{r} = R\hat{r}$.
    
    Now we simply define a priori new polarization tensors from the outer product of these basis vectors:
    \begin{align}
        \begin{cases}
            e^{\hat{r}+}_{ij} \equiv \hat{\theta}_i\hat{\theta}_j - \hat{\phi}_i\hat{\phi}_j \longleftrightarrow \textbf{e}^{\hat{r}+} \equiv \hat{\theta}\otimes\hat{\theta} - \hat{\phi}\otimes\hat{\phi}, \\
                e^{\hat{r}\times}_{ij} \equiv \hat{\phi}_i\hat{\theta}_j + \hat{\theta}_i\hat{\phi}_j \longleftrightarrow \textbf{e}^{\hat{r}\times} \equiv \hat{\phi}\otimes\hat{\theta} + \hat{\theta}\otimes\hat{\phi}.
        \end{cases} \label{eqn:general polarization tensors}
    \end{align}
    Notice that equations~\ref{eqn:general polarization tensors} are actually just a generalization of equations~\ref{eqn:basic polarization tensors}, where we let $\hat{x}\rightarrow \hat{\theta}$, $\hat{y}\rightarrow \hat{\phi}$, and $\hat{z}\rightarrow \hat{r}$.  The definitions give these new polarization tensors all of the required properties.  They are traceless, transverse along the $\hat{r}$-axis, and reduce to our original $\hat{z}$-axis polarization tensors~\ref{eqn:basic polarization tensors} when $\theta=\phi=0$.  The final step is a simply replacement.  Since these polarization tensors~\ref{eqn:general polarization tensors} are a generalization of the original polarization tensors~\ref{eqn:basic polarization tensors}, we can substitute these new tensors into the expressions in the metric perturbation from before, \ref{eqn: primed metric perturbation} and \ref{eqn: e plus/cross prime}.
    
\begin{figure}
    \centering
      \begin{subfigure}[t]{0.48\linewidth}
      \centering
        \includegraphics[width=\linewidth]{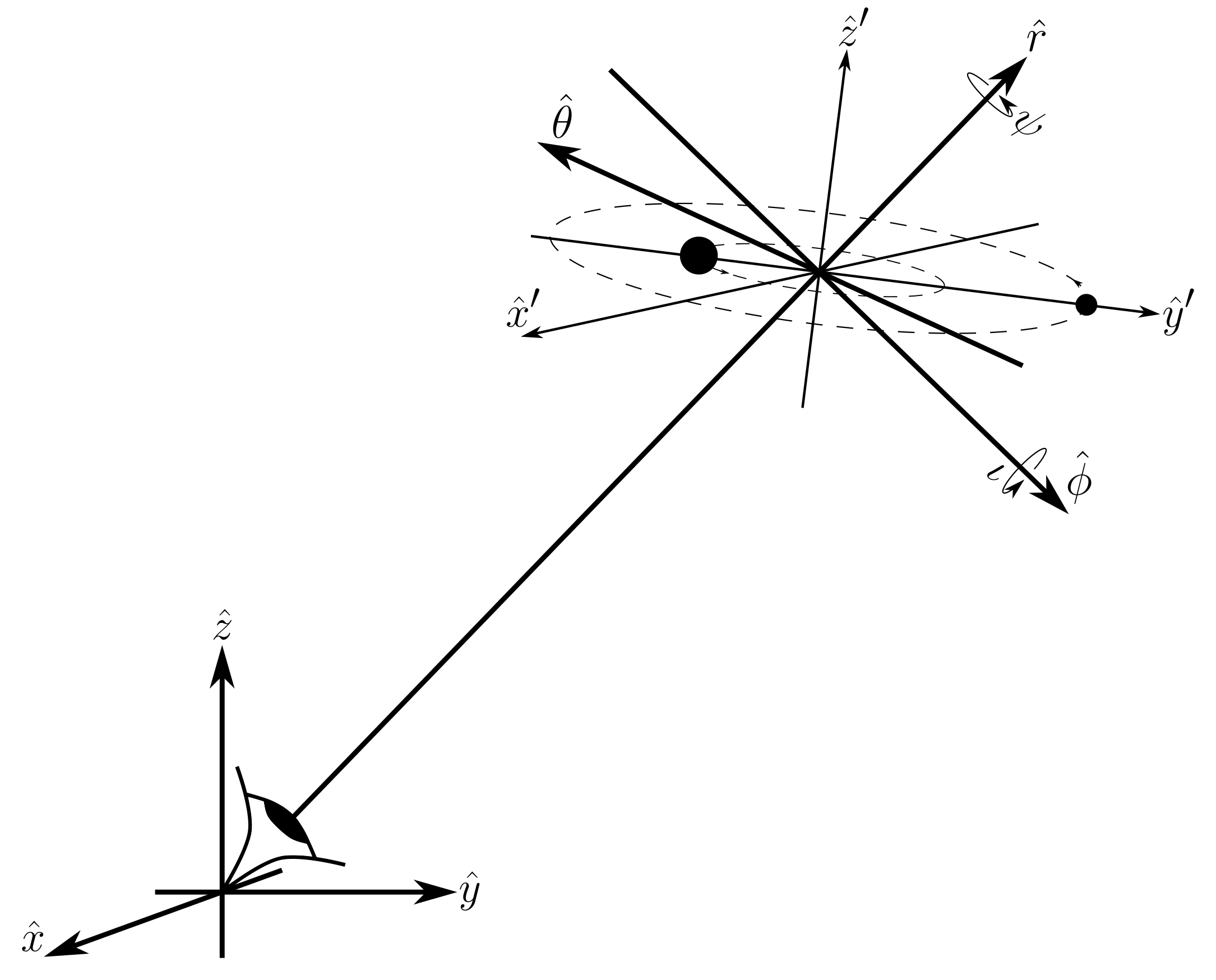}
        \caption{Fully generalized geometrical positioning of a gravitational wave source with respect to an observer.  Angles $\theta$ and $\phi$ denote source sky position, and $\iota$ and $\psi$ indicate source inclination and polarization.  Source appears ``face-on'' if $\iota=0$, and ``edge-on'' if $\iota=\pi/2$, and ``face-off'' if $\iota=\pi$.}
        \label{fig:general source orientation}
      \end{subfigure}
      \hfill
      \begin{subfigure}[t]{0.48\linewidth}
      \centering
        \includegraphics[width=\linewidth]{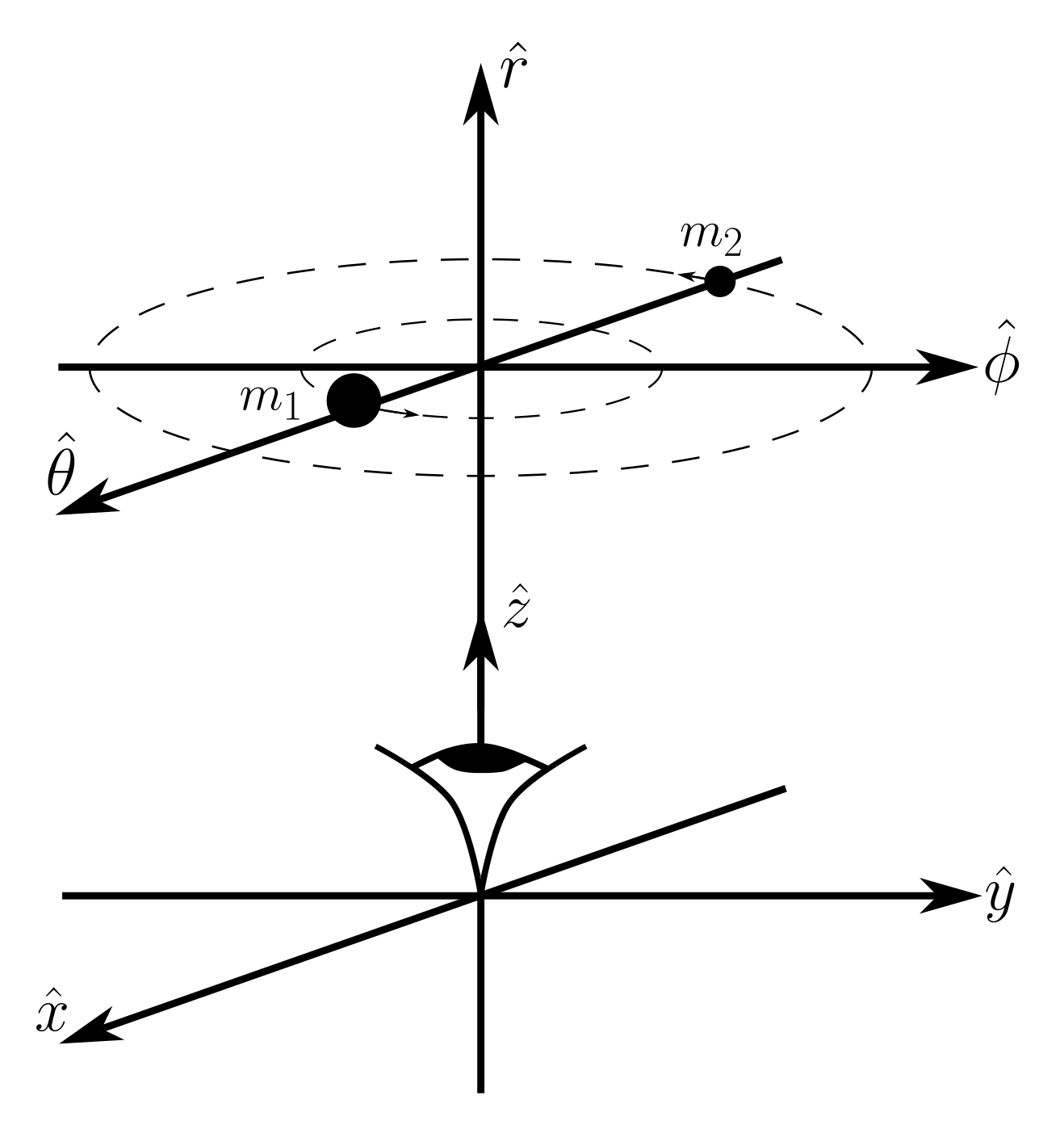}
        \caption{The source basis vectors are chosen so that when $\theta = \phi = 0$, then $\hat{r}=\hat{z}$, $\hat{\theta}=\hat{x}$, and $\hat{\phi}=\hat{y}$.  In this case, if the source is not inclined (i.e. $\iota=0$), then the observer will see the binary orbiting in the counter-clockwise direction.}
        \label{fig:example general source orientation}
      \end{subfigure}
    \caption[Binary System: Position]{}
    \label{fig:general source orientation full figure}
    \end{figure}

    The final fully general result for our problem of interest can now be expressed compactly in a couple of different ways:
    \begin{align}
    h^{TT\hat{r}}_{ij} &= e^{\hat{r}+}_{ij}H_{+} + e^{\hat{r}\times}_{ij}H_{\times} = e^{\hat{r}\textsc{A}}_{ij}H_{\textsc{A}}, \nonumber \\
    &= E^{\hat{r}+}_{ij}h_{+} + E^{\hat{r}\times}_{ij}h_{\times} = E^{\hat{r}\textsc{A}}_{ij}h_{\textsc{A}} \qquad \mathrm{for}\quad \textsc{A} \in [+, \times] , \label{eqn: general metric perturbation} \\
    &\mathrm{where}\quad \begin{cases}
            H_{+} \equiv h_{+}\frac{1}{2}\left(1 + \cos^2(\iota)\right)\cos(2\psi) - h_{\times}\cos(\iota)\sin(2\psi), \\
            H_{\times} \equiv h_{+}\frac{1}{2}\left(1 + \cos^2(\iota)\right)\sin(2\psi) + h_{\times}\cos(\iota)\cos(2\psi),
            \end{cases} \tag{\ref{eqn: h prime} r} \\
        &\qquad\quad \ \ \begin{cases}
            E^{\hat{r}+}_{ij} \equiv \frac{1}{2}\left(1 + \cos^2(\iota)\right) \left[\cos(2\psi)e^{\hat{r}+}_{ij} + \sin(2\psi)e^{\hat{r}\times}_{ij} \right], \\
            E^{\hat{r}\times}_{ij} \equiv \cos(\iota) \left[-\sin(2\psi)e^{\hat{r}+}_{ij} + \cos(2\psi)e^{\hat{r}\times}_{ij} \right],
            \end{cases} \label{eqn: E plus/cross general} \\
        &\qquad\quad \ \ \begin{cases}
            h_{+}(t) \equiv -h(t) \cos\big(2\Theta(t)\big), \\
            h_{\times}(t) \equiv -h(t) \sin\big(2\Theta(t)\big), \tag{\ref{eqn: h plus and cross} r}
            \end{cases}\\[2pt]
        &\hspace{1.8cm} h(t) \equiv \frac{4(G\mathcal{M})^{5/3}}{c^4 R}\omega(t)^{2/3} , \tag{\ref{eqn: amplitude of metric perturbation} r}\\[2pt]
        &\qquad\quad \ \ \begin{cases}
            e^{\hat{r}+}_{ij} \equiv \hat{\theta}_i\hat{\theta}_j - \hat{\phi}_i\hat{\phi}_j, \\
            e^{\hat{r}\times}_{ij} \equiv \hat{\phi}_i\hat{\theta}_j + \hat{\theta}_i\hat{\phi}_j ,
            \end{cases} \tag{\ref{eqn:general polarization tensors} r} \\
        &\qquad\quad \ \ \begin{cases}
            \begin{tabular}{c l c r}
                $\hat{r}$ &$= \big[\sin(\theta)\cos(\phi),$ &$\sin(\theta)\sin(\phi),$ &$\cos(\theta) \big] , $  \\
                $\hat{\theta}$ &$= \big[\cos(\theta)\cos(\phi),$ &$\cos(\theta)\sin(\phi),$ &$\sin(\theta) \big] , $  \\
                $\hat{\phi}$ &$= \big[-\sin(\phi),$ &$\cos(\phi),$ &$0 \big] , $  
            \end{tabular}
            \end{cases}  \tag{\ref{eqn:source basis vectors} r} 
    \end{align}
    As was the case before, we can either look at this as a transformation to the original $h_{+}$ and $h_{\times}$ terms, or a transformation to the polarization tensors.  We can test that these expressions are indeed a generalization of the original problem with all of the required properties.  The metric perturbation is finally in the TT-gauge now for a gravitational wave propagating along the $\hat{r}$-axis.  It is transverse to the direction of propagation of the gravitational wave and traceless, both of which follow since the polarization tensors are transverse and traceless, and the metric perturbation directly decomposes into a linear combination of these tensors.  The sky angles $\theta$ and $\phi$ point toward the location of the source, and the angles $\iota$ and $\psi$ indicate the source's own orientation.  All of this is shown in Figure~\ref{fig:general source orientation full figure}.  Furthermore, our motivation for this approach and these definitions stems from our comfort and familiarity with spherical coordinate unit basis vectors.  

%---------------------------------------------------------------------------------
%---------------------------------------------------------------------------------    
    \section{Source Parameters and a Word about Notation}
    
    We take a moment here simply to point out the physical significance behind the notation we have employed so far, as it will greatly simplify our work later and lead to solutions which are written in a way that are easier to understand physically.
    
    A circular binary system produces gravitational waves which in the zeroth order Newtonian regime is described by a total of eight parameters: 
    \begin{equation*}
        \mathrm{Source} \ \mathrm{Parameters:} \quad \{R, \theta, \phi, \iota, \psi, \theta_0, \mathcal{M}, \omega_0 \} \quad \longrightarrow \quad \begin{cases}
            \mathrm{Extrinsic:} \quad &\{R, \theta, \phi, \iota, \psi, \theta_0\} \\
            \mathrm{Intrinsic:} \quad &\{\mathcal{M}, \omega_0 \}
        \end{cases}
    \end{equation*}
    (Note that $\theta_0$ and $\omega_0$ will be shown explicitly in the next section~\ref{sec: two frequency models}, but they appear in $\Theta(t)$ and $\omega(t)$ above).
    
    Of these eight parameters, the angles $\theta$, $\phi$, $\iota$, and $\psi$ are the binary's ``geometrical orientation'' angles - they describe the orientation of the orbit of the binary, and the binary's location on the sky in reference to the Earth's frame.  The parameter $\theta_0$ we'll think of as the initial phase of the binary, although it was pointed out that this is also the third Euler angle for the binary's orientation.  The rest of the parameters - $R$, $\mathcal{M}$, $\theta_0$, and $\omega_0$ - influence only the strain $h$ of the gravitational wave.  Lastly, it should be pointed out too that the strain is also influenced by the $\theta$ and $\phi$, because the strain is evaluated at the retarded time, and the retarded time will depend on these two parameters (as discussed in Section~\ref{sec: Time Retardation}).
    
    In terms of notation, as we mentioned before we can look at the metric perturbation in equation~\ref{eqn: general metric perturbation} in either of two ways.  In this paper we prefer to use the notation $h^{TT\hat{r}}_{ij} = E^{\hat{r}\textsc{A}}_{ij}h_{\textsc{A}}$ because it groups all of the geometrical orientation and location angles into the definition of the polarization tensor, and keeps the strain in the simple form that we started with in equation~\ref{eqn: h plus and cross}.  We will point out that many sources seem to prefer a notation which more closely resembles the alternative form of writing $h^{TT\hat{r}}_{ij} = e^{\hat{r}\textsc{A}}_{ij}H_{\textsc{A}}$ \citep[see for example,][]{CC_main_paper,PPTA_cw_2014,NG_11yr_cw}.  Perhaps the only disadvantage of this is that while the polarization tensor is a function only of the sky position angles, the strain $H_\textsc{A}$ becomes a mixture of both $h_+$ and $h_{\times}$.  But since:
    \begin{equation*}
        e^{\hat{r}\textsc{A}}_{ij}H_{\textsc{A}} = E^{\hat{r}\textsc{A}}_{ij}h_{\textsc{A}},
    \end{equation*}
    we can use these two notations interchangeably as is most convenient.
    
    Also as an aside, we point out that when reviewing pulsar timing literature, one should pay close attention to the use of notation and definitions of quantities, namely the vectors that go into building the polarization tensors.  Often one will find the use of the vectors $\hat{\Omega}$, $\hat{m}$, and $\hat{n}$ for the source basis vectors (instead of $\hat{r}$, $\hat{\theta}$, and $\hat{\phi}$ as we have used.  What becomes tricky and sometimes frustrating, however, is that these vectors won't always be defined in the same way.  Typically they differ in the direction they are pointed, i.e. by a $\pm$ sign \citep[see for example,][]{Arzoumanian_2014, chamberlin2015, taylor2020}.  Noting these differences is crucial, as they will necessarily introduce $\pm$ signs in the definitions of the polarization tensors and antenna patterns (see Section~\ref{subsec:The Antenna Response}).  
    
    Mathematically this is perfectly fine, in the sense that all of the math is still consistent when working within a given choice of notation/definitions.  However, the conceptualization quantities such as angular parameters between notations may differ.  This is why we propose and explain the motivation of the notation and conventions established here in our work, as they simply adopt the often used and familiar spherical coordinate basis vectors as their building point.

%---------------------------------------------------------------------------------
%---------------------------------------------------------------------------------    
    \section{Two Models:  Monochromatism and Frequency Evolution}\label{sec: two frequency models}
    
    There are two important models used for computing the evolution of the metric perturbation and the timing residual.  In short, the frequency evolution model is a more generalized version of the timing residual, and reduces to the monochromatic model in the appropriate limit (described below).
    
    Here we call attention to $t_0$, the ``fiducial time'' for our two models defined below.  As the model's reference time, we define $t_0$ as the time at which $\theta_0$ and $\omega_0$ are measured: \begin{align}
        \Theta(t_0) \equiv \theta_0 , \nonumber \\
        \omega(t_0) \equiv \omega_0 , 
    \end{align}
    We can choose any time to be the fiducial time of our model, but perhaps the two most comfortable choices are either $t_0 = 0$, or $t_0=-\frac{R}{c}$.  Conceptually choosing $t_0=0$ means these parameters are the Earth frame ``present-day'' values of the orbital phase and frequency of the binary (when our experiment first begins), and choosing $t_0=-\frac{R}{c}$ are the ``retarded-day'' values, that is the values of these parameters that produced the gravitational waves which are currently arriving at the Earth (again, at the time our experiment first begins).  Later in this work we will choose the fiducial time to be $t_0 = -\frac{R}{c}$.

%---------------------------------------------------------------------------------
%---------------------------------------------------------------------------------    
        \subsection{Monochromatism}\label{subsec: two frequency models - mono}
        
        In the most simple model of the gravitational waves emitted by a binary system, one assumes no energy is lost from the emission of gravitational waves.  Under this assumption, the binary's energy will remain conserved, the orbit will not coalesce (it has ``infinite'' coalescence time), and hence the orbital period/frequency will remain constant.  This assumption means that the gravitational waves emitted by the binary system will have a constant frequency for all time.  Hence the name ``monochromatic'' gravitational waves.
        
        For the monochromatic gravitational wave model, the orbital frequency and phase are given by:
        \begin{align}
        \omega(t) =& \omega_0 , \label{eqn:monochrome freq} \\ 
        \Theta(t) =& \theta_0 + \omega_0(t - t_0) . \label{eqn:monochrome phase}
        \end{align}

%---------------------------------------------------------------------------------
%---------------------------------------------------------------------------------        
        \subsection{Frequency Evolution}\label{subsec: two frequency models - freq evo}
        
        More realistically, however, energy in the binary system is lost due to the emission of gravitational wave radiation.  This means that over time the binary will collapse and coalesce.  As the binary collapses, the orbital frequency will increase, meaning that the gravitational wave frequency will also increase.  This evolution towards higher frequencies is called ``frequency chirping.''
        
        Under assumption~\ref{as:slow}, gravitational wave radiation from a binary system causes the frequency of the orbit to change as:
        \begin{equation}
            \dot{\omega} = \frac{d\omega}{dt} = \frac{96}{5}\left(\frac{G\mathcal{M}}{c^3}\right)^{5/3}\omega^{11/3} .
        \label{eqn: omega dot}
        \end{equation}
        As the orbit coalesces, the orbital frequency increases without bound.  Formally we define the time the binary coalesces to be when it reaches infinite orbital frequency, that is $\omega(t) \rightarrow \infty$ as $t \rightarrow \tau_c$.  We can compute the time of coalescence of the binary by integrating equation~\ref{eqn: omega dot}:
        \begin{equation*}
            \int\limits^{\tau_c}_{t_0} dt = \int\limits^{\infty}_{\omega_0} \frac{5}{96}\left(\frac{c^3}{G\mathcal{M}}\right)^{5/3} \omega^{-11/3} d\omega,
        \end{equation*}
        which gives:
        \begin{align}
            \tau_c &= t_0 + \Delta\tau_c , \nonumber \\
            &\mathrm{where}\quad \Delta\tau_c \equiv \frac{5}{256}\left(\frac{c^3}{G\mathcal{M}}\right)^{5/3} \frac{1}{\omega_0^{8/3}} .
        \label{eqn: time to coalescence}
        \end{align}
        Here in our notation $\tau_c$ is the ``time of coalescence'' and $\Delta\tau_c \equiv \tau_c - t_0$ is the ``time of coalescence measured from our chosen fiducial time.''  We make the pedantic distinction of this here as the difference between the two will become more significant later on.  This difference is represented in Figure~\ref{fig: PTA chronology PW}, and the dependence of $\Delta \tau_c$ on the chirp mass and orbital frequency parameters can be seen in the contour plot in Figure~\ref{fig: DTauc & F contours}.
        
        For arbitrary times, we can repeat the integration:
        \begin{equation*}
            \int\limits^{t}_{t_0} dt = \int\limits^{\omega}_{\omega_0} \frac{5}{96}\left(\frac{c^3}{G\mathcal{M}}\right)^{5/3} \omega^{' \ -11/3} d\omega^{'},
        \end{equation*}
        which leads us to an expression for the orbital frequency as a function of time for our binary system:
        \begin{equation}
            \omega(t) = \omega_0\left[1 - \frac{t-t_0}{\Delta\tau_c} \right]^{-3/8} .
        \label{eqn:freq evolution freq}
        \end{equation}
        As for the phase of the binary, we know $\frac{d\Theta}{dt} = \omega(t)$.  We can integrate this expression over time, or similary we can first write $\omega(t) = \frac{d\Theta}{d\omega}\frac{d\omega}{dt}$, so using equation~\ref{eqn: omega dot} again we can write $\frac{d\Theta}{d\omega} = \frac{5}{96}\left(\frac{c^3}{G\mathcal{M}}\right)^{5/3}\omega^{-8/3}$ and integrate this for arbitary $\omega$:
        \begin{equation*}
            \int\limits^{\Theta}_{\theta_0}d\Theta^{'} = \int\limits^{\omega}_{\omega_0} \frac{5}{96}\left(\frac{c^3}{G\mathcal{M}}\right)^{5/3}\omega^{' \ -8/3} d\omega^{'}, 
        \end{equation*}
        which leads us to an expression for the orbital phase as a function of time for our binary system (expressed here in a number of useful forms):
        \begin{align}
            \Theta(t) &= \theta_0 + \theta_c\left[ 1 - \left(\frac{\omega(t)}{\omega_0}\right)^{-5/3} \right] , \nonumber \\
            &= \theta_0 + \theta_c\left[ 1 - \left(1 - \frac{t-t_0}{\Delta\tau_c}\right)^{5/8} \right] , \nonumber \\
            &= \theta_0 + \frac{1}{32}\left(\frac{c^3}{G\mathcal{M}}\right)^{5/3}  \left[ \omega_0^{-5/3} - \omega(t)^{-5/3} \right] ,  \label{eqn:freq evolution phase}
        \end{align}
        where we define the quantity:
        \begin{equation}
            \theta_c \equiv \frac{8}{5}\Delta\tau_c \omega_0 = \frac{1}{32}\left(\frac{c^3}{G\mathcal{M}\omega_0}\right)^{5/3} . 
        \label{eqn: thetac}
        \end{equation}
        This is a rather important quantity itself, and conceptually it is the number of radians that the binary will sweep out before it coalesces (i.e. $\frac{\theta_c}{2\pi}$ is the number of orbital revolutions before the system coalesces).  That is, in the limit the binary coalesces where $\omega(t) \rightarrow \infty$, we see that $\Theta \rightarrow \theta_0 + \theta_c$.  For this reason we will refer to $\theta_c$ as the ``coalescence angle.''  Remember, both $\Delta \tau_c$ and $\theta_c$ are measured with respect to the chosen fiducial time.
        
        Note that in the ``large coalescence time'' limit that $\Delta\tau_c \rightarrow \infty$, the frequency evolution model (equations~\ref{eqn:freq evolution freq} and \ref{eqn:freq evolution phase}) reduces to the monochromatic model (equations~\ref{eqn:monochrome freq} and \ref{eqn:monochrome phase}) as we would expect.  In practice, the monochromatic limit for pulsar timing experiments merely requires that the gravitational wave frequency stays nearly constant as pulsar's radiation travels between the pulsar and the Earth, so that $\Delta \tau_c \gg \left(1-\hat{r}\cdot\hat{p}\right)\frac{L}{c}$ (see ahead to equation~\ref{eqn: Res(t) pw freq evo frequency phase E and P}).  However in this work since we are specifically investigating and classifying different models based on their mathematical differences, we will use the term ``monochromatic'' specifically when working under the assumption of Section~\ref{subsec: two frequency models - mono}.
        
        Finally, using this information we can also express equation~\ref{eqn: omega dot} as:
        \begin{align}
            \dot{\omega} &= \frac{d\omega}{dt} = \frac{96}{5}\left(\frac{G\mathcal{M}}{c^3}\right)^{5/3}\omega^{11/3} , \nonumber \\
            &= \frac{3}{8}\frac{\omega^{11/3}}{\Delta\tau_c \omega_0^{8/3}} ,  \nonumber \\
            &= \frac{3}{8}\frac{\omega}{\tau_c - t} .
        \label{eqn: omega dot alt versions}
        \end{align}

\cleardoublepage
\chapter{The Continuous Wave Timing Residual}\label{ch: The Continuous Wave Timing Residual}

A binary system of two massive objects orbiting each other will, under general relativity, cause the metric perturbation to vary in time sinusoidally, and this perturbation will propagate away from the source as a wave.  The result is that spacetime will stretch and compress as the wave propagates through the universe, and will change the path (geodesic) that external objects move along as the wave interferes with them.  For gravitational wave experiments, it is especially important to try and understand how the gravitational wave will affect null geodesics, i.e. the path that light moves along, because we can design experiments that can actually test and observe this effect on light.

With this in mind we now turn to the question of using pulsars to design a gravitational wave experiment.  When observed with radio telescopes here on Earth, highly regularly spinning pulsars act as incredibly precise clocks, with each pulse acting as a clock tick.  Now consider the effect of a gravitational wave on the photons in a pulse of light beamed from a pulsar, as it travels from the pulsar towards our observatories here on Earth.  In the absence of any gravitational wave disturbances, the photons would move along an essentially flat spacetime background as they traveled from the pulsar to the Earth.  Since the spacetime the photons traverse is static, the interval between the light pulses will remain constant, or in other words, the pulsar's pulse period $T$ will remain constant as observed on Earth.  However, a gravitational wave will cause that flat spacetime to perturb sinusoidally, and in this way the photon spacetime path from the pulsar to the Earth will change.  The question we want to ask is how is the observed period of the pulsar affected by a gravitational wave, and can we find an observable that allows us to turn this theoretical observation into an experimental one?

For the work below we will make the following assumptions:\\

\begin{minipage}{0.9\linewidth}
\fbox {
    \parbox{\linewidth}{
    \textbf{Assumptions:} Observed Pulsar Period
    \begin{enumerate}
        \item The universe is flat and static.  Therefore the expression for the retarded time is the familiar: $t_\mathrm{ret} = t - \frac{|\vec{x}-\vec{x}^{'}|}{c}$ . \label{as: flat static retarded time}
        \item Earth is at the center of our coordinate system.
        \item The pulsar is located at a fixed angular sky position and distance from the Earth.
        \item The local background spacetime between the pulsar and the Earth is flat Minkowski, ${\eta_{\mu \nu} = \mathrm{diag}(-c^2, \ 1, \ 1, \ 1)}$. \label{as: local background Minkowski}
        \item Transverse-traceless (``TT'') gauge. 
        \item The antenna patterns are assumed to remain constant over the Earth-pulsar baseline.
        \item The pulsar's rotational period timescale must be appropriately small in comparison to the timescale of the gravitational wave.  Specifically we have two case requirements: \label{as: pulsar period timescale}
        \begin{arrowlist}
            \item \underline{Monochromatic Source}:  $\omega_0 T \ll 1$ ,
            \item \underline{Frequency Evolving Source}:  $\frac{T}{\Delta \tau_c} \ll 1$ .
        \end{arrowlist}
    \end{enumerate}
    }
}
\end{minipage}\\ \\

%---------------------------------------------------------------------------------
%---------------------------------------------------------------------------------        
    \section{The Observed Pulsar Period}
    
    So the first question we ask is, what is the path of a photon traveling from the pulsar to the Earth?\footnote{The derivation presented here follows the logic presented in \cite{maggiore_2018}.}  To determine this we begin with the spacetime interval and our metric, a flat background spacetime with a gravitational wave perturbation coming from some binary source, as given by equation~\ref{eqn: general metric perturbation}:
    \begin{equation}
        ds^2 = -(c dt)^2 + \left[ \eta_{ij} + h^{TT\hat{r}}_{ij} \right] dx^i dx^j .
    \label{eqn: spacetime interval}
    \end{equation}
    It's important to remember here that the metric perturbation $h^{TT\hat{r}}_{ij}$ is a function of the retarded time, which itself is a function of $t$ and $\vec{x}$, that is, our notation for the functional dependence on $t_\mathrm{ret}$, $t$, and $\vec{x}$ can be indicated as:
    \begin{equation}
        h^{TT\hat{r}}_{ij} \equiv h^{TT\hat{r}}_{ij}\left(t_{\mathrm{ret}}\right) \equiv h^{TT\hat{r}}_{ij}\big(t_{\mathrm{ret}}(t,\vec{x})\big) \equiv h^{TT\hat{r}}_{ij}\left(t,\vec{x}\right) .
    \label{eqn: h as function of tret}
    \end{equation}
    Remembering this order, that the metric perturbation is a function of the retarded time, which itself depends on the field point of interest $\vec{x}$, will become important shortly, which is why we make special emphasis of it here before continuing.
    
    Let the pulsar be at a fixed angular position given by $\theta_p$ and $\phi_p$; the vector pointing from the Earth towards the pulsar will be denoted as $\hat{p}$:
    \begin{equation}
        \hat{p} \equiv \big[\sin(\theta_p)\cos(\phi_p),\ \sin(\theta_p)\sin(\phi_p),\ \cos(\theta_p) \big]
        \label{eqn: p hat}
    \end{equation}
    The photon will travel along the radial spatial path connecting the pulsar and the Earth (hence $d\theta=d\phi=0$), so using spherical coordinates we can write $dx^i = \hat{p}^i dr$.  Moreover, since we are considering the motion of a photon, $ds = 0$.  So equation~\ref{eqn: spacetime interval} becomes:
    \begin{align}
        (c dt)^2 &= \left[ \eta_{ij} + h^{TT\hat{r}}_{ij} \right] \hat{p}^i\hat{p}^j dr^2 ,  \nonumber \\
                 &= \left[ 1 + \hat{p}^i\hat{p}^j h^{TT\hat{r}}_{ij} \right] dr^2 , \nonumber \\
                 &= \left[ 1 + \hat{p}^i\hat{p}^j E^{\hat{r}\textsc{A}}_{ij} h_{\textsc{A}} \right] dr^2  \quad\quad \longrightarrow \quad \pm \ dr = \frac{c dt}{\left[1+\hat{p}^i\hat{p}^j E^{\hat{r}\textsc{A}}_{ij} h_{\textsc{A}} \right]^{1/2}} , \label{eqn: photon spacetime interval}
    \end{align}
    where the second line just comes from writing out the Einstein notation: $\hat{p}^i\hat{p}^j\eta_{ij} = (\hat{p}^x)^2 + (\hat{p}^y)^2 + (\hat{p}^z)^2 = |\hat{p}|^2 = 1$.  (Or alternatively, we can write the spatial part of the Minkowski metric as a decomposed sum of the outer products of the unit Cartesian vectors $\eta_{ij} \equiv \left(\hat{x}_i \hat{x}_j \right) + \left(\hat{y}_i \hat{y}_j \right) + \left(\hat{z}_i \hat{z}_j \right)$ and then recognize that this is just the dot product of $\hat{p}$ with itself: $\hat{p}^i\hat{p}^j\eta_{ij} = |\hat{p}|^2 = 1$).  In the third line we switch to the notation established and motivated in equation~\ref{eqn: general metric perturbation}, and note that here the quantity $\hat{p}^i\hat{p}^j E^{\hat{r}\textsc{A}}_{ij}$ is function only of the source's orientation and sky location angles, and importantly, doesn't have any time dependence.  We will find later that this combines with another term to become what we will call the ``antenna response'' function (discussed in detail in Section~\ref{subsec:The Antenna Response}).
    
    The $\pm$ from the square root denotes a radially outbound ($+$) versus radially inbound ($-$) photon.  We now integrate both sides of this expression to get the photon's path.  The photon leaves the pulsar at the ``emitted'' time $t_\mathrm{em}$ at a distance $L$ from the Earth, and travels radially inbound arriving at the Earth at the ``observed'' time $t_\mathrm{obs}$:
    \begin{align*}
        L = -\int\limits^0_L dr 
          = \int\limits^{t_{\mathrm{obs}}}_{t_{\mathrm{em}}} \frac{c dt}{\left[ 1 + \hat{p}^i\hat{p}^j E^{\hat{r}\textsc{A}}_{ij} h_{\textsc{A}} \right]^{1/2}} 
          &\approx \int\limits^{t_{\mathrm{obs}}}_{t_{\mathrm{em}}}  \left[ 1 - \frac{1}{2}\hat{p}^i\hat{p}^j E^{\hat{r}\textsc{A}}_{ij} h_{\textsc{A}} \right] c dt,  \\
          &= c\left(t_\mathrm{obs} - t_\mathrm{em}\right) -\frac{1}{2}\hat{p}^i\hat{p}^j E^{\hat{r}\textsc{A}}_{ij}\int\limits^{t_{\mathrm{obs}}}_{t_{\mathrm{em}}} h_{\textsc{A}}\left(t, \vec{x} \right) c dt .
    \end{align*}
    We can see in the final equality here that to ``zeroth order'' in the metric perturbation the path that the photon takes is $L \approx c\left( t_\mathrm{obs} - t_\mathrm{em}\right)$, or $t_\mathrm{obs} \approx t_\mathrm{em} + L/c$,  which is exactly as we would expect for flat unperturbed spacetime.  Since we are only interested in the solution to first order in the metric perturbation, we can use this to replace the upper limit of integration and then evaluate the metric perturbation along the zeroth order path (the overall integrand is still first order).  We can also use this to write the spatial path of our photon to zeroth order as:
    \begin{equation}
        \vec{x}_0(t) \equiv \left[ct_\mathrm{em} + L - ct\right]\hat{p} \approx \left[ct_\mathrm{obs} - ct\right]\hat{p} .
        \label{eqn:zeroth order photon path}
    \end{equation}
    This describes the desired path - i.e. $\vec{x}_0 = L\hat{p}$ at the time the photon leaves the pulsar and $\vec{x}_0 = 0$ at the time it arrives at the Earth.  Now we can write:
    \begin{equation}
        t_\mathrm{obs} \approx t_\mathrm{em} + \frac{L}{c} + \frac{1}{2}\hat{p}^i\hat{p}^j E^{\hat{r}\textsc{A}}_{ij}\int\limits^{t_{\mathrm{em}}+L/c}_{t_\mathrm{em}}h_{\textsc{A}}\big(t, \vec{x}_0(t) \big) dt .
    \label{eqn: tobs light pulse}
    \end{equation}
    
    So we have considered the path that a photon will travel between the pulsar and the Earth, and have found an expression equation~\ref{eqn: tobs light pulse} which gives the difference in emitted and observed times of the photon.  Next imagine the pulsar emits a flash, then rotates once and emits a second flash.  We'll denote the true period of the pulsar as $T$, but now want to find what is the observed period here on Earth.  We repeat the exact same steps above for a photon one period later.  Now the emitted time is $t_\mathrm{em}^{'} = t_\mathrm{em}+T$ and the observed time is $t_\mathrm{obs}^{'}$.  The the spatial path then becomes:
    \begin{align}
        \vec{x}^{'}_0(t) &\equiv \vec{x}_0(t,t_\mathrm{em} \rightarrow t_\mathrm{em}^{'}) = \left[ct_\mathrm{em} + cT + L - ct\right]\hat{p} = \left[ct_\mathrm{em} + L - c(t-T)\right]\hat{p} , \nonumber \\
        & = \vec{x}_0(t-T) .
    \label{eqn:zeroth order photon path prime}
    \end{align}
    The expression one period later changes to:
    \begin{align}
        t_\mathrm{obs}^{'} &\approx t_\mathrm{em} + T + \frac{L}{c} + \frac{1}{2}\hat{p}^i\hat{p}^j E^{\hat{r}\textsc{A}}_{ij}\int\limits^{t_{\mathrm{em}}+T+L/c}_{t_{\mathrm{em}+T}} h_{\textsc{A}}\big(t, \vec{x}_0(t-T) \big) dt , \nonumber \\
        & = t_\mathrm{em} + T + \frac{L}{c} + \frac{1}{2}\hat{p}^i\hat{p}^j E^{\hat{r}\textsc{A}}_{ij}\int\limits^{t_{\mathrm{em}}+L/c}_{t_{\mathrm{em}}} h_{\textsc{A}}\big(t+T, \vec{x}_0(t) \big) dt ,
    \label{eqn: tobs_prime light pulse}
    \end{align}
    where in the second line we performed the substitution/coordinate-shift $\Tilde{t}=t-T$.  This shifts the limits of integration, and since the label of the coordinate of integration in an integral is arbitrary, we can let $\Tilde{t} \rightarrow t$.  The integrals in expressions~\ref{eqn: tobs light pulse} and~\ref{eqn: tobs_prime light pulse} now only differ by a factor of $T$ in the time-coordinate, but \textit{not} in the spatial path.  So the difference between equation~\ref{eqn: tobs light pulse} and~\ref{eqn: tobs_prime light pulse} is the observed pulsar pulse period here on Earth:
    \begin{align}
        T_\mathrm{obs} \equiv t^{'}_\mathrm{obs} - t_\mathrm{obs} &\approx T + \Delta T, \nonumber \\
        \mathrm{where} \quad \Delta T &= \frac{1}{2}\hat{p}^i\hat{p}^j E^{\hat{r}\textsc{A}}_{ij}\int\limits^{t_{\mathrm{em}}+L/c}_{t_{\mathrm{em}}} \bigg[h_{\textsc{A}}\big(t+T, \vec{x}_0(t) \big) - h_{\textsc{A}}\big(t, \vec{x}_0(t) \big) \bigg]dt ,
    \label{eqn: observed pulsar period}
    \end{align}
    with equation~\ref{eqn: observed pulsar period} being the general expression for the resulting perturbation of a gravitational wave on the observed pulsar period.
    
    In Section~\ref{sec:The Observable Residual} we will explain that a more useful quantity to consider is the fractional change in the pulsar's period, so dividing equation~\ref{eqn: observed pulsar period} by $T$ we write:
    \begin{align}
        \frac{\Delta T}{T} &= \frac{1}{2}\hat{p}^i\hat{p}^j E^{\hat{r}\textsc{A}}_{ij}\int\limits^{t_{\mathrm{em}}+L/c}_{t_{\mathrm{em}}} \left[ \frac{h_{\textsc{A}}\big(t+T, \vec{x}_0(t) \big) - h_{\textsc{A}}\big(t, \vec{x}_0(t)\big)}{T} \right]dt , \nonumber \\
        &= \frac{1}{2}\hat{p}^i\hat{p}^j E^{\hat{r}\textsc{A}}_{ij}\int\limits^{t_{\mathrm{em}}+L/c}_{t_{\mathrm{em}}} \left[ \frac{h_{\textsc{A}}\big(t+T, \vec{x} \big) - h_{\textsc{A}}\big(t, \vec{x}\big)}{T} \right]\Bigg\rvert_{\vec{x}=\vec{x}_{0}(t)} dt , \nonumber \\
        &\approx \frac{1}{2}\hat{p}^i\hat{p}^j E^{\hat{r}\textsc{A}}_{ij}\int\limits^{t_{\mathrm{em}}+L/c}_{t_{\mathrm{em}}} \frac{\partial h_{\textsc{A}}\left(t, \vec{x}\right)}{\partial t} \Bigg\rvert_{\vec{x}=\vec{x}_{0}(t)} dt .
    \label{eqn: Delta T / T limit}
    \end{align}
    The approximation of the integrand that we made in the final line is crucial, and  makes use of the definition of a derivative.  However, to arrive at this approximation we must consider the two fundamentally different cases of interest.  Regardless if the source is truly monochromatic or if its frequency is evolving, the result here is the same, but the approximation statements will be different as shown in the box below.\\
    
    \begin{minipage}{0.9\linewidth}
    \fbox {
        \parbox{\linewidth}{
        \underline{\textbf{Monochromatic Source}}\newline
        For a monochromatic source we see that from the functional form of the phase, frequency, and strain (Section~\ref{subsec: two frequency models - mono} and equation~\ref{eqn: amplitude of metric perturbation}), $t$ will always appear in the combination $\omega_0 t$, a dimensionless quantity, in our equations.  Let's use this to non-dimensionalize our expression so that we can properly take the limit definition of our derivative.  It is tempting to take the limit definition of our derivative of the integrand in equation~\ref{eqn: Delta T / T limit} in terms of $T$ alone, but this is a dimensional quantity, and therefore this statement would not be sufficiently rigorous.  So let:
        \begin{equation*}
            \Tilde{t} \equiv \omega_0 t \quad \implies \Tilde{T} = \omega_0 T .
        \end{equation*}
        This means we can write:
        \begin{align*}
            \frac{h_{\textsc{A}}\big(t+T, \vec{x} \big) - h_{\textsc{A}}\big(t, \vec{x}\big)}{T} &= \frac{h_{\textsc{A}}\big(\Tilde{t}+\Tilde{T}, \vec{x} \big) - h_{\textsc{A}}\big(\Tilde{t}, \vec{x}\big)}{T} \times \frac{\omega_0}{\omega_0},
        \end{align*}
        }
    }
    \end{minipage}
    
    \begin{minipage}{0.9\linewidth}
    \fbox {
        \parbox{\linewidth}{
        \begin{align*}
            \hspace{4.2cm} &= \left[ \frac{h_{\textsc{A}}\big(\Tilde{t}+\Tilde{T}, \vec{x} \big) - h_{\textsc{A}}\big(\Tilde{t}, \vec{x}\big)}{\Tilde{T}} \right]\omega_0 , \\
            &\underset{\Tilde{T} \ll 1}{\approx} \left[ \frac{\partial h_{\textsc{A}}\big(\Tilde{t}\big)}{\partial \Tilde{t}}\right] \omega_0, \\
            &= \left[ \frac{\partial t}{\partial \Tilde{t}}\frac{\partial h_{\textsc{A}}\big(t\big)}{\partial t}\right] \omega_0, \\
            &= \left[ \frac{1}{\omega_0}\frac{\partial h_{\textsc{A}}\big(t\big)}{\partial t}\right] \omega_0, \\
            &= \frac{\partial h_{\textsc{A}}\big(t\big)}{\partial t} .
        \end{align*}
        The first line is true because again, $t$ and $T$ appear everywhere in our expression for $h_\textsc{A}$ in the combinations $\omega_0 t$ and $\omega_0 T$.  In the third line we have applied our         
        derivative definition approximation, namely that in the limit $\Tilde{T}\rightarrow 0$ the approximation becomes an exact equality.  And in the fourth line we apply the chain rule to write the derivative with respect to $t$.  So the crucial assumption for a monochromatic source is that we require $\Tilde{T}=\omega_0 T \ll 1$.  For our sources of interest, SMBHBs, the orbital period will be on the order of months to decades, so the orbital angular frequency $\omega_0$ will be on the order of $10^{-7}$ to $10^{-9}$ Hz.  And the pulsar rotation period will be on the order of milliseconds.  So $\omega_0 T \sim \mathcal{O}(10^{-10})$ or so, which indeed will be a very small quantity, so this approximation is safe. \\
        
        \underline{\textbf{Frequency Evolving Source}}\newline
        For a source whose orbit evolves with time, we see that from the functional form of the phase, frequency, and strain (Section~\ref{subsec: two frequency models - freq evo} and equation~\ref{eqn: amplitude of metric perturbation}), $t$ will always
        }
    }
    \end{minipage}
    
    \begin{minipage}{0.9\linewidth}
    \fbox {
        \parbox{\linewidth}{
        appear in the combination $\frac{t}{\Delta \tau_c}$ in our equations, which is again a dimensionless quantity.  For the same reasons as in the monochromatic cases, let:
        \begin{equation*}
            \Tilde{t} \equiv \frac{t}{\Delta \tau_c} \quad \implies \Tilde{T} = \frac{T}{\Delta \tau_c} .
        \end{equation*}
        Once again, we carry through the same series of steps:
        \begin{align*}
            \frac{h_{\textsc{A}}\big(t+T, \vec{x} \big) - h_{\textsc{A}}\big(t, \vec{x}\big)}{T} &= \frac{h_{\textsc{A}}\big(\Tilde{t}+\Tilde{T}, \vec{x} \big) - h_{\textsc{A}}\big(\Tilde{t}, \vec{x}\big)}{T} \times \frac{1/\Delta \tau_c}{1/\Delta \tau_c}, \\
            &= \left[ \frac{h_{\textsc{A}}\big(\Tilde{t}+\Tilde{T}, \vec{x} \big) - h_{\textsc{A}}\big(\Tilde{t}, \vec{x}\big)}{\Tilde{T}} \right]\frac{1}{\Delta \tau_c}, \\
            &\underset{\Tilde{T} \ll 1}{\approx} \left[ \frac{\partial h_{\textsc{A}}\big(\Tilde{t}\big)}{\partial \Tilde{t}}\right]\frac{1}{\Delta \tau_c}, \\
            &= \left[ \frac{\partial t}{\partial \Tilde{t}}\frac{\partial h_{\textsc{A}}\big(t\big)}{\partial t}\right]\frac{1}{\Delta \tau_c}, \\
            &= \left[ \Delta \tau_c\frac{\partial h_{\textsc{A}}\big(t\big)}{\partial t}\right] \frac{1}{\Delta \tau_c}, \\
            &= \frac{\partial h_{\textsc{A}}\big(t\big)}{\partial t}.
        \end{align*}
        So the crucial assumption for a source whose frequency does evolve over time is that we require $\Tilde{T}=\frac{T}{\Delta \tau_c} \ll 1$.  Again for our sources of interest, SMBHBs, the time of coalescence (measured from the fiducial time) will typically be on the order of kiloyears all of the way to gigayears.  Combined with a millisecond pulsar period this requirement again will normally be safely met.  This might only become a problem if we are considering an agressively evolving source, wherein we actually capture the moment of coalescence.  If this is the case then this assumption would break down.
        }
    }
    \end{minipage} \\ \\
    
    Next, let's consider the functional form of this integrand expression in equation~\ref{eqn: Delta T / T limit}.  From equations~\ref{eqn: h plus and cross},~\ref{eqn: amplitude of metric perturbation}, and~\ref{eqn: h as function of tret} the functional dependence of the metric perturbation on time looks like: $h_{\textsc{A}} = h_{\textsc{A}}\Bigg(h\Big(\omega\big(t_\mathrm{ret}(t,\vec{x})\big)\Big), \Theta\big(t_\mathrm{ret}(t,\vec{x})\big)\Bigg)$.  Furthermore, under assumption~\ref{as: flat static retarded time}, $\frac{\partial t_\mathrm{ret}}{\partial t} = 1$.  Therefore taking the time derivative of the metric perturbation, the chain rule yields:
    \begin{align}
        \frac{\partial h_\textsc{A}\left(t_\mathrm{ret}\left(t,\vec{x}\right)\right)}{\partial t} &= \frac{d h_{\textsc{A}}\left(t_\mathrm{ret}\right)}{d t_\mathrm{ret}} \frac{\partial t_\mathrm{ret}}{\partial t} =  \frac{d h_{\textsc{A}}\left(t_\mathrm{ret}\right)}{d t_\mathrm{ret}} = \frac{d h_{\textsc{A}}\left(h\left(\omega\left(t_\mathrm{ret}\right)\right), \Theta\left(t_\mathrm{ret}\right)\right)}{d t_\mathrm{ret}}, \nonumber \\
        &= \frac{\partial h_{\textsc{A}}(h,\Theta)}{\partial h} \frac{\partial h(\omega)}{\partial \omega} \frac{d \omega}{d t_\mathrm{ret}} + \frac{\partial h_{\textsc{A}}(h,\Theta)}{\partial \Theta} \frac{d \Theta}{d t_\mathrm{ret}} , \nonumber \\
        & =  \left(\frac{h_\textsc{A}}{h}\right) \left(\frac{2}{3}\frac{h}{\omega}\right) \dot{\omega}(t_\mathrm{ret}) + \bigg(2 h_\textsc{A}\left(h,  \Theta + \frac{\pi}{4} \right)\bigg)\omega(t_\mathrm{ret})  , \nonumber \\
        &=  \frac{2}{3} h_\textsc{A}(t_\mathrm{ret}) \frac{\dot{\omega}(t_\mathrm{ret})}{\omega(t_\mathrm{ret})} + 2 h_{\textsc{A}}\left(\omega(t_\mathrm{ret}), \Theta(t_\mathrm{ret}) + \frac{\pi}{4} \right)\omega(t_\mathrm{ret}) ,
    \label{eqn:partial h partial t chain rule}
    \end{align}
    where in the third line here we have used the notation described in the ``Notational Aside'' box.  The ``dot'' derivative $\dot{\omega}$ here is a derivative with respect to $t_\mathrm{ret}$.
    
    We see here by equation~\ref{eqn:partial h partial t chain rule} that if we carefully consider the order of the derivatives and then evaluate the expression along the photon path $\vec{x}=\vec{x}_0(t)$, we can say (again see the ``Notational Aside'' box below):
    \begin{align}
        \frac{\partial h_\textsc{A}\left(t_\mathrm{ret}\left(t,\vec{x}\right)\right)}{\partial t}\Bigg\rvert_{\vec{x}=\vec{x}_{0}(t)} &= \ \ \frac{d h_{\textsc{A}}\left(t_\mathrm{ret}\right)}{d t_\mathrm{ret}}\Bigg\rvert_{t_\mathrm{ret}=t_\mathrm{ret}^0} , \nonumber \\
        &= \ \ \frac{2}{3} h_\textsc{A}(t^0_\mathrm{ret}) \frac{\dot{\omega}(t^0_\mathrm{ret})}{\omega^0} + 2 h_{\textsc{A}}\left(\omega^0, \Theta^0 + \frac{\pi}{4} \right)\omega^0 , \nonumber  \\ 
        &= \ \ \frac{d h_\textsc{A}\left(t^0_\mathrm{ret}\right)}{d t^0_\mathrm{ret}} .
    \label{eqn:partial h partial t evaluated at x0}
    \end{align}
    In words, taking the time derivative of $h_\textsc{A}=h_\textsc{A}(t)$ and then evaluating the result at $\vec{x}=\vec{x}_0(t)$ is equivalent to just writing $h_\textsc{A}=h_\textsc{A}\left(t^0_\mathrm{ret}\right)$ and taking a derivative of that with respect to $t^0_\mathrm{ret}$.  This will end up being a very important result later in Section~\ref{sec: Plane-Wave Formalism}.
    
    With each of these established notations we can write the integrand of the fractional change in the pulsar period in a number of useful forms, all of which we will exploit later in our studies:
    \begin{align}
        \frac{\Delta T}{T}(t_\mathrm{obs}) &\approx \frac{1}{2}\hat{p}^i\hat{p}^j E^{\hat{r}\textsc{A}}_{ij} \int\limits^{t_u}_{t_l} \frac{\partial h_\textsc{A}\left(t_\mathrm{ret}\left(t,\vec{x}\right)\right)}{\partial t}\Bigg\rvert_{\vec{x}=\vec{x}_{0}(t)} dt , \nonumber \\
        &= \frac{1}{2}\hat{p}^i\hat{p}^j E^{\hat{r}\textsc{A}}_{ij} \int\limits^{t_u}_{t_l} \frac{d h_\textsc{A}\left(t^0_\mathrm{ret}\right)}{d t^0_\mathrm{ret}} dt , \nonumber \\
        &= \frac{1}{2}\hat{p}^i\hat{p}^j E^{\hat{r}\textsc{A}}_{ij} \int\limits^{t_u}_{t_l} \Bigg[ \frac{2}{3} h_\textsc{A}(t^0_\mathrm{ret}) \frac{\dot{\omega}(t^0_\mathrm{ret})}{\omega^0} + 2 h_\textsc{A}\bigg(\omega^0, \Theta^0 + \frac{\pi}{4}\bigg)\omega^0  \Bigg] dt , \label{eqn: Delta T / T} \\
        &\mathrm{where} \quad \begin{cases}
            \begin{tabular}{l l l}
                $t_u$ &= $t_\mathrm{em} + \frac{L}{c}$ & $\approx t_\mathrm{obs}$ , \\
                $t_l$ &= $t_\mathrm{em}$ & $\approx t_\mathrm{obs} - \frac{L}{c}$ .
            \end{tabular}
        \end{cases} \nonumber
    \end{align}
    
    \begin{minipage}{0.9\linewidth}
    \fbox {
        \parbox{\linewidth}{ \textbf{Notational Aside}\newline
        The following property of the sine and cosine functions will be quite useful to us in terms of keeping our notation ``clean'' and convenient throughout this paper:
        \begin{align}
            &\begin{cases}
                \sin\left(2\Theta\right) &= -\cos\Big(2\left(\Theta + \frac{\pi}{4}\right)\Big) , \\
                                         &= \quad \cos\Big(2\left(\Theta - \frac{\pi}{4}\right)\Big) , \\[10pt]
                \cos\left(2\Theta\right) &= \quad \sin\Big(2\left(\Theta + \frac{\pi}{4}\right)\Big) , \\
                                         &= -\sin\Big(2\left(\Theta -\frac{\pi}{4}\right)\Big) .
            \end{cases}
        \label{eqn: useful sine/cosine identities}
        \end{align}
        }
    }
    \end{minipage}
    
    \begin{minipage}{0.9\linewidth}
    \fbox {
        \parbox{\linewidth}{
        These relations allow us to write:
        \begin{equation}
            \frac{\partial h_\textsc{A}(h,\Theta)}{\partial \Theta} = \left\{\begin{tabular}{r l r} $2h\sin\left(2\Theta\right)$ & $= -2h\cos\left(2\Theta + \frac{\pi}{2}\right)$ & $(+)$ \\[5pt] $-2h\cos\left(2\Theta\right)$ & $= -2h\sin\left(2\Theta + \frac{\pi}{2}\right)$ & $(\times)$
            \end{tabular}\right\} = 2h_\textsc{A}\left(\Theta + \frac{\pi}{4}\right) .
        \label{eqn: useful partial h partial Theta}
        \end{equation}
        
        This notation is motivated in the sense that it will allow us to continually express our formulae in terms of the original $h_\textsc{A}$ expressions in equation~\ref{eqn: h plus and cross}, with the inclusion of an additional phase factor.\\
        
        Additionally, we define the following streamlined notation:
        \begin{equation}
            \begin{cases}
                t^0_\mathrm{ret} &\equiv t_\mathrm{ret}\big(\vec{x}=\vec{x}_0(t)\big) ,\\
                \omega^0 &\equiv \omega\left(t^0_\mathrm{ret}\right) ,\\
                \Theta^0 &\equiv \Theta\left(t^0_\mathrm{ret}\right) .
            \end{cases}
        \label{eqn: t^0_ret, w^0, Theta^0 notation definition}
        \end{equation}
        In words this is the retarded time, the orbital frequency, and the orbital phase as a function of time \textit{along} the photon's path.
        }
    }
    \end{minipage} \\ \\
    
    Here we emphasize the limits of integration.  The motion of the photon as it travels from the pulsar to the Earth can be referenced in terms of the emitted time or the observation time.  So far we have explained this derivation with reference to the emitted time since it is perhaps more natural.  But for more conceptual convenience going forward we will express the limits with respect to the observation time.  Again, since overall we are working on a solution which is good only to first order in the metric perturbation, we can use the zeroth order path of the photon to interchange the limits in this way (recall back to equation~\ref{eqn: tobs light pulse}, higher order corrections would introduce additional factors of the metric perturbation).  Additionally we make it explicitly clear that that the metric perturbation $h_\textsc{A}$ is still a function of \textit{both} the orbital frequency (which appears in the amplitude, equation~\ref{eqn: amplitude of metric perturbation}) and the phase.  We do this because other derivations in the literature seem to ignore the amplitude's dependence on the frequency or don't properly explain the assumptions they are making about the amplitude's frequency dependence.
    
    Once again, this expression is the fractional change in the period with time due to the presence of a gravitational wave.  As a reminder, conceptually we are integrating along the photon's path between the pulsar and the Earth.  This is achieved by integrating over the dummy time variable $t$ from the time the photon leaves the pulsar, $t_\mathrm{obs}-L/c$, to the time it arrives at the Earth and is observed, $t_\mathrm{obs}$.  Hence the result is a function of the observation time - i.e. it is the observed fractional change in the period of the pulsar at some given observation time.

%---------------------------------------------------------------------------------
%---------------------------------------------------------------------------------    
    \section{The Observable: The Pulsar Timing Residual} \label{sec:The Observable Residual}
    
    It may seem like the expression for $\frac{\Delta T}{T}$ in equation~\ref{eqn: Delta T / T} is a quantity which we could try to observe experimentally.  However, as we will see in the coming sections when we actually explicitly solve this equation, the resulting quantity will still be extraordinarily small.  For example, jumping ahead to Section~\ref{sec: Plane-Wave Formalism} and considering the solution of $\frac{\Delta T}{T}$ in the plane-wave formalism in equation~\ref{eqn: Delta T / T pw}, aside from the geometrical terms out front we see that the solution is effectively on the order of the metric perturbation itself, $\mathcal{O}\left(h\right)$.  If the period of our pulsar is something like $1 \mathrm{ms}$ and the magnitude of the metric perturbation produced by our SMBHB is something like $10^{-17}$, then we would be looking at trying to measure a fluctuation in the period of our otherwise extremely regular pulsar clock ticks on the order of $\Delta T \sim 10^{-20} \mathrm{s}$!  So this itself is not useful to us experimentally, but it does get us one step closer to an experimental quantity which we can measure.~\footnote{See the box at the end of this section for a discussion of the difference between measuring the pulsar periods vs. pulse time-of-arrivals.}
    
    Consider for a moment the integral of this quantity:
    \begin{equation*}
        \int \frac{\Delta T}{T}\left( t_\mathrm{obs}\right) dt_\mathrm{obs}.
    \end{equation*}
    What does this represent physically, and is it useful to us?  For concreteness let's imagine the following toy problem.  There is a clock which we are observing, which we expect has a true period of $T = 2 \mathrm{s}$, but some unknown physical effect is causing that clock period to shift by a constant amount such that $\Delta T \left(t_\mathrm{obs}\right) = 3 \mathrm{s}$.  We are going to study this clock with a reference clock of our own for some amount of ``observation time.''  Even if the observed clock doesn't ``tick'' at the same rate as ours, as clocks they should still be synchronized in that a given time interval on the observed clock should always correspond to the same interval of time on our reference clock.
    
    \begin{figure}
        \centering
        \includegraphics[width=0.8\linewidth]{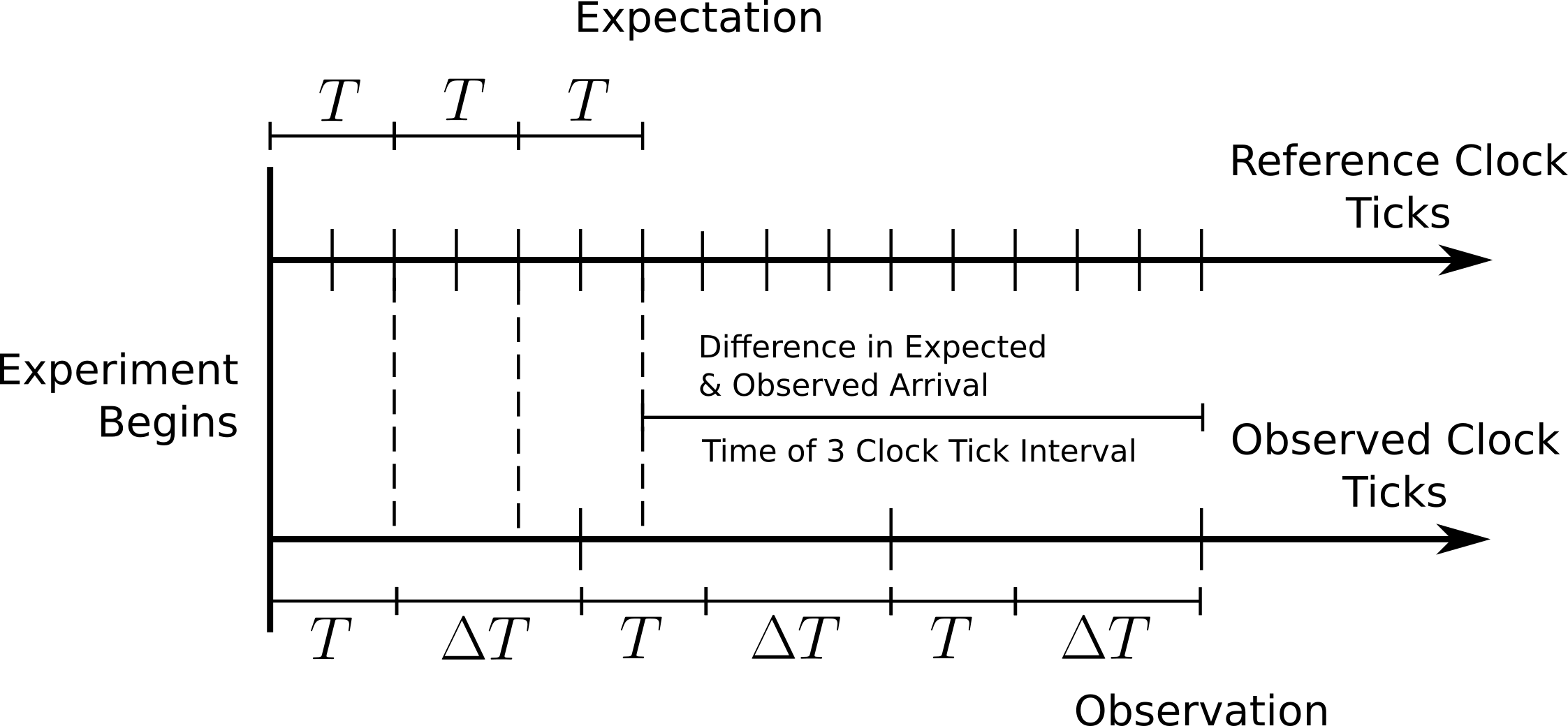}
        \caption[Toy Clock Experiment: Period Shift]{A toy experiment.  Imagine an observed clock with period $T=2 \mathrm{s}$, with some effect which causes its period to shift by $\Delta T = 3 \mathrm{s}$.  An interval of time lasting $6 \mathrm{s}$ our time we expect to be recorded by three periods on the observed clock if they are synchronized.  However, we observe three periods (``clock ticks'') lasting $15 \mathrm{s}$, i.e. $9 \mathrm{s}$ longer than our expectation.}
        \label{fig: example timing residual}
    \end{figure}
    Imagine we begin our experiment as shown in Figure~\ref{fig: example timing residual}.  Focus on our observation of the time interval lasting three observed clock ticks.  If we expect the period of this observed clock to be $T=2 \mathrm{s}$ our time, then we expect three periods of the observed clock to measure out $t=6 \mathrm{s}$.  However, in this example we observe this interval to last for $t = 15 \mathrm{s}$.  So the difference in the expected time interval and the observed time interval of the third clock tick is $9 \mathrm{s}$.  This is also what our expression tells us.  If we integrate the fractional change in the period of the observed clock over the interval of interest from our time of $t = 0 \mathrm{s}$ to $t = 6 \mathrm{s}$, then this gives us the difference in the expected and observed times for the three ticks:
    \begin{equation*}
        \int\limits^{6 \mathrm{s}}_0 \frac{\Delta T}{T}\left( t_\mathrm{obs}\right) dt_\mathrm{obs} = \int\limits^{6 \mathrm{s}}_0 \frac{3 \mathrm{s}}{2\mathrm{s}}dt_\mathrm{obs} = \frac{3}{2} (6 \mathrm{s}) = \quad 9 \mathrm{s} \quad = 15 \mathrm{s} - 6 \mathrm{s}.
    \end{equation*}
    In fact, this quantity tells us that for any interval of time that is observed and measured:
    \begin{equation*}
        \int\limits \frac{\Delta T}{T}\left( t_\mathrm{obs}\right) dt_\mathrm{obs} = \frac{3}{2} t ,
    \end{equation*}
    so if I expect to measure an interval of time $t$ on the observed clock, I will actually observe it to be different from my expectation by $\frac{3}{2}t$.  For example,
    \begin{align*}
        \int\limits^{2 \mathrm{s}}_0 \frac{\Delta T}{T}\left( t_\mathrm{obs}\right) dt_\mathrm{obs} = \quad 3 \mathrm{s} \quad = 5 \mathrm{s} - 2 \mathrm{s}, \\
        \int\limits^{4 \mathrm{s}}_0 \frac{\Delta T}{T}\left( t_\mathrm{obs}\right) dt_\mathrm{obs} = \quad 6 \mathrm{s} \quad = 10 \mathrm{s} - 4 \mathrm{s}, \\
        \int\limits^{4 \mathrm{s}}_{1 \mathrm{s}} \frac{\Delta T}{T}\left( t_\mathrm{obs}\right) dt_\mathrm{obs} = \quad 4.5 \mathrm{s} \quad = 7.5 \mathrm{s} - 3 \mathrm{s}.
    \end{align*}
    (In the last example here, if our clocks are in sync then the $3 \mathrm{s}$ time interval between $1 \mathrm{s}$ and $4 \mathrm{s}$ on our own reference clock would correspond to $1.5$ periods on our observed clock.  But here we observe $1.5$ periods to be $7.5 \mathrm{s}$ long, or $4.5 \mathrm{s}$ longer than expected.)
    
    This is a very simple toy model, but it illustrates another key concept.  Say at this point we come to some realization that the intrinsic period of the observed clock $T$ is actually $T = 5 \mathrm{s}$, and that there is no unknown physical effect causing that clock period to shift like we originally suspected.  Then if we change our model to reflect this, the observed clock's measurement of time will now always match our expectation (e.g. we will now expect three periods on the observed clock to correspond to a time interval of $15 \mathrm{s}$, which is what we indeed observe).  Moreover, since $\Delta T \left(t_\mathrm{obs}\right) = 0$ now our integral quantity will always be zero, i.e. there will be no difference between any measured time interval on our reference clock and on our observed clock.
    
    We also gain more insight on this quantity by writing it in the following way, using the definition that $\Delta T \equiv T_\mathrm{obs} - T$:
    \begin{equation*}
        \int \frac{\Delta T}{T}\left( t_\mathrm{obs}\right) dt_\mathrm{obs} = \int \frac{T_\mathrm{obs}\left( t_\mathrm{obs}\right) - T}{T} dt_\mathrm{obs} = \int \left[\frac{T_\mathrm{obs}\left( t_\mathrm{obs}\right)}{T} - 1\right] dt_\mathrm{obs} = \int \frac{T_\mathrm{obs}\left( t_\mathrm{obs}\right)}{T} dt_\mathrm{obs} - t .
    \end{equation*}
    The very final term $t$ on the right hand side is just the time interval we would expect to the observed clock to measure if it was synchronized with our reference clock.  We also know that our term on the left hand side is the difference between the observed and expected measured time intervals on the observed clock, so this means the quantity $\int \frac{T_\mathrm{obs}}{T}\left( t_\mathrm{obs}\right) dt_\mathrm{obs}$ is the actual observed time interval.  We can see this directly from our earlier examples too.  From the first example for the interval $t = 0 \mathrm{s}$ to $t = 6 \mathrm{s}$, the expected time is simply $t=6 \mathrm{s}$, and the observed time is $\int\limits^{6 \mathrm{s}}_0\frac{T_\mathrm{obs}}{T}\left( t_\mathrm{obs}\right) dt_\mathrm{obs} = \int\limits^{6 \mathrm{s}}_0\frac{5 \mathrm{s}}{2 \mathrm{s}}\left( t_\mathrm{obs}\right) dt_\mathrm{obs} = \frac{5}{2}(6 \mathrm{s}) = 15\mathrm{s}$, which is what we had earlier.
    
    Therefore, since this quantity is a residual of the observed and expected time intervals measured by our observed clock, we call it the ``timing residual,'' which can be written as:
    \begin{align}
        &\begin{array}{rll}
            \mathrm{Res}(t) &= \mathlarger{\int}\frac{\Delta T}{T}\left(t_\mathrm{obs}\right) dt_\mathrm{obs} &= \ \mathrm{Obs}(t) - \mathrm{Exp}(t) , \\
            \mathrm{Res}(t,t_0) &= \mathlarger{\int}\limits^t_{t_0} \frac{\Delta T}{T}\left(t_\mathrm{obs}\right) dt_\mathrm{obs} &= \ \mathrm{Obs}(t,t_0) - \mathrm{Exp}(t,t_0),
        \end{array} \label{eqn:timing residual} \\
        &\mathrm{where} \quad \begin{cases}
            \mathrm{Exp}(t,t_0) = t - t_0 , \\
            \mathrm{Obs}(t,t_0) = \mathlarger{\int}\limits^t_{t_0} \frac{T_\mathrm{obs} (t_\mathrm{obs})}{T} dt_\mathrm{obs} .
        \end{cases}
    \end{align}
    Here we denote $\mathrm{Obs}(t)$ as the ``observed time interval'' and $\mathrm{Exp}(t)$ as the ``expected time interval'' of the observed clock.  If we want to measure the residual time intervals beginning from a specific time, then we specify $t_0$ in the definite integral as our initial start time.  Therefore the residual at the time $t_0$ will always be zero because this is the start of the time interval and the residual measures the deviation away from this specific time.  Otherwise the indefinite integral gives us the general expression for how the residual changes over time, without ``zeroing'' it out at a specific time $t_0$.  If we want to know the difference in the observed time and the expected time for a specific interval, then we need to explicitly give those limits.  Another example of a $\Delta T$ which varies sinusoidally in time, as we would expect for a time shift caused by a gravitational wave, is given in Figure~\ref{fig: example timing residual 2}.
    \begin{figure}
    \centering
      \begin{subfigure}[t]{0.8\linewidth}
      \centering
        \includegraphics[width=1\linewidth]{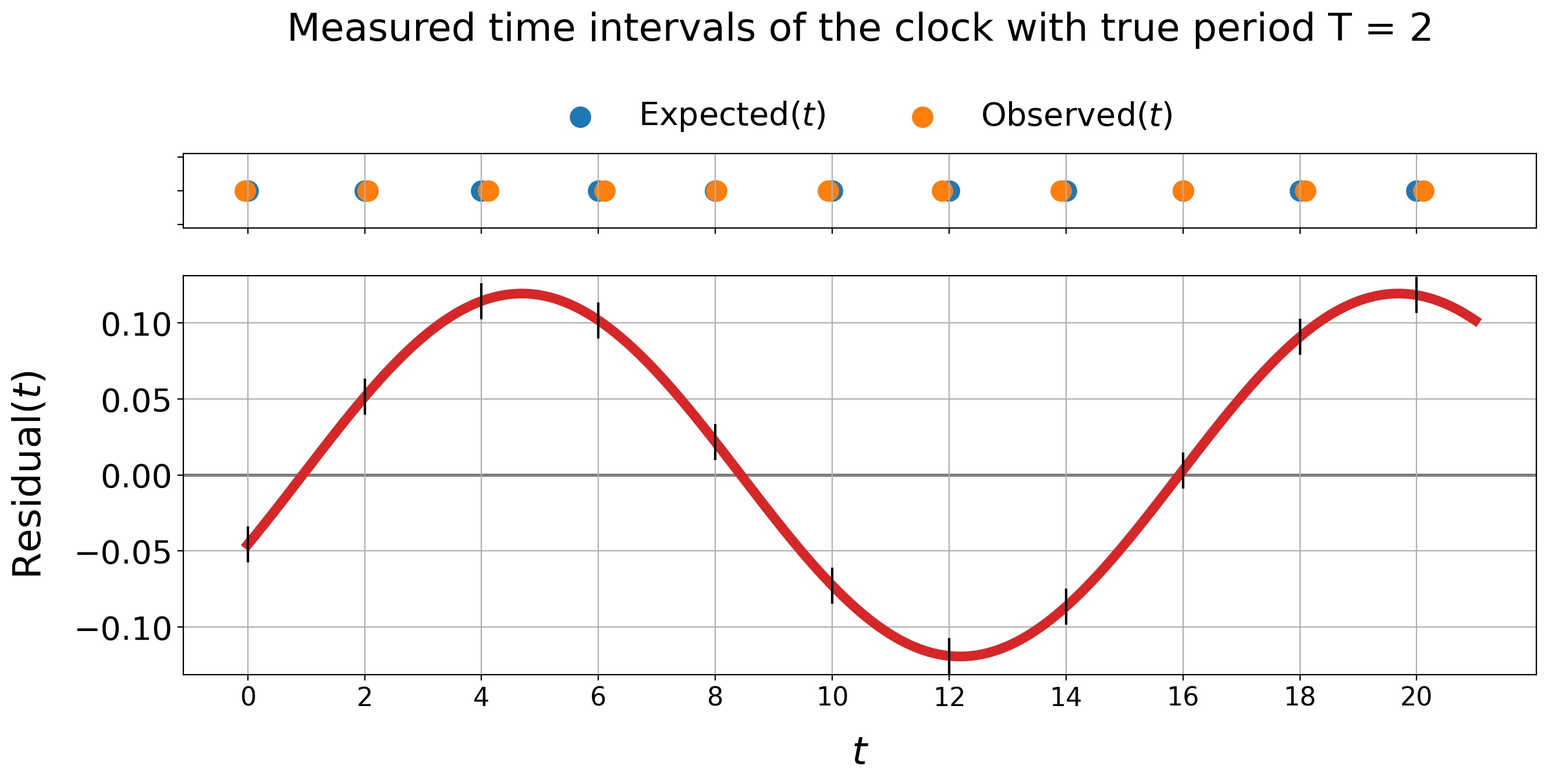}
        \caption{Example of $\mathrm{Res}(t)$.  In this case, the measured time intervals spend equal amounts of time being shorter than expected as they do longer than expected.}
        \label{fig:Res(t) example}
        \vspace{0.75cm}
        \includegraphics[width=1\linewidth]{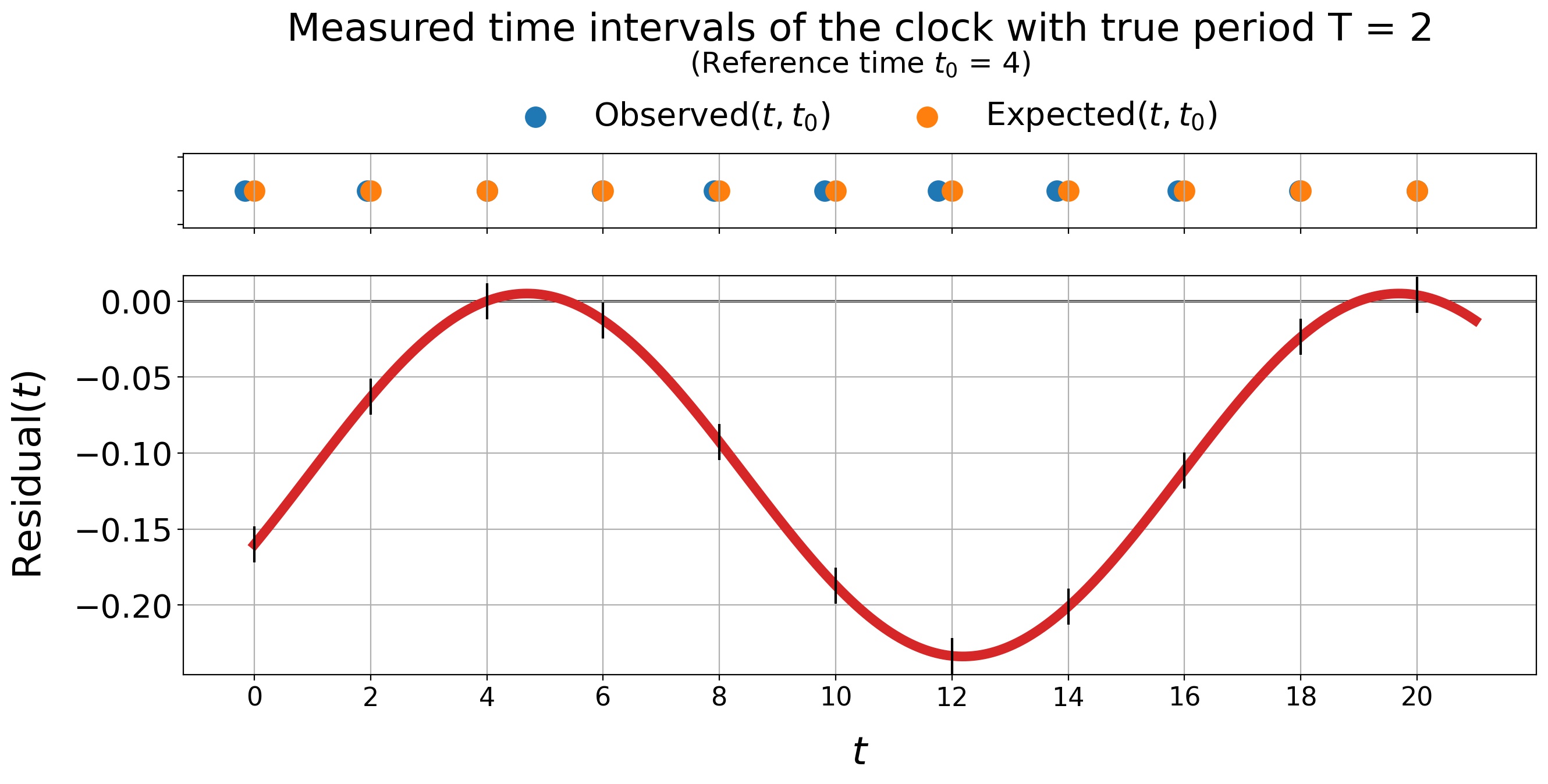}
        \caption{Example of $\mathrm{Res}(t,t_0)$, zeroed at $t = t_0 = 4$.  In this case, nearly all measured time intervals will appear to be shorter than we would expect them to be.}
        \label{fig:Res(t,t0) example}
      \end{subfigure}
    \caption[Toy Clock Experiment: Timing Residual 1]{Here is another toy example of our experiment, where the true period of the clock we are watching is $T=2$, and $\Delta T (t) = A \sin\left(2\pi t/T_\mathrm{gw} + \theta_0 \right)$, with the parameters set to $A=0.1$, $T_\mathrm{gw}=15$, and $\theta_0=3\pi /8$.  Both figures are the same data set, but show how the measured time intervals can change when we ``zero'' the expected and observed time intervals at a specific time $t_0$.  Above each timing residual is a visualization of the arrivals of the observed and the expected clock ticks, for the times $t$ indicated on the horizontal axis.}
    \label{fig: example timing residual 2}
    \end{figure}
    
    Physically, we can understand what is happening as follows.  Although the fluctuations of the pulsar's period $\Delta T$ are too small to measure directly, over time, after every pulsar period, these period shifts begin to stack.  If the values of $\Delta T > 0$, after each pulsar period the additional time added pushes the expected time-of arrival (commonly referred to as a ``TOA'') and the observed time-of-arrival of the subsequent ticks further and further apart.  If the values of $\Delta T < 0$, the pulsar periods are shorter than normal so the observed TOA of subsequent ticks comes earlier than expected.  So while an individual $\Delta T$ may not be measurable, this measurement allows us to exploit the cumulative effect of the pulsar's period shift over a long period of time.  If enough of these $\Delta T$'s stack on top of each other over a long enough sequence of clock ticks, then the resulting shift in time on the observed clock will become measurable.
    
    This brings us to another crucial insight into the feasibility of making this measurement possible.  Ideally, we want the period of our observed clock, which in this case is our pulsar, to be much smaller than the period of our induced $\Delta T$ shift, which is the period of our gravitational wave, that is:
    \begin{equation*}
    T_\mathrm{gw} > T_\mathrm{clock} \quad \longleftrightarrow \quad 1 > \omega_\mathrm{gw} T_\mathrm{clock} .
    \end{equation*}
    This is because longer gravitational wave periods mean more time for the $\Delta T$ shifts to constructively add on top of each other after each period of the clock.  This will give more time for the observed time intervals to increase with respect to the expected time intervals, and thus will make the timing residuals larger.  If the observed clock period and the gravitational wave period are close in magnitude to each other, then fewer clock cycles $\Delta T$ shifts will be able to compound, and the timing residual will be smaller.  In fact we see this in our earlier assumption for a monochromatic source, assumption~\ref{as: pulsar period timescale}.  So this is the key physical motivation in turning this theoretical effect (an otherwise immeasurable deviation to a pulsar's clock period) into one which can be measured experimentally.  An example of this is shown in Figure~\ref{fig: example timing residuals vs delta T}.
    \begin{figure}
        \centering
        \includegraphics[width=0.9\linewidth]{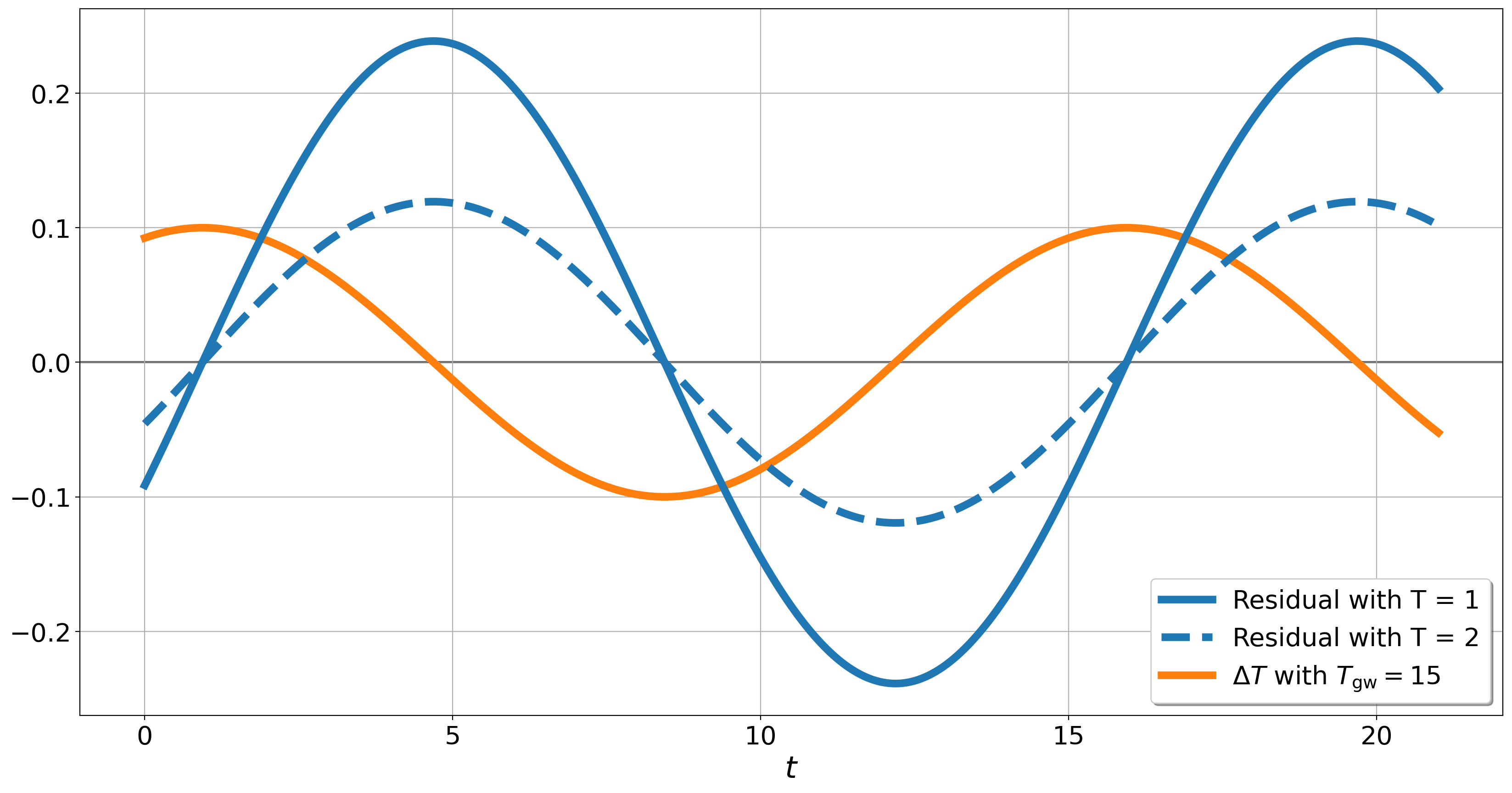}
        \caption[Toy Clock Experiment: Timing Residual 2]{Continuation of the example shown in Figure~\ref{fig: example timing residual 2}.  Here we plot two timing residuals from two clocks with different intrinsic periods.  What this aims to show is that ideally we need to use a clock (i.e. pulsar) which has a period $T$ much smaller than the period $T_\mathrm{gw}$ of the gravitational wave inducing the $\Delta T$ shift, so that more clock cycles will allow the $\Delta T$ shifts to \textit{constructively} add and make the timing residuals larger in magnitude, to the point where they can be measured (even if the $\Delta T$ shift itself is too small to be measurable).  The smaller the $T$ (compared to $T_\mathrm{gw}$), the greater the magnitude of the resulting residual.}
        \label{fig: example timing residuals vs delta T}
    \end{figure}
    
    Now we can design our experiment:
    %  EMAIL DESCRIPTION (2019) OF EXPERIMENT FROM STEPHEN TAYLOR:
        % — Point your telescope at a pulsar. Observe it for 20-30 minutes, recording many, many pulses. The pulses are phase-folded to get a very clean average pulse for that observation.
        % — Convolve the observed pulse with a long-timescale-averaged template pulse for that pulsar. Compute the phase offset of the observation from the template.
        % — Multiply the phase by the period to get a time offset for that observation.
        % — Add the time offset to the time-stamp of a distinctive feature of the pulse (either the peak or leading edge). That gives you the TOA for that pulse.
        % — Repeat ad infinitum for a time-series of TOAs.
    \begin{itemize}
        \item Observe a ``clock tick'' from a pulsar and record the time that the tick occurred, using some master reference clock here on Earth.
        \item Make subsequent observations like this of the pulsar's clock ticks over the course of years to decades.  This will be our $\mathrm{Obs}(t)$ data.
        \item For a highly regular pulsar, we will have a good idea of what the pulsar's clock period $T$ is from one observation.  However, if a gravitational wave is in fact adding some $\Delta T$ to it, we don't know how many cycles may have passed where period deviations have stacked on top of the true period.  Therefore we average a long timescale period for the pulsar from our collected data.  With this period we can then compute our $\mathrm{Exp}(t)$ data.
        \item Given the observed time interval data and the expected time interval data we can now compute the residual from their difference.
        \item If there is nothing affecting the period of our pulsar then experimentally we would expect the residual to simply look like white noise centered about $\mathrm{Res}=0$.  If, however, there is something affecting the period and our observations are sensitive enough to detect it, then we should see some deviation away from zero in the residual over time.  We can then compare the data to our model for a timing residual induced by a gravitational wave.  For example, this analysis could be done under a Bayesian framework in order to try and estimate the parameters of the source from the data itself.
        \item This entire process is iterative.  As we collect more data over longer observation timescales, we both increase our $\mathrm{Obs}(t)$ data and refine our $\mathrm{Exp}(t)$ data, which in turn refines our $\mathrm{Res}(t)$ data and hopefully leads to more accurate measurements of the gravitational wave parameters.
    \end{itemize} 
    
    \vspace{0.5cm}
    \begin{minipage}{0.9\linewidth}
    \fbox {
        \parbox{\linewidth}{ \textbf{Measuring Pulsar Periods vs. TOAs}~\citep{michael_TOAvsPeriod}\newline
        One potential for misconception when discussing the timing of a pulsar is how well astronomers can currently measure pulsar periods vs. the actual TOAs of pulsar pulses.  The precision to which we can measure the periods of many millisecond pulsars is truly astonishing - down to period uncertainties around the order of \textit{attoseconds}!  However, our current ability to measure TOAs (and hence, timing residuals) of pulsars is currently around the order of hundreds of nanoseconds~\citep{NG_11yr_data}. \\
        
        It may seem strange that despite our incredible ability to measure a pulsar's period, our uncertainties on the arrival times of the actual pulses from a pulsar are around nine orders of magnitude larger!  If we know a pulsar's period $T$ down to a $\sigma_T$ uncertainty of order attoseconds, then given our previous discussion, why is it perhaps not possible to measure the period shift $\Delta T$ induced by a gravitational wave directly?  Why is it we must try to measure the integrated effect of these period shifts, i.e. the timing residuals, instead?  The reason for this discrepancy lies in \textit{how} we make the two different measurements themselves. \\
        
        Because millisecond pulsars are spinning so rapidly, when we point our telescopes at them and record their light pulses, we observe many pulsar period cycles in a relatively short amount of time.  In order to measure the pulsar's \textit{period}, all we basically need to do is divide the total observation time by the number of pulsar pulses that occurred during that time.  And this type of observation is easily repeatable - every time we look back at that pulsar we can repeat this
        }
    }
    \end{minipage}
    
    \begin{minipage}{0.9\linewidth}
    \fbox {
        \parbox{\linewidth}{
        measurement, counting more pulses over more time.  Notice that by doing this we are not really trying to time the exact moment each of those pulses arrived.  All we care about is how many pulses arrived in the time we were observing the pulsar - we simply need to be able to see that there is a pulse, and then count that pulse.  Calculating the pulsar's period by dividing the observation window by the number of pulses effectively averages everything out, giving us the \textit{average} period.  Maybe there are gravitational waves causing those pulses individually to arrive slightly sooner (some $T-\Delta T$) or slightly later (some $T+\Delta T$) than we would expect, but this measurement is averaging that out and giving us the underlying ``true'' period to within some (insanely small) $\sigma_T$ uncertainty. \\
        
        But that said, if during any of those observations we want to know the \textit{exact time} every one of those pulsar light pulses arrived, then we need to have extremely good temporal resolution on all of the individual pulsar light pulses themselves.  Given the averaging procedure we may now know the true underlying period for that pulsar, but being able to say just how much sooner or how much later all of the pulsar pulses are appearing as compared to the expected theoretical TOA is more difficult.  This is why our current capabilities make this measurement many orders of magnitude larger than measuring simply the pulsar's underlying period.
        }
    }
    \end{minipage} \\

%---------------------------------------------------------------------------------
%---------------------------------------------------------------------------------    
    \section{Time Retardation}\label{sec: Time Retardation}
\begin{figure}
        \centering
        \includegraphics[width=0.8\linewidth]{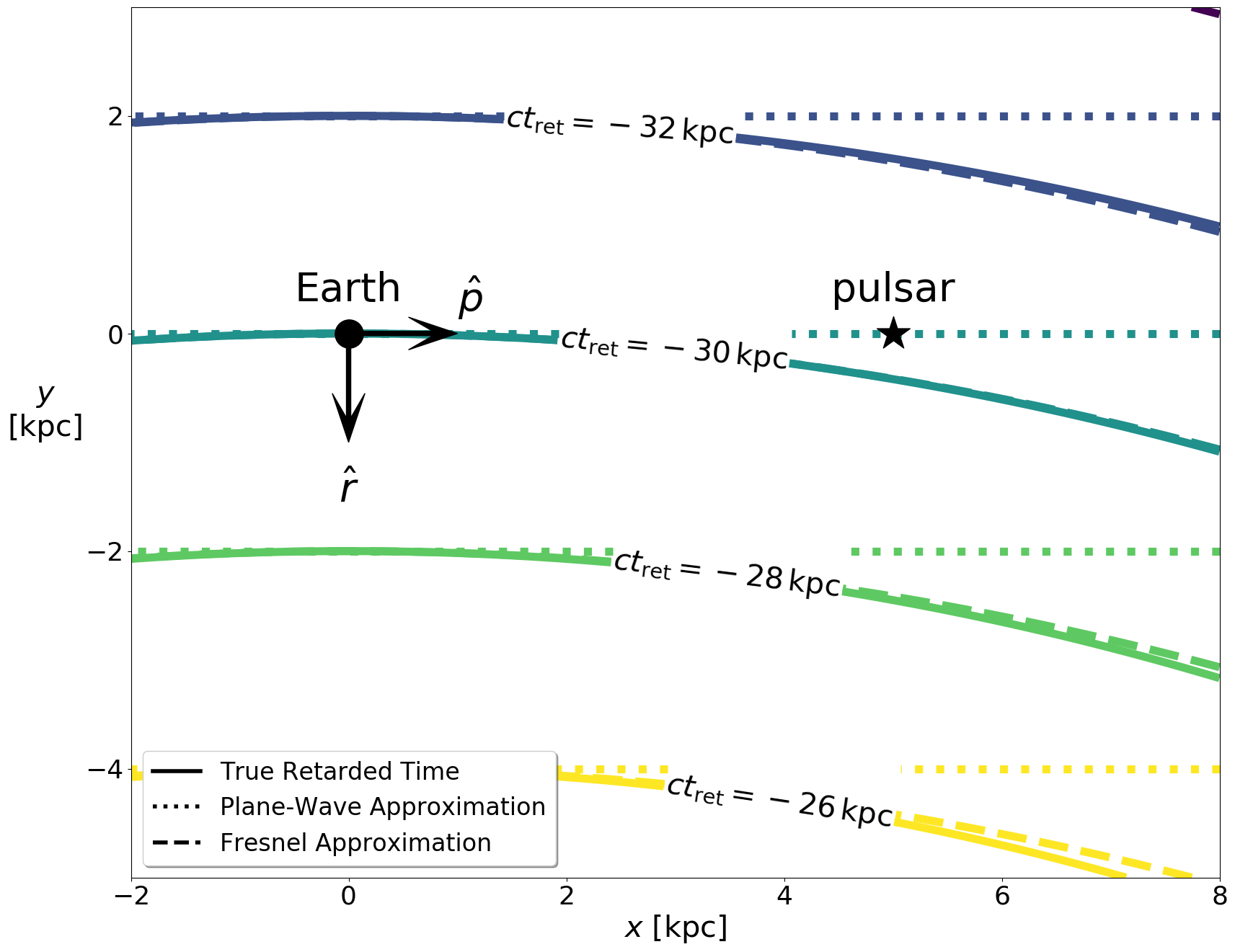}
        \caption[Time Retardation Approximations]{An example contour diagram of the retarded time, in the source-Earth-pulsar plane at fixed time $t=0$.  Labels on the contours give the value of $t_\mathrm{ret}$ multiplied by the speed of light.  Thus these values indicated for our fixed $t=0$ the distance each contour line is from the source.  The approximations indicate up to what term is included in the expansion in equation~\ref{eqn:retarded time}.  Since the lines show constant $t_\mathrm{ret}$, this traces the curvature of the wavefront.  For illustrative purposes that exaggerate and show the differences in the approximation regimes, we set the pulsar at a distance $L=5$ kpc and the source at $R=30$ kpc from the Earth.  A source with frequency $\omega_0=1$~nHz along with this pulsar would have a Fresnel number $F=27.3$ in this example.  As we can see here, the curvature of the wavefront means the wave arrives at the pulsar's location \textit{later} than what is predicted by the plane-wave approximation.}
        \label{fig: tret contours}
    \end{figure}

    For a static flat universe, the retarded time is:
    \begin{align}
        t_\mathrm{ret} = t - \frac{|\vec{x} - \vec{x}^{'}|}{c} &= t - \frac{1}{c}\sqrt{|\vec{x}^{'}|^2 - 2\vec{x}^{'}\cdot\vec{x} + |\vec{x}|^2} , \nonumber \\
        &= t - \frac{|\vec{x}^{'}|}{c}\sqrt{1 - 2\hat{x}^{'}\cdot\hat{x}\left(\frac{|\vec{x}|}{|\vec{x}^{'}|}\right) + \left(\frac{|\vec{x}|}{|\vec{x}^{'}|}\right)^2} , \nonumber \\
        &= t - \frac{|\vec{x}^{'}|}{c}\left[1 - \hat{x}^{'}\cdot\hat{x}\left(\frac{|\vec{x}|}{|\vec{x}^{'}|}\right) + \frac{1}{2}\left(1-\left(\hat{x}^{'}\cdot\hat{x}\right)^2\right)\left(\frac{|\vec{x}|}{|\vec{x}^{'}|}\right)^2 + \ldots \right], \nonumber \\
        &= t \ \ - \underbrace{\frac{|\vec{x}^{'}|}{c}}_{\mathrm{``Far} \ \mathrm{Field"}} \  + \ \ \underbrace{\left(\hat{x}^{'}\cdot\hat{x}\right)\frac{|\vec{x}|}{c}}_{\mathrm{``Plane-Wave"}} \ \ - \ \ \underbrace{\frac{1}{2}\left(1-\left(\hat{x}^{'}\cdot\hat{x}\right)^2\right)\frac{|\vec{x}|}{c}\frac{|\vec{x}|}{|\vec{x}^{'}|}}_{\mathrm{``Fresnel"}} \ \ + \ \ \ldots 
    \label{eqn:retarded time}
    \end{align}
    where $\vec{x}^{'}$ is the wave source's position and $\vec{x}$ is the field point of interest.  The expansion we have made here is good when the source is much further away than our field point.  It also gives us a way of categorizing different wave solution approximation regimes, which will ultimately be the crux of our work to come, so gaining insight into this is critical.  Keeping successive terms after the $t$ in this expression, the first term is our ``far field'' regime, the second term is the ``plane-wave'' regime, and the third term is the ``Fresnel'' regime.  One way to visually interpret the different approximation regimes is by plotting contours of constant $t_\mathrm{ret}$, as shown in Figure~\ref{fig: tret contours}.  The shape of the contours trace the shape of the wavefront as it propagates away from the source.  The Fresnel term gives the first order curvature of the physical wavefront.  Conceptually, it introduces an additional time delay to the arrival time of the wavefront predicted by the plane-wave approximation (see again Figure~\ref{fig: tret contours}).  Our goal is to understand when this curvature term/time delay will become significant in our timing residual model.
    
    As mentioned earlier, from equation~\ref{eqn:source basis vectors} we define $\vec{x}^{'} \equiv R \hat{r}$ as the source location.  In our problem we want to follow the path of a photon along the radial line connecting the Earth to the pulsar, i.e. along the direction $\hat{p}$.  Therefore, the field point of interest we will always be working with will be some vector that points along this line, so we can write $\vec{x} = x \hat{p}$ (where $x \geq 0$).  (Namely the path we are interested in is given by equation~\ref{eqn:zeroth order photon path} which between our limits of integration is always positive).  Hence we can write:
    \begin{align}
        t_\mathrm{ret}(t,x\hat{p}) &= t - \frac{1}{c}\sqrt{R^2 - 2\left(\hat{r}\cdot\hat{p}\right)Rx + x^2} , \nonumber\\
        &= t  - \frac{R}{c}  +  \left(\hat{r}\cdot\hat{p}\right)\frac{x}{c}  -  \frac{1}{2}\left(1-\left(\hat{r}\cdot\hat{p}\right)^2\right)\frac{x}{c}\frac{x}{R}  +  \ldots .
    \label{eqn: tret along Earth-pulsar baseline}
    \end{align}
    
    In pulsar timing, typically most sources are approximated as far enough from the Earth-pulsar baseline that gravitational waves arrive as plane-waves.  However, we are interested in seeing under what circumstances the effects of the wavefront's curvature from the Fresnel regime may begin to be important in the computation of the timing residual.  
    
    Namely, the plane-wave and Fresnel regimes become increasingly different away from the coordinate system origin, so let's consider the maximum size of our detector, which is the full Earth-pulsar distance $\vec{x} = L\hat{p}$:
    \begin{equation}
        t_\mathrm{ret}(t,L\hat{p}) = t \ \ - \ \underbrace{\frac{R}{c}}_{\mathrm{``Far} \ \mathrm{Field"}} \ + \ \ \underbrace{\left(\hat{r}\cdot\hat{p}\right)\frac{L}{c}}_{\mathrm{``Plane-Wave"}} \ \ - \ \ \underbrace{\frac{1}{2}\left(1-\left(\hat{r}\cdot\hat{p}\right)^2\right)\left(\frac{L}{c}\right)\left(\frac{L}{R}\right)}_{\mathrm{``Fresnel"}} \ \ + \ \ \ldots
    \label{eqn: tret along Earth-pulsar baseline at x=L}
    \end{equation}
    Now in general a wave solution behaves sinusoidally, oscillating with some angular frequency $\omega$.  For our order of magnitude estimate let's consider a monochromatic wave - then the wave solution evaluated at the retarded time will behave like:
    \begin{equation*}
        \sin(\omega_\mathrm{gw} t_\mathrm{ret}) \sim \sin\Bigg( \omega_\mathrm{gw} t - \omega_\mathrm{gw}\frac{R}{c} + \left(\hat{r}\cdot\hat{p}\right)\omega_\mathrm{gw}\frac{L}{c} - \frac{1}{2}\left(1-\left(\hat{r}\cdot\hat{p}\right)^2\right)\omega_\mathrm{gw}\left(\frac{L}{c}\right)\left(\frac{L}{R}\right)   \Bigg) .
    \end{equation*}
    
    Since sine functions are cyclic on the interval from $0 \rightarrow 2\pi$, the Fresnel term will become significant if it is an appreciable fraction of $2\pi$, so let's write:
    \begin{align*}
        2\pi &\sim \frac{1}{2}\left(1-\left(\hat{r}\cdot\hat{p}\right)^2\right)\omega_\mathrm{gw}\left(\frac{L}{c}\right)\left(\frac{L}{R}\right) , \\
        &\equiv \left(1-\left(\hat{r}\cdot\hat{p}\right)^2\right)\pi F ,
    \end{align*}
    or more roughly if we ignore all of the geometric terms and factors of order unity in this expression we can write this condition as roughly:
    \begin{align}
        F &\sim 1 , \label{eqn: Fresnel number condition} \\
        &\mathrm{where} \quad F \equiv \frac{L^2}{\lambda_\mathrm{gw} R} = \frac{\omega_0}{\pi}\frac{L}{c}\frac{L}{R} , \label{eqn: Fresnel number} \\
        &\mathrm{and} \quad \lambda_\mathrm{gw} \equiv \frac{\pi c}{\omega_0} . \label{eqn: gw wavelength}
    \end{align}
    
    Equation~\ref{eqn: gw wavelength} comes from the fact that the gravitational wave travels at the speed of light, so $\lambda_\mathrm{gw} = c T = 2\pi c / \omega_\mathrm{gw}$, and the gravitational wave frequency is twice the orbital frequency of the source, so $\omega_\mathrm{gw} = 2 \omega_0$.  (Again, we emphasize here this is a rough approximation to our original statement, which doesn't worry about factors of order unity).  The quantity $F$ is the familiar ``Fresnel number'' from diffraction theory, so when $F \sim \mathcal{O}(1)$, this suggests that our plane-wave approximation may be less accurate and using the Fresnel approximation would be required.  In fact, it is because of this connection to the Fresnel number that we have been calling this term in the retarded time expansion the ``Fresnel'' term.  Hence in the sections to come we will develop the formalisms for both regimes and study how the timing residuals in these regimes differ.  To see visually the dependence of $F$ on the orbital frequency, source distance, and pulsar distance parameters, see the contour plot in Figure~\ref{fig: DTauc & F contours}.
    
    Of course it is still important to remember that this condition equation~\ref{eqn: Fresnel number condition} should just be used to get a quick sense of which regime may be more dominant - the geometric terms which we ignored can become important, but we will discuss those in detail in the coming sections. \\
    
    \begin{minipage}{0.9\linewidth}
    \fbox {
        \parbox{\linewidth}{ \textbf{Natural Plane-Wave Limit}\newline
        Note there is an important limit here wherein higher order terms past the plane-wave term in equation~\ref{eqn: tret along Earth-pulsar baseline at x=L} naturally vanish.  Writing out additional terms in the Taylor expansion in equation~\ref{eqn: tret along Earth-pulsar baseline}, we would find that in addition to the ``Fresnel'' term, all higher terms contain a factor of $\left(1 -\left(\hat{r}\cdot\hat{p}\right)^2\right)$.  This means when $\hat{r}\cdot\hat{p} = \pm 1$, all terms except for the plane-wave term vanish - hence we refer to this as the ``natural plane-wave limit.'' \\
        
        Physically, this corresponds to the pulsar and the source being either ``aligned'' ($+1$) or ``anti-aligned'' ($-1$) with the Earth.  We can see why all effects of curvature of the wavefront would not be noticeable in this limit simply from considering Figure~\ref{fig: tret contours}.  All points along the vertical line that runs through the Earth marker (pointing along the arrow) ``see'' the incoming wavefront as a plane - there is no curvature along the wavefront in this precise geometrical alignment.  This can also be seen in Figure~\ref{fig: antenna pattern example}, where no induced change in the pulsar's period would occur in either of these alignments.
        }
    }
    \end{minipage} \\ \\
    
    Finally in our problem, in order to solve for the timing residual we will need to evaluate the retarded time along the path of the photon as it travels between the pulsar and the Earth, given by equation~\ref{eqn:zeroth order photon path}.  Inserting this into equation~\ref{eqn: tret along Earth-pulsar baseline} gives (using our notation in equation~\ref{eqn: t^0_ret, w^0, Theta^0 notation definition}):
    \begin{align}
        t_\mathrm{ret}^0 &= t - \frac{R}{c}\sqrt{1 - 2\hat{r}\cdot\hat{p}\left(\frac{ct_\mathrm{obs} -ct}{R}\right) + \left(\frac{ct_\mathrm{obs} -ct}{R}\right)^2} , \nonumber \\
        &= t \ - \frac{R}{c} \ +  \left(\hat{r}\cdot\hat{p}\right)\frac{ct_\mathrm{obs} -ct}{c} \ -  \frac{1}{2}\left(1-\left(\hat{r}\cdot\hat{p}\right)^2\right)\frac{\left(ct_\mathrm{obs} -ct\right)^2}{cR} \ + \ \ldots .
      \label{eqn:tret 0}
    \end{align}

%---------------------------------------------------------------------------------
%---------------------------------------------------------------------------------    
    \section{Four Pulsar Timing Regimes}\label{sec: four pulsar timing regimes}
    
    To date, pulsar timing literature has been primarily considered timing models derived assuming that the PTA receives plane-waves coming from distant sources.  They have also been derived assuming that the frequency of the gravitational wave is monochromatic, or that in some cases the frequency evolves slightly in the thousands to tens of thousands of years it takes light to travel from the pulsar to the Earth.  We can classify these as two separate model regimes, ``plane-wave monochromatic'' and ``plane-wave frequency evolution.''
    
\begin{figure}
        \centering
        \includegraphics[width=0.9\linewidth]{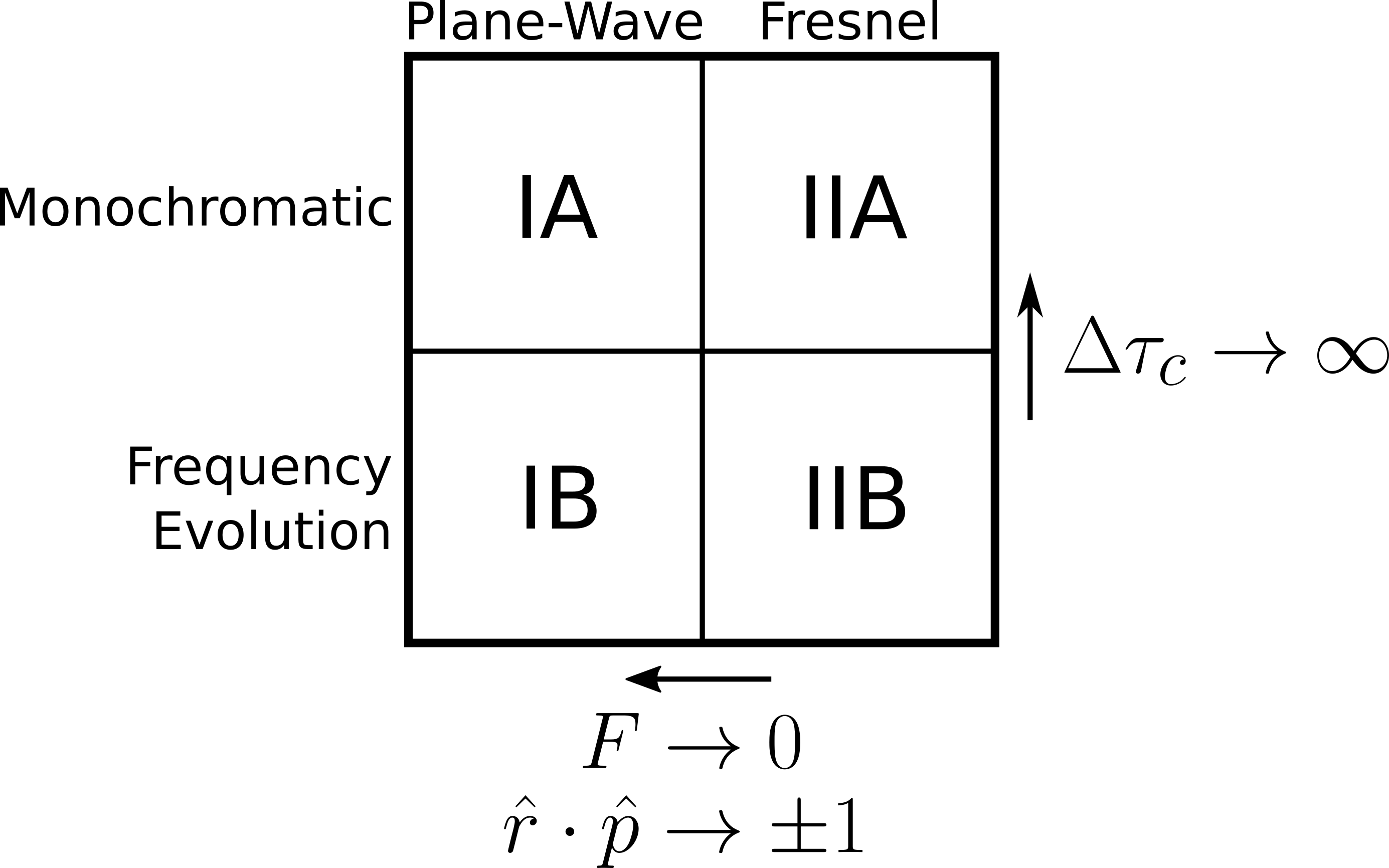}
        \caption[Gravitational Wave Timing Regimes Classification]{My classification of the gravitational wave regimes coming from sources of interest.  They increase in generality from left-to-right and top-to-bottom.  The models reduce from frequency evolution to monochromatism in the large coalescence time limit ($\Delta \tau_c \rightarrow \infty$), and from Fresnel to plane-wave in either the small Fresnel number limit ($F \rightarrow 0$) or the natural plane-wave limit ($\hat{r}\cdot\hat{p} \rightarrow \pm 1$).  The plane-wave regime has been well studied and used in pulsar timing literature, and in my work I add a new ``Fresnel'' regime.}
        \label{fig: 4 regimes}
    \end{figure}

    One of the primary goals of my work has been to further generalize these models by removing the plane-wave assumption.  In my work I add two new and complimentary timing regimes to the literature, represented in Figure~\ref{fig: 4 regimes}.  These regimes increase in generality from left-to-right and top-to-bottom.  The new regime I have labeled the ``Fresnel'' regime as we will show it becomes important for significant Fresnel numbers.  To this I have derived an analytic formula for the timing residuals in the Fresnel monochromatic regime, and I propose a physically motivated conjecture as to what the most general timing model should be, the Fresnel frequency evolution regime.  The Fresnel frequency evolution regime recovers all of the previously predicted results by each of the other three regimes in the appropriate limits.  In general, the frequency evolution regime reduces to the monochromatic regime in the large coalescence time limit $\Delta \tau_c \rightarrow \infty$ (see Section~\ref{sec: two frequency models}).  And the Fresnel regime reduces back to the plane-wave regime in either the natural plane-wave limit $\hat{r}\cdot\hat{p} = \pm 1$ or the small Fresnel number limit $F\rightarrow 0$.
    
    The key to this derivation of the Fresnel formalism is that I keep out to the Fresnel term in the expansion of the retarded time in equation~\ref{eqn:retarded time}.  Conceptually, we account for additional time delay of a curved gravitational wavefront (as compared to a plane wavefront) along the path of a photon traveling from the pulsar to the Earth (as the example in Figure~\ref{fig: tret contours} shows).  This will slightly alter the frequency and phase of the timing residual in the pulsar term, which later we will show can produce a measurable effect.
    
    It is important to realize that this derivation and formalism does \textit{not} correct for wavefront curvature effects in the amplitude or the geometrical orientation terms like the antenna patterns in Section~\ref{subsec:The Antenna Response}.  In reality, the pulsar-source distance is different from the Earth-source distance $R$.  This means that the amplitude of the metric perturbation (equation~\ref{eqn: amplitude of metric perturbation}) should be different for the photon when it leaves the pulsar versus when it arrives at the Earth.  Furthermore, since the pulsar is at a different location, the pulsar-source geometry will be different, and hence a photon leaving the pulsar will be affected by a metric perturbation with different orientation angles $\iota$ and $\psi$ (and $\theta_0$ since that is also an Euler angle as discussed in Section~\ref{sec: generalized source orientation & location}).  We could Taylor expand $|\vec{x}-\vec{x}^{'}|$ in the small parameter $\frac{L}{R}$ and look for $\mathcal{O}\left(\frac{L}{R}\right)$ corrections in the orientation geometry as well, but all of these corrections will directly affect the amplitude of the metric perturbation and timing residual.  For a typical $L = 1$kpc pulsar, even a source at $R = 1$Mpc would only give a correction of order $\mathcal{O}(0.001)$.  In general we just don't expect any of these geometrical curvature effects to produce any measurable effect in the timing residual.
    
    However, corrections of order $\mathcal{O}\left(\frac{L}{R}\right)$ in the frequency and phase of the metric perturbation and timing residual are different.  This is primarily because the phase of a sinusoidal function wraps around the interval $[0, 2\pi)$.  Additionally, as we discussed in Section~\ref{sec: Time Retardation}, corrections of order $\mathcal{O}\left(\frac{L}{R}\right)$ in the retarded time combine with factors $\omega_\mathrm{gw}\frac{L}{c}$ to produce net corrections of the order of the Fresnel number.  Overall, we find that these corrections on a $2\pi$-interval can produce measurable effects in the overall timing residual.  It is one of the main goals of my work to study when and how measureable these effects will be in this new Fresnel regime.

%---------------------------------------------------------------------------------
%---------------------------------------------------------------------------------    
    \section{Plane-Wave Formalism (I)}\label{sec: Plane-Wave Formalism}
    
    We start by keeping only terms in the plane-wave regime in equation~\ref{eqn:tret 0}:
    \begin{align}
        t_\mathrm{ret}^0 &\approx t - \frac{R}{c} + \left(\hat{r}\cdot\hat{p}\right)(t_\mathrm{obs}-t) , \nonumber \\
        &= \left[ -\frac{R}{c} + \left(\hat{r}\cdot\hat{p}\right)t_\mathrm{obs}\right] + \left(1 - \hat{r}\cdot\hat{p}\right)t .
    \label{eqn:tret 0 pw}
    \end{align}
    The integral in equation~\ref{eqn: Delta T / T} is over the dummy time variable $t$.  The explicit dependence on $t$ in that expression is in $t^0_\mathrm{ret}$, which in this section is given in equation~\ref{eqn:tret 0 pw}.  Again remember, conceptually our time variable $t$ is defined for the interval $t_\mathrm{obs}-L/c \leq t \leq t_\mathrm{obs}$ which gives us the path of the photon as it traverses between the pulsar and the Earth.
    
    As we found in equation~\ref{eqn: Delta T / T}, there are a number of useful ways to express the fractional change in the pulsar's period.  We could approach solving the integral using any of these forms.  However, in the plane-wave formalism, it turns out the solution is rather remarkably simple and elegant once we make the following critical observation.  Let's change the integration variable from $t \rightarrow t^0_\mathrm{ret}$ using equation~\ref{eqn:tret 0 pw}, which tells us $\frac{dt_\mathrm{ret}^0}{dt} = \left(1 - \hat{r}\cdot\hat{p}\right)$.  This one statement is actually the crux of the entire plane-wave formalism - because this derivative is just a constant, it can immediately be pulled out of the integral.  The result is:
    \begin{align}
        \frac{\Delta T}{T}(t_\mathrm{obs}) &= \frac{1}{2}\hat{p}^i\hat{p}^j E^{\hat{r}\textsc{A}}_{ij} \int\limits^{t_\mathrm{obs}}_{t_\mathrm{obs}-L/c} \frac{d h_\textsc{A}(t^0_\mathrm{ret})}{d t^0_\mathrm{ret}} dt , \nonumber \\
        &= \frac{1}{2}\hat{p}^i\hat{p}^j E^{\hat{r}\textsc{A}}_{ij} \int\limits^{t_\mathrm{obs}-R/c}_{t_\mathrm{obs}-R/c-\left(1 - \hat{r}\cdot\hat{p}\right)L/c} \frac{d h_\textsc{A}(t^0_\mathrm{ret})}{d t^0_\mathrm{ret}} \frac{dt^0_\mathrm{ret}}{\left(1 - \hat{r}\cdot\hat{p}\right)} ,\nonumber \\
        &= \frac{1}{2}\frac{\hat{p}^i\hat{p}^j E^{\hat{r}\textsc{A}}_{ij}}{\left(1 - \hat{r}\cdot\hat{p}\right)} \int\limits^{h_\textsc{A}\left(t_\mathrm{obs}-R/c\right)}_{h_\textsc{A}\left(t_\mathrm{obs}-R/c-\left(1 - \hat{r}\cdot\hat{p}\right)L/c\right)} dh_\textsc{A} , \nonumber \\
        &\equiv \frac{1}{2} F^{\textsc{A}} \int\limits^{h_\textsc{A}\left(t_\mathrm{obs}-R/c\right)}_{h_\textsc{A}\left(t_\mathrm{obs}-R/c-\left(1 - \hat{r}\cdot\hat{p}\right)L/c\right)} dh_\textsc{A} .
    \label{eqn: solving pw Delta T / T}
    \end{align}
    The function $F^\textsc{A}$ is named the antenna response function and is discussed in detail in the next section~\ref{subsec:The Antenna Response}.  In this final expression we see that all dependence on the actual functional form of the metric perturbation $h_\textsc{A}$ actually drops out of the integration.    The importance of this statement is that it means the fractional change in the pulsar's period does not depend on the intermediate motion of the photon between the pulsar and the Earth, only on those two endpoints of its journey - when the pulsar's light first left the pulsar, and when it finally arrived at the Earth.  These endpoints we denote with ``E'' and ``P'' subscripts.  So the timing residual in the plane-wave regime breaks apart into an ``Earth term'' and a ``pulsar term.''  Furthermore, this result is independent of whether or not the source is a monochromatic or a frequency evolving gravitational wave.  Therefore, completing the integration we have have the general result:
    \begin{align}
        \frac{\Delta T}{T}(t_\mathrm{obs}) &= \frac{1}{2} F^\textsc{A}\Bigg[ h_\textsc{A}\left(t_\mathrm{obs}-\frac{R}{c}\right) - h_\textsc{A}\left(t_\mathrm{obs}-\frac{R}{c}-\left(1 - \hat{r}\cdot\hat{p}\right)\frac{L}{c}\right) \Bigg] , \nonumber \\
        &\equiv \frac{1}{2} F^\textsc{A}\bigg[ h_\textsc{A}\Big(\omega^0_E,\Theta^0_E\Big) - h_\textsc{A}\Big(\omega^0_P,\Theta^0_P\Big) \bigg] , \label{eqn: Delta T / T pw}\\
        &\mathrm{where}\quad\begin{cases}
            \omega^0_E &\equiv \omega\Big(t_\mathrm{obs}-\frac{R}{c}\Big) , \\
            \omega^0_P &\equiv \omega\Big(t_\mathrm{obs}-\frac{R}{c}-\left(1 - \hat{r}\cdot\hat{p}\right)\frac{L}{c}\Big) , \\
            \Theta^0_E &\equiv \Theta\Big(t_\mathrm{obs}-\frac{R}{c}\Big) , \\
            \Theta^0_P &\equiv \Theta\Big(t_\mathrm{obs}-\frac{R}{c}-\left(1 - \hat{r}\cdot\hat{p}\right)\frac{L}{c}\Big) .
        \end{cases}\label{eqn: Delta T / T pw w^0_EP and Theta^0_EP}
    \end{align}
    
    Now we want to solve for the timing residual equation~\ref{eqn:timing residual} by integrating the fractional change in the pulsar's period equation~\ref{eqn: Delta T / T pw}:
    \begin{equation}
        \mathrm{Res}(t) = \frac{1}{2} F^\textsc{A} \int \bigg[ h_\textsc{A}\Big(\omega^0_E,\Theta^0_E\Big) - h_\textsc{A}\Big(\omega^0_P,\Theta^0_P\Big) \bigg] dt_\mathrm{obs}.
    \label{eqn: solving pw Res}
    \end{equation}
    At this point depending on which model we are investigating, there will be a couple of important differences (primarily conceptual) in the solution, so we will treat the two cases separately.

%---------------------------------------------------------------------------------
%---------------------------------------------------------------------------------    
        \subsection{The Plane-Wave Antenna Response} \label{subsec:The Antenna Response}
        The quantity $F^\textsc{A} \equiv \frac{\hat{p}^i\hat{p}^j E^{\hat{r}\textsc{A}}_{ij}}{\left(1-\hat{r}\cdot\hat{p}\right)}$ has an important physical interpretation.  Using the definitions $E^{\hat{r}\textsc{A}}_{ij}$ and $e^{\hat{r}\textsc{A}}_{ij}$ from equation~\ref{eqn: E plus/cross general} and~\ref{eqn:general polarization tensors}, and changing from Einstein notation to dot product notation we can write:
        \begin{align}
            \hspace{-1.8cm}\begin{cases}
                \begin{tabular}{l l l l}
                    $f^{+}$ &$\equiv \frac{\hat{p}^i\hat{p}^j e^{\hat{r}+}_{ij}}{\left(1-\hat{r}\cdot\hat{p}\right)}$ &$= \frac{\hat{p}^i\hat{p}^j\hat{\theta}_i\hat{\theta}_j - \hat{p}^i\hat{p}^j\hat{\phi}_i\hat{\phi}_j}{\left(1-\hat{r}\cdot\hat{p}\right)}$ &$= \frac{\left(\hat{p}\cdot\hat{\theta}\right)^2 - \left(\hat{p}\cdot\hat{\phi}\right)^2}{\left(1-\hat{r}\cdot\hat{p}\right)}$ , \\
                    $f^{\times}$ &$\equiv \frac{\hat{p}^i\hat{p}^j e^{\hat{r}\times}_{ij}}{\left(1-\hat{r}\cdot\hat{p}\right)}$ &$= \frac{\hat{p}^i\hat{p}^j\hat{\phi}_i\hat{\theta}_j + \hat{p}^i\hat{p}^j\hat{\theta}_i\hat{\phi}_j}{\left(1-\hat{r}\cdot\hat{p}\right)}$ &$= \frac{2\left(\hat{p}\cdot\hat{\theta}\right)\left(\hat{p}\cdot\hat{\phi}\right)}{\left(1-\hat{r}\cdot\hat{p}\right)}$ ,
                \end{tabular}
            \end{cases} \label{eqn: antenna f}
        \end{align}
        \vspace{-1cm}
        \begin{align}
            \begin{cases}
                \begin{tabular}{l l l}
                    $F^+$ &$\equiv \frac{\hat{p}^i\hat{p}^j E^{\hat{r}+}_{ij}}{\left(1-\hat{r}\cdot\hat{p}\right)}$ &$= \frac{1}{2}\left(1+\cos^2(\iota)\right) \left[ \cos(2\psi) f^+ + \sin(2\psi) f^\times \right]$ , \\
                    $F^\times$ &$\equiv \frac{\hat{p}^i\hat{p}^j E^{\hat{r}\times}_{ij}}{\left(1-\hat{r}\cdot\hat{p}\right)}$ &$= \cos(\iota) \left[ -\sin(2\psi) f^+ + \cos(2\psi) f^\times \right]$ .
                \end{tabular}
            \end{cases} \label{eqn: antenna F}
        \end{align}
        These are purely geometrical functions which depend on the source location ($\theta$ and $\phi$) and orientation ($\iota$ and $\psi$) in relation to the pulsar location ($\theta_p$ and $\phi_p$).  We can think of the Earth-pulsar baseline as an ``antenna'' for our experiment.  These functions are therefore an antenna response that then determines how sensitive $\frac{\Delta T}{T}$ in equation~\ref{eqn: Delta T / T pw} and $\mathrm{Res}(t)$ in equation~\ref{eqn: solving pw Res} will be to the geometrical alignment of our pulsar and source.  We derive these relations specifically from the plane-wave regime.  Later in Section~\ref{sec: Fresnel Formalism} we will find it is not as easy to write the antenna response cleanly in this way in the Fresnel formalism, but that asymptotically the Fresnel formalism also has the same antenna response as the plane-wave formalism.
        
\begin{figure}
            \centering
            \includegraphics[width=0.8\linewidth]{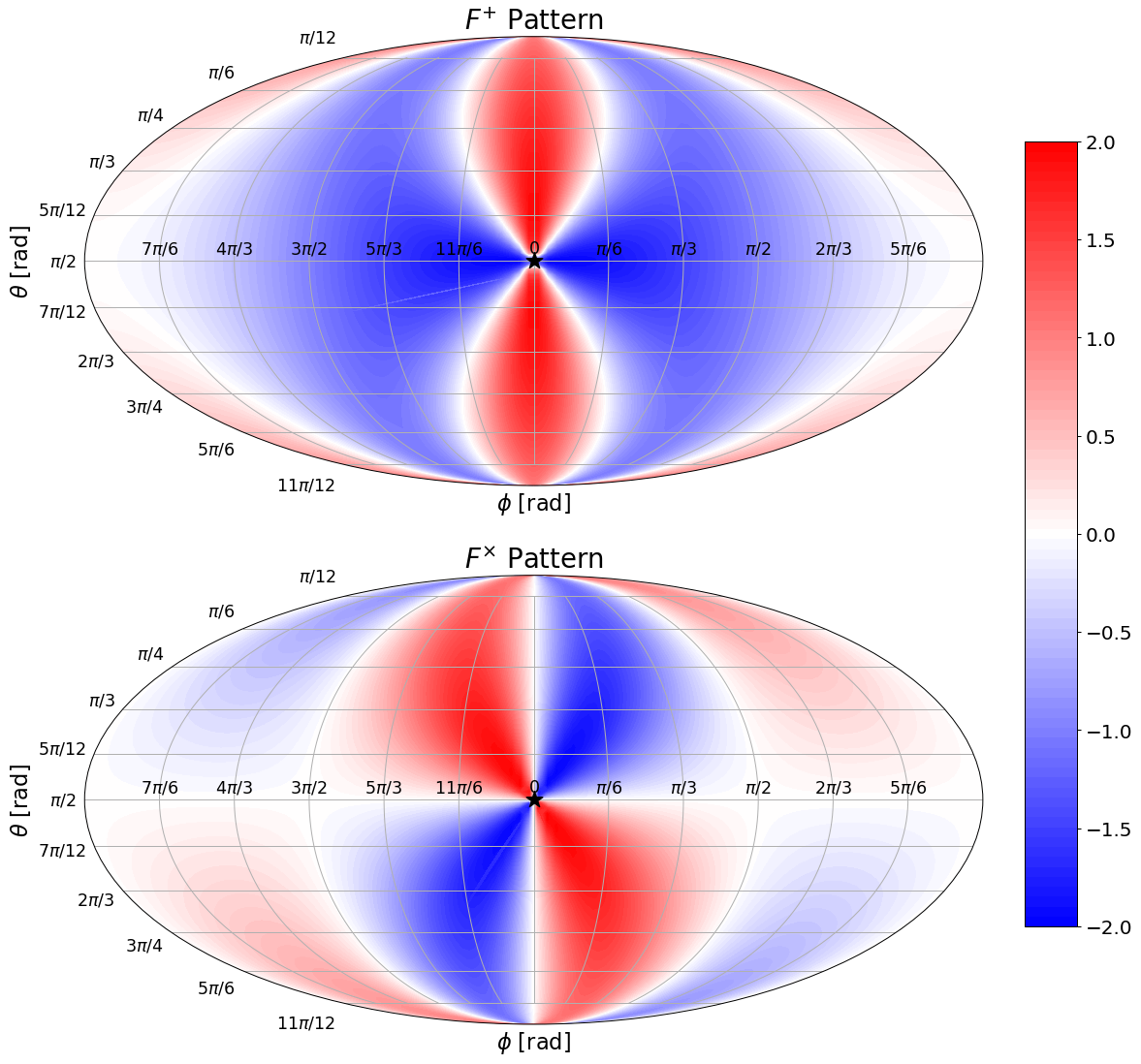}
            \caption[PTA Antenna Patterns Example]{An example of the antenna patterns as a function of sky position angles $(\theta,\phi)$ of a source with $\iota=\psi=0$ (face-on, not polarized).  Note this corresponds to example~\ref{ex: antenna response case F=f} in Section~\ref{subsec:The Antenna Response}, so really these plots are also of $f^+$ and $f^\times$.  The pulsar has been fixed at $\theta_p=\frac{\pi}{2}$ and $\phi_p = 0$, as indicated in the figure by the star.  The quadrupolar pattern of the gravitational wave's effect can clearly be seen in the antenna patterns.  Here red and blue simply indicate regions of the sky that would apply opposite signs to the fractional period shift and timing residual quantities.  So for example, a source in a blue region would appear ``blueshifted'' with respect to the \textit{same source} at the \textit{same time} which appears in the red region (which would appear ``redshifted,'' respectively).  Of course, in reality the gravitational waves cause this change to vary sinusoidally in time in the fractional period shift and the timing residual, so really here the blue and red colors indicate oppositely induced changes in these quantities at any given time.  The white regions show where the antenna would not be sensitive to detecting a source.}
            \label{fig: antenna pattern example}
        \end{figure}

        An example of this antenna response as a function of source sky position is shown in Figure~\ref{fig: antenna pattern example}.  It is important to note that the definitions of these response functions in equations~\ref{eqn: antenna f} and~\ref{eqn: antenna F} are mathematically undefined when the source and pulsar are perfectly aligned, i.e. when $\hat{r}\cdot\hat{p} = +1$.  (Note that if $\hat{r}\cdot\hat{p} = \pm 1$, then $\hat{p}\cdot\hat{\theta} \equiv \hat{p}\cdot\hat{\phi} \equiv 0$ because by definition of the basis vectors, $\hat{\theta}$ and $\hat{\phi}$ are orthogonal to $\hat{r}$).  However, the expressions for the fractional period shift and timing residual will remain well defined.  Looking first at the fractional period shift equation~\ref{eqn: Delta T / T pw}, we see that in the limit $\hat{r}\cdot\hat{p} \rightarrow 1$ the term in square brackets goes to zero.  It turns out that this goes to zero fast enough to kill the diverging antenna response, so the overall fractional period shift approaches zero as the source and pulsar become perfectly aligned.  We will see in the next section~\ref{subsec: Monochromatic Model (PW)} that the same thing happens to the timing residual in this limit.  And in the limit that the source and pulsar become anti-aligned, $\hat{r}\cdot\hat{p} = -1$, all of these expressions, including the antenna response, are well defined and also go to zero.
        
        Finally, here are a few important case examples:
        \begin{enumerate}
            \item If the source is both face-on $\iota=0$ and $\psi=0$, then $F^+ = f^+$ and $F^\times = f^\times$ (see Figure~\ref{fig: antenna pattern example}). \label{ex: antenna response case F=f}
            \item If the source is both face-off $\iota=\pi$ and $\psi=0$, then $F^+ = f^+$ and $F^\times = -f^\times$.  Physically, the only difference in this case as compared to case~\ref{ex: antenna response case F=f} is that the binary source now appears to orbit in the reverse direction.  The plus pattern is invariant to this orbital direction reversal, but the cross pattern becomes ``mirrored'' (considering Figure~\ref{fig: antenna pattern example} again, $F^+$ remains the same, but the blue and red areas in $F^\times$ are swapped).
            \item If $\iota = \pi/2$ (edge-on) and $\psi = 0$, then $F^+ = \frac{1}{2}f^+$ and $F^\times = 0$, meaning only the plus polarization of the gravitational wave is affecting the change in the pulsar's period.
            \item If $\iota = \pi/2$ (edge-on) and $\psi = \pi/4$, then $F^+ = \frac{1}{2}f^\times$ and $F^\times = 0$, meaning only the cross polarization of the gravitational wave is affecting the change in the pulsar's period.
        \end{enumerate}
         
%---------------------------------------------------------------------------------
%---------------------------------------------------------------------------------    
        \subsection{Plane-Wave, Monochromatic (IA)}\label{subsec: Monochromatic Model (PW)}
        
        For a monochromatic gravitational wave, the orbital frequency remains constant and hence the amplitude of the metric perturbation loses its time-dependence (equation~\ref{eqn: amplitude of metric perturbation}).  Explicitly in this model (see equations~\ref{eqn:monochrome freq} and~\ref{eqn:monochrome phase}) equation~\ref{eqn: Delta T / T pw w^0_EP and Theta^0_EP} becomes:
        \begin{equation}
            \underset{t_0 \ = \ -\frac{R}{c}}{\mathrm{choosing}}\quad\begin{cases}
                \omega^0_E &= \omega^0_P \quad \equiv \omega_0 , \\
                \Theta^0_E &= \theta_0 + \omega_0 t_\mathrm{obs} , \\
                \Theta^0_P &= \theta_0 +\omega_0\Big(t_\mathrm{obs}-\left(1-\hat{r}\cdot\hat{p}\right)\frac{L}{c}\Big) .
            \end{cases}
        \label{eqn: Delta T / T pw w^0_EP and Theta^0_EP mono}
        \end{equation}
        Here we have also chosen our model's fiducial time, primarily motivated by convenience here.  As described earlier in Section~\ref{sec: two frequency models}, let's pick $t_0 = -\frac{R}{c}$.  This will cancel out some factors of $\frac{R}{c}$ in these expressions, but also it will keep our formula here consistent with our later formulae, wherein we will make this same choice.
        
        We use one last change of variable furnished by equation~\ref{eqn: Delta T / T pw w^0_EP and Theta^0_EP mono} to write the timing residual equation~\ref{eqn: solving pw Res} as:  
        \begin{align}
            \mathrm{Res}(t) &= \frac{1}{2} F^\textsc{A} \left[ \int h_\textsc{A}\left(\omega_0,\Theta_E^0\right) dt_\mathrm{obs} \quad - \int h_\textsc{A}\left(\omega_0,\Theta_P^0\right) dt_\mathrm{obs} \right] , \nonumber \\
            &=\frac{1}{2} F^\textsc{A} \left[ \int h_\textsc{A}\left(\omega_0,\Theta_E^0\right) \frac{d\mathrm{\Theta^0_E}}{\omega_0} \quad - \int h_\textsc{A}\left(\omega_0,\Theta_P^0\right) \frac{d\mathrm{\Theta^0_P}}{\omega_0} \right] .   \label{eqn: solving pw mono Res}
        \end{align}
        Because the orbital frequency is constant it can be pulled out of the integral, and the remaining integration in equation~\ref{eqn: solving pw mono Res} then becomes the simple integration of the sine and cosine terms in eqn~\ref{eqn: h plus and cross}.  Using the helpful identities in equation~\ref{eqn: useful sine/cosine identities}, the final result can be expressed compactly as:
        \begin{align}
            \mathrm{Res}(t) &= \frac{F^\textsc{A}}{4\omega_0} \bigg[ h_\textsc{A}\Big(\Theta_E-\frac{\pi}{4}\Big) - h_\textsc{A}\Big(\Theta_P-\frac{\pi}{4}\Big) \bigg] , \label{eqn: Res(t) pw mono} \\
            &\mathrm{for}\quad \textsc{A} \in [+, \times], \nonumber \\
        \nonumber \\
            &\underset{\left(t_0 \ = \ -\frac{R}{c}\right)}{\mathrm{where}}\quad\quad \begin{cases}
                \Theta_E \equiv \Theta\Big(t-\frac{R}{c}\Big) &\equiv \theta_0 + \omega_0 t , \\
                \Theta_P \equiv \Theta\Big(t-\frac{R}{c}-\left(1-\hat{r}\cdot\hat{p}\right)\frac{L}{c}\Big) &\equiv \theta_0 + \omega_0\Big(t-\left(1-\hat{r}\cdot\hat{p}\right)\frac{L}{c}\Big) ,
                \end{cases}\label{eqn: Res(t) pw mono phase E and P} \\
            &\hspace{2cm}\begin{cases}
                h_{+} \equiv -h_0 \cos\big(2\Theta\big), \\
                h_{\times} \equiv -h_0 \sin\big(2\Theta\big),
            \end{cases} \label{eqn: h0 plus and cross} \\
            &\hspace{2cm}h_0 \equiv \frac{4(G\mathcal{M})^{5/3}}{c^4 R}\omega_0^{2/3} , \label{eqn: amplitude 0 of metric perturbation} \\
            &\hspace{2cm}\begin{cases}
                \begin{tabular}{l l l}
                    $f^{+}$ &$\equiv \frac{\hat{p}^i\hat{p}^j e^{\hat{r}+}_{ij}}{\left(1-\hat{r}\cdot\hat{p}\right)}$ &$= \frac{\left(\hat{p}\cdot\hat{\theta}\right)^2 - \left(\hat{p}\cdot\hat{\phi}\right)^2}{\left(1-\hat{r}\cdot\hat{p}\right)}$ , \\
                    $f^{\times}$ &$\equiv \frac{\hat{p}^i\hat{p}^j e^{\hat{r}\times}_{ij}}{\left(1-\hat{r}\cdot\hat{p}\right)}$ &$= \frac{2\left(\hat{p}\cdot\hat{\theta}\right)\left(\hat{p}\cdot\hat{\phi}\right)}{\left(1-\hat{r}\cdot\hat{p}\right)}$ ,
                \end{tabular}
            \end{cases} \tag{\ref{eqn: antenna f} r} \\
            &\hspace{2cm}\begin{cases}
                \begin{tabular}{l l l}
                    $F^+$ &$\equiv \frac{\hat{p}^i\hat{p}^j E^{\hat{r}+}_{ij}}{\left(1-\hat{r}\cdot\hat{p}\right)}$ &$= \frac{1}{2}\left(1+\cos^2(\iota)\right) \left[ \cos(2\psi) f^+ + \sin(2\psi) f^\times \right]$ , \\
                    $F^\times$ &$\equiv \frac{\hat{p}^i\hat{p}^j E^{\hat{r}\times}_{ij}}{\left(1-\hat{r}\cdot\hat{p}\right)}$ &$= \cos(\iota) \left[ -\sin(2\psi) f^+ + \cos(2\psi) f^\times \right]$ ,
                \end{tabular}
            \end{cases} \tag{\ref{eqn: antenna F} r} \\
            &\hspace{2cm}\begin{cases}
                \begin{tabular}{r l l r}
                    $\hat{r}$ &$= \big[\sin(\theta)\cos(\phi),$ &$\sin(\theta)\sin(\phi),$ &$\cos(\theta) \big]$, \\
                    $\hat{\theta}$ &$= \big[\cos(\theta)\cos(\phi),$ &$\cos(\theta)\sin(\phi),$ &$\sin(\theta) \big]$, \\
                    $\hat{\phi}$ &$= \big[-\sin(\phi),$ &$\cos(\phi),$ &$0 \big]$, \\
                    $\hat{p}$ &$= \big[\sin(\theta_p)\cos(\phi_p),$ &$\sin(\theta_p)\sin(\phi_p),$ &$\cos(\theta_p) \big]$.
                \end{tabular}
                \end{cases} \tag{\ref{eqn:source basis vectors}, \ref{eqn: p hat} r} 
        \end{align}
        In this model regime, looking at eqn~\ref{eqn: Res(t) pw mono phase E and P} we see that the Earth and pulsar terms have the same frequencies.  However, their difference is in the phase offset between the two.  Comparing both of these expressions, we see that the pulsar term $\Theta_p$ differs in phase by $-\omega_0\left(1-\hat{r}\cdot\hat{p}\right)\frac{L}{c}$ compared to the Earth term $\Theta_E$.
        
        As mentioned in Section~\ref{subsec:The Antenna Response}, the timing residual expression equation~\ref{eqn: Res(t) pw mono} remains well defined in both limits $\hat{r}\cdot\hat{p}\rightarrow \pm 1$ (which in Section~\ref{sec: Fresnel Formalism} we will refer to as the ``natural plane-wave'' limits), despite the fact that the antenna response is not well defined in the limit that $\hat{r}\cdot\hat{p}\rightarrow + 1$.  As the system becomes anti-aligned $\hat{r}\cdot\hat{p}\rightarrow -1$ the antenna patterns $F^\textsc{A}\rightarrow 0$ and this kills the entire timing residual.  And as the system becomes aligned $\hat{r}\cdot\hat{p}\rightarrow + 1$ the pulsar phase $\Theta_P \rightarrow \Theta_E$ and the difference in the metric perturbations at the Earth and pulsar goes to zero fast enough to also kill the entire timing residual.  Conceptually the photons ``surf'' the gravitational waves in this second case, and thus the photons is never ``feel'' the gravitational wave itself alter their own motion between the pulsar and Earth.
        
        Note that in terms of our notation, we are writing the residual as a function $t$, so we are changing notation in equation~\ref{eqn: Delta T / T pw w^0_EP and Theta^0_EP mono} from $t_\mathrm{obs} \rightarrow t$ through the integration.  It is also important to realize that in this regime the source chirp mass $\mathcal{M}$ and distance $R$ parameters are fully degenerate in the timing residual - they only appear in the combination $\frac{\mathcal{M}^{5/3}}{R}$ with each other in the amplitude terms here (see equation~\ref{eqn: amplitude 0 of metric perturbation}).  Therefore we would not be able to measure these parameters independently through actual observations (only this specific combination of the two).
        
        After applying a couple of trigonometric identities (namely, $\cos(x) - \cos(x+a) \equiv 2\sin\left(\frac{a}{2}\right)\sin\left(\frac{a}{2}+x\right)$, and similar for the sine identity), an alternative but equivalent way of writing equation~\ref{eqn: Res(t) pw mono} is:
        \begin{align}
            \mathrm{Res}(t) &= A_{c(\mathrm{IA})} \ F^\textsc{A} \ s_{E,\textsc{A}}  \qquad \mathrm{for}\quad \textsc{A} \in [+, \times], \label{eqn: Res(t) pw mono - alternative expression} \\[4pt]
            &\mathrm{where}\quad \begin{cases}
                A_{c(\mathrm{IA})} \equiv \frac{h_{c(\mathrm{IA})}}{4\omega_0} , \\
                h_{c(\mathrm{IA})} \equiv 2 h_0 \sin\left(\omega_0\left(1-\hat{r}\cdot\hat{p}\right)\frac{L}{c}\right) ,
                \end{cases}\label{eqn: Res(t) pw mono characteristic strain and amplitude} \\
            &\qquad\quad \ \ \begin{cases}    
                s_{E,+} \equiv \sin\left(2\Theta_E - \frac{\pi}{2}-\omega_0\left(1-\hat{r}\cdot\hat{p}\right)\frac{L}{c}\right) , \\
                s_{E,\times} \equiv -\cos\left(2\Theta_E - \frac{\pi}{2}-\omega_0\left(1-\hat{r}\cdot\hat{p}\right)\frac{L}{c}\right) .
                \end{cases}
        \end{align}    
        The benefit of this notation is that it allows us to define in equation~\ref{eqn: Res(t) pw mono characteristic strain and amplitude} a ``characteristic strain'' $h_{c(\mathrm{IA})}$ and ``characteristic timing residual amplitude'' $A_{c(\mathrm{IA})}$ for the IA regime.  These characteristic terms will be useful when comparing this model to the heuristic Fresnel monochromatic model IIA, in Section~\ref{subsec: Monochromatic Model (Fresnel)}.

%---------------------------------------------------------------------------------
%---------------------------------------------------------------------------------    
        \subsection{Plane-Wave, Frequency Evolution (IB)}\label{subsec: Frequency Evolution (PW)}
        
        Here we will make the following assumptions: \\
        
        \begin{minipage}{0.9\linewidth}
        \fbox {
            \parbox{\linewidth}{
            \textbf{Assumptions:} Frequency Evolving Timing Residuals
            \begin{enumerate}
                \item The observation time scale is much, much less than the time of coalescence (measured from the fiducial time): $t_\mathrm{obs} \ll \Delta \tau_c$ \label{as: t_obs << tau_c}
            \end{enumerate}
            }
        }
        \end{minipage}\\ \\
        
        The frequency evolution model of a gravitational wave described in Section~\ref{subsec: two frequency models - freq evo} is more realistic as it captures the changing frequency of the orbit of the binary.  Conceptually, as the photon first leaves the pulsar it will feel gravitational waves of some initial frequency which were emitted by the source.  As the pulsar-Earth baseline is on the order of kiloparsecs in distance, the photon will then take thousands of years before it arrives at the Earth.  During this time, the binary will lose energy due to the emission of gravitational waves and its orbit will collapse, increasing the overall frequency of the emitted gravitational waves with time.  So as the photon travels toward the Earth, it will continuously experience gravitational waves of higher frequencies.  However, crucially, as we found in equation~\ref{eqn: Delta T / T pw} and discussed, in the plane-wave formalism the result of the fractional change in the pulsar's period (and subsequently, in the observed timing residual), only depends on the frequency the photon experiences at the two endpoints, when it initially leaves the pulsar and when it finally arrives at Earth.  In the monochromatic model, these frequencies are the same since the orbit does not coalesce, but in this model the thousands of years travel can potentially alter the frequency felt by the photon at the endpoints in a significant way, as well will see here.  Therefore in this model, the Earth and pulsar terms will not only have different phases as we saw in the plane-wave monochromatic regime, but also different frequencies.
        
        Explicitly in this model (see equations~\ref{eqn:freq evolution freq} and~\ref{eqn:freq evolution phase}) equation~\ref{eqn: Delta T / T pw w^0_EP and Theta^0_EP} becomes:
        \begin{equation}
            \underset{t_0 \ = \ -\frac{R}{c}}{\mathrm{choosing}}\quad\begin{cases}
                \omega^0_E &= \omega_0\left[1 - \frac{t_\mathrm{obs}}{\Delta\tau_c}\right]^{-3/8} , \\
                \omega^0_P &= \omega_0\left[1 - \frac{t_\mathrm{obs}-\left(1-\hat{r}\cdot\hat{p}\right)\frac{L}{c}}{\Delta\tau_c}  \right]^{-3/8} , \\
                \Theta^0_E &= \theta_0 + \theta_c\left[ 1 - \left(1 - \frac{t_\mathrm{obs}}{\Delta\tau_c}\right)^{5/8} \right] , \\
                \Theta^0_P &= \theta_0 + \theta_c\left[ 1 - \left(1 - \frac{t_\mathrm{obs} - \left(1-\hat{r}\cdot\hat{p}\right)\frac{L}{c}}{\Delta\tau_c}\right)^{5/8} \right] .
            \end{cases}
        \label{eqn: Delta T / T pw w^0_EP and Theta^0_EP freq evo}
        \end{equation}
        Again here we have made the choice to set the fiducial time $t_0 = -\frac{R}{c}$.  This means that below in equation~\ref{eqn: Delta T / T pw w^0_EP and Theta^0_EP freq evo approx t0=-R/c} when we specify the value of $\omega_0$, we are directly specifying the value of the frequency of the gravitational wave affecting the Earth end of the photon's journey, which is arguably a conceptual convenience.
        
        Integrating the timing residual in equation~\ref{eqn: solving pw Res} would be difficult using the frequency evolution model since the orbital frequency is no longer a constant, but a function of $t_\mathrm{obs}$.  However, this is where we make our final important assumption.  Fortunately, while the frequency evolution model allows the frequency of the gravitational wave to evolve appreciably in the thousands of years it takes the photon to travel from the pulsar to the Earth, we still make the assumption that on the observation time scales of our experiment itself (which lasts on the order of years to decades) the frequency remains constant.  This means we can Taylor expand the functions in the small parameter $\frac{t_\mathrm{obs}}{\Delta \tau_c}$ equation~\ref{eqn: Delta T / T pw w^0_EP and Theta^0_EP freq evo}, the result being:
        \begin{equation}
            \underset{t_0 \ = \ -\frac{R}{c}}{\mathrm{choosing}}\quad\begin{cases}
                \begin{tabular}{ l l l }
                    $\omega^0_E$ & $\approx \omega_0$ & $\equiv \omega_{0E}$ , \\
                    $\omega^0_P$ & $\approx \omega_0\left[1 + \frac{\left(1-\hat{r}\cdot\hat{p}\right)\frac{L}{c}}{\Delta\tau_c}  \right]^{-3/8}$ & $\equiv \omega_{0P}$ , \\
                    $\Theta^0_E$ & $\approx \theta_0 + \omega_0t_\mathrm{obs}$ & $\equiv \theta_{0E} + \omega_{0E}t_\mathrm{obs}$ , \\
                    $\Theta^0_P$ & $\approx \left[\theta_0 + \theta_c\left(1-\left[1+ \frac{\left(1-\hat{r}\cdot\hat{p}\right)\frac{L}{c}}{\Delta\tau_c}\right]^{5/8}\right)\right] + \omega_{0P}t_\mathrm{obs}$ & $\equiv \theta_{0P} + \omega_{0P}t_\mathrm{obs}$ ,
                \end{tabular}
            \end{cases} \label{eqn: Delta T / T pw w^0_EP and Theta^0_EP freq evo approx t0=-R/c}
        \end{equation}
        In the orbital frequency we only keep to the zeroth order term because again we are assuming it doesn't evolve in time on our observation time scales, and in the phase we keep out to linear order in $\frac{t_\mathrm{obs}}{\Delta \tau_c}$ (note the $\Delta \tau_c$ cancels in the algebra leaving it to linear order in $t_\mathrm{obs}$ as we see here).
        
        Therefore, on \textit{observation time scales}, the gravitational wave affecting the endpoints of the pulsar-Earth baseline is approximately monochromatic (comparing equations~\ref{eqn: Delta T / T pw w^0_EP and Theta^0_EP mono} and~\ref{eqn: Delta T / T pw w^0_EP and Theta^0_EP freq evo approx t0=-R/c} we see they are of the same basic form, a constant frequency and a phase linearly increasing in time).  This allows us to employ the same change of variables that we did when solving equation~\ref{eqn: solving pw mono Res} in the monochromatic formalism.  The only difference here is that now when we consider equation~\ref{eqn: Delta T / T pw w^0_EP and Theta^0_EP freq evo approx t0=-R/c}, we see that there are \textit{two} frequencies, an initial ``pulsar'' frequency and an initial ``Earth'' frequency, corresponding to the respective frequencies of the gravitational waves which arrived at those two endpoints as the photon made its journey from the pulsar to the Earth (when we initially begin our experiment).  In the monochromatic model those two frequencies were of course the same, because in that model the frequency never changes, but here the difference between the two frequencies is the result of the slowly changing orbital frequency over the thousands of years travel for the photon between the pulsar and Earth.
        
        Therefore making the appropriate change of variables through equation~\ref{eqn: Delta T / T pw w^0_EP and Theta^0_EP freq evo approx t0=-R/c} (mirroring what we did earlier in the monochromatic case with equation~\ref{eqn: solving pw mono Res}), now we can write equation~\ref{eqn: solving pw Res} as:
        \begin{align}
            \mathrm{Res}(t) &\approx \frac{1}{2} F^\textsc{A} \left[ \int h_\textsc{A}\left(\omega_{0E},\Theta_E^0\right) dt_\mathrm{obs} \quad - \int h_\textsc{A}\left(\omega_{0P},\Theta_P^0\right) dt_\mathrm{obs} \right] , \nonumber \\
            &\approx \frac{1}{2} F^\textsc{A} \left[ \int h_\textsc{A}\left(\omega_{0E},\Theta_E^0\right) \frac{d\mathrm{\Theta^0_E}}{\omega_{0E}} \quad - \int h_\textsc{A}\left(\omega_{0P},\Theta_P^0\right) \frac{d\mathrm{\Theta^0_P}}{\omega_{0P}} \right] .   \label{eqn: solving pw freq evo Res}
        \end{align}
        Just like we found before, the orbital frequencies in each integral are constants so they can be pulled out of the integration.  And like before, the remaining integration in equation~\ref{eqn: solving pw freq evo Res} then becomes the simple integration of the sine and cosine terms in equation~\ref{eqn: h plus and cross}.  Using the helpful identities in equation~\ref{eqn: useful sine/cosine identities}, the final result can be expressed compactly as:
        \begin{align}
            \mathrm{Res}(t) &= \frac{F^\textsc{A}}{4} \left[ \frac{h_\textsc{A}\Big(\omega_{0E},\Theta_E-\frac{\pi}{4}\Big)}{\omega_{0E}} - \frac{h_\textsc{A}\Big(\omega_{0P},\Theta_P-\frac{\pi}{4}\Big)}{\omega_{0P}} \right] , \label{eqn: Res(t) pw freq evo} \\
            &\mathrm{for}\quad \textsc{A} \in [+, \times], \nonumber \\
        \nonumber \\
            &\underset{\left(t_0 \ = \ -\frac{R}{c}\right)}{\mathrm{where}}\quad\begin{cases}
                \omega_{0E} &\equiv \omega_0 , \\
                \omega_{0P} &\equiv \omega_0\left[1 + \frac{\left(1-\hat{r}\cdot\hat{p}\right)\frac{L}{c}}{\Delta\tau_c}  \right]^{-3/8} , \\
                \theta_{0E} &= \theta_0 , \\
                \theta_{0P} &= \theta_0 + \theta_c\left(1-\left[1+ \frac{\left(1-\hat{r}\cdot\hat{p}\right)\frac{L}{c}}{\Delta\tau_c}\right]^{5/8}\right) , \\
                \Theta_E & = \theta_{0E} + \omega_{0E}t , \\
                \Theta_P & = \theta_{0P} + \omega_{0P}t ,
                \end{cases} \label{eqn: Res(t) pw freq evo frequency phase E and P} \\
            &\hspace{1.6cm}\begin{cases}
                \Delta\tau_c \equiv \frac{5}{256}\left(\frac{c^3}{G\mathcal{M}}\right)^{5/3} \frac{1}{\omega_0^{8/3}} , \\
            \theta_c \equiv \frac{8}{5}\Delta\tau_c \omega_0 ,
            \end{cases} \tag{\ref{eqn: time to coalescence}, \ref{eqn: thetac} r} \\
            &\hspace{1.6cm}\begin{cases}
                h_{+} \equiv -h(\omega) \cos\big(2\Theta\big), \\
                h_{\times} \equiv -h(\omega) \sin\big(2\Theta\big),
            \end{cases} \tag{\ref{eqn: h plus and cross} r} \\
            &\hspace{1.6cm}h(\omega) \equiv \frac{4(G\mathcal{M})^{5/3}}{c^4 R}\omega^{2/3} , \tag{\ref{eqn: amplitude of metric perturbation} r}\\
            &\hspace{1.6cm}\begin{cases}
                \begin{tabular}{l l l}
                    $f^{+}$ &$\equiv \frac{\hat{p}^i\hat{p}^j e^{\hat{r}+}_{ij}}{\left(1-\hat{r}\cdot\hat{p}\right)}$ &$= \frac{\left(\hat{p}\cdot\hat{\theta}\right)^2 - \left(\hat{p}\cdot\hat{\phi}\right)^2}{\left(1-\hat{r}\cdot\hat{p}\right)}$ , \\
                    $f^{\times}$ &$\equiv \frac{\hat{p}^i\hat{p}^j e^{\hat{r}\times}_{ij}}{\left(1-\hat{r}\cdot\hat{p}\right)}$ &$= \frac{2\left(\hat{p}\cdot\hat{\theta}\right)\left(\hat{p}\cdot\hat{\phi}\right)}{\left(1-\hat{r}\cdot\hat{p}\right)}$ ,
                \end{tabular}
            \end{cases} \tag{\ref{eqn: antenna f} r} \\
            &\hspace{1.6cm}\begin{cases}
                \begin{tabular}{l l l}
                    $F^+$ &$\equiv \frac{\hat{p}^i\hat{p}^j E^{\hat{r}+}_{ij}}{\left(1-\hat{r}\cdot\hat{p}\right)}$ &$= \frac{1}{2}\left(1+\cos^2(\iota)\right) \left[ \cos(2\psi) f^+ + \sin(2\psi) f^\times \right]$ , \\
                    $F^\times$ &$\equiv \frac{\hat{p}^i\hat{p}^j E^{\hat{r}\times}_{ij}}{\left(1-\hat{r}\cdot\hat{p}\right)}$ &$= \cos(\iota) \left[ -\sin(2\psi) f^+ + \cos(2\psi) f^\times \right]$ ,
                \end{tabular}
            \end{cases} \tag{\ref{eqn: antenna F} r} \\
            &\hspace{1.6cm}\begin{cases}
                \begin{tabular}{r l l r}
                    $\hat{r}$ &$= \big[\sin(\theta)\cos(\phi),$ &$\sin(\theta)\sin(\phi),$ &$\cos(\theta) \big]$, \\
                    $\hat{\theta}$ &$= \big[\cos(\theta)\cos(\phi),$ &$\cos(\theta)\sin(\phi),$ &$\sin(\theta) \big]$, \\
                    $\hat{\phi}$ &$= \big[-\sin(\phi),$ &$\cos(\phi),$ &$0 \big]$, \\
                    $\hat{p}$ &$= \big[\sin(\theta_p)\cos(\phi_p),$ &$\sin(\theta_p)\sin(\phi_p),$ &$\cos(\theta_p) \big]$.
                \end{tabular}
                \end{cases} \tag{\ref{eqn:source basis vectors}, \ref{eqn: p hat} r} 
        \end{align}
        Once again note that in terms of our notation, we are writing the residual as a function $t$, so we change notation in equation~\ref{eqn: Delta T / T pw w^0_EP and Theta^0_EP mono} from $t_\mathrm{obs} \rightarrow t$ through the integration.  As explained earlier, from equation~\ref{eqn: Res(t) pw freq evo frequency phase E and P} we can now see explicitly how the Earth and pulsar terms differ in this regime in both phase and frequency.  Because the Earth and pulsar terms combine in equation~\ref{eqn: Res(t) pw freq evo}, in these cases where frequency evolution is non-negligible we expect this to physically produce ``beats'' in the timing residual data that we observe.
        
        Note, in this regime the degeneracy of the source chirp mass $\mathcal{M}$ and distance $R$ parameters is now broken since $\mathcal{M}$ now appears in the phase and frequency of the pulsar term as well as the amplitude, and not in the same combination with $R$ everywhere.  So in principle a loud enough observed source with significant frequency evolution should allow for independent measurements of both of these source parameters.
        
        Also, as a sanity check, we can see that this result equation~\ref{eqn: Res(t) pw freq evo} reduces to the monochromatic result equation~\ref{eqn: Res(t) pw mono} in the appropriate limit, as we would expect.  For the source to produce monochromatic gravitational waves, the time of coalescence must be infinite.  Therefore, we take the limit that $\Delta \tau_c \rightarrow \infty$, and see that this immediately implies $\omega_{0P} \rightarrow \omega_0$ as we would expect (now both endpoints, pulsar and Earth, feel the same frequency).  But it also implies $\theta_c \rightarrow \infty$, so we must be careful when taking the limit of $\theta_{0P}$ in order to do it correctly.  In the large $\Delta \tau_c$ limit we can Taylor expand $\theta_{0P}$ and note the following cancellations:
        \begin{align*}
            \theta_{0P} &= \theta_0 + \theta_c\left(1-\left[1+ \frac{\left(1-\hat{r}\cdot\hat{p}\right)\frac{L}{c}}{\Delta\tau_c}\right]^{5/8}\right) , \\
            &= \theta_0 + \frac{8}{5}\Delta\tau_c \omega_0\left(1-\left[1+ \frac{\left(1-\hat{r}\cdot\hat{p}\right)\frac{L}{c}}{\Delta\tau_c}\right]^{5/8}\right) , \\
            &\approx \theta_0 + \frac{8}{5}\Delta\tau_c \omega_0\left(1-\left[1+ \frac{5}{8}\frac{\left(1-\hat{r}\cdot\hat{p}\right)\frac{L}{c}}{\Delta\tau_c}\right]\right) , \\
            &= \theta_0 - \omega_0\left(1-\hat{r}\cdot\hat{p}\right)\frac{L}{c} .
        \end{align*}
        Therefore in this limit we have $\Theta_P \rightarrow \theta_0 + \omega_0\Big(t - \left(1-\hat{r}\cdot\hat{p}\right)\frac{L}{c}\Big)$.  This along with the expression for $\Theta_E$ now matches equation~\ref{eqn: Res(t) pw mono phase E and P}.

\begin{figure}
            \centering
            \includegraphics[width=1\linewidth]{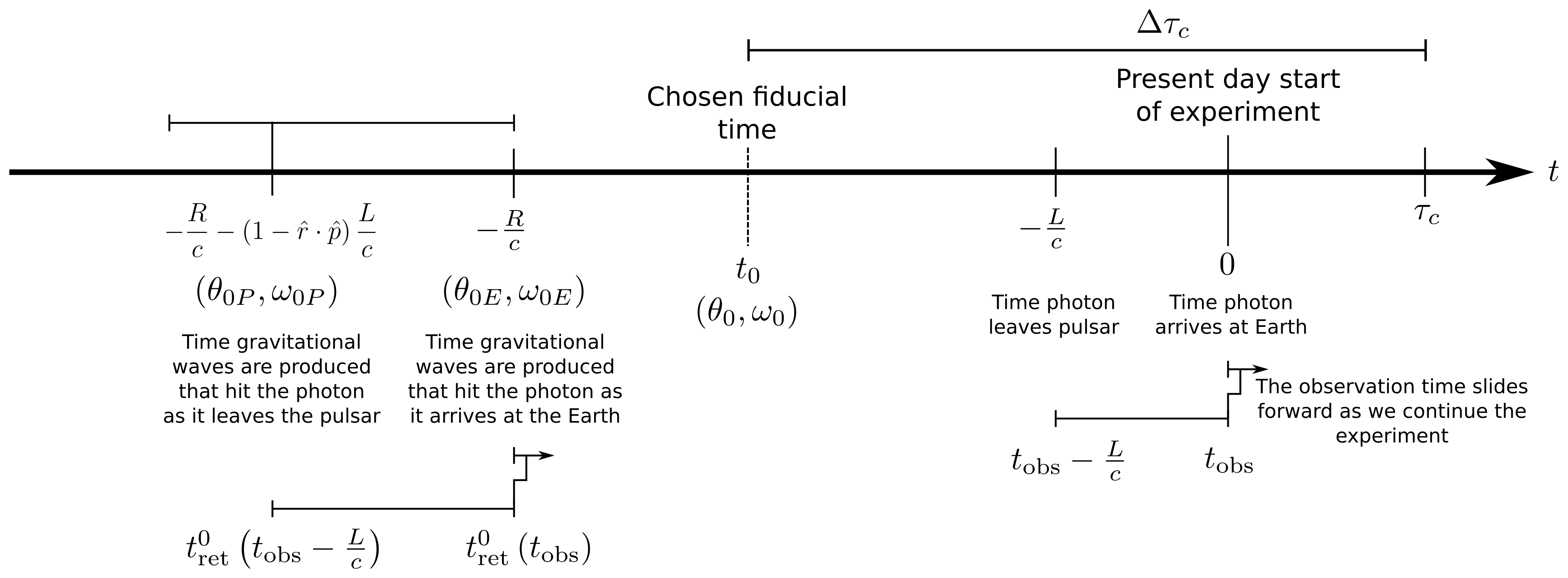}
            \caption[Time Scales Chronology]{Here we give a conceptual chronological summary of the important times and scales involved in the calculation of our timing residuals.  This also shows the terminology we are using for the various measured quantities of the source's orbital phase and frequency, with the time they correspond to.  The fiducial time can be chosen at any time, and in this work we choose it to be at $t_0 = -\frac{R}{c}$, which means that $\theta_{0E} = \theta_0$ and $\omega_{0E} = \omega_0$.  The coalescence time $\tau_c$ is measured from present day, whereas $\Delta \tau_c$ is measured from the fiducial time.  We begin the experiment at $t=0$, and at this time our measurement of the pulsar timing residual depends on the initial observation time $t_\mathrm{obs}$ and the time the photon first left the pulsar, $t_\mathrm{obs} - \frac{L}{c}$.  Furthermore we show how these times are mapped to the retarded times corresponding to when the gravitational waves were produced, $t^0_\mathrm{ret}\left(t_\mathrm{obs}\right)$ and $t^0_\mathrm{ret}\left(t_\mathrm{obs}-\frac{L}{c}\right)$.  These time scales are continually integrated as we observe the pulsar for the duration of our experiment, so we can think of them as ``sliders'' of fixed width that move forward in time.  The time at which the gravitational waves are produced that hit the photon as it leaves the pulsar depends on the Earth-pulsar-source geometry (indicated by the $\hat{r}\cdot\hat{p}$ term).  It will fall somewhere within the range $t=-\frac{R}{c}$ (if aligned) to $t=-\frac{R}{c}-2\frac{L}{c}$ (if anti-aligned).  Finally we see the retardation delay caused by a plane-wave gravitational wavefront.  In the case of the Fresnel formalism, additional curvature terms would factor in here.  Namely we see in our heuristic model in Section~\ref{subsec: Monochromatic Model (Fresnel)} that the time the gravitational waves produced which hit the photon as it leaves the pulsar would become approximately $t=-\frac{R}{c}-\left(1-\hat{r}\cdot\hat{p}\right)\frac{L}{c} -\frac{1}{2}\left(1-\left(\hat{r}\cdot\hat{p}\right)^2\right)\frac{L}{c}\frac{L}{R}$. }
            \label{fig: PTA chronology PW}
        \end{figure}

        A chronological summary of events and times is shown in Figure~\ref{fig: PTA chronology PW}.  While this figure is specifically for the plane-wave formalism, it also holds in general for the Fresnel formalism, which would only slightly alter the time the gravitational waves are produced that hit the photon as it leaves the pulsar.

%---------------------------------------------------------------------------------
%---------------------------------------------------------------------------------    
    \section{Fresnel Formalism (II)}\label{sec: Fresnel Formalism}
    For the work below we will make the following assumptions: \\
    
    \begin{minipage}{0.9\linewidth}
    \fbox {
        \parbox{\linewidth}{
        \textbf{Assumptions:} Fresnel Formalism
        \begin{enumerate}
            \item We are not working in a plane-wave limit.  The Fresnel formalism will reduce to the plane-wave formalism in the following limits: \label{as: fresnel pw limits}
            \begin{enumerate}
                \item $F \rightarrow 0$.  As the Fresnel number between the pulsar and source decreases, the Fresnel effects become more and more negligible.
                \item $\hat{r}\cdot\hat{p} \rightarrow \pm 1$.  These are the natural plane-wave limits.  In this case the source becomes perfectly aligned/anti-aligned with the pulsar, which means there are geometrically no curvature effects possible in this precise alignment. \label{as: fresnel no alignment}
            \end{enumerate}
            \item Corrections in the geometrical orientation components which affect the amplitude of the timing residual, namely the antenna patterns, which are due to the curvature of a gravitational wave are negligible.  Only corrections which directly affect the phase and frequency of these quantities are significant. \label{as: no geometry curvature}
        \end{enumerate}
        }
    }
    \end{minipage}\\ \\
    
    The key to deriving this model is that we keep out to the Fresnel term in the expansion of the retarded time in equation~\ref{eqn:retarded time}.  Conceptually, we account for additional time delay of a curved gravitational wavefront (as compared to a plane wavefront) along the path of a photon traveling from the pulsar to the Earth (as the example in Figure~\ref{fig: tret contours} shows).  This will slightly alter the frequency and phase of the timing residual in the pulsar term, which we will show later can produce measurable effects.  It is important to realize that this derivation and formalism does not correct for wavefront curvature effects in the geometrical orientation terms like the antenna patterns, as stated in assumption~\ref{as: no geometry curvature}.  We expect those corrections to introduce curvature effects of order $\mathcal{O}\left(\frac{L}{R}\right)$ in the \textit{amplitudes} of the terms that appear in the timing residuals, which we do not expect to be measurable.  Curvature corrections to the phase and frequency terms we will show later can be measured.
    
    Now we return to equation~\ref{eqn:tret 0} and this time we keep out to the Fresnel regime and drop all higher order terms after that.  The result can be written in the form:
    \begin{align}
        t^0_\mathrm{ret} &\approx \left(t_\mathrm{obs}-\frac{R}{c}\right) - \frac{R}{c}\left(1-\hat{r}\cdot\hat{p}\right)u - \frac{1}{2}\frac{R}{c}\left(1-\left(\hat{r}\cdot\hat{p}\right)^2\right)u^2 , \nonumber\\
        &\mathrm{where}\quad u \equiv \frac{ct_\mathrm{obs}-ct}{R} .
    \label{eqn:tret 0 fr}
    \end{align}
    The key difference here this time is that if we try to perform the same change of variables as we did before in equation~\ref{eqn: solving pw Delta T / T}, the $\frac{\partial t^0_\mathrm{ret}}{\partial t}$ term will no longer be a constant (it will be a function of $t$), and so we won't be able to pull it out of the integral.  Instead let's go back to equation~\ref{eqn: Delta T / T} and look at the third line, using the definition of $u$ to change variables and write the integral as the following:
    \begin{align}
          \frac{\Delta T}{T} &= \frac{1}{2}\hat{p}^i\hat{p}^j E^{\hat{r}\textsc{A}}_{ij} \int\limits^{t_\mathrm{obs}}_{t_\mathrm{obs}-\frac{L}{c}} \Bigg[ \frac{2}{3} h_\textsc{A}(t^0_\mathrm{ret}) \frac{\dot{\omega}(t^0_\mathrm{ret})}{\omega^0} + 2 h_\textsc{A}\bigg(\omega^0, \Theta^0 + \frac{\pi}{4}\bigg)\omega^0  \Bigg] dt , \nonumber \\
          &= \frac{1}{2}\hat{p}^i\hat{p}^j E^{\hat{r}\textsc{A}}_{ij} \int\limits^{0}_{L/R} \Bigg[ \frac{2}{3} h_\textsc{A}(t^0_\mathrm{ret}) \frac{\dot{\omega}(t^0_\mathrm{ret})}{\omega^0} + 2 h_\textsc{A}\bigg(\omega^0, \Theta^0 + \frac{\pi}{4}\bigg)\omega^0  \Bigg] \frac{-R}{c}du , \nonumber \\
          &= +\frac{1}{2}\hat{p}^i\hat{p}^j E^{\hat{r}\textsc{A}}_{ij} \frac{R}{c} \int\limits^{L/R}_{0} \Bigg[ \frac{2}{3} h_\textsc{A}(t^0_\mathrm{ret}) \frac{\dot{\omega}(t^0_\mathrm{ret})}{\omega^0} + 2 h_\textsc{A}\bigg(\omega^0, \Theta^0 + \frac{\pi}{4}\bigg)\omega^0  \Bigg] du .
    \label{eqn: solving fr Delta T / T}
    \end{align}

%---------------------------------------------------------------------------------
%---------------------------------------------------------------------------------    
        \subsection{Fresnel, Monochromatic (IIA)}\label{subsec: Monochromatic Model (Fresnel)}
        
        For a monochromatic gravitational wave the orbital frequency is a constant $\omega^0 \equiv \omega\left(t^0_\mathrm{ret}\right) \equiv \omega_0$ (from equation~\ref{eqn:monochrome freq}), which means $\dot{\omega} \equiv 0$, $h(t) = h_0$ (equation~\ref{eqn: amplitude 0 of metric perturbation}), and the first term in square brackets in equation~\ref{eqn: solving fr Delta T / T} vanishes.  For the second term, let's take a look at the explicit expressions for both the plus and cross terms.  Using equation~\ref{eqn: useful partial h partial Theta} we can write them as:
        \begin{align}
            2 h_\textsc{A}\bigg(\Theta^0 + \frac{\pi}{4}\bigg) &= \left\{\begin{tabular}{r l}
                    $2 h_0 \sin\left(2\Theta^0\right)$ & $= 2 h_0 \sin\left(2\Theta\left(t^0_\mathrm{ret}(u)\right)\right)$ , \\[8pt]
                    $-2 h_0 \cos\left(2\Theta^0\right)$ & $= -2 h_0 \cos\left(2\Theta\left(t^0_\mathrm{ret}(u)\right)\right)$ ,
                \end{tabular}\right. \nonumber \\[10pt]
                &= \left\{\begin{tabular}{l r}
                    $= 2 h_0 \sin\left(\textsf{A} - \textsf{B}u - \textsf{C}u^2\right)$ , & $(+)$ \\[8pt]
                    $= -2 h_0 \cos\left(\textsf{A} - \textsf{B}u - \textsf{C}u^2\right)$ , & $(\times)$
                \end{tabular}\right. \label{eqn:partial h partial Theta at x0 in u}\\[10pt]        
            &\underset{\left(t_0 \ = \ -\frac{R}{c}\right)}{\mathrm{where}}\quad \left\{\begin{tabular}{l l l}
            $\textsf{A}$ & $\equiv 2\theta_0 + 2\omega_0 t_\mathrm{obs}$ , & \\[8pt]
            $\textsf{B}$ & $\equiv 2\omega_0\frac{R}{c}\left(1-\hat{r}\cdot\hat{p}\right)$ & $= \frac{2\textsf{C}}{\left(1+\hat{r}\cdot\hat{p}\right)}$ , \\[8pt]
            $\textsf{C}$ & $\equiv \omega_0\frac{R}{c}\left(1-\left(\hat{r}\cdot\hat{p}\right)^2\right)$ & $= \omega_0\frac{R}{c}\left(1-\hat{r}\cdot\hat{p}\right)\left(1+\hat{r}\cdot\hat{p}\right)$ , \\
             & & $= \frac{\left(1+\hat{r}\cdot\hat{p}\right)\textsf{B}}{2}$ .
            \end{tabular}\right. \label{eqn: f A B C expressions}
        \end{align}
        Once again we make the familiar choice that $t_0=-\frac{R}{c}$.  Now we can write equation~\ref{eqn: solving fr Delta T / T} in the following compact form:
        \begin{align}
            \frac{\Delta T}{T} &= \hat{p}^i\hat{p}^j E^{\hat{r}\textsc{A}}_{ij}\omega_0\frac{R}{c} h_0 g_\textsc{A} , \\
            &\mathrm{where}\quad \begin{cases}
            g_{+} \equiv \int\limits^{L/R}_0 \sin\left(\textsf{A} - \textsf{B}u - \textsf{C}u^2\right) du ,\\
            g_{\times} \equiv -\int\limits^{L/R}_0 \cos\left(\textsf{A} - \textsf{B}u - \textsf{C}u^2\right) du .
            \end{cases}\label{eqn: g integrals}
        \end{align}
        As a sanity check, note that if we drop the $u^2$ term and carry out the integration in equation~\ref{eqn: g integrals}, we arrive back at the expression in equation~\ref{eqn: solving pw mono Res} and hence we do indeed regain the plane-wave monochromatic approximation equation~\ref{eqn: Res(t) pw mono} as we would expect.
        
        By appealing to complex methods, both the integrals in equation~\ref{eqn: g integrals} can be solved simultaneously by taking the real and imaginary parts of the solution of the following expression:
        \begin{align}\int\limits_0^{L/R} \exp{\bigg[i\Big(\textsf{A}-\textsf{B}u-\textsf{C}u^2\Big)\Bigg]}du &= \int\limits_0^{L/R} \exp{\left[i\left(\left(\textsf{A}+\frac{\textsf{B}^2}{4\textsf{C}}\right) - \frac{\left(2\textsf{C}u+\textsf{B}\right)^2}{4\textsf{C}}\right)\right]}du , \nonumber \\
        &\hspace{-1cm}\equiv \int\limits_0^{L/R} \exp{\Bigg[i\bigg(\Phi - \frac{\pi}{2}x^2\bigg)\Bigg]}du, \nonumber \\
        &\hspace{-1cm}= \sqrt{\frac{\pi}{2\textsf{C}}} e^{i\Phi} \int\limits_{\eta_1}^{\eta_2} e^{-i\frac{\pi}{2}x^2} dx , \nonumber \\
        &\hspace{-1cm}= \sqrt{\frac{\pi}{2\textsf{C}}} e^{i\Phi} \left[\int\limits_{0}^{\eta_2} e^{-i\frac{\pi}{2}x^2} dx -\int\limits_{0}^{\eta_1} e^{-i\frac{\pi}{2}x^2} dx \right], \nonumber \\
        &\hspace{-1cm}= \sqrt{\frac{\pi}{2\textsf{C}}}\Big(\cos(\Phi) + i\sin(\Phi)\Big) \Big[ C(\eta_2) - iS(\eta_2) - C(\eta_1) + iS(\eta_1) \Big] , \label{eqn: exp^(i A B C)}  \\
        \mathrm{where}\quad &\begin{cases}
            \Phi \equiv \textsf{A} + \frac{\textsf{B}^2}{4\textsf{C}} , \\[8pt]
            \eta_1 \equiv \frac{\textsf{B}}{\sqrt{2\pi\textsf{C}}} , \\[8pt]
            \eta_2 \equiv \eta_1 \Big[1 + \frac{2\textsf{C}}{B}\left(\frac{L}{R}\right) \Big] ,
            \end{cases}
        \end{align}
        where we make the substitution $\sqrt{\frac{\pi}{2}}x = \frac{2\textsf{C}u + \textsf{B}}{2\sqrt{\textsf{C}}}$.  Note that the only required assumption here is that $\textsf{C} > 0$.  Given the form of $\textsf{C}$ in equation~\ref{eqn: f A B C expressions} this will only fail if the gravitational wave source and the pulsar are perfectly aligned or anti-aligned, such that $\hat{r}\cdot\hat{p}=\pm 1$.  However, if this is the case and the source and pulsar are perfectly aligned or anti-aligned, then the quadratic term in $u$ would vanish, and the problem would once again be a plane-wave problem (hence assumption~\ref{as: fresnel no alignment}).
        
        Here we note that we can write our solution using the following definitions of the well known Fresnel integrals:
        \begin{align}
            \begin{cases}
            S(\eta) \equiv \int\limits^\eta_0 \sin\left(\frac{\pi}{2} x^2\right) dx \quad \xrightarrow[|\eta| \gg 1]{\mathrm{asymptotic}} \quad \frac{\mathrm{sgn}(\eta)}{2} - \frac{1}{\pi \eta}\cos\left(\frac{\pi}{2}\eta^2\right) , \\
            C(\eta) \equiv \int\limits^\eta_0 \cos\left(\frac{\pi}{2} x^2\right) dx \quad \xrightarrow[|\eta| \gg 1]{\mathrm{asymptotic}} \quad \frac{\mathrm{sgn}(\eta)}{2} + \frac{1}{\pi \eta}\sin\left(\frac{\pi}{2}\eta^2\right) ,
            \end{cases}
        \label{equation: Fresnel integrals}
        \end{align}
        Taking the real and imaginary parts of equation~\ref{eqn: exp^(i A B C)} gives us the solutions to equation~\ref{eqn: g integrals}:
        \begin{align}
            &\begin{cases}
            g_{+} = \sqrt{\frac{\pi}{2\textsf{C}}} \bigg[ \Big\{C\left(\eta_2\right)-C\left(\eta_1\right)\Big\}\sin(\Phi) - \Big\{S\left(\eta_2\right)-S\left(\eta_1\right)\Big\}\cos(\Phi) \bigg] , \\
            g_{\times} = -\sqrt{\frac{\pi}{2\textsf{C}}} \bigg[ \Big\{C\left(\eta_2\right)-C\left(\eta_1\right)\Big\}\cos(\Phi) + \Big\{\left(\eta_2\right)-S\left(\eta_1\right)\Big\}\sin(\Phi) \bigg] , 
            \end{cases} \label{eqn: g integrals solved}
        \end{align}
        
        Now we can solve equation~\ref{eqn:timing residual} to get the timing residual.  The convenient thing here is that the only term that depends on $t_\mathrm{obs}$ is $\Phi = \Phi\big(\textsf{A}\left(t_\mathrm{obs}\right)\big)$, which we can use to write $\frac{d\Phi}{dt_\mathrm{obs}}=\frac{d\Phi}{d\textsf{A}}\frac{d\textsf{A}}{dt_\mathrm{obs}} = 2\omega_0$ and perform the following change of variables in the integration:
        \begin{equation}
            \mathrm{Res}(t) = \int \left[\hat{p}^i\hat{p}^j E^{\hat{r}\textsc{A}}_{ij}\omega_0\frac{R}{c} h_0 g_\textsc{A}\right] dt_\mathrm{obs} = \hat{p}^i\hat{p}^j E^{\hat{r}\textsc{A}}_{ij}\omega_0\frac{R}{c} h_0 \int g_\textsc{A} \frac{d\mathrm{\Phi}}{2\omega_0} .
        \end{equation}
        Finishing out the integration of $g_\textsc{A}$ from equation~\ref{eqn: g integrals solved} and tidying up the algebra (and again, integration changes $t_\mathrm{obs}\rightarrow t$), the entire result can be expressed as:
        \begin{align}
            \mathrm{Res}(t) &= \hat{p}^i\hat{p}^j E^{\hat{r}\textsc{A}}_{ij} \sqrt{\frac{\pi R/c}{8\omega_0\left(1-\left(\hat{r}\cdot\hat{p}\right)^2\right)}} \Bigg[  \Big\{C\left(\eta_2\right)-C\left(\eta_1\right)\Big\} h_\textsc{A}\left(\Theta^{'}\right) \Bigg. \nonumber \\
            &\hspace{6.5cm}+ \ \Bigg.\Big\{S\left(\eta_2\right)-S\left(\eta_1\right)\Big\} h_\textsc{A}\left(\Theta^{'}-\frac{\pi}{4}\right) \Bigg] , \label{eqn: Res(t) fresnel mono} \\
            &\mathrm{for}\quad \textsc{A} \in [+, \times] , \nonumber \\
        \nonumber \\
            &\underset{\left(t_0 \ = \ -\frac{R}{c}\right)}{\mathrm{where}}\quad \Theta^{'} \equiv \Theta\left(t+\frac{1}{2}\frac{R}{c}\frac{\left(1-\hat{r}\cdot\hat{p}\right)}{\left(1+\hat{r}\cdot\hat{p}\right)} \right) \equiv \theta_0 + \omega_0\left(t+\frac{1}{2}\frac{R}{c}\frac{\left(1-\hat{r}\cdot\hat{p}\right)}{\left(1+\hat{r}\cdot\hat{p}\right)} \right) , \\
            &\hspace{1.5cm}\begin{cases}
                \eta_1 \equiv \sqrt{\frac{2\omega_0 R/c \left(1-\hat{r}\cdot\hat{p}\right)}{\pi \left(1+\hat{r}\cdot\hat{p}\right)}} , \\
                \eta_2 \equiv \eta_1 \Big[1+\left(1+\hat{r}\cdot\hat{p}\right)\left(\frac{L}{R}\right)\Big]  ,
            \end{cases} \\
            &\hspace{1.5cm}\begin{cases}
                h_{+} \equiv -h_0 \cos\big(2\Theta\big), \\
                h_{\times} \equiv -h_0 \sin\big(2\Theta\big),
            \end{cases} \tag{\ref{eqn: h0 plus and cross} r} \\
            &\hspace{1.5cm} h_0 \equiv \frac{4(G\mathcal{M})^{5/3}}{c^4 R}\omega_0^{2/3}  , \tag{\ref{eqn: amplitude 0 of metric perturbation} r}\\
            &\hspace{1.5cm}\begin{cases}
                \begin{tabular}{l l l}
                    $f^{+}$ &$\equiv \frac{\hat{p}^i\hat{p}^j e^{\hat{r}+}_{ij}}{\left(1-\hat{r}\cdot\hat{p}\right)}$ &$= \frac{\left(\hat{p}\cdot\hat{\theta}\right)^2 - \left(\hat{p}\cdot\hat{\phi}\right)^2}{\left(1-\hat{r}\cdot\hat{p}\right)}$ , \\
                    $f^{\times}$ &$\equiv \frac{\hat{p}^i\hat{p}^j e^{\hat{r}\times}_{ij}}{\left(1-\hat{r}\cdot\hat{p}\right)}$ &$= \frac{2\left(\hat{p}\cdot\hat{\theta}\right)\left(\hat{p}\cdot\hat{\phi}\right)}{\left(1-\hat{r}\cdot\hat{p}\right)}$ ,
                \end{tabular}
            \end{cases} \tag{\ref{eqn: antenna f} r} \\
            &\hspace{1.5cm}\begin{cases}
                \begin{tabular}{l l l}
                    $F^+$ &$\equiv \frac{\hat{p}^i\hat{p}^j E^{\hat{r}+}_{ij}}{\left(1-\hat{r}\cdot\hat{p}\right)}$ &$= \frac{1}{2}\left(1+\cos^2(\iota)\right) \left[ \cos(2\psi) f^+ + \sin(2\psi) f^\times \right]$ , \\
                    $F^\times$ &$\equiv \frac{\hat{p}^i\hat{p}^j E^{\hat{r}\times}_{ij}}{\left(1-\hat{r}\cdot\hat{p}\right)}$ &$= \cos(\iota) \left[ -\sin(2\psi) f^+ + \cos(2\psi) f^\times \right]$ ,
                \end{tabular}
            \end{cases} \tag{\ref{eqn: antenna F} r} \\
            &\hspace{1.5cm}\begin{cases}
                \begin{tabular}{r l l r}
                    $\hat{r}$ &$= \big[\sin(\theta)\cos(\phi),$ &$\sin(\theta)\sin(\phi),$ &$\cos(\theta) \big]$, \\
                    $\hat{\theta}$ &$= \big[\cos(\theta)\cos(\phi),$ &$\cos(\theta)\sin(\phi),$ &$\sin(\theta) \big]$, \\
                    $\hat{\phi}$ &$= \big[-\sin(\phi),$ &$\cos(\phi),$ &$0 \big]$, \\
                    $\hat{p}$ &$= \big[\sin(\theta_p)\cos(\phi_p),$ &$\sin(\theta_p)\sin(\phi_p),$ &$\cos(\theta_p) \big]$.
                \end{tabular}
                \end{cases} \tag{\ref{eqn:source basis vectors}, \ref{eqn: p hat} r} 
        \end{align}
        
        The Fresnel monochromatic model equation~\ref{eqn: Res(t) fresnel mono} approaches the plane-wave monochromatic model equation~\ref{eqn: Res(t) pw mono} as we would expect in the small Fresnel number limit ($F \rightarrow 0$) and the natural plane-wave limit ($\hat{r}\cdot\hat{p} \rightarrow \pm 1$).  However, it is also important to remember that the plane-wave monochromatic model itself converges to zero in \textit{both} of the natural plane-wave limits, as discussed in Section~\ref{subsec: Monochromatic Model (PW)}.  Thus both of these effects result in zero timing residual for these two precise alignments, and hence the natural plane-wave limit will not be of much interest to us going forward in this study.
        
        Additionally, in this regime the degeneracy of the source chirp mass $\mathcal{M}$ and distance $R$ parameters is also broken, so in principle a loud enough observed source with a significant Fresnel number with the pulsar should allow for independent measurements of both of these source parameters.

        \paragraph{Asymptotic Model}\mbox{}

        Also of interest here we note the result when considering the asymptotic expansion of the Fresnel integrals.  Both $\eta_2, \ \eta_1 \gg 1$, so long as we are not looking at a source that is nearly perfectly aligned or anti-aligned in the plane-wave limit (where $\hat{r}\cdot\hat{p} \approx \pm 1$).  In this case, after replacing the Fresnel integrals with their asymptotic expansions given in equation~\ref{equation: Fresnel integrals}, the result simplifies to:
        \begin{align}
            \mathrm{Res}(t) \approx \frac{F^\textsc{A}}{4\omega_0} \Bigg[  &\Bigg\{\frac{1}{1+\left(1+\hat{r}\cdot\hat{p}\right)\left(\frac{L}{R}\right)}\sin\bigg(\frac{\pi}{2}\eta_2^2\bigg)-\sin\bigg(\frac{\pi}{2}\eta_1^2\bigg)\Bigg\} h_\textsc{A}\left(\Theta^{'}\right) \nonumber \\
            &+  \Bigg\{\frac{-1}{1+\left(1+\hat{r}\cdot\hat{p}\right)\left(\frac{L}{R}\right)}\cos\bigg(\frac{\pi}{2}\eta_2^2\bigg)+\cos\bigg(\frac{\pi}{2}\eta_1^2\bigg)\Bigg\} h_\textsc{A}\left(\Theta^{'}-\frac{\pi}{4}\right) \Bigg] , \nonumber \\[10pt]
            \approx \frac{F^\textsc{A}}{4\omega_0} \Bigg[&\Bigg\{\sin\bigg(\frac{\pi}{2}\eta_2^2\bigg)-\sin\bigg(\frac{\pi}{2}\eta_1^2\bigg)\Bigg\} h_\textsc{A}\left(\Theta^{'}\right) \Bigg. \nonumber \\
            &\hspace{0.5cm}+ \Bigg. \Bigg\{-\cos\bigg(\frac{\pi}{2}\eta_2^2\bigg)+\cos\bigg(\frac{\pi}{2}\eta_1^2\bigg)\Bigg\} h_\textsc{A}\left(\Theta^{'}-\frac{\pi}{4}\right) \Bigg] . \label{eqn: Res(t) fresnel mono asymptotic}
        \end{align}
        In the final line we dropped the $\mathcal{O}\left(\frac{L}{R}\right)$ factors in the amplitude that came from the asymptotic expansion, and now we see the Fresnel formalism solution more closely resembles that of the plane-wave solution, as they both share the same dominant amplitude term.

        \paragraph{Heuristic Model}\mbox{}
        
        Our careful study and derivation of the timing residual in both the plane-wave and Fresnel regimes reveals some interesting insights into the underlying physics in these models.  In the previous Sections~\ref{sec: Plane-Wave Formalism} and~\ref{sec: Fresnel Formalism} we have solved for the timing residual model including higher order terms in the expansion of the retarded times.
        
        In the limit that the source is further away, we know that the Fresnel formalism should reduce to the plane-wave formalism.  And looking at the plane-wave solution of the timing residual in equation~\ref{eqn: Res(t) pw mono} we see that it has two terms which are evaluated at the endpoints of our photon's journey.  Evaluating equation~\ref{eqn:tret 0 pw} at the start of the photon's motion $t_\mathrm{obs} - L/c$ and the end of the photon's motion $t_\mathrm{obs}$ ultimately leads us to the two times that the phase terms are evaluated at in equation~\ref{eqn: Res(t) pw mono phase E and P}, namely $t$ and $t-\left(1-\hat{r}\cdot\hat{p}\right)\frac{L}{c}$ (given we choose $t_0 = -\frac{R}{c}$).
        
        Referring again to Figure~\ref{fig: tret contours}, we know the biggest potential difference between the Fresnel and plane-wave regimes will come when evaluating the retarded time at the pulsar's position, since that is at the largest spatial separation along the Earth-pulsar baseline.  So these two observations lead us to make the following guess: perhaps we can model the timing residual by using the plane-wave formalism, but modify the retarded time expression at the pulsar endpoint to include the higher order correction terms in the retarded time.
        
        To try this, first note that, a priori, if we write the following expression:
        \begin{align}
            \overline{t}_\mathrm{ret} &\equiv t - \frac{L}{c} - \frac{1}{c}\sqrt{R^2 - 2\left(\hat{r}\cdot\hat{p}\right)RL + L^2} , \nonumber \\
            &= t - \frac{L}{c} - \frac{R}{c}\left[1-\left(\hat{r}\cdot\hat{p}\right)\left(\frac{L}{R}\right) + \frac{1}{2}\left(1-\left(\hat{r}\cdot\hat{p}\right)^2\right)\left(\frac{L}{R}\right)^2 + \ldots \right] , \nonumber \\
            &= t \quad - \frac{R}{c}  \quad - \left(1-\hat{r}\cdot\hat{p}\right)\frac{L}{c} \quad - \frac{1}{2}\left(1-\left(\hat{r}\cdot\hat{p}\right)^2\right)\frac{L}{c}\frac{L}{R} \quad + \ldots .
        \label{eqn: tret model}
        \end{align}
        that the first three terms in this expression, when later combined with our model evaluated at $t_0 = -\frac{R}{c}$, will recover the retarded time factor for the pulsar term in the plane-wave expansion equation~\ref{eqn: Res(t) pw mono phase E and P}.  And further given what we have established previously it suggests that this fourth term is the Fresnel regime correction for the pulsar term.
        
        These ideas lead us on the basis of physical interpretation of this problem only, to propose the following model for the monochromatic regime timing residual of a pulsar which includes higher order corrections of the effect of the gravitational wavefront's curvature coming from the retarded time:
        \begin{align}
            \overline{\mathrm{Res}}(t) &\equiv \frac{F^\textsc{A}}{4\omega_0} \bigg[ h_\textsc{A}\Big(\Theta_E-\frac{\pi}{4}\Big) - h_\textsc{A}\Big(\overline{\Theta}_P-\frac{\pi}{4}\Big) \bigg] , \label{eqn: Res(t) fresnel mono heuristic} \\
            &\mathrm{for}\quad \textsc{A} \in [+, \times], \nonumber \\
            \nonumber \\
            &\underset{\left(t_0 \ = \ -\frac{R}{c}\right)}{\mathrm{where}}\quad \begin{cases}
                \Theta_E \equiv \Theta\Big(t-\frac{R}{c}\Big) &\equiv \theta_0 + \omega_0 t , \\
                \overline{\Theta}_P \equiv \Theta\big(\bar{t}_\mathrm{ret}\big) &\approx \Theta\Bigg(t - \frac{R}{c}  - \left(1-\hat{r}\cdot\hat{p}\right)\frac{L}{c} - \frac{1}{2}\left(1-\left(\hat{r}\cdot\hat{p}\right)^2\right)\frac{L}{c}\frac{L}{R}\Bigg) , \\
                &\equiv \theta_0 + \omega_0\Bigg(t-\left(1-\hat{r}\cdot\hat{p}\right)\frac{L}{c} - \frac{1}{2}\left(1-\left(\hat{r}\cdot\hat{p}\right)^2\right)\frac{L}{c}\frac{L}{R}\Bigg) .
                \end{cases}\label{eqn: Res(t) fresnel mono heuristic phase E and P}
        \end{align}
        The overbar notation used here and below is simply to distinguish that, unlike the previous timing residual models, this has not been analytically derived, but rather proposed.
        
        This proposed model has the required property that it reduces to the plane-wave formalism in the limits given in assumption~\ref{as: fresnel pw limits} in Section~\ref{subsec: Monochromatic Model (Fresnel)}.  It takes the general form of the plane-wave monochromatic formalism (equation~\ref{eqn: Res(t) pw mono}) but with the corrections motivated by the Fresnel regime study.  \textit{It is critical to understand that this formula is not a mathematically derived formula for the timing residual}.  It is merely meant to be a model which attempts to accurate encapsulate the basic physics principles behind the true timing residual model motivated from our previous sections.  As we showed, the real Fresnel regime solution equation~\ref{eqn: Res(t) fresnel mono} is more complicated.  But from a practical standpoint, we found that this formula is much easier to study numerically in our Fisher analysis in Chapter~\ref{ch: Fisher matrix analysis}.
        
\begin{figure}
            \centering
            \includegraphics[width=0.9\linewidth]{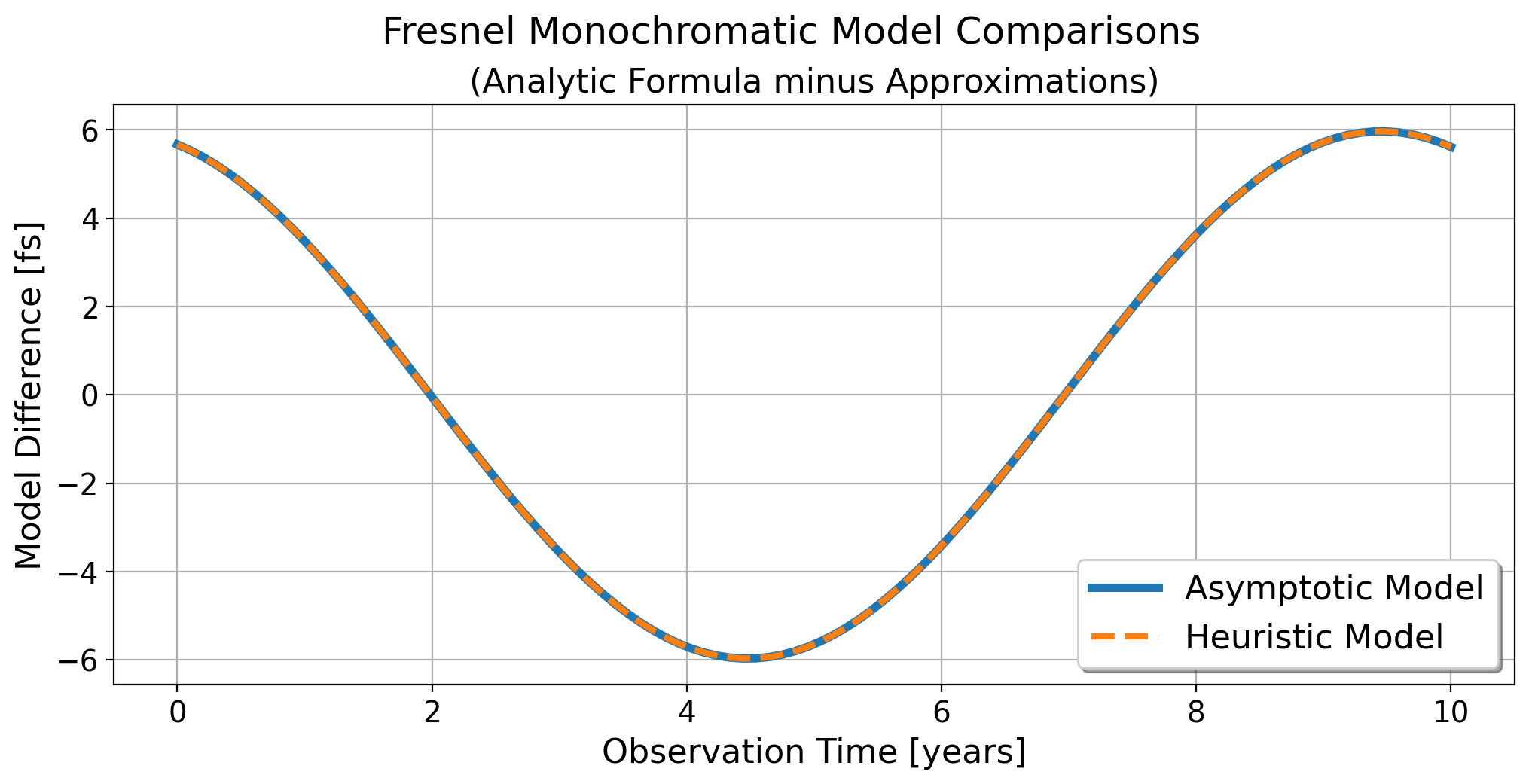}
            \caption[Monochromatic Timing Residual Models Comparison]{An example of the difference between the monochromatic timing residual models in the Fresnel formalism.  Here we take the difference between the analytically derived model equation~\ref{eqn: Res(t) fresnel mono}, and the two approximations we discussed, the asymptotic model (equation~\ref{eqn: Res(t) fresnel mono asymptotic}) and the heuristic model (equation~\ref{eqn: Res(t) fresnel mono heuristic}).  Both approximations accurately recover the analytic formula, with the difference here on the order of femtoseconds.  This is well below current pulsar timing precision, which is around $100$ns~\citep{NG_11yr_data}.  For this example the following sets of parameters were used: $\left\{R,\theta,\phi,\iota,\psi,\theta_0,\mathcal{M},\omega_0\right\}=\left\{100 \ \mathrm{Mpc}, \frac{\pi}{2} \ \mathrm{rad}, 0 \ \mathrm{rad}, 0 \ \mathrm{rad}, 0 \ \mathrm{rad}, 0 \ \mathrm{rad}, 10^7  \ M_\odot, 10 \ \mathrm{nHz}\right\}$ and $\left\{L, \theta_p, \phi_p\right\} = \left\{20 \ \mathrm{kpc}, 0 \ \mathrm{rad}, 0 \ \mathrm{rad}\right\}$, which gives the source a coalescence time $\Delta\tau_c = 2.0$ Gyr, and a Fresnel number of $F=1.31$ with this pulsar.}
            \label{fig: fresnel model comparisons}
        \end{figure}

        As a model we can still test it to see how accurately it matches the known analytic solution equation~\ref{eqn: Res(t) fresnel mono}, and it turns out, both the asymptotic and the heuristic models do exceptionally well at predicting the behavior of the timing residual.  Figure~\ref{fig: fresnel model comparisons} shows an example of the difference between the analytic and asymptotic formulae, and the analytic and the heuristic formulae.  For this example, the difference in models is on the order of femtoseconds, which is well below the precision with which we can currently measure timing residuals.  But perhaps even more importantly, we see that these two approximation models, equations~\ref{eqn: Res(t) fresnel mono asymptotic} and~\ref{eqn: Res(t) fresnel mono heuristic} match each other, which further validates our heuristic understanding of the underlying physics affecting the timing residual model.
        
        Alternatively, we can decompose equation~\ref{eqn: Res(t) fresnel mono heuristic} into the sum of a plane-wave part and a correction part due to the curvature of the gravitational wavefront, similar to what was done in~\citet{DF_main_paper}:
        \begin{align}
            \overline{\mathrm{Res}}(t) &\equiv \mathrm{Res}_{(\mathrm{IA})}(t) + \mathrm{Res}_{(\mathrm{IIA,correction})}(t) , \label{eqn: Res(t) fresnel mono approx - alternative expression} \\[4pt]
            &\mathrm{where}\quad \mathrm{Res}_{(\mathrm{IIA,correction})}(t) \equiv \frac{F^\textsc{A}}{4\omega_0} \bigg[ h_\textsc{A}\Big(\Theta_P-\frac{\pi}{4}\Big) - h_\textsc{A}\Big(\overline{\Theta}_P-\frac{\pi}{4}\Big) \bigg] , \nonumber \\
            &\hspace{3.45cm} = A_{c(\mathrm{IIA,correction})} \ F^\textsc{A} \ s_{P,\textsc{A}} \label{eqn: Res(t) fresnel mono - alternative expression correction term} \\[4pt]
            &\qquad\quad \ \begin{cases}
            A_{c(\mathrm{IIA,correction})} \equiv \frac{h_{c(\mathrm{IIA,correction})}}{4\omega_0} , \\
            h_{c(\mathrm{IIA,correction})} \equiv 2 h_0 \sin\left(\frac{\omega_0}{2}\left(1-\left(\hat{r}\cdot\hat{p}\right)^2\right)\frac{L}{c}\frac{L}{R}\right) ,
            \end{cases} \label{eqn: Res(t) fr mono correction characteristic strain and amplitude} \\            
            &\qquad\quad \ \begin{cases}    
            s_{P,+} \equiv \sin\left(2\Theta_P - \frac{\pi}{2} - \frac{\omega_0}{2}\left(1-\left(\hat{r}\cdot\hat{p}\right)^2\right)\frac{L}{c}\frac{L}{R}\right) , \\
            s_{P,\times} \equiv -\cos\left(2\Theta_P - \frac{\pi}{2} - \frac{\omega_0}{2}\left(1-\left(\hat{r}\cdot\hat{p}\right)^2\right)\frac{L}{c}\frac{L}{R}\right) .
            \end{cases}\label{eqn: Res(t) fr mono correction "r" terms}
        \end{align}
        Here $\mathrm{Res}_{(\mathrm{IA})}(t)$ is the plane-wave monochromatic model from Section~\ref{subsec: Monochromatic Model (PW)}, and equation~\ref{eqn: Res(t) fresnel mono - alternative expression correction term} is the correction to the plane-wave regime due to the wavefront curvature from the Fresnel terms.  Note that $\Theta_P$ in equations~\ref{eqn: Res(t) fresnel mono - alternative expression correction term} and~\ref{eqn: Res(t) fr mono correction "r" terms} comes from the IA model equation~\ref{eqn: Res(t) pw mono phase E and P}.  Using the same trick as in Section~\ref{subsec: Monochromatic Model (PW)} we can express this correction in terms of its own ``characteristic'' strain and timing residual amplitudes $ h_{c(\mathrm{IIA,correction})}$ and $A_{c(\mathrm{IIA,correction})}$.  This provides a very convenient interpretation and comparison of the IA and IIA regimes.  Since the IIA model is identical to the IA model with an additional correction due to the curvature of the gravitational wavefront, and since we now have expressions for the characteristic amplitudes of both the IA timing residual as well as this correction term, we can define the ratio of the amplitude of the correction to the amplitude of the plane-wave part as:
        \begin{equation}
            \rho \equiv \left|\frac{A_{c(\mathrm{IIA,correction})}}{A_{c(\mathrm{IA})}}\right| = \left|\frac{h_{c(\mathrm{IIA,correction})}}{h_{c(\mathrm{IA})}}\right| = \left|\frac{\sin\left(\frac{\omega_0}{2}\left(1-\left(\hat{r}\cdot\hat{p}\right)^2\right)\frac{L}{c}\frac{L}{R}\right)}{\sin\left(\omega_0\left(1-\hat{r}\cdot\hat{p}\right)\frac{L}{c}\right)}\right| .
        \end{equation}
        This ratio compares the relative size of the Fresnel correction in the IIA model to the plane-wave contribution, and is a similar quantity again to what~\citeauthor{DF_main_paper} calculated in their work.  If the value of $\rho \sim \mathcal{O}(1)$ or larger, then the Fresnel terms contribute a very significant correction to the otherwise plane-wave monochromatic model.  Together with the Fresnel number itself (equation~\ref{eqn: Fresnel number}), $F$ and $\rho$ are two useful metrics in evaluating the significance of the curvature corrections to our models.

%---------------------------------------------------------------------------------
%---------------------------------------------------------------------------------    
        \subsection{Fresnel, Frequency Evolution - Conjecture (IIB)}\label{subsec: Frequency Evolution CONJECTURE (Fresnel)}
        
        Given the understanding of the previous timing regimes, we now propose the most general timing residual model to date, one which includes the Fresnel corrections, allows for frequency evolution, and still recovers all of the previous three regimes in the appropriate limits.  This is not analytically derived - a complete and careful derivation along the lines of what we have done previously is still needed here.  However, our proposed conjecture here is motivated by the reasoning explained from our insights of the heuristic model in Section~\ref{subsec: Monochromatic Model (Fresnel)}.  This model makes the same assumptions as in the plane-wave frequency evolution regime (Section~\ref{subsec: Frequency Evolution (PW)}) and the Fresnel monochromatic regime (Section~\ref{subsec: Monochromatic Model (Fresnel)}).
        
        In the monochromatic formalism we generalized the plane-wave solution by modifying the retarded time in the pulsar term to include the Fresnel correction term, which encoded information about the curvature of the wave.  We propose doing the same thing with the frequency evolution formalism.  We begin with the plane-wave frequency evolution regime timing residual equation~\ref{eqn: Res(t) pw freq evo}, and we modify the pulsar term's retarded time using equation~\ref{eqn: tret model} as discussed in the Fresnel monochromatic case.  This will in turn modify both the orbital frequency of the pulsar term $\overline{\omega}_{0P}$ as well as the phase $\overline{\theta}_{0P}$ from equation~\ref{eqn: Res(t) pw freq evo frequency phase E and P}.  The model can be expressed compactly as:
        \begin{align}
            \overline{\mathrm{Res}}(t) &= \frac{F^\textsc{A}}{4} \left[ \frac{h_\textsc{A}\Big(\omega_{0E},\Theta_E-\frac{\pi}{4}\Big)}{\omega_{0E}} - \frac{h_\textsc{A}\Big(\overline{\omega}_{0P},\overline{\Theta}_P-\frac{\pi}{4}\Big)}{\overline{\omega}_{0P}} \right] , \label{eqn: Res(t) fr freq evo} \\
            &\mathrm{for}\quad \textsc{A} \in [+, \times], \nonumber \\
        \nonumber \\
            &\underset{\left(t_0 \ = \ -\frac{R}{c}\right)}{\mathrm{where}}\quad \begin{cases}
                \omega_{0E} &\equiv \omega_0 , \\
                \overline{\omega}_{0P} &\equiv \omega_0\left[1 + \frac{\left(1-\hat{r}\cdot\hat{p}\right)\frac{L}{c} + \frac{1}{2}\left(1-\left(\hat{r}\cdot\hat{p}\right)^2\right)\frac{L}{c}\frac{L}{R}}{\Delta\tau_c}  \right]^{-3/8} , \\
                \theta_{0E} &= \theta_0 , \\
                \overline{\theta}_{0P} &= \theta_0 + \theta_c\left(1-\left[1+ \frac{\left(1-\hat{r}\cdot\hat{p}\right)\frac{L}{c} + \frac{1}{2}\left(1-\left(\hat{r}\cdot\hat{p}\right)^2\right)\frac{L}{c}\frac{L}{R}}{\Delta\tau_c}\right]^{5/8}\right) , \\
                \Theta_E & = \theta_{0E} + \omega_{0E}t , \\
                \overline{\Theta}_P & = \overline{\theta}_{0P} + \overline{\omega}_{0P}t ,
                \end{cases}\label{eqn: Res(t) fresnel freq evo phase E and P} \\
            &\hspace{1.6cm}\begin{cases}
                \Delta\tau_c \equiv \frac{5}{256}\left(\frac{c^3}{G\mathcal{M}}\right)^{5/3} \frac{1}{\omega_0^{8/3}} , \\
            \theta_c \equiv \frac{8}{5}\Delta\tau_c \omega_0 ,
            \end{cases} \tag{\ref{eqn: time to coalescence}, \ref{eqn: thetac} r} \\
            &\hspace{1.6cm}\begin{cases}
                h_{+} \equiv -h(\omega) \cos\big(2\Theta\big), \\
                h_{\times} \equiv -h(\omega) \sin\big(2\Theta\big),
            \end{cases} \tag{\ref{eqn: h plus and cross} r} \\
            &\hspace{1.6cm}h(\omega) \equiv \frac{4(G\mathcal{M})^{5/3}}{c^4 R}\omega^{2/3} , \tag{\ref{eqn: amplitude of metric perturbation} r}\\
            &\hspace{1.6cm}\begin{cases}
                \begin{tabular}{l l l}
                    $f^{+}$ &$\equiv \frac{\hat{p}^i\hat{p}^j e^{\hat{r}+}_{ij}}{\left(1-\hat{r}\cdot\hat{p}\right)}$ &$= \frac{\left(\hat{p}\cdot\hat{\theta}\right)^2 - \left(\hat{p}\cdot\hat{\phi}\right)^2}{\left(1-\hat{r}\cdot\hat{p}\right)}$ , \\
                    $f^{\times}$ &$\equiv \frac{\hat{p}^i\hat{p}^j e^{\hat{r}\times}_{ij}}{\left(1-\hat{r}\cdot\hat{p}\right)}$ &$= \frac{2\left(\hat{p}\cdot\hat{\theta}\right)\left(\hat{p}\cdot\hat{\phi}\right)}{\left(1-\hat{r}\cdot\hat{p}\right)}$ ,
                \end{tabular}
            \end{cases} \tag{\ref{eqn: antenna f} r} \\
            &\hspace{1.6cm}\begin{cases}
                \begin{tabular}{l l l}
                    $F^+$ &$\equiv \frac{\hat{p}^i\hat{p}^j E^{\hat{r}+}_{ij}}{\left(1-\hat{r}\cdot\hat{p}\right)}$ &$= \frac{1}{2}\left(1+\cos^2(\iota)\right) \left[ \cos(2\psi) f^+ + \sin(2\psi) f^\times \right]$ , \\
                    $F^\times$ &$\equiv \frac{\hat{p}^i\hat{p}^j E^{\hat{r}\times}_{ij}}{\left(1-\hat{r}\cdot\hat{p}\right)}$ &$= \cos(\iota) \left[ -\sin(2\psi) f^+ + \cos(2\psi) f^\times \right]$ ,
                \end{tabular}
            \end{cases} \tag{\ref{eqn: antenna F} r} \\
            &\hspace{1.6cm}\begin{cases}
                \begin{tabular}{r l l r}
                    $\hat{r}$ &$= \big[\sin(\theta)\cos(\phi),$ &$\sin(\theta)\sin(\phi),$ &$\cos(\theta) \big]$, \\
                    $\hat{\theta}$ &$= \big[\cos(\theta)\cos(\phi),$ &$\cos(\theta)\sin(\phi),$ &$\sin(\theta) \big]$, \\
                    $\hat{\phi}$ &$= \big[-\sin(\phi),$ &$\cos(\phi),$ &$0 \big]$, \\
                    $\hat{p}$ &$= \big[\sin(\theta_p)\cos(\phi_p),$ &$\sin(\theta_p)\sin(\phi_p),$ &$\cos(\theta_p) \big]$.
                \end{tabular}
                \end{cases} \tag{\ref{eqn:source basis vectors}, \ref{eqn: p hat} r} 
        \end{align}
        This model behaves as expected in the previously established limits, recovering all previous regimes.
        
        And just like we did with the heuristic Fresnel monochromatic model in Section~\ref{subsec: Monochromatic Model (Fresnel)}, we can also decompose this formula into a plane-wave part and a correction part due to the curvature of the gravitational wavefront:
        \begin{align}
            \overline{\mathrm{Res}}(t) &\equiv \mathrm{Res}_{(\mathrm{IB})}(t) + \mathrm{Res}_{(\mathrm{IIB,correction})}(t) , \label{eqn: Res(t) fresnel freq evo - alternative expression} \\[4pt]
            &\mathrm{where}\quad \mathrm{Res}_{(\mathrm{IIB,correction})}(t) \equiv \frac{F^\textsc{A}}{4} \left[ \frac{h_\textsc{A}\Big(\omega_{0P},\Theta_P-\frac{\pi}{4}\Big)}{\omega_{0P}} - \frac{h_\textsc{A}\Big(\overline{\omega}_{0P},\overline{\Theta}_P-\frac{\pi}{4}\Big)}{\overline{\omega}_{0P}} \right] . \label{eqn: Res(t) fresnel freq evo - alternative expression correction term}
        \end{align}
        Here $\mathrm{Res}_{(\mathrm{IB})}(t)$ is the plane-wave frequency evolution model from Section~\ref{subsec: Frequency Evolution (PW)}, and equation~\ref{eqn: Res(t) fresnel freq evo - alternative expression correction term} is the correction to the plane-wave regime due to the wavefront curvature from the Fresnel terms.  Note that $\omega_{0P}$ and $\Theta_P$ in equation~\ref{eqn: Res(t) fresnel freq evo - alternative expression correction term} come from the IB model equation~\ref{eqn: Res(t) pw freq evo frequency phase E and P}.
        
        Finally, like in the Fresnel monochromatic and plane-wave frequency evolution regimes, there are no degeneracies between any of the source parameters.  In the specific case of the parameters $\mathcal{M}$ and $R$ both appear in combination in the amplitude terms (equation~\ref{eqn: amplitude of metric perturbation}), and both appear independently in the pulsar term phase and frequency functions (frequency evolution effects bring in the chirp mass, while Fresnel corrections bring in the source distance).  So in principle a source well described by this model with significant frequency evolution and Fresnel numbers among the pulsars in the PTA should allow for independent measurements of all source parameters.

%---------------------------------------------------------------------------------
%---------------------------------------------------------------------------------    
    \section{Regime \& Model Comparisons}\label{sec: regime & model comparisons}
    
    Figure~\ref{fig: model-regime comparison} shows four examples that highlight the differences in these timing model regimes, and Figure~\ref{fig: source timing regimes example} emphasizes just how different the predicted timing results can be depending on which model you use.  It is important to keep in mind the limitations and assumptions under which each of these timing models are derived when making direct comparisons.  The models can give significantly different and unreliable results if they are being applied to a regime in which they cannot accurately account for that regime's physics.  For example, all four models should give the same predictions for a source with high coalescence time and low Fresnel number since they all can describe the physics of a plane-wave monochromatic source.  However, they should all predict different timing residuals for a source with low coalescence time and high Fresnel number since each of the models (apart from model IIB) have assumptions built into them which prevent them from describing all of the physics in this scenario (that is IA and IB can't describe the effects of a curved wavefront, and IIA can't describe the effects of frequency evolution).
    
    This has lead us to consider the following question, ``by how much will our parameter estimations be affected when using the wrong model regime for a particular source?''  Consider, for example, a particular source has both significant Fresnel and frequency evolution effects that contribute to the timing residual it induces on our PTA.  It would be important to know if using the IB model (which doesn't capture those Fresnel effects) for parameter estimation would result in estimates of the \textit{wrong} source parameters, and by how much they are wrong.  At the time of completing my PhD work and finishing this dissertation I had begun a preliminary investigation into this question, so this may result in the topic of future studies beyond the work presented here.
    
\begin{figure}
        \centering
        \includegraphics[width=1\linewidth]{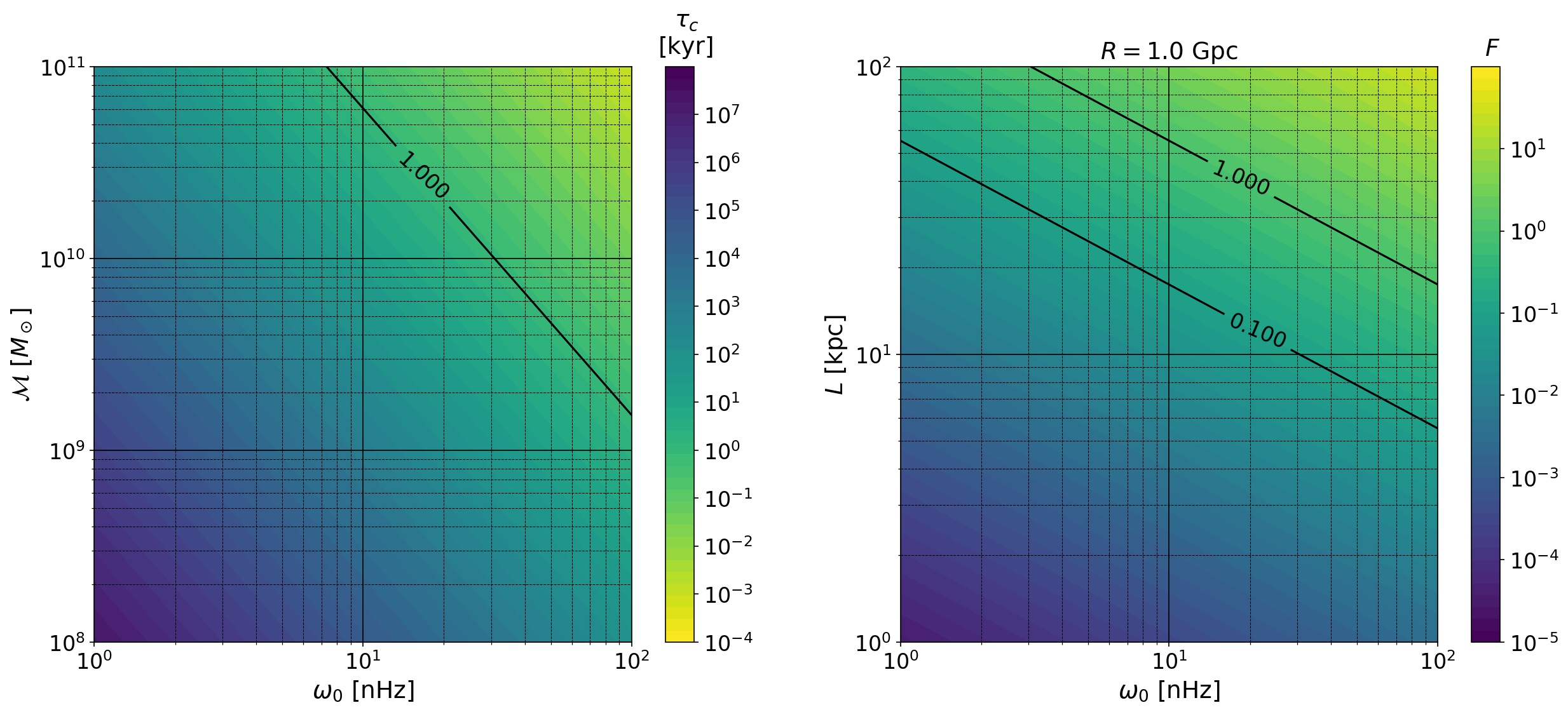}
        \caption[Coalescence Time and Fresnel Number Contours]{Contour plots of the coalescence time $\Delta \tau_c$ and Fresnel number $F$.  We expect the frequency evolution regime to become significant for smaller coalescence times, where the chirp mass $\mathcal{M}$ and orbital frequency $\omega_0$ are sufficiently high.  Similarly the Fresnel regime should become significant for higher Fresnel numbers.  Here as an example we fix the source distance $R$ of a gigaparsec and look at contours in the pulsar distance $L$ and orbital frequency. }
        \label{fig: DTauc & F contours}
    \end{figure}

    The four model regimes can be further subdivided into different categories given the parameter dependencies of the coalesence time and the Fresnel number (equations~\ref{eqn: time to coalescence} and~\ref{eqn: Fresnel number}), as shown in Table~\ref{tab: sub regimes}.  The parameters $\left\{\mathcal{M}, \omega_0\right\}$ control the transition between the monochromatic and frequency evolution regimes, and the parameters $\left\{R, \omega_0, L\right\}$ control the transition between the plane-wave and Fresnel regimes (see Figure~\ref{fig: 4 regimes}).  Figure~\ref{fig: DTauc & F contours} shows contour plots of both the coalescence time $\Delta \tau_c$ and Fresnel number $F$ in terms of their parameter dependencies.  In general these can help give us a sense of where in parameter space we can expect a transition from one of our timing regimes to another.
    \begin{table}
        \centering
        \begin{tabular}{c|c}
        \hline
          \begin{tabular}{c} large coalescence time limit \\ $\left(\Delta\tau_c \ \rightarrow \ \infty \right)$\end{tabular}  &  \begin{tabular}{c} small Fresnel number limit \\ $\left(F \ \rightarrow \ 0 \right)$\end{tabular} \\
          \hline
          \begin{tabular}{l l} low mass limit & $(\mathcal{M}\rightarrow 0)$ \\ low orbital frequency limit & $(\omega_0\rightarrow 0)$\end{tabular}  &  \begin{tabular}{l l} far source limit & $(R\rightarrow \infty)$ \\ low orbital frequency limit & $(\omega_0\rightarrow 0)$ \\ small detector limit & $(L\rightarrow 0)$\end{tabular}\\
        \hline
        \end{tabular}
        \caption[Gravitational Wave Timing Model Sub-Regimes]{The timing residual model regimes are divided by the coalescence time and Fresnel number limits.  These transitions can be expressed in terms of the model parameter limits indicated here.}
        \label{tab: sub regimes}
    \end{table}

    Consequently, the orbital frequency parameter has the most direct influence on the regimes - increasing the orbital frequency pushes the models towards \textit{both} the Fresnel and frequency evolution regimes, and vice versa.  When studying these pulsar timing models, special consideration should be taken in this regard, because studying systems at higher frequencies not only suggests that the physical effects due to frequency evolution could become important, but that also the effects of curvature as well.  These physical effects will change the timing model simultaneously, which means understanding the the full Fresnel frequency evolution model is very important, because it has the potential to predict very different timing residuals than the currently used plane-wave frequency evolution model.
    
    Furthermore, it also gives us motivation to look for and time pulsars at greater distances.  Since the Fresnel number scales as $L^2$, increasing the pulsar distance is the easiest way to push the timing model into the Fresnel regime.  This gives us motivation to look for and time pulsars at greater distances in our PTAs, since it is the one factor that we have more direct control over in enabling us to explore the Fresnel regime in pulsar timing experiments.  We explore this idea in much greater detail in Chapter~\ref{ch:results - first investigation}.
    \begin{figure}
        \centering
        \includegraphics[width=\linewidth]{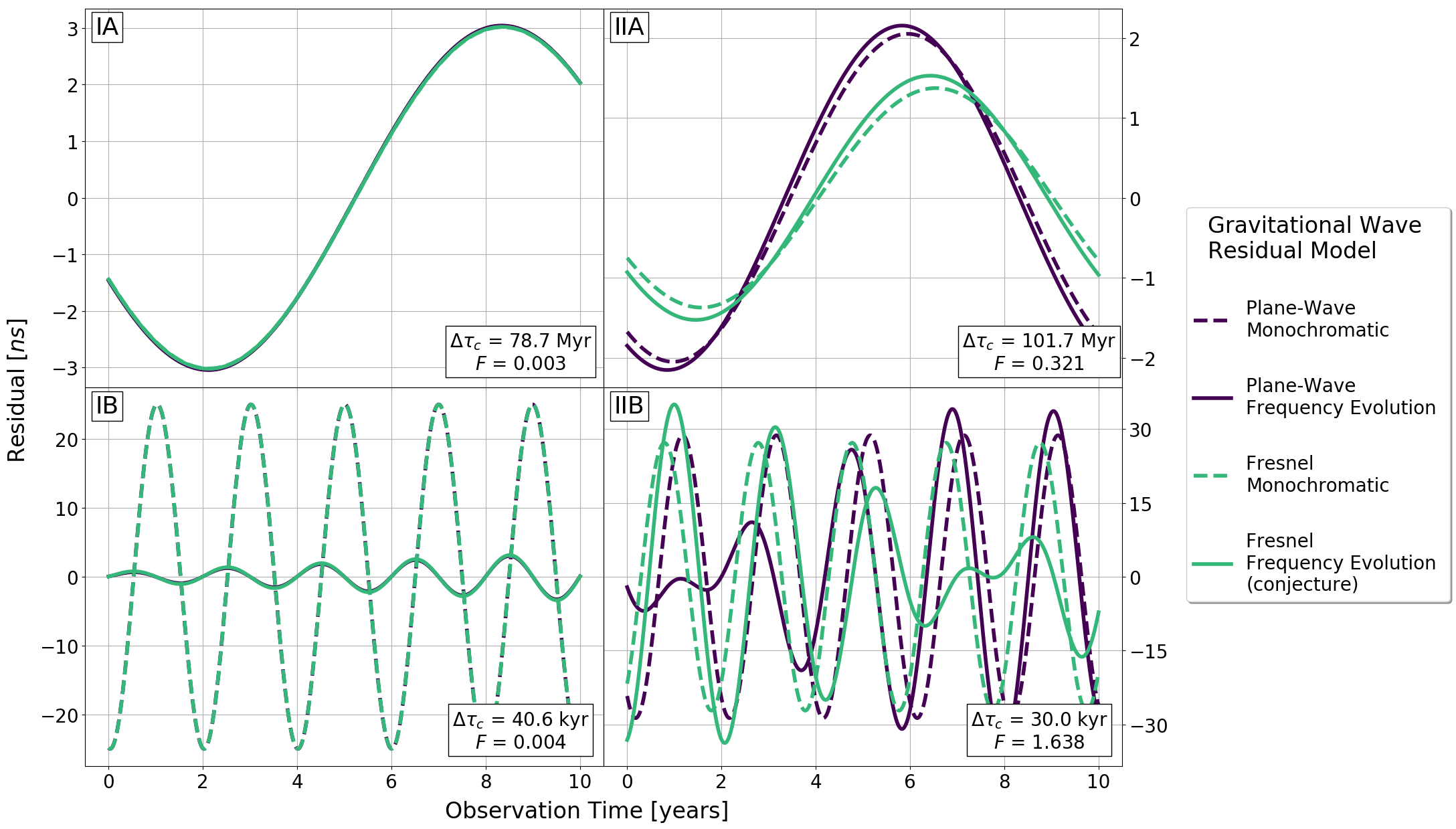}
        \caption[Gravitational Wave Timing Model Regime Comparisons 1]{To emphasize regime differences we plot the gravitational wave timing residuals for a single set of parameters as predicted by each of the four models.  Each panel is labelled by the approximate regime the source is in based on its physical properties (coalescence time and Fresnel number).  In all cases: $\left\{\theta,\phi,\iota,\psi,\theta_0,\theta_p,\phi_p\right\}=\left\{\frac{\pi}{3}, \ 0, \ \frac{\pi}{4}, \ \frac{\pi}{4}, \ 0, \ 0, \ 0\right\} \mathrm{rad}$.  \\[5pt]
        \textbf{(IA)} \underline{Low Fresnel number, high coalescence time}: As a sanity check, we find that all four models do correctly converge on the plane-wave monochromatic regime prediction as expected. Here: $\left\{R,\mathcal{M},\omega_0, L\right\}=\left\{50 \ \mathrm{Mpc}, \ 1\times10^8 \ M_\odot, \ 8 \ \mathrm{nHz}, \ 0.8 \ \mathrm{kpc}\right\}$.  \\[5pt]
        \textbf{(IB)}  \underline{Low Fresnel number, low coalescence time}: The monochromatic models separate from the frequency evolution models since they cannot capture the significant change in the orbital evolution of the source.  Additionally, since the Fresnel number is not significant, the Fresnel models converge on the plane-wave models. Here: $\left\{R,\mathcal{M},\omega_0,L\right\}=\left\{100 \ \mathrm{Mpc}, \ 5\times10^8 \ M_\odot, \ 50 \ \mathrm{nHz}, \ 0.5 \ \mathrm{kpc}\right\}$.  \\[5pt]
        \textbf{(IIA)}  \underline{High Fresnel number, high coalescence time}: The plane-wave models separate from the Fresnel models since they cannot capture the significant effect of the wavefront's curvature.  Additionally, since frequency evolution is not significant, those models converge on the monochromatic model predictions. Here: $\left\{R,\mathcal{M},\omega_0, L\right\}=\left\{50 \ \mathrm{Mpc}, \ 6\times10^7 \ M_\odot, \ 10 \ \mathrm{nHz}, \ 7 \ \mathrm{kpc}\right\}$.  \\[5pt] }
        \label{fig: model-regime comparison}
    \end{figure}
    \begin{figure}
    % https://tex.stackexchange.com/questions/487728/pagebreak-in-caption-within-a-beginfigure-environment
        \ContinuedFloat
        \caption[]{ \textit{Continued} \\[5pt]
        \textbf{(IIB)}  \underline{High Fresnel number, low coalesence time}: Here the monochromatic models can no longer capture the interesting physics such as the beat frequency due to the frequency evolution of the source, and the plane-wave models can't reliably account for the additional effects that wavefront curvature introduces.  All four models begin to predict very different results for the same set of parameters due to their respective limitations and underlying assumptions, and the most reliable model becomes the full Fresnel frequency evolution model. Here: $\left\{R,\mathcal{M},\omega_0,L\right\}=\left\{100 \ \mathrm{Mpc}, \ 6\times10^8 \ M_\odot, \ 50 \ \mathrm{nHz}, \ 10 \ \mathrm{kpc}\right\}$.    }
    \end{figure}

    \begin{figure}
        \centering
        \includegraphics[width=1\linewidth]{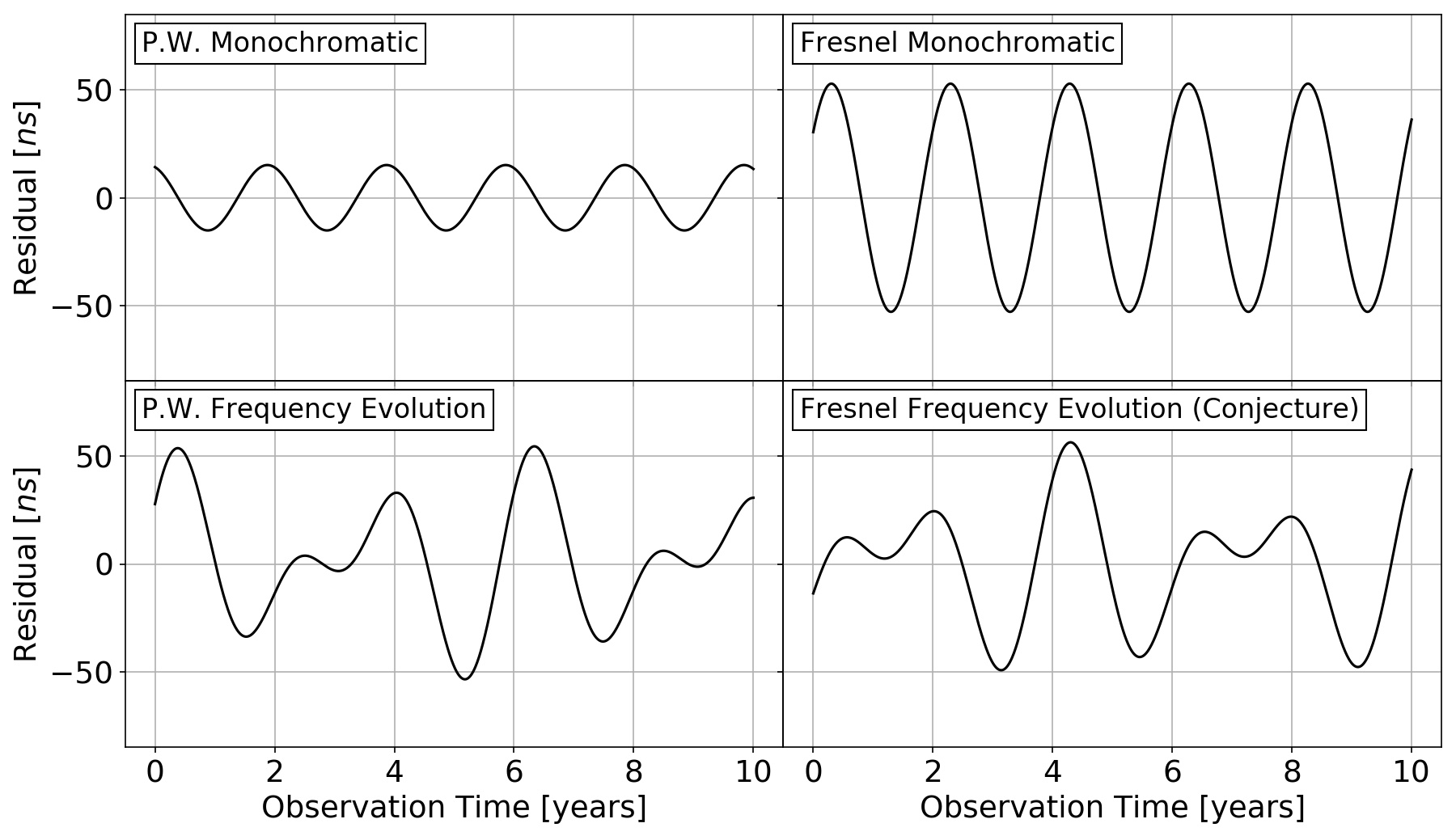}
        \caption[Gravitational Wave Timing Model Regime Comparisons 2]{Here we compute the predicted timing residual for Source 2 (Table~\ref{tab: source parameters reference}) at $R=100$ Mpc and the pulsar J1747-4036 (given the values in Table~\ref{tab: pta reference}, pulsar \#40) using the four pulsar timing regime models.  This source has a coalescence time of $12.793$ kyr, and the Fresnel number of this source and pulsar is $F=0.837$.  This is meant to show just how different the results can be in the predicted timing models for a given source and pulsar across each regime.  It is important to consider the strength of the frequency evolution in the signal and how large the Fresnel number is, in order to decide the accuracy of each model regime's predicted results, since certain regimes have physical limitations.  For example, the monochromatic models are insensitive to any frequency evolution over the pulsar-Earth baseline that the photon travels, which may otherwise produce a beat frequency in the signal (as seen in the two frequency evolution regimes here).  And the plane-wave regimes are not sensitive to Fresnel corrections in the phase, which may otherwise both shift the phase, produce different amplitudes, and further change the underlying beat frequency.}
        \label{fig: source timing regimes example}
    \end{figure}

%---------------------------------------------------------------------------------
%---------------------------------------------------------------------------------   
    \section{Regime Search Parameters \& Degeneracies}\label{sec:regime degeneracies}
        
    Each of the model regimes which we have discussed allow for the measurement of different sets of source parameters experimentally depending on which parameters are degenerate in the model.  However certain parameters are degenerate or highly covariant with others in certain regimes, which has motivated us to parametrize the model in specific ways.  For reference, Table~\ref{tab: model regime degeneracies} indicates the parameters in each of our model regimes.
    \begin{table}[ht]
        \centering
        \begin{tabular}{ c|c|c } 
        \hline
                 &  Plane-Wave  &  Fresnel \\ 
                 \hline
                 Monochrome  &  ${\vec{s}_\mathrm{IA} = \left\{A_\mathrm{E,res}, \theta, \phi, \iota, \psi, \theta_0, \omega_0\right\}}$  &  $\vec{s}_{\mathrm{IIA}} = \vec{s}_{\mathrm{IA}} \cup \left\{R\right\}$ \\
                 \hline
                 Frequency Evolution  &  $\vec{s}_{\mathrm{IB}} = \vec{s}_{\mathrm{IA}} \cup \left\{\mathcal{M}\right\}$  &  $\vec{s}_{\mathrm{IIB}} = \vec{s}_{\mathrm{IA}} \cup \left\{R,\mathcal{M}\right\}$  \\
                 \hline
        \end{tabular}
        \caption[Model Regime Degeneracies]{Our base parametrization choice for the timing residual models in each regime, based on the model degeneracies and covariances.  The parameter $A_\mathrm{E,res}$ is the ``Earth term timing residual amplitude'' (equation~\ref{eqn: earth term amplitude}).  In general, frequency evolution allows the direct measurement of chirp mass $\mathcal{M}$ from the frequency $\omega(t)$ and phase $\Theta(t)$ terms in the timing residual models, and Fresnel corrections allow for direct measurement of source distance $R$ from these frequency and phase terms.  Fisher matrices with alternative parametrization choices were computed through the Jacobian transformation in equation~\ref{eqn: fisher coordinate transformation}.}
        \label{tab: model regime degeneracies}
    \end{table}
    
    Starting with the most basic model, the plane-wave monochromatic regime, the source coordinate distance $R$ and chirp mass $\mathcal{M}$ are fully degenerate, appearing only in the amplitude of the timing residual (equation~\ref{eqn: Res(t) pw mono}).  We use equations~\ref{eqn: Res(t) pw mono}, \ref{eqn: h0 plus and cross}, and~\ref{eqn: amplitude 0 of metric perturbation} to introduce the following parameter:
    \begin{equation}
        A_\mathrm{E,res} \equiv \frac{h_0}{4 \omega_0} = \frac{\left(G\mathcal{M}\right)^{5/3}}{c^4 R \omega_0^{1/3}} ,
    \label{eqn: earth term amplitude}
    \end{equation}
    which is the ``Earth term timing residual amplitude,'' a similar quantity to characteristic amplitudes (equations~\ref{eqn: Res(t) pw mono characteristic strain and amplitude} and~\ref{eqn: Res(t) fr mono correction characteristic strain and amplitude}).  Therefore in the plane-wave monochromatic regime, this Earth term timing residual amplitude parameter replaces the chirp mass and source distance parameters.  This new parameter also removes the orbital frequency $\omega_0$ from the timing residual amplitude terms in our model.  Therefore $\omega_0$ is only being measured from its contribution to the time evolution of the phase of the wave (equation~\ref{eqn: Res(t) pw mono phase E and P}), not from its contribution to the timing residual amplitude.

    As we discussed in the previous sections the $R$-$\mathcal{M}$ degeneracy is broken in the other three regimes due to either frequency evolution or Fresnel curvature effects.  However, we found that if we removed the $A_{E,\mathrm{res}}$ parameter from these models and included $R$, $\mathcal{M}$, and $\omega_0$ each separately in the amplitude of the timing residual, it introduced strong covariances between the three parameters which made it very difficult to measure each of them independently.  This is because the combination of $\frac{\mathcal{M}}{R\omega_0^{1/3}}$ appears in both the Earth and pulsar terms of every timing residual measured, independent of the pulsar being timed, which makes these parameters covariant.  We need the differences caused by frequency evolution and Fresnel corrections that are due to different pulsar distances and locations across the sky to help reduce the covariances between these three parameters.  Therefore, in every regime we continue to include the Earth term timing residual amplitude $A_{E,\mathrm{res}}$ as its own separate parameter.  To clarify, this does mean that in the two frequency evolution regimes IB and IIB, we wrote the amplitude of the pulsar term as the combination of parameters $A_{P,\mathrm{res}} = A_{E,\mathrm{res}} \left(\frac{\omega_0}{\omega_{0P}}\right)^{1/3}$ for the plane-wave regime, and $A_{P,\mathrm{res}} = A_{E,\mathrm{res}} \left(\frac{\omega_0}{\overline{\omega}_{0P}}\right)^{1/3}$ for the Fresnel regime.  The other parameters $R$, $\mathcal{M}$, and $\omega_0$ are then measured from the other components of the model where they appear.
    
    From this we see one novel result of this study is that the Fresnel monochrome regime allows for the recovery of the coordinate distance even if the gravitational wave is effectively monochrome.  Furthermore, this study shows that going into the full Fresnel frequency evolution regime allows separate measurements of both the coordinate distance and the chirp mass from the frequency and phase terms in the timing residual.  These key points are investigated in detail in Section~\ref{sec: Measuring Chirp Mass from a Monochromatic Source}.
\cleardoublepage
\chapter{Fisher Matrix Analysis}\label{ch: Fisher matrix analysis}

The Fisher matrix provides a powerful tool for analyzing the predictive capabilities of a Bayesian-based model of some physical process.  At its core, the beauty and purpose of the Fisher matrix formalism is that it gives a ``best-case-scenario'' of how well the model parameters will be measured and constrained by the experiment itself, given the inherent uncertainties in making the observations, before the experiment is even performed.  An excellent motivational summary of the Fisher matrix formalism is:

\begin{displayquote}
    ``The whole point of the Fisher matrix formalism is to predict how well the experiment will be able to constrain the model parameters, \textit{before doing the experiment} and in fact without even simulating the experiment in any detail.  We can then forecast the results of different experiments and look at tradeoffs such as precision versus cost.  In other words, we can engage in \textit{experimental design}.
    
    ...The beauty of the Fisher matrix approach is that there is a simple prescription for setting up the Fisher matrix \textit{knowing only your model and your measurement uncertainties}; and that under certain standard assumptions, the Fisher matrix  is the inverse of the covariance matrix.  So all you have to do is set up the Fisher matrix then invert it to obtain the covariance matrix (that is, the uncertainties on your model parameters)... the inverse of the Fisher matrix is the best you can \textit{possibly} do given the information content of your experiment.''
    
\rightline{--- David Wittman (\citeauthor{wittman_Fisher_for_beginners})}
\end{displayquote}

Here we will develop the derivation of the Fisher matrix itself from Bayesian principles, and apply it to our pulsar timing model.

%---------------------------------------------------------------------------------
%---------------------------------------------------------------------------------
    \section{High Statistical Limit Applied to Bayes' Theorem}\label{sec:high stat limit of Bayes}
    
    From Bayes' Theorem we know:
    \begin{equation}
        \underbrace{p\left(\vec{\theta}\middle|\vec{X}\right)}_{\mathrm{``posterior"}} = \frac{\overbrace{p\left(\vec{X}\middle|\vec{\theta}\right)}^{\mathrm{``likelihood"}}\overbrace{p\left(\vec{\theta}\right)}^{\mathrm{``prior"}}}{\underbrace{p\left(\vec{X}\right)}_{\mathrm{``evidence"}}} ,
    \label{eqn: Bayes theorem}
    \end{equation}
    where $\vec{\theta}$ are the model parameters and $\vec{X}$ are the observed random variables (i.e. the data/evidences).  Conceptually, the ``evidence'' is the total probability of the data $\vec{X}$ regardless of any other information.  The ``prior'' represents what we know about the parameters $\vec{\theta}$ before we see the data, and the ``likelihood'' represents what we have learned from the observations in our experiment.  In this work we will use the notation $p\left(\vec{X}\middle|\vec{\theta}\right) \equiv \mathscr{L}\left(\vec{X}\middle|\vec{\theta}\right)$ to help more quickly differentiate between posterior and the likelihood.
    
    Often when applying Bayesian statistics to a model the evidence is something that is unknown.  However, since it is only a normalizing constant (which normalizes integration of the posterior to $1$), the actual shape and ``information'' contained in the posterior distribution comes entirely from the likelihood and the prior.  This is why it is often written as shorthand that:
    \begin{equation*}
        \mathrm{posterior} \propto \mathrm{likelihood}\times\mathrm{prior} .
    \end{equation*}
    Once we have the posterior distribution we can normalize it numerically in the end if we desire, hence often we don't have a direct need to compute $p\left(\vec{X}\right)$ ahead of time.
    
    Typically we find ourselves performing an experiment where we record observations of some quantity.  This quantity we try to predict ahead of time by mathematically describing it using a model which may be dependent on a number of parameters, our $\vec{\theta}$.  However, we often also do not know the true values of the parameters governing the thing we are observing in our experiment.  So the problem, then, is to figure out how to infer and estimate these parameters $\vec{\theta}$ given a set of observation data $\vec{X}$ from our experiment, and this can be done through a Fisher matrix analysis.
    
    Now, the key behind the Fisher matrix and this problem is rooted in how we would expect our experiment to behave if we were we able to repeat the experiment over and over for many realizations (tending towards \textit{infinity}).  From Bayes' Theorem we see that the posterior and the likelihood are directly proportional to each other, and in this so called ``high statistical limit'' or ``asymptotic limit'' of repeating our experiment many times we expect the posterior will tend to depend less and less on the prior information.

    Moreover, often the tendency is that asymptotically the posterior will become Gaussian distributed~\citep{bernstein}.  These ideas come from the Bernstein-von Mises theorem, and require certain assumptions such as Cromwell's rule~\citep{cromwell}, which says that the prior probability on a parameter should never be fully $0$ or $1$.  The rationale on this assumption is that if the parameter is fixed absolutely at always being observed or never being observed, then the posterior can't be influenced by the actual data and observations taken into the model described by the likelihood.
    
    The Fisher matrix is by no means a perfect estimator, and this type of analysis does have its limitations which must be considered.  One is the role of the previously mentioned prior information.  It is not universally true that in the high asymptotic limit the prior information will have \textit{no} influence on the resulting posterior.  In fact, the prior can be quite informative.  One key idea is that in the region of interest of parameter space we are focusing on (near the true values), the prior information should remain approximately constant/unchanging~\citep{michele_fisher}.  This is because, as will be shown in the next sections, the Fisher matrix is effectively a second-order Taylor expansion of the likelihood function which is being used to approximate the posterior.  Because this quadratic approximation describes the local ``curvature'' in parameter space about the true parameter values, the Fisher matrix should remain decent so long as the curvature of the prior is less than the curvature of the likelihood.
    
\begin{figure}
    \centering
      \begin{subfigure}[t]{0.48\linewidth}
      \centering
        \includegraphics[width=1\linewidth]{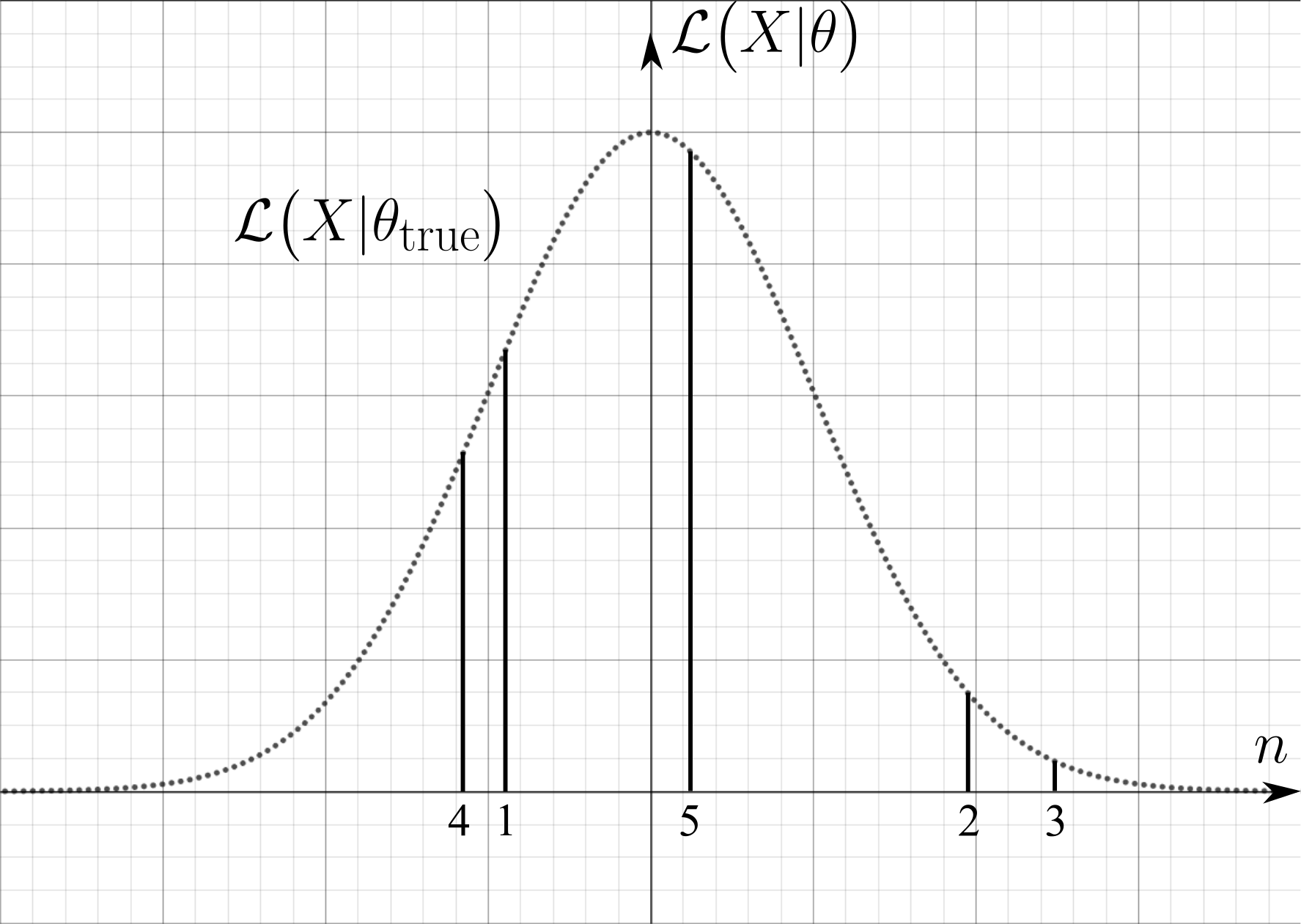}
        \caption{If we happen to choose the correct parameter parameter $\theta$ in our model, then the likelihood distribution will peak for zero noise.  Every time we repeat the experiment the noise will be different, so we will compute a different likelihood, as what is being exemplified by the first five iterations of our experiment here.  Eventually, if we repeat this experiment many many times, we will see the full distribution's shape.}
        \label{fig: MLE gaussian 1}
      \end{subfigure}
      \hfill
      \begin{subfigure}[t]{0.48\linewidth}
      \centering
        \includegraphics[width=1\linewidth]{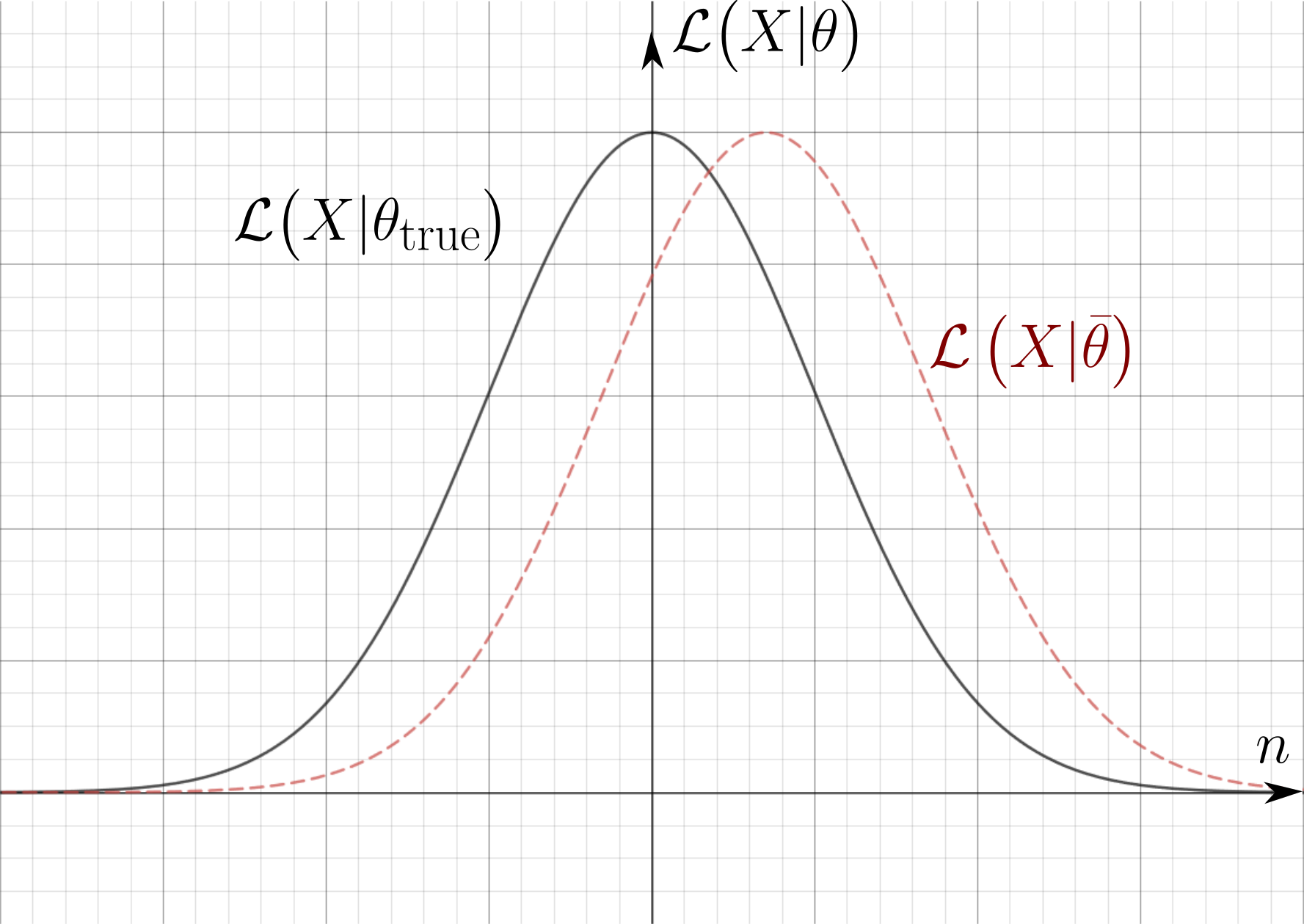}
        \caption{If we choose a different parameter $\Bar{\theta}\neq\theta_\mathrm{true}$, then the distribution will peak for some non-zero value of the noise.}
        \label{fig: MLE gaussian 2}
      \end{subfigure}
    \caption[Likelihood Function Visualization]{The likelihood function $\mathscr{L}\left(X\middle|\theta\right)$ gives us a ``goodness-of-fit'' distribution of our data $X$ for a given set of parameter $\theta$ in our model.  The data is a random variable since it is a true ``signal'' plus some random noise $n$.}
    \label{fig: MLE visualization}
    \end{figure}

    So with these ideas in mind we are effectively looking for a maximum likelihood estimation (MLE), that is a way of predicting the values of $\vec{\theta}$ in parameter space which maximize our likelihood.  But something important to always remind ourselves of (which we emphasize explicitly here) is that our observations $\vec{X}$ are random variables (R.V.s).  Now what we expect is that the observations contain some ``true'' value produced by nature (the thing we are really \textit{trying} to observe), combined with some inherent noise (also produced by nature... thanks, nature).  In other words $\vec{X} = \vec{X}_\mathrm{true} + \vec{n}$, where $\vec{n}$ is the noise in each measurement.  \textit{This} is what makes our observation an R.V. - here $\vec{X}$ is an R.V. because $\vec{n}$ is an R.V.  What this means is that when we perform an experiment, even if we were using the correct model with the true parameters for the thing being observed, $\vec{\theta}_\mathrm{true}$, the likelihood won't necessarily be at its maximum value.  If the noise is exactly zero, then the likelihood is indeed maximized, but every time we repeat the experiment there will be different random noise added into the observation (as shown in Figure~\ref{fig: MLE gaussian 1}).  The random noise spreads the likelihood out into a distribution in terms of the noise.  If we happen to pick the wrong set of parameters, some $\Bar{\theta} \neq \theta_\mathrm{true}$, then the distribution will shift and peak for some non-zero value of the noise, as shown in Figure~\ref{fig: MLE gaussian 2}.  This is why we must treat the problem statistically, and consider what is asymptotically the most likely set of parameters for our model, and what is our statistical uncertainty of their measurements?
    
%---------------------------------------------------------------------------------
%---------------------------------------------------------------------------------
    \section{Conceptually Understanding the Fisher Information}\label{sec: Conceptually Understanding the Fisher Information}
    
    In order to focus on motivating all of the logic and ideas which go into understanding and deriving the Fisher matrix~\citep{darkenergy_fisher}, we consider here first on a very simple example where we imagine that we perform an experiment which collects one datum $X_\mathrm{obs}$ with Gaussian uncertainty $\sigma$, and our functional model for this observation only depends on one parameter $X(\theta)$.  The ideas here will extend to a fully multidimensional data and parameter-space, but for the moment we keep it simple so as to focus entirely on the logic.
    
    We begin by defining a Chi-squared test as one way of trying to quantify a sense of the ``goodness-of-fit'' of our model (given its parameter) to the observed data:
    \begin{equation}
        \chi^2 \equiv \frac{\Tilde{X}^2}{\sigma^2} = \frac{\Big(X(\theta) - X_\mathrm{obs}\Big)^2}{\sigma^2} ,
    \end{equation}
    where $X_\mathrm{obs} = X_\mathrm{true} + n$ with inherent noise $n$.  If this model and its parameters is correct and really does describe the physics of our experiment, then the likelihood should be a Gaussian, which can be written in terms of our Chi-squared expression:
    \begin{equation}
        \mathscr{L}\left(X\middle|\theta\right) = \frac{1}{\sqrt{2\pi}\sigma}\exp\left[-\frac{1}{2}\chi^2\right] .
    \label{eqn: 1D likelihood(X|theta)}
    \end{equation}
    Now as we discussed in Section~\ref{sec:high stat limit of Bayes}, our goal is to infer the parameters $\theta$ from our observation $X_\mathrm{obs}$.  Due to the inherent noise in our observations, $X_\mathrm{obs}$ is an R.V., which means that simply minimizing the Chi-squared value, i.e. maximizing our likelihood, for this one realization of our experiment won't necessarily give us the ``true'' parameters.  This would only happen if the noise were \textit{exactly} zero, otherwise the noise pushes our observation away from the true $X_\mathrm{true}$ (see Figure~\ref{fig: MLE gaussian 1}).
    
    From Bayes' theorem equation~\ref{eqn: Bayes theorem} we know the posterior is proportional to the likelihood, $p\left(\theta\middle|X\right) \propto \mathscr{L}\left(X\middle|\theta\right)$.  Additionally we may or may not know prior information about our parameter which we could supply to help inform our experiment.  As mentioned already in Section~\ref{sec:high stat limit of Bayes}, in the approximation we are about to make we are sort of assuming that the prior is not very informative over the small region of parameter space that we are going to focus in on.  Therefore let's focus on the high statistical limit of just our Chi-squared value in the likelihood - what is the value we \textit{expect} it to be as we repeat the experiment ad infinitum?  Mathematically, we are going to solve for the expectation value $\left\langle \chi^2 \right\rangle$ in order to ask what is the expected likelihood.
  
    In addition to this, however, we are also interested in focusing our attention on what is the expectation of our Chi-squared and likelihood near the true parameter $\theta_\mathrm{true}$, the true value of $\theta$ which we are trying to infer from our data.  Hence first we Taylor expand Chi-squared for small $\Delta \theta = \theta - \theta_\mathrm{true}$ about the true parameter as:
    \begin{equation}
        \chi^2 =\quad \chi^2\bigg\rvert_{\theta_\mathrm{true}} \quad+\quad \frac{d}{d\theta}\left[\chi^2\right]\bigg\rvert_{\theta_\mathrm{true}}\Delta \theta \quad+\quad \frac{1}{2}\frac{d^2}{d\theta^2}\left[\chi^2\right]\bigg\rvert_{\theta_\mathrm{true}}\Delta \theta^2 \quad+\quad \mathcal{O}\left(\Delta \theta^3\right) .
    \label{eqn: chi-squared taylor}
    \end{equation}
    Next we want to take the expectation value of this.  As a reminder, the expectation value is over our R.V., which here is $X_\mathrm{obs}$ or $n$ (or $\Tilde{X}$).  Since we are treating these continuously, the expectation value weights the argument against the probability density we have, which in this case is our likelihood.  Furthermore, we are going to take the expectation value \textit{evaluated at the true parameter} $\theta_\mathrm{true}$.  This is because again, we want to focus our attention on what happens to Chi-squared near the true parameter over an (ideally) infinite number of experiment iterations (see Figure~\ref{fig: MLE gaussian 2}).  So let the following notation mean:
    \begin{align}
        \langle \ \cdot \ \rangle = \int\limits^\infty_{-\infty} \left[ \ \cdot \ \right] \ \ \mathscr{L}\left(X\middle|\theta_\mathrm{true}\right) dX_\mathrm{obs} &= \int\limits^\infty_{-\infty} \left[ \ \cdot \ \right] \ \ \frac{1}{\sqrt{2\pi}\sigma}\exp\left[-\frac{1}{2}\chi\left(X_\mathrm{obs},\theta_\mathrm{true}\right)^2\right] dX_\mathrm{obs} , \nonumber \\
        &= \int\limits^\infty_{-\infty} \left[ \ \cdot \ \right] \ \ \frac{1}{\sqrt{2\pi}\sigma}\exp\left[-\frac{1}{2}\chi\left(n,\theta_\mathrm{true}\right)^2\right] dn , \nonumber \\
        &= \int\limits^\infty_{-\infty} \left[ \ \cdot \ \right] \ \ \frac{1}{\sqrt{2\pi}\sigma}\exp\left[-\frac{1}{2}\chi\left(\Tilde{X}\left(\theta_\mathrm{true}\right)\right)^2\right] d\!\!\left[\Tilde{X}\left(\theta_\mathrm{true}\right)\right] ,
    \end{align}
    where the three integral expressions here are all equivalent since they are all directly related to each other through a simple change of variables given their definitions.  So now we can take the expectation value of equation~\ref{eqn: chi-squared taylor} and write:
    \begin{equation}
        \left\langle\chi^2\right\rangle = \ \ \left\langle\chi^2\bigg\rvert_{\theta_\mathrm{true}}\right\rangle \ \ +\ \ \left\langle\frac{d}{d\theta}\left[\chi^2\right]\bigg\rvert_{\theta_\mathrm{true}}\right\rangle\Delta \theta \ \ +\ \ \frac{1}{2}\left\langle\frac{d^2}{d\theta^2}\left[\chi^2\right]\bigg\rvert_{\theta_\mathrm{true}}\right\rangle\Delta \theta^2 \ \ +\ \ \mathcal{O}\left(\Delta \theta^3\right) .
    \end{equation}
    Working this out piece-by-piece we have for the first two terms:
    \begin{align}
        \left\langle\chi^2\bigg\rvert_{\theta_\mathrm{true}}\right\rangle \hspace{0.85cm}&= \int\limits^\infty_{-\infty} \left[\frac{\Tilde{X}\left(\theta_\mathrm{true}\right)^2}{\sigma^2}\right] \frac{1}{\sqrt{2\pi}\sigma}\exp\left[-\frac{1}{2}\chi\left(\theta_\mathrm{true}\right)^2\right] d\!\!\left[\Tilde{X}\left(\theta_\mathrm{true}\right)\right] , \nonumber \\
        \hspace{0.85cm}&= \frac{1}{\sigma^2}\frac{1}{\sqrt{2\pi}\sigma}\int\limits^\infty_{-\infty} \Tilde{X}\left(\theta_\mathrm{true}\right)^2 \exp\left[-\frac{1}{2}\frac{\Tilde{X}\left(\theta_\mathrm{true}\right)^2}{\sigma^2}\right] d\!\!\left[\Tilde{X}\left(\theta_\mathrm{true}\right)\right] , \nonumber \\
        &= \frac{1}{\sigma^2}\frac{1}{\sqrt{2\pi}\sigma}\sqrt{2\pi}\sigma^3 \equiv 1 ,
    \end{align}
    \begin{align}
        \left\langle\frac{d}{d\theta}\left[\chi^2\right]\bigg\rvert_{\theta_\mathrm{true}}\right\rangle &= \int\limits^\infty_{-\infty} \left[\frac{2}{\sigma^2}\Tilde{X}\left(\theta_\mathrm{true}\right)\frac{\partial}{\partial\theta}\left[\Tilde{X}\right]\bigg\rvert_{\theta_\mathrm{true}}\right] \frac{1}{\sqrt{2\pi}\sigma}\exp\left[-\frac{1}{2}\chi\left(\theta_\mathrm{true}\right)^2\right] d\!\!\left[\Tilde{X}\left(\theta_\mathrm{true}\right)\right] , \nonumber \\
        &= \frac{2}{\sigma^2} \frac{\partial}{\partial\theta}\Big[X\left(\theta_\mathrm{true}\right)\Big] \frac{1}{\sqrt{2\pi}\sigma}  \int\limits^\infty_{-\infty} \Tilde{X}\left(\theta_\mathrm{true}\right) \exp\left[-\frac{1}{2}\frac{\Tilde{X}\left(\theta_\mathrm{true}\right)^2}{\sigma^2}\right] d\!\!\left[\Tilde{X}\left(\theta_\mathrm{true}\right)\right] , \nonumber \\
        &\equiv 0 .
    \label{eqn: first order term Taylor expansion of expectation of Chi-squared 1D}
    \end{align}
    The result of this shows us that the first non-zero terms of our expectation of Chi-squared near the true parameter value is:
    \begin{align}
        \left\langle\chi^2\right\rangle \quad &=\quad 1 \quad+\quad F \Delta\theta^2 \quad+\quad \mathcal{O}\left(\Delta \theta^3\right) , \nonumber \\
        &\mathrm{where}\quad F = F\left(\theta_\mathrm{true}\right) \equiv \frac{1}{2}\left\langle\frac{d^2}{d\theta^2}\left[\chi^2\right]\bigg\rvert_{\theta_\mathrm{true}}\right\rangle .
    \label{eqn: Taylor expansion of expectation Chi-squared}
    \end{align}
    We can make a conceptual understanding of this in the following way.  When there is no noise, Chi-squared is zero at $\theta_\mathrm{true}$.  However, the average or expected value of Chi-squared evaluated at the true parameter $\left\langle\chi^2\big\rvert_{\theta_\mathrm{true}}\right\rangle$ when weighted against the likelihood probabilities of $\chi^2$ which are \textit{also} evaluated at the true parameter (this is important) is $1$.  Next looking locally around $\theta_\mathrm{true}$ for small deviations $\mathcal{O}\left(\Delta\theta\right)$, we find the that linear term $\Delta \theta$ vanishes, which tells us we are at an extremum here in our parameter space.  However, the quadratic $\Delta\theta^2$ term does not vanish and can be computed, and since it is quadratic this quantity tells us the local curvature around our extremum.  We call this quantity the ``Fisher information,'' denoted by $F$, and it is a value which depends on the true parameter $\theta_\mathrm{true}$.
    
    So we have used the likelihood evaluated at the true parameter $\theta_\mathrm{true}$ to calculate the expectation value of our model's Chi-squared for small deviations off of the true parameter.  This is our high statistical limit, and now we take that and use it to give us our \textit{new high-statistical limit likelihood}, and hence also our high statistical limit approximation of our posterior:
    \begin{align}
        p\left(\theta\middle|X\right) \propto \mathscr{L}\left(X\middle|\theta\right) = \textsf{C}\exp\left[-\frac{1}{2}\left\langle \chi^2 \right\rangle\right] &= \textsf{C}\exp\left[-\frac{1}{2} \Big(1 + F\Delta\theta^2\Big) \right] , \nonumber \\
        &= \Bar{\textsf{C}} \exp\left[-\frac{1}{2} F\Delta\theta^2\right] ,
    \label{eqn: high stat limit - the new likelihood and posterior}
    \end{align}
    where notice that $\textsf{C}$ is our new normalization constant for this probability density (here we just haven't computed it explicity), and $\Bar{\textsf{C}}$ has simply absorbed the additional factor of $\exp\left[-\frac{1}{2}\right]$ which is just a constant.  Since this is a Gaussian still, the new normalization constant would be $\Bar{\textsf{C}} = \sqrt{\frac{F}{2\pi}}$.  Note that in this asymptotic approximation, we have dropped $\mathcal{O}\left(\Delta\theta^3\right)$ terms.
    
    Consider what we have done here.  The posterior is a distribution on the parameters (given the data), while the likelihood is a distribution of the data (given the parameters).  However, through Bayes' theorem and a high statistical limit, averaging over realizations of the data itself, we now view the likelihood as a function of the \textit{parameters}.  The result is a Gaussian dependent on this new Fisher information, which is itself a function of the true parameter values.  So the Fisher analysis takes us from a Gaussian likelihood in data-space to a Gaussian posterior in parameter-space.  Looking at equation~\ref{eqn: high stat limit - the new likelihood and posterior}, the inverse of that Fisher information is the variance on $\Delta\theta$, that is $F^{-1} = \sigma^2$.  So inverting the Fisher information here would tell us how large our uncertainty is on the parameter $\theta$, centered on the true value of $\theta_\mathrm{true}$.
    
    The final crucial key to this comes from the Cram\'er-Rao theorem~\citep{cramer-rao}, which states that the variance produced by some unbiased estimator will be at least as high as the inverse of this Fisher information.  Or in other words, the Fisher information gives the ``best-case-scenario'' for how well we can measure the parameter in our model, which is what makes it such an important quantity.
    
    To get equation~\ref{eqn: high stat limit - the new likelihood and posterior} we took the high-statistical limit of Chi-squared about our true parameter, then used that to define our new likelihood in the high statistical limit.  Note though we can cast the definition of the Fisher information in a more general form rather than the form given in equation~\ref{eqn: Taylor expansion of expectation Chi-squared} as follows.  Start with the likelihood function equation~\ref{eqn: 1D likelihood(X|theta)}.  Then take the negative second derivative of the natural log of that quantity:
    \begin{align*}
        -\frac{\partial^2}{\partial\theta^2}\ln{\mathscr{L}\left(X\middle|\theta\right)} &= -\frac{\partial^2}{\partial\theta^2}\left[\ln{\frac{1}{\sqrt{2\pi}\sigma}} - \frac{1}{2}\chi^2\right] , \nonumber \\
        &=\frac{1}{2}\frac{\partial^2}{\partial\theta^2}\chi^2 .
    \end{align*}
    Next evaluate this expression at $\theta=\theta_\mathrm{true}$, take the expectation value of both sides to integrate over the random variable, and drop $\mathcal{O}\left(\Delta\theta^3\right)$ terms:
    \begin{align*}
        -\left\langle \frac{\partial^2}{\partial\theta^2}\ln{\mathscr{L}\left(X\middle|\theta_\mathrm{true}\right)} \right\rangle &=\frac{1}{2}\frac{\partial^2}{\partial\theta^2} \left\langle\chi^2\right\rangle , \nonumber \\
        &=\frac{1}{2}\frac{\partial^2}{\partial\theta^2} \left[1 + F\Delta\theta^2 + \mathcal{O}\left(\Delta\theta^3\right) \right] , \nonumber \\
        &=\frac{1}{2}\frac{\partial^2}{\partial\theta^2} \left[1 + F\left(\theta - \theta_\mathrm{true}\right)^2 + \mathcal{O}\left(\Delta\theta^3\right) \right] , \nonumber \\
        &\approx F .
    \end{align*}
    Hence we can see that, assuming again we are considering sufficiently small deviations away from the true parameter $\Delta\theta = \theta - \theta_\mathrm{true}$, in addition to equation~\ref{eqn: Taylor expansion of expectation Chi-squared}, we can say our Fisher information is:
    \begin{equation}
        F = -\left\langle \frac{\partial^2}{\partial\theta^2}\ln{\mathscr{L}\left(X\middle|\theta_\mathrm{true}\right)} \right\rangle .
    \label{eqn: Fisher information "definition"}
    \end{equation}
    Most sources say this is the definition of the Fisher information, which we can choose it to be, but the logic has this built in assumption regarding small $\Delta\theta$ which we want to point out from our logic and motivated derivation here.

%---------------------------------------------------------------------------------
%---------------------------------------------------------------------------------
    \section{Generalizing from the Fisher Information to the Fisher Matrix}
    
    Now that we have the conceptual understanding of the ideas underlying the Fisher information, here we present the fully generalized version.  Before we considered one datum $X_\mathrm{obs}$ and a functional model to describe it which only depended on one parameter $\theta$.  From our analysis, we found our ``Fisher information'' was a scalar value that described the local curvature in our 1-D parameter space around the true parameter.
    
    However, now suppose we collect many data, represented by a $d$-dimensional vector $\vec{X}_\mathrm{obs}$, each with Gaussian $\sigma$ uncertainties, and construct a model for that data which depends on multiple parameters, represented by a $k$-dimensional vector $\vec{\theta}$.  In our work and derivation here we are assuming that there are no covariances between observed data measurements - while each data measurement may carry its own uncertainty, it is not correlated with another data measurement.  Finally, in our notation here, we will use indices $(a,b,c)$ in our data space, and indices $(i,j)$ in our parameter space.  We will use both vector/matrix notation and Einstein index notation (and implied summation conventions) in our work below.
    
    As we did before, let's define a multidimensional goodness-of-fit Chi-squared quantity for this observed data set and our model:
    \begin{align}
        \chi^2 &\equiv \vec{\Tilde{X}}^T \bm{\Sigma}^{-1} \vec{\Tilde{X}} = \Sigma^{-1}_{ab}\Tilde{X}^a\Tilde{X}^b = \frac{1}{\sigma^2_c}\left(\Tilde{X}^c\right)^2 = \frac{1}{\sigma^2_c}\bigg(X^c\left(\vec{\theta}\right) - X^c_\mathrm{obs}\bigg)^2 , \label{eqn: chi squared multivariate with variance matrix} \\
        &\mathrm{where}\quad \Sigma^{-1}_{ab} \equiv \frac{1}{\sigma^2_c}\delta^c_{ab} , \nonumber \\
        &\mathrm{and}\hspace{0.7cm} \delta^c_{ab} \equiv\begin{cases} 1, \quad a=b=c \\ 0, \quad\text{else} \end{cases} \nonumber
    \end{align}
    Here $\bm{\Sigma}$ is our covariance matrix for the data observations, however, as we noted already we are assuming here there are no covariances between data measurements (this is really just a ``variance matrix'').  Note that our definition of the delta notation here is just a clever way of expressing that $\bm{\Sigma} = \mathrm{diag}\left(\sigma^2_1, \sigma^2_2, \dots, \sigma^2_d \right)$ is just a diagonal matrix where each element on the diagonal can be a different value.  We still note that $\vec{X}_\mathrm{obs} = \vec{X}_\mathrm{true} + \vec{n}$, where the noise makes this a multidimensional R.V.  Again, if our model is correct and does truly describe the physics of our experiment, then we expect the likelihood should be a (now $d$-dimensional multivariate) Gaussian, and written in terms of this Chi-squared quantity we can express it as:
    \begin{equation}
        \mathscr{L}\left(\vec{X}\middle|\vec{\theta}\right) = \frac{1}{\sqrt{(2\pi)^d\mathrm{det}(\bm{\Sigma})}}\exp\left[-\frac{1}{2}\chi^2\right] ,
    \label{eqn: DD likelihood(X|theta)}
    \end{equation}
    where $\mathrm{det}(\bm{\Sigma}) = \displaystyle\prod_{c}^{d} \sigma_c^2 = \left(\displaystyle\prod_{c}^{d} \sigma_c \right)^2$.  Following the same logic trail as we did before, now we perform a $k$-dimensional Taylor expansion of Chi-squared about $\vec{\theta}_\mathrm{true}$ which looks like:
    \begin{equation}
        \chi^2 =\ \  \chi^2\bigg\rvert_{\vec{\theta}_\mathrm{true}} \ \ +\ \  \frac{\partial}{\partial\theta_i}\left[\chi^2\right]\bigg\rvert_{\vec{\theta}_\mathrm{true}}\Delta \theta^i \ \ +\ \  \frac{1}{2}\frac{\partial^2}{\partial\theta_i\theta_j}\left[\chi^2\right]\bigg\rvert_{\vec{\theta}_\mathrm{true}}\Delta\theta^i\Delta\theta^j \ \ +\ \  \mathcal{O}\left(\Delta \theta^3\right) ,
    \label{eqn: multidimensional chi-squared taylor}
    \end{equation}
    and then take the expectation value of this expression, where the expectation value now looks like:
    \begin{align}
        \langle \ \cdot \ \rangle &= \int\limits^\infty_{-\infty} \left[ \ \cdot \ \right] \ \ \mathscr{L}\left(\vec{X}\middle|\vec{\theta}_\mathrm{true}\right) d^dX_\mathrm{obs} , \nonumber \\
        &= \int\limits^\infty_{-\infty} \left[ \ \cdot \ \right] \ \ \frac{1}{\sqrt{(2\pi)^d\mathrm{det}(\bm{\Sigma})}}\exp\left[-\frac{1}{2}\chi\left(\vec{\Tilde{X}}\left(\vec{\theta}_\mathrm{true}\right)\right)^2\right] d^d\!\!\left[\Tilde{X}\left(\vec{\theta}_\mathrm{true}\right)\right] .
    \end{align}
    Remember, we are still evaluating the expectation value at the \textit{true} parameter values $\vec{\theta}_\mathrm{true}$. The resulting integrals are certainly uglier, but keeping careful collection of like terms they all work out in the same way as before.  The zeroth order term looks like: \\
    
    \begin{align}
        \left\langle\chi^2\bigg\rvert_{\vec{\theta}_\mathrm{true}}\right\rangle &= \int\limits^\infty_{-\infty} \left[\frac{\left(\Tilde{X}^c\left(\vec{\theta}_\mathrm{true}\right)\right)^2}{\sigma^2_c}\right] \frac{1}{\sqrt{(2\pi)^d\mathrm{det}(\bm{\Sigma})}}\exp\left[-\frac{1}{2}\frac{\left(\Tilde{X}^c\left(\vec{\theta}_\mathrm{true}\right)\right)^2}{\sigma^2_c}\right] d^d\!\!\left[\Tilde{X}\left(\vec{\theta}_\mathrm{true}\right)\right] , \nonumber \\
        &= \int\limits^\infty_{-\infty} \frac{\left(\Tilde{X}^c\left(\vec{\theta}_\mathrm{true}\right)\right)^2}{\sigma^2_c} \frac{1}{\sqrt{2\pi}\sigma_c}\exp\left[-\frac{1}{2}\frac{\left(\Tilde{X}^c\left(\vec{\theta}_\mathrm{true}\right)\right)^2}{\sigma^2_c}\right] d\!\!\left[\Tilde{X}^c\left(\vec{\theta}_\mathrm{true}\right)\right]  \nonumber \\
        & \qquad \times \ \ \int\limits^\infty_{-\infty}  \frac{1}{\sqrt{(2\pi)^{d-1}\displaystyle\prod_{a}^{d-1} \sigma_a^2}}\exp\left[-\frac{1}{2}\frac{\left(\Tilde{X}^a\left(\vec{\theta}_\mathrm{true}\right)\right)^2}{\sigma^2_a}\right] d^{d-1}\!\!\left[\Tilde{X}\left(\vec{\theta}_\mathrm{true}\right)\right] , \nonumber \\
        &= 1 \times 1 = 1 .
    \label{eqn: expectation chi squared multivariate taylor first term}
    \end{align}
    Here we have separated out like terms, the $\Tilde{X}^c$ integral from the rest of the $\Tilde{X}^a$ for $a\neq c$ integrals (so note that the implied sum doesn't apply to the $c$ index in the third to last line above).  As for the first order term, we can see that when we take the first partial:
    \begin{equation*}
        \frac{\partial}{\partial\theta_i}\left[\chi^2\right] = 2\frac{\Tilde{X}^c}{\sigma^2_c}\frac{\partial X^c}{\partial\theta_i} = \left(\frac{2}{\sigma^2_c}\frac{\partial X^c}{\partial\theta_i}\right)\Tilde{X}^c ,
    \end{equation*}
    we will end up with an odd Gaussian in $\Tilde{X}^c$ like we did in equation~\ref{eqn: first order term Taylor expansion of expectation of Chi-squared 1D}, which will kill off the first order term (as we would expect, if we are sitting at an extremum like we found in the 1D case).  So our result now looks like:
    \begin{align}
        \left\langle\chi^2\right\rangle \quad &=\quad 1 \quad+\quad F_{ij} \Delta\theta^i\Delta\theta^j \quad+\quad \mathcal{O}\left(\Delta \theta^3\right) , \nonumber \\
        &\mathrm{where}\quad F_{ij} = F_{ij}\left(\vec{\theta}_\mathrm{true}\right) \equiv \frac{1}{2}\left\langle\frac{\partial^2}{\partial\theta_i\partial\theta_j}\left[\chi^2\right]\bigg\rvert_{\vec{\theta}_\mathrm{true}}\right\rangle .
    \label{eqn: Taylor expansion of expectation Chi-squared multidimensional}
    \end{align}
    Now rather than just a scalar quantity we have a full matrix, referred to as the ``Fisher (Information) Matrix,'' whose elements are given by eqn~\ref{eqn: Taylor expansion of expectation Chi-squared multidimensional}.  Also like before, following the same steps and logic that lead up to equation~\ref{eqn: Fisher information "definition"} (remember, $\Delta\theta = \theta - \theta_\mathrm{true}$):
    \begin{align*}
        -\left\langle \frac{\partial^2}{\partial\theta_i\partial\theta_j}\ln{\mathscr{L}\left(\vec{X}\middle|\vec{\theta}_\mathrm{true}\right)} \right\rangle &=\frac{1}{2}\frac{\partial^2}{\partial\theta_i\partial\theta_j} \left\langle\chi^2\right\rangle , \\
        &=\frac{1}{2}\frac{\partial^2}{\partial\theta_i\partial\theta_j} \left[1 + F_{ab}\Delta\theta^a\Delta\theta^b + \mathcal{O}\left(\Delta\theta^3\right) \right] , \\
        &=\frac{1}{2}\frac{\partial}{\partial\theta_i} \left[F_{ab}\delta^{aj}\Delta\theta^b + F_{ab}\Delta\theta^a\delta^{bj} + \mathcal{O}\left(\Delta\theta^2\right) \right] , \\
        &=\frac{1}{2}\frac{\partial}{\partial\theta_i} \left[F_{jb}\Delta\theta^b + F_{aj}\Delta\theta^a + \mathcal{O}\left(\Delta\theta^2\right) \right] , \\
        &=\frac{1}{2}\left[F_{ji} + F_{ij} + \mathcal{O}\left(\Delta\theta^2\right) \right] , \\
        &\approx F_{ij} ,
    \end{align*}
    where in the second-to-last line we used the fact that the Fisher matrix, as we can see from equation~\ref{eqn: Taylor expansion of expectation Chi-squared multidimensional} is a symmetric matrix, so $F_{ji} = F_{ij}$.  This means that we can define (approximately as we pointed out in the previous section) that:
    \begin{equation}
        F_{ij} = -\left\langle \frac{\partial^2}{\partial\theta_i\partial\theta_j}\ln{\mathscr{L}\left(\vec{X}\middle|\vec{\theta}_\mathrm{true}\right)} \right\rangle .
    \end{equation}
    Like with what we found in equation~\ref{eqn: high stat limit - the new likelihood and posterior}, in the high-statistical limit we expect now that our posterior distribution should look like a $k$-dimensional multivariate Gaussian distribution described using the Fisher matrix as:
    \begin{align}
        p\left(\vec{\theta}\middle|\vec{X}\right) &\sim \sqrt{\frac{\mathrm{det}(\bm{F})}{(2\pi)^k}} \exp\left[-\frac{1}{2}F_{ij}\Delta\theta^i\Delta\theta^j\right] = \sqrt{\frac{\mathrm{det}(\bm{F})}{(2\pi)^k}} \exp\left[-\frac{1}{2} \Delta\vec{\theta}^T \bm{F} \Delta\vec{\theta} \right] , \nonumber \\
        &=\sqrt{\frac{1}{(2\pi)^k\mathrm{det}(\bm{C})}} \exp\left[-\frac{1}{2}C^{-1}_{ij}\Delta\theta^i\Delta\theta^j\right] = \sqrt{\frac{1}{(2\pi)^k\mathrm{det}(\bm{C})}} \exp\left[-\frac{1}{2} \Delta\vec{\theta}^T \bm{C}^{-1} \Delta\vec{\theta} \right] .
    \label{eqn: Fisher/covariance posterior}
    \end{align}
    Here $\bm{C}$ is the parameter covariance matrix, which as we can see is the inverse of the Fisher matrix, $\bm{F}^{-1}=\bm{C}$.  (In the 1D case, we found the inverse of the Fisher information was the variance, $F^{-1}=\sigma^2$, the natural generalization of that is that now we are talking in terms of matrices).  So inverting the Fisher matrix would give us all of the covariances between each of the model parameters, and hence how much uncertainty we could expect when trying to measure these parameters in an experiment.  Once again, this Fisher analysis takes us from a multivariate Gaussian likelihood in data-space to a multivariate Gaussian posterior in parameter-space.
    
    Finally, we note that all of the results presented in this generalization faithfully reduce to the 1D case (in both data and parameters) that we gave in Section~\ref{sec: Conceptually Understanding the Fisher Information}.

%---------------------------------------------------------------------------------
%---------------------------------------------------------------------------------
    \section{Two Common Fisher Matrix Forms}
    
    In the literature the Fisher Matrix is often defined\footnote{Again, this is still a sort of \textit{approximate} definition, with the small $\Delta\theta$ assumption built into it as pointed out in Section~\ref{sec: Conceptually Understanding the Fisher Information}.} in one of two forms:
    \begin{align}
        F_{ij}\left(\vec{\theta}_\mathrm{true}\right) &= -\left\langle \frac{\partial^2}{\partial\theta_i\partial\theta_j}\ln{\mathscr{L}\left(\vec{X}\middle|\vec{\theta}_\mathrm{true}\right)} \right\rangle , \nonumber \\
        &=  \Bigg\langle \left( \frac{\partial}{\partial\theta_i}\ln{\mathscr{L}\left(\vec{X}\middle|\vec{\theta}_\mathrm{true}\right)}\right) \left( \frac{\partial}{\partial\theta_j}\ln{\mathscr{L}\left(\vec{X}\middle|\vec{\theta}_\mathrm{true}\right)}\right) \Bigg\rangle.
    \label{eqn: Fisher matrix}
    \end{align}
    Here we prove that the second form of the Fisher matrix as stated here is equivalent to the first.
    
    These two forms give two ways of describing what the Fisher Matrix is.  In the first form in equation~\ref{eqn: Fisher matrix}, it is the ``expected value of the observed information,'' where the observed information is the negative Hessian matrix of the log-likelihood.  In the second form in equation~\ref{eqn: Fisher matrix}, it is the ``variance of the score,'' where the score is the gradient of the log-likelihood with respect to the parameter vector (and the variance is the expectation value of a squared deviation of an R.V.).
    
    Thinking of the Fisher matrix as the expected value of the observed information is helpful, since when deriving the Fisher matrix we emphasized how by Taylor expanding $\chi^2$ about the true parameters $\theta_\mathrm{true}$ and taking the expectation value of that also around $\theta_\mathrm{true}$, we were localizing ourselves to the behavior of the function right around its maximum, that is its curvature right around its maximum (which is effectively what our Fisher matrix is).  The Hessian matrix \textit{is} the local curvature of a function of multiple variables, hence the natural conceptual connection. \\
    
    \begin{minipage}{0.9\linewidth}
    \fbox {
        \parbox{\linewidth}{
        \underline{\textbf{Proof}}
    
        First note:
        \begin{align}
            -\frac{\partial^2}{\partial\theta_i\partial\theta_j}\ln{\mathscr{L}\left(\vec{X}\middle|\vec{\theta}\right)} &= -\frac{\partial}{\partial\theta_i}\left[\frac{\partial}{\partial\theta_j}\ln{\mathscr{L}\left(\vec{X}\middle|\vec{\theta}\right)}\right] , \nonumber \\
            &= -\frac{\partial}{\partial\theta_i}\left[\frac{1}{\mathscr{L}\left(\vec{X}\middle|\vec{\theta}\right)} \frac{\partial}{\partial\theta_j}\mathscr{L}\left(\vec{X}\middle|\vec{\theta}\right)\right] , \nonumber \\
            &= \frac{1}{\mathscr{L}\left(\vec{X}\middle|\vec{\theta}\right)^2}\left( \frac{\partial}{\partial\theta_i}\mathscr{L}\left(\vec{X}\middle|\vec{\theta}\right) \right)\left( \frac{\partial}{\partial\theta_j}\mathscr{L}\left(\vec{X}\middle|\vec{\theta}\right) \right) \nonumber \\
            &\hspace{2.5cm}-\ \  \frac{1}{\mathscr{L}\left(\vec{X}\middle|\vec{\theta}\right)}\frac{\partial^2}{\partial\theta_i\partial\theta_j}\mathscr{L}\left(\vec{X}\middle|\vec{\theta}\right) , \nonumber \\
            &= \left( \frac{\partial}{\partial\theta_i}\ln{\mathscr{L}\left(\vec{X}\middle|\vec{\theta}\right)} \right)\left( \frac{\partial}{\partial\theta_j}\ln{\mathscr{L}\left(\vec{X}\middle|\vec{\theta}\right)} \right) \nonumber \\
            &\hspace{2.5cm}-\ \  \frac{1}{\mathscr{L}\left(\vec{X}\middle|\vec{\theta}\right)}\frac{\partial^2}{\partial\theta_i\partial\theta_j}\mathscr{L}\left(\vec{X}\middle|\vec{\theta}\right) .
        \label{eqn: 2 fisher forms proof first step}
        \end{align}
        Next we take the expectation value of both sides of this equation, and consider what happens in that process to the second term on the RHS:
        \begin{align}
            \left\langle \frac{1}{\mathscr{L}\left(\vec{X}\middle|\vec{\theta}\right)}\frac{\partial^2}{\partial\theta_i\partial\theta_j}\mathscr{L}\left(\vec{X}\middle|\vec{\theta}\right) \right\rangle &= \int\limits^\infty_{-\infty} \frac{1}{\mathscr{L}\left(\vec{X}\middle|\vec{\theta}\right)}\left(\frac{\partial^2}{\partial\theta_i\partial\theta_j}\mathscr{L}\left(\vec{X}\middle|\vec{\theta}\right)\right) \mathscr{L}\left(\vec{X}\middle|\vec{\theta}\right) d^nx , \nonumber
        \end{align}
        }
    }
    \end{minipage}   
        
    \begin{minipage}{0.9\linewidth}
    \fbox {
        \parbox{\linewidth}{
        \begin{align}
            &\hspace{1cm}=\frac{\partial^2}{\partial\theta_i\partial\theta_j} \int\limits^\infty_{-\infty} \mathscr{L}\left(\vec{X}\middle|\vec{\theta}\right) d^nx , \nonumber \\
            &\hspace{1cm}= \frac{\partial^2}{\partial\theta_i\partial\theta_j} (1) \equiv 0 .
        \end{align}
        This leaves us with the following expression after taking the expectation value of equation~\ref{eqn: 2 fisher forms proof first step}:
        \begin{equation*}
            -\left\langle \frac{\partial^2}{\partial\theta_i\partial\theta_j}\ln{\mathscr{L}\left(\vec{X}\middle|\vec{\theta}\right)} \right\rangle = \left\langle \left( \frac{\partial}{\partial\theta_i}\ln{\mathscr{L}\left(\vec{X}\middle|\vec{\theta}\right)} \right)\left( \frac{\partial}{\partial\theta_j}\ln{\mathscr{L}\left(\vec{X}\middle|\vec{\theta}\right)} \right) \right\rangle .
        \end{equation*}
        
        \hfill \textit{Q.E.D.}
        }
    }
    \end{minipage}\\ \\

%---------------------------------------------------------------------------------
%---------------------------------------------------------------------------------
    \section{Adding Prior Knowledge to the Fisher Matrix}\label{sec: adding prior to fisher matrix}
    
    As we stated at the start of Section~\ref{sec:high stat limit of Bayes}, the Fisher analysis makes that assumption that in the high-statistical limit of repeating our experiment many times,the effect of the prior on the posterior should become less significant.  Because we are effectively ``zooming'' in on a local region of parameter space when we perform our Taylor expansion approximation, one of the main ideas is that the local curvature of the prior should remain effectively constant, meaning only the curvature of the likelihood is what is important and contributing to the posterior.  These assumptions and this derivation then leads to our Fisher matrix, whose inverse gives us the best-case-scenario for how well we can measure the parameters in our model.  Hence the Fisher matrix itself is effectively a best-case-scenario measurement estimate of the parameters with no prior knowledge built into its derivation.
    
    In practice, however, we may be performing an experiment where we do \textit{know} additional information about the parameters (perhaps something related to their physical nature), which can help further constrain our best-case-scenario, and make it even more precise (\citeauthor{wittman_Fisher_for_beginners}).  In this case we may want to build back in some of that prior knowledge, and ``update'' our Fisher matrix approximation.  To do this we can choose to take our high-statistical likelihood and multiply by an additional prior.  If that prior knowledge is Gaussian distributed, we can express our updated Fisher matrix posterior as:
    \begin{align}
        p\left(\vec{\theta}\middle|\vec{X}\right) &\sim \textsf{C} \exp\left[-\frac{1}{2} \Delta\vec{\theta}^T \bm{F} \Delta\vec{\theta} \right] \times \exp\left[-\frac{1}{2} \Delta\vec{\theta}^T \textbf{Cov}^{-1} \Delta\vec{\theta} \right] , \nonumber \\
        &= \textsf{C} \exp\left[-\frac{1}{2} \Delta\vec{\theta}^T \Big(\bm{F} + \textbf{Cov}^{-1}\Big) \Delta\vec{\theta} \right],
    \label{eqn: Fisher/covariance posterior with added prior}
    \end{align}
    where $\textbf{Cov}$ is some new known prior covariance matrix, and $\textsf{C}$ is the new normalizing factor.  So in an analysis, if we have the Fisher matrix for our model and experiment, we can include additional prior constraints by adding to the Fisher matrix the inverted Covariance matrix of our prior.  Inverting this new Fisher matrix, $\bm{F}^{\bm{'}} = \Big(\bm{F} + \textbf{Cov}^{-1}\Big)$, would give us even more constrained covariances on our experiment's parameters, $\textbf{Cov}^{\bm{'}} = \Big(\bm{F} + \textbf{Cov}^{-1}\Big)^{-1}$.
    
    In reality, often we may find that from prior knowledge we can constrain some of our parameters, but not all of them.  This means we cannot write a covariance matrix $\textbf{Cov}$ with the same dimensions as our Fisher matrix $\bm{F}$, which can  be inverted $\textbf{Cov}^{-1}$, because some of the columns/rows would contain only zeros.  Hence we can't simply write down equation~\ref{eqn: Fisher/covariance posterior with added prior}.  Instead in practice we simply add the covariances directly to the appropriate elements of the Fisher matrix.

%---------------------------------------------------------------------------------
%---------------------------------------------------------------------------------
    \section{The Timing Residual Fisher Matrix}
    
    Now let's apply this Fisher matrix analysis to our timing residual model, in order to predict how well pulsar timing experiments will be able to constrain the model parameters of our different timing models (from Section~\ref{sec: four pulsar timing regimes}).  Once again, since the Fisher matrix is the inverse of the parameter covariance matrix, $\bf{F} = \bf{C}^{-1}$, inverting the Fisher matrix gives us the covariances of all of our parameters, which tells us how well we should be able to measure them experimentally.
    
    Let our observed pulsar timing residual data (indexed by $a$, $b$, and $c$ below) be represented by a $d$-dimensional R.V., $\overrightarrow{\mathrm{Res}} = \left\{ \mathrm{Res}_c : c=1,2,\ldots d \right\}$.  As we did previously, let the parameters of the timing residual model in Table~\ref{tab: model regime degeneracies} (indexed by $i,j$) be described by a $k$-dimensional $\vec{\theta}$.  Therefore the true timing residual we are looking for is burried in random Gaussian noise, $\overrightarrow{\mathrm{Res}} = \overrightarrow{\mathrm{Res}}_\mathrm{true} + \vec{n}$, which is what makes the observation an R.V.  The timing data for every pulsar at every observation can then be modeled as a function of $\vec{\theta}$ by choosing a timing residual model from Section~\ref{sec: four pulsar timing regimes}, organized into the vector $\overrightarrow{\mathrm{Res}}\left(\vec{\theta}\right)$.\footnote{Note that in our studies, we will simulate our own data sets rather than use real data.  Therefore we must ``inject'' a source by simulating our own ``true'' timing residual.  This is done by taking our desired injection parameters $\theta_\mathrm{true}$ and inputting them into our desired model function, that is $\overrightarrow{\mathrm{Res}}_\mathrm{true} = \overrightarrow{\mathrm{Res}}\left(\vec{\theta}_\mathrm{true}\right)$.  The goal is then to see how well we can recover our injected parameters.}  Let's assume our measurements of the timing residual have uncorrelated Gaussian timing uncertainties $\sigma$, with covariance matrix given by $\bm{\Sigma} = \mathrm{diag}\left(\sigma^2_1, \sigma^2_2, \dots, \sigma^2_d \right)$, as shown in equation~\ref{eqn: chi squared multivariate with variance matrix}.
    
    Once again using implied Einstein summation notation as we did earlier, our Chi-squared goodness-of-fit quantity can be written as:
    \begin{equation}
        \chi^2 \equiv \overrightarrow{\widetilde{\mathrm{Res}}}^T \bm{\Sigma}^{-1} \overrightarrow{\widetilde{\mathrm{Res}}} = \Sigma^{-1}_{ab}\widetilde{\mathrm{Res}}^a\widetilde{\mathrm{Res}}^b = \frac{1}{\sigma^2_c} \left(\widetilde{\mathrm{Res}}^c\right)^2 = \frac{1}{\sigma^2_c}\bigg( \mathrm{Res}^c\left(\vec{\theta}\right) - \mathrm{Res}^c \bigg)^2 ,
    \end{equation}
    where $\overrightarrow{\widetilde{\mathrm{Res}}} \equiv \overrightarrow{\mathrm{Res}}\left(\vec{\theta}\right) - \overrightarrow{\mathrm{Res}}$.  One of the key ideas here is that because our noise $\vec{n}$ is Gaussian distributed, and because we are modeling our data as the true signal plus this noise (with no correlations between timing data), then our timing residual data is also Gaussian distributed.  Therefore our likelihood should be a Gaussian written in terms of this Chi-squared quantity as:
    \begin{align}
        \mathscr{L}\left(\overrightarrow{\mathrm{Res}}\mid\vec{\theta}\right) &= \frac{1}{\sqrt{(2\pi)^d \mathrm{det}(\bm{\Sigma})}}\exp{\left[-\frac{1}{2} \chi^2 \right]} , \nonumber \\
        &= \frac{1}{\sqrt{(2\pi)^d\mathrm{det}(\bm{\Sigma})}}  \exp{\Bigg[-\frac{1}{2}\frac{1}{\sigma^2_c}\left( \mathrm{Res}^c\left(\vec{\theta}\right) - \mathrm{Res}^c \right)^2 \Bigg]} , \nonumber \\
        &= \frac{1}{\sqrt{(2\pi)^d\mathrm{det}(\bm{\Sigma})}}  \exp{\Bigg[-\frac{1}{2}\frac{1}{\sigma^2_c}\left( \widetilde{\mathrm{Res}}^c \right)^2 \Bigg]} .
    \label{eqn: likelihood function}
    \end{align}
    Now that we have our likelihood function defined, we need to compute the Fisher matrix.  We will use the first form of equation~\ref{eqn: Fisher matrix} for our derivation.  First note the following:
    \begin{align*}
        \ln{\mathscr{L}\left(\overrightarrow{\mathrm{Res}}\mid\vec{\theta}_\mathrm{true}\right)} &= -\frac{1}{2}\ln\left[(2\pi)^d\mathrm{det}(\bm{\Sigma})\right] - \frac{1}{2}\frac{1}{\sigma^2_c} \left(\widetilde{\mathrm{Res}}^c\right)^2 , \\
        \frac{\partial}{\partial\theta_j}\ln{\mathscr{L}\left(\overrightarrow{\mathrm{Res}}\mid\vec{\theta}_\mathrm{true}\right)} &= 0 - \frac{1}{2}\frac{1}{\sigma^2_c}2\left(\widetilde{\mathrm{Res}}^c\right)\frac{\partial\widetilde{\mathrm{Res}}^c}{\partial\theta_j} = - \frac{1}{\sigma^2_c}\left(\widetilde{\mathrm{Res}}^c\right)\frac{\partial\mathrm{Res}^c}{\partial\theta_j} , \\
        -\frac{\partial^2}{\partial\theta_i\partial\theta_j}\ln{\mathscr{L}\left(\overrightarrow{\mathrm{Res}}\mid\vec{\theta}_\mathrm{true}\right)} &= +\frac{1}{\sigma^2_c}\left[ \frac{\partial\mathrm{Res}^c}{\partial\theta_i}\frac{\partial\mathrm{Res}^c}{\partial\theta_j} + \widetilde{\mathrm{Res}}^c\frac{\partial^2\mathrm{Res}^c}{\partial\theta_i\partial\theta_j}  \right] ,
    \end{align*}
    where all of the residual functions in these expressions on the RHS are evaluated at the true set of parameters $\vec{\theta}_\mathrm{true}$ (like what is shown explicitly on the LHS).  Now we want to take the high-statistical limit and average over the R.V., which here is $\widetilde{\mathrm{Res}}\left(\vec{\theta}_\mathrm{true}\right)$, remembering that the expectation value is also evaluated at the true parameter.  So the resulting integral that we want to solve for is:
    \begin{align}
         F_{ij}\left(\vec{\theta}_\mathrm{true}\right) &= -\left\langle \frac{\partial^2}{\partial\theta_i\partial\theta_j}\ln{\mathscr{L}\left(\overrightarrow{\mathrm{Res}}\mid\vec{\theta}_\mathrm{true}\right)} \right\rangle ,\nonumber \\
        &= \int\limits^\infty_{-\infty} \frac{1}{\sigma^2_c}\left[ \frac{\partial\mathrm{Res}^c}{\partial\theta_i}\frac{\partial\mathrm{Res}^c}{\partial\theta_j} + \widetilde{\mathrm{Res}}^c\frac{\partial^2\mathrm{Res}^c}{\partial\theta_i\partial\theta_j}  \right]\bigg\rvert_{\vec{\theta}_\mathrm{true}} \mathscr{L}\left(\overrightarrow{\mathrm{Res}}\mid\vec{\theta}_\mathrm{true}\right)d^d\!\!\left[\widetilde{\mathrm{Res}}\left(\vec{\theta}_\mathrm{true}\right)\right] .
    \end{align}
    Now we see there are two separate integrals we need to evaluate.  Importantly, note that $\frac{\partial\mathrm{Res}^c\left(\vec{\theta}_\mathrm{true}\right)}{\partial\theta_i}$ and $\frac{\partial^2\mathrm{Res}^c\left(\vec{\theta}_\mathrm{true}\right)}{\partial\theta_i\partial\theta_j}$ are just constants which can be pulled out of their respective integrals.  Therefore the integrals that we need to now solve, using equation~\ref{eqn: likelihood function}, are:
    \begin{equation*}
        \int\limits^\infty_{-\infty} \mathscr{L}\left(\overrightarrow{\mathrm{Res}}\mid\vec{\theta}_\mathrm{true}\right)d^d\!\!\left[\widetilde{\mathrm{Res}}\left(\vec{\theta}_\mathrm{true}\right)\right] \equiv 1 ,
    \end{equation*}
    \begin{align*}
        &\int\limits^\infty_{-\infty} \widetilde{\mathrm{Res}}^c\left(\vec{\theta}_\mathrm{true}\right) \mathscr{L}\left(\overrightarrow{\mathrm{Res}}\mid\vec{\theta}_\mathrm{true}\right)d^d\!\!\left[\widetilde{\mathrm{Res}}\left(\vec{\theta}_\mathrm{true}\right)\right] , \nonumber \\
        &\hspace{2cm}= \int\limits^\infty_{-\infty} \widetilde{\mathrm{Res}}^c\left(\vec{\theta}_\mathrm{true}\right) \frac{1}{\sqrt{(2\pi)^d\mathrm{det}(\bm{\Sigma})}}  \exp{\Bigg[-\frac{1}{2}\frac{1}{\sigma^2_c}\left( \widetilde{\mathrm{Res}}^c\left(\vec{\theta}_\mathrm{true}\right) \right)^2 \Bigg]}  d\!\!\left[\widetilde{\mathrm{Res}}\left(\vec{\theta}_\mathrm{true}\right)\right] , \nonumber \\
        &\hspace{2cm}= \frac{1}{\sqrt{(2\pi)^d\mathrm{det}(\bm{\Sigma})}}  \int\limits^\infty_{-\infty} \widetilde{\mathrm{Res}}^c\left(\vec{\theta}_\mathrm{true}\right)   \exp{\Bigg[-\frac{1}{2}\frac{1}{\sigma^2_c}\left( \widetilde{\mathrm{Res}}^c\left(\vec{\theta}_\mathrm{true}\right) \right)^2 \Bigg]}  d\!\!\left[\widetilde{\mathrm{Res}}^c\left(\vec{\theta}_\mathrm{true}\right)\right] \nonumber \\
        & \hspace{5.5cm} \times \ \ \int\limits^\infty_{-\infty} \exp{\Bigg[-\frac{1}{2}\frac{1}{\sigma^2_a}\left( \widetilde{\mathrm{Res}}^a\left(\vec{\theta}_\mathrm{true}\right) \right)^2 \Bigg]}  d^{d-1}\!\!\left[\widetilde{\mathrm{Res}}\left(\vec{\theta}_\mathrm{true}\right)\right] \equiv 0 .
    \end{align*}
    The second integral is killed off because of the odd Gaussian integral over $\widetilde{\mathrm{Res}}^c\left(\vec{\theta}_\mathrm{true}\right)$.  Note that in this particular integral the notation is similar as to that back in equation~\ref{eqn: expectation chi squared multivariate taylor first term}.  We have separated out like terms, the $\widetilde{\mathrm{Res}}^c\left(\vec{\theta}_\mathrm{true}\right)$ integral from the rest of the $\widetilde{\mathrm{Res}}^a\left(\vec{\theta}_\mathrm{true}\right)$ for $a\neq c$ integrals (so note that the implied sum doesn't apply to the $c$ index in the second to last line above).
    
    This concludes our work and gives us the final Fisher matrix for our multivariate Gaussian timing residual model:
    \begin{equation}
         F_{ij}\left(\vec{\theta}_\mathrm{true}\right) = \sum\limits_c \frac{1}{\sigma^2_c}\left(\frac{\partial\mathrm{Res}^c}{\partial\theta_i}\frac{\partial\mathrm{Res}^c}{\partial\theta_j}\right)\bigg\rvert_{\vec{\theta}_\mathrm{true}} .
    \label{eqn: timing residual fisher matrix}
    \end{equation}
    As a reminder, the $c$ indices range over all $d$ data that we collect in our experiment (for every pulsar at every observation time), and the $i$,$j$ indices range over the number of $k$ parameters in our model.  Using the Fisher matrix we can estimate the posterior distribution of our parameters:
    \begin{align}
        p\left(\vec{\theta}\middle|\vec{X}\right) &\sim \sqrt{\frac{\mathrm{det}(\bm{F})}{(2\pi)^k}} \exp\left[-\frac{1}{2}F_{ij}\Delta\theta^i\Delta\theta^j\right] = \sqrt{\frac{\mathrm{det}(\bm{F})}{(2\pi)^k}} \exp\left[-\frac{1}{2} \Delta\vec{\theta}^T \bm{F} \Delta\vec{\theta} \right] , \nonumber \\
        &=\sqrt{\frac{1}{(2\pi)^k\mathrm{det}(\bm{C})}} \exp\left[-\frac{1}{2}C^{-1}_{ij}\Delta\theta^i\Delta\theta^j\right] = \sqrt{\frac{1}{(2\pi)^k\mathrm{det}(\bm{C})}} \exp\left[-\frac{1}{2} \Delta\vec{\theta}^T \bm{C}^{-1} \Delta\vec{\theta} \right] ,
    \tag{\ref{eqn: Fisher/covariance posterior} r}
    \end{align}
    where $\Delta\vec{\theta} \equiv \vec{\theta}-\vec{\theta}_\mathrm{true}$.  At this point inverting the Fisher matrix will give us an estimate of the parameter covariance matrix, which is our goal.
    
    Please note that in our work below, we are considering the case where all of the timing uncertainty measurements in our experiment are the same:
    \begin{equation}
         F_{ij}\left(\vec{\theta}_\mathrm{true}\right) = \sum\limits_c \frac{1}{\sigma^2}\left(\frac{\partial\mathrm{Res}^c}{\partial\theta_i}\frac{\partial\mathrm{Res}^c}{\partial\theta_j}\right)\bigg\rvert_{\vec{\theta}_\mathrm{true}} .
    \tag{\ref{eqn: timing residual fisher matrix} common $\sigma$}
    \end{equation}

%---------------------------------------------------------------------------------
%---------------------------------------------------------------------------------
    \section{Breaking the Timing Residual Fisher Matrix into Submatrices}\label{sec: breaking the timing residual fisher matrix into submatrices}
    
    The main parameters of interest in this model are the source parameters.  However, distances to most pulsars are not known to a high degree of accuracy.  For instance numerous pulsars
    %, including ones timed by NANOGrav, 
    %
    have parallax uncertainties on the order of 100pc~\citep{NG_11yr_data, pulsar_parallax2019}, which is much larger than a typical gravitational wavelength $\lambda_\mathrm{gw}$ for our sources of interest (see equation~\ref{eqn: gw wavelength}).  For this reason, previous studies have thought to include the pulsar distances as free parameters in addition to the source parameters in their analyses~\citep{CC_main_paper}.  This allows us to use the gravitational wave data to also help measure the distances to pulsars in our PTA.  When including pulsar distances as model parameters, the Fisher matrix takes a symmetric block-matrix form, separating into a source parameter only symmetric matrix $\bm{F^S}$, a pulsar distance parameter only symmetric matrix $\bm{F^L}$, and a matrix with cross terms $\bm{F}^{\bm{SL}}$. Dividing the model parameters into ``source'' parameters and ``pulsar distance'' parameters, $\vec{\theta} = \left[\vec{s}, \vec{L}\right]$, equation~\ref{eqn: timing residual fisher matrix} then becomes (for common $\sigma$):
    \begin{align}
        \begin{cases}
            F^S_{ij} &= \sum\limits_c \frac{1}{\sigma^2}\left(\frac{\partial \mathrm{Res}^c}{\partial s_i}\right) \left(\frac{\partial \mathrm{Res}^c}{\partial s_j}\right) , \\[8pt]
            F^{SL}_{ij} &= \sum\limits_c \frac{1}{\sigma^2}\left(\frac{\partial \mathrm{Res}^c}{\partial s_i}\right) \left(\frac{\partial \mathrm{Res}^c}{\partial L_j}\right) , \\[8pt]
            F^L_{ij} &= \sum\limits_c \frac{1}{\sigma^2}\left(\frac{\partial \mathrm{Res}^c}{\partial L_i}\right)^2 \delta_{ij} , \\[10pt]
            \bm{F} &= \left[\begin{array}{c|ccc}
                \\[-5pt]
                \bm{F^S}                         & \cdots & \bm{F}^{\bm{SL}}  & \cdots \\[8pt] \hline
                \vdots                                & \ddots &                        & 0 \\
                \left(\bm{F}^{\bm{SL}}\right)^T  &        & \bm{F^L}          & \\
                \vdots                                & 0      &                        & \ddots
            \end{array}\right] .
        \end{cases}
    \label{eqn: Fisher block matrix}
    \end{align}

    For our Fisher analysis we used the PYTHON SymPy package to first symbolically write and manipulate our models from Chapter~\ref{ch: The Continuous Wave Timing Residual}.  Inverting the Fisher matrix was accomplished through singular value decomposition (SVD).  We checked that the condition number of each Fisher matrix was below $10^{14}$ before performing the inversion~\citep{condition_number_source}.  This requirement helped ensure that the SVD procedure accurately calculated the inverse matrix - if the condition number exceeded this value, then we did not calculate the covariance matrix.  As a final check for accuracy the computed covariance matrix was multiplied against its original Fisher matrix, and the result subtracted from the identity matrix.  This was checked against an ``error threshold'' matrix defined by:
    \begin{equation}
        \bm{I} - \bm{F}\bm{C} < \epsilon \left[\begin{array}{cc} 1 & \cdots \\ \vdots & \ddots \\ \end{array}\right] ,
    \end{equation}
    where $\epsilon$ was our ``error threshold.''  For all of the results presented in Chapter~\ref{ch:results - first investigation}, the value of $\epsilon \leq 0.01$ (and in almost all cases $\leq 10^{-4}$).

%---------------------------------------------------------------------------------
%---------------------------------------------------------------------------------
    \section{Fisher Coordinate Transformation}\label{sec:fisher coordinate transformation}
        
    Once we have chosen some parametrization for our model $\vec{\theta}$ and found its Fisher matrix $\bm{F}\left(\vec{\theta}\right)$, we can change the parametrization choice to some different $\vec{\theta^{'}}$ and rather easily find the corresponding new Fisher matrix $\bm{F}^{\bm{'}}\left(\vec{\theta^{'}}\right)$.  Let our new parameters be continuously differentiable functions of the old parameters, i.e. let $\vec{\theta^{'}} = \vec{\theta^{'}}\left(\vec{\theta}\right)$.  Then we can write:
    \begin{equation}
      \frac{\partial}{\partial \theta^{'}_i} = \frac{\partial \theta_m}{\partial \theta^{'}_i}\frac{\partial}{\partial \theta_m} = J_{mi}\frac{\partial}{\partial \theta_m} ,
    \label{eqn: fisher coordinate transformation - Jacobian}
    \end{equation}
    where $\bm{J}$ is the Jacobian matrix.  This means that using our definition of the Fisher matrix (either the original definition equation~\ref{eqn: Fisher matrix} or the result when applied to our model, equation~\ref{eqn: timing residual fisher matrix}) we can express the new Fisher matrix $\bm{F}^{\bm{'}}\left(\vec{\theta^{'}}\right)$, replace the partial derivatives in that expression (which are with respect to $\theta^{'}_i$ and $\theta^{'}_j$ now), and write the result in terms of the old Fisher matrix.  Explicitly with equation~\ref{eqn: timing residual fisher matrix} this looks like:
    \begin{equation*}
        F^{'}_{ij}\left(\vec{\theta^{'}}\right) = \sum_c \frac{1}{\sigma^2} \frac{\partial \mathrm{Res}^c}{\partial \theta^{'}_i} \frac{\partial \mathrm{Res}^c}{\partial \theta^{'}_j} = \sum_c \frac{1}{\sigma^2} J_{mi} \frac{\partial \mathrm{Res}^c}{\partial \theta_m} J_{nj} \frac{\partial \mathrm{Res}^c}{\partial \theta_n} = J_{mi} J_{nj} F_{mn}\left(\vec{\theta}\right) .
    \end{equation*}
    (Note if we repeat this with the original definition in equation~\ref{eqn: Fisher matrix}, the expectation value is an integral over the random variable $\vec{X}$, so we can simply pull the Jacobian matrix outside of the expectation values once we make the change of variables in the partial derivatives, since the Jacobian is a function of the model parameters, not the random variable).  In general, a coordinate transformation of the Fisher matrix model parameters can be written as:
    \begin{align}
        &F^{'}_{ij}\left(\vec{\theta^{'}}\right) = J_{mi} J_{nj} F_{mn}\left(\vec{\theta}\right) \quad \longleftrightarrow \quad \bm{F}^{\bm{'}} = \bm{J}^T \bm{F} \bm{J} , \nonumber \\
        &\mathrm{where}\quad J_{ij} \equiv \frac{\partial \theta_i}{\partial \theta^{'}_j} .
    \label{eqn: fisher coordinate transformation}
    \end{align}

\cleardoublepage
%---------------------------------------------------------------------------------
%---------------------------------------------------------------------------------
\chapter{Pulsar Distance Wrapping Problem}\label{ch:L-wrapping problem}

A common problem in parameter estimation from continuous wave pulsar timing residuals is the ``pulsar distance wrapping problem'' \citep{CC_main_paper,Ellis_2013}.  The pulsar distance affects the phase of the timing residual terms (see equations~\ref{eqn: Res(t) pw mono phase E and P}, \ref{eqn: Res(t) pw freq evo frequency phase E and P}, \ref{eqn: Res(t) fresnel mono heuristic phase E and P}, and~\ref{eqn: Res(t) fresnel freq evo phase E and P}).  Since the phase wraps around the periodic interval $[0,2\pi)$, if the pulsar distance is an unrestricted free parameter, it can become difficult or even impossible to recover in parameter estimation.  This is because it is often easy to find arbitrarily large values of $L$ which cause the phase to wrap around the $2\pi$-interval and yield the same timing residual.

Interestingly, it is also because the phase wraps around $2\pi$ that we can even try to measure $R$ from the Fresnel formalism in the first place!  This the key idea behind why it is even possible to study Fresnel corrections, and why they aren't always just completely negligible.  As discussed in more detail in Section~\ref{sec: four pulsar timing regimes}, corrections of $\mathcal{O}\left(\frac{L}{R}\right)$ to the amplitude of the timing residual model would likely be impossible to measure, however, when they appear in the \textit{phase} they can become appreciable over the $2\pi$-interval.

In the monochromatic regimes, we can analytically compute the size of the deviation in $L$ that would still give the same timing residual.  First let's focus on the plane-wave monochromatic regime.  Looking at the form of the plane-wave timing residual we see that from equations~\ref{eqn: Res(t) pw mono}, \ref{eqn: Res(t) pw mono phase E and P}, and~\ref{eqn: h0 plus and cross}, and the cyclic nature of sine and cosine, that $2\Theta_p = 2\Theta_p + 2\pi n$ for $n\in \mathbb{Z}$.  In other words, any $2\pi$ multiple added onto the phase would result in the exact same timing residual.  So let's now ask the question, if we take the true value of the pulsar distance $L$ that appears in $\Theta_p$ and add some $\Delta L$ value to it, how large would this deviation need to be in order to cause the phase to wrap and to thus give the same timing residual?  Or in other words, defining $L^{'} \equiv L + \Delta L_n$, 
%for what $\Delta L_n$ does $2\Theta_p\left(L^{'}\right) = 2\Theta_p\Big(L\Big)$?
%
for what $\Delta L_n$ does $2\Theta_p\left(L^{'}\right) = 2\Theta_p\Big(L\Big) + 2\pi n$?

From equation~\ref{eqn: Res(t) pw mono phase E and P} this question becomes:
\begin{align*}
    2\theta_0 + 2\omega_0\left(t-\left(1-\hat{r}\cdot\hat{p}\right)\frac{L^{'}}{c}\right) &= 2\theta_0 + 2\omega_0\left(t-\left(1-\hat{r}\cdot\hat{p}\right)\frac{L}{c}\right) + 2\pi n , \\
    \therefore\quad -\omega_0\left(1-\hat{r}\cdot\hat{p}\right)\frac{L+\Delta L_n}{c} &= -\omega_0\left(1-\hat{r}\cdot\hat{p}\right)\frac{L}{c} + \pi n , \\
    &\hspace{-2cm} \longrightarrow \quad \Delta L_n = n \frac{\pi c}{\omega_0\left(1-\hat{r}\cdot\hat{p}\right)} = n \frac{\lambda_\mathrm{gw}}{\left(1-\hat{r}\cdot\hat{p}\right)} .
\end{align*}
Note that in the final step we absorb a negative factor into $n$, since this is just any positive or negative integer, and we recognize from equation~\ref{eqn: gw wavelength} that this is a scaled multiple of the gravitational wavelength.  We can build our intuition of this result in the following way.  For a monochromatic gravitational wave, the gravitational wavelength remains constant for the entire time the photon travels from the pulsar to the Earth, since there is no frequency evolution in this case to cause the wavelength to be different at the pulsar versus at the Earth.  So physically the timing residual in the monochromatic regime only depends on the difference in the phase of the gravitational wave strain at the pulsar when the photon leaves, and the phase at the Earth when the photon arrives.  We can see this looking at equations~\ref{eqn: Res(t) pw mono} and~\ref{eqn: Res(t) pw mono phase E and P}.  If the source and pulsar are perpendicular to each other, such as what is depicted in Figure~\ref{fig: tret contours}, then $\Delta L_n = n \lambda_\mathrm{gw}$, so moving our pulsar closer or further away from the source by $n$ multiples of the gravitational wavelength itself simply cycles the strain phase by that same multiple of the gravitational wavelength, resulting in no change to the timing residual.

However, the geometrical alignment of our system also plays into this expression as a scaling factor that either stretches or compresses the total distance we need to move along the gravitational wavelength to return the phase to its original value.  As we can see, in the limit that $\hat{r}\cdot\hat{p} \rightarrow 1$, that is the source-pulsar become aligned, $\Delta L_n \rightarrow \infty$.  Similarly, as $\hat{r}\cdot\hat{p} \rightarrow -1$, that is the source-pulsar become anti-aligned, $\Delta L_n \rightarrow n \frac{\lambda_\mathrm{gw}}{2}$.  This result tells us something interesting and helpful, that as the source and pulsar become more aligned with each other, the cyclic degeneracy of the timing residual in the monochromatic regime actually vanishes naturally!  Furthermore, the smallest possible $\Delta L$ we could shift our pulsar by is half a gravitational wavelength, for $n=1$ in an anti-aligned system.

The reason this becomes the ``pulsar-wrapping problem'' is because as stated in Section~\ref{sec: breaking the timing residual fisher matrix into submatrices}, most pulsar distances are not measured to a high degree of accuracy.  For this reason we include the pulsar distances as free-parameters in our model.  However, without any restrictions on the searched values of $L$ in our analyses, any $L^{'} = L + \Delta L_n$ selected will be just as ``correct'' mathematically as the true $L$ value that we want to find.  But if we can measure the distances to our pulsars down to uncertainties on the order of $\Delta L_n$, then this knowledge can be added as prior knowledge to our parameter estimation search (such as described in Section~\ref{sec: adding prior to fisher matrix}).  Ideally we would like to be able to measure all pulsar distances uncertainties $\sigma_L < \Delta L_1$, to ensure that our parameter search never wraps over the phase simply by choosing too large of a value of $L$.

This poses an experimental challenge by today's standards.  Most pulsars in our PTA are on the order of kiloparsecs in distance, but for typical sources of interest ($\omega_0 \sim 1-100$nHz), the gravitational wavelength $\lambda_\mathrm{gw}$ can range roughly between tens of parsecs to subparsec distances.  Moreover, the geometric scaling factor can either help or hurt us in this sense.  If our pulsar happens to be within 90\textdegree \ of our source ($\hat{r}\cdot\hat{p} \rightarrow 1$), then the condition relaxes and we don't need as tight of an experimental uncertainty.  But if our pulsar happens to be more than 90\textdegree \ from our source ($\hat{r}\cdot\hat{p} \rightarrow -1$), then we require even more precision in our measurement of $L$.

We can perform this same analysis in the Fresnel monochromatic regime by using our heuristic model equations~\ref{eqn: Res(t) fresnel mono heuristic} and~\ref{eqn: Res(t) fresnel mono heuristic phase E and P}.  Once again, asking for what $\Delta L_n$ does %$2\Theta_p\left(L^{'}\right) = 2\Theta_p\Big(L\Big)$,
${2\Theta_p\left(L^{'}\right) = 2\Theta_p\Big(L\Big) + 2\pi n}$,
and for convenience defining $\textsc{A} \equiv \left(1-\hat{r}\cdot\hat{p}\right)$ and $\textsc{B} \equiv \frac{1}{2}\left(1-\left(\hat{r}\cdot\hat{p}\right)^2\right)$, we find:
\begin{align*}
    2\theta_0 + 2\omega_0\left( t - \frac{R}{c}  - \textsc{A}\frac{L^{'}}{c} - \textsc{B}\frac{L^{'}}{c}\frac{L^{'}}{R} \right) &= 2\theta_0 + 2\omega_0\left( t - \frac{R}{c}  - \textsc{A}\frac{L}{c} - \textsc{B}\frac{L}{c}\frac{L}{R} \right) + 2\pi n , \\
    \therefore\quad -2\omega_0\textsc{A}\frac{L+\Delta L_n}{c} - 2\omega_0\textsc{B}\frac{\left(L + \Delta L_n\right)^2}{c R} &= -2\omega_0\textsc{A}\frac{L}{c} -2\omega_0\textsc{B}\frac{L^2}{c R} + \pi n , \\
    \therefore\quad \textsc{B}\Delta L_n^2 + \left(\textsc{A}R +2\textsc{B}L\right)\Delta L_n + n\frac{\pi cR}{\omega_0} &= 0 , \\
    &\hspace{-2cm} \longrightarrow \quad \Delta L_n = \frac{-\left(\textsc{A}R +2\textsc{B}L\right) \pm \sqrt{\left(\textsc{A}R +2\textsc{B}L\right)^2 + 4\textsc{B}n\frac{\pi cR}{\omega_0}}}{2\textsc{B}} ,
\end{align*}
where once again, just to keep consistency with our steps compared to what we did above in the plane-wave monochromatic case, we absorb an extra negative factor into $n$.

\begin{figure}
    \centering
    \includegraphics[width=1\linewidth]{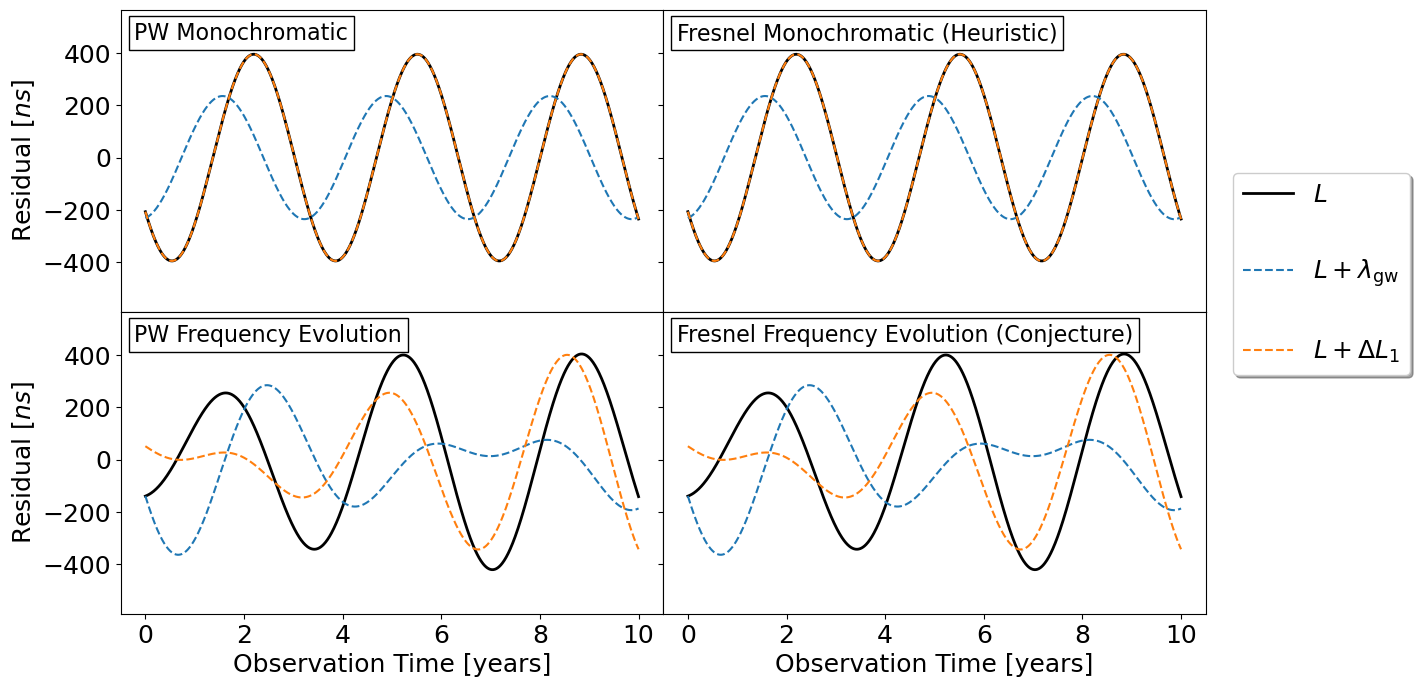}
    \caption[Pulsar Distance Wrapping Problem Example]{Here we show visually what the pulsar distance wrapping problem does to the timing residual.  These are the calculated timing residuals for each of our four regimes using Source 1 from Table~\ref{tab: source parameters reference} set at $R=1$~Gpc and Pulsar 1 from Table~\ref{tab: pta reference}.  Here $\Delta L_1$ is the wrapping cycle calculated from the plane-wave monochromatic formula given in equation~\ref{eqn:Delta L in monochromatic regimes}.  In this example, the relative difference between the $\Delta L_1$ values calculated using the plane-wave vs. Fresnel cases is very small ($\sim \mathcal{O}\left(10^{-7}\right)$), which is why the plane-wave $\Delta L_1$ value here works visually just as well even in the Fresnel monochromatic regime.  As we can see, adding just a gravitational wavelength $\lambda_\mathrm{gw}$ to the pulsar distance doesn't give back the same timing residual, but accounting for the relative positions between the source and pulsar as given by equation~\ref{eqn:Delta L in monochromatic regimes} does in the monochromatic regimes.  In the frequency evolving regimes, the wrapping cycle degeneracy is mathematically broken since the source's frequency is changing with time.}
    \label{fig: L wrapping example}
\end{figure}

In summary, for the monochromatic regimes, any $L^{'} = L \pm \Delta L_n$ such that:
\begin{equation}
\Delta L_n \equiv \begin{cases}
    n \frac{\pi c}{\omega_0\left(1-\hat{r}\cdot\hat{p}\right)} = n \frac{\lambda_\mathrm{gw}}{\left(1-\hat{r}\cdot\hat{p}\right)} , \\
    &\hspace{-6cm}\text{(Plane-Wave, monochromatic)} \\[15pt]
    -\left(\frac{R}{\left(1+\hat{r}\cdot\hat{p}\right)} + L\right) + \sqrt{ \left(\frac{R}{\left(1+\hat{r}\cdot\hat{p}\right)} + L\right)^2 + n\frac{2\pi cR}{\omega_0 \left(1-\left(\hat{r}\cdot\hat{p}\right)^2\right)}} , \\[10pt]
    \qquad = -\left(\frac{R}{\left(1+\hat{r}\cdot\hat{p}\right)} + L\right) + \sqrt{ \left(\frac{R}{\left(1+\hat{r}\cdot\hat{p}\right)} + L\right)^2 + n\frac{2\lambda_\mathrm{gw}R}{\left(1-\left(\hat{r}\cdot\hat{p}\right)^2\right)}} , \\
    &\hspace{-6cm} \text{(\textit{Heuristic} Fresnel, monochromatic)}
\end{cases}
\label{eqn:Delta L in monochromatic regimes}
\end{equation}
for $n \in \mathbb{Z}$ will give the same timing residual as the timing residual evaluated at the true $L$.  We will refer to $\Delta L_1$ for the \textit{plane-wave monochromatic case} as ``one wrapping cycle.''  When solving the quadratic equation in the Fresnel monochromatic case, we only consider the positive solution (discussed further in the box below).  As with the Fresnel formalism in general, this result for $\Delta L_n$ does not hold for the natural plane-wave limit.  In Table~\ref{tab: pta reference} we provide for reference the values of the wrapping cycle $\Delta L_1$ proxy (from the plane-wave monochromatic formula) for each pulsar in our PTA with Source 1 from Table~\ref{tab: source parameters reference}.  A visual of this pulsar distance wrapping problem is shown in Figure~\ref{fig: L wrapping example}. \\

\begin{minipage}{0.9\linewidth}
\fbox {
    \parbox{\linewidth}{ \textbf{Limit Check}\newline
    As a sanity check, we do find that the $\Delta L_n$ result in the Fresnel monochromatic case does reduce to the plane-wave monochromatic formula in the limit $R \rightarrow \infty$ (assuming $\hat{r}\cdot\hat{p} \neq \pm 1$).  Note that if we consider the negative solution that comes from the quadratic expression for the Fresnel monochromatic case, then take it's limit $R\rightarrow \infty$, we find:
    \begin{equation*}
        \Delta L_n \approx -\frac{2R}{\left(1+\hat{r}\cdot\hat{p}\right)} - n\frac{\lambda_\mathrm{gw}}{\left(1-\hat{r}\cdot\hat{p}\right)} .
    \end{equation*}
    This expression in general produces an extremely large negative $\Delta L_n$ value (such that the overall $L^{'}$ would be negative, which is un-physical), unless an extremely large negative value of $n$ were chosen.  This is why we choose to discard and only consider the positive solution above.
    }
}
\end{minipage} \\

We only perform this analysis in the monochromatic regimes.  This same approach doesn't work in the frequency evolution regimes because they are time dependent.  The frequency of the gravitational wave (and hence its wavelength) is naturally different at the pulsar and the Earth locations.  Shifting the pulsar by some $\Delta L$ value will not only shift the phase of the pulsar term, but also the frequency of the pulsar term, which in turn will affect the amplitude of that term.  This does not happen if the wave is monochromatic, which is what made that analysis possible.

These relations give us a proxy of how well our observational uncertainties on the pulsar distances $\sigma_L$ need to be if we are to include them as free parameters in our models.  Namely $\sigma_L \sim \Delta L_1$, one wrapping cycle, which means that the uncertainty on our pulsar distance measurements needs to be on the order of the wavelength of the gravitational wave $\lambda_\mathrm{gw}$ that we are trying to measure.

\begin{figure}
\centering
  \begin{subfigure}[t]{0.48\linewidth}
  \centering
    \includegraphics[width=0.75\linewidth]{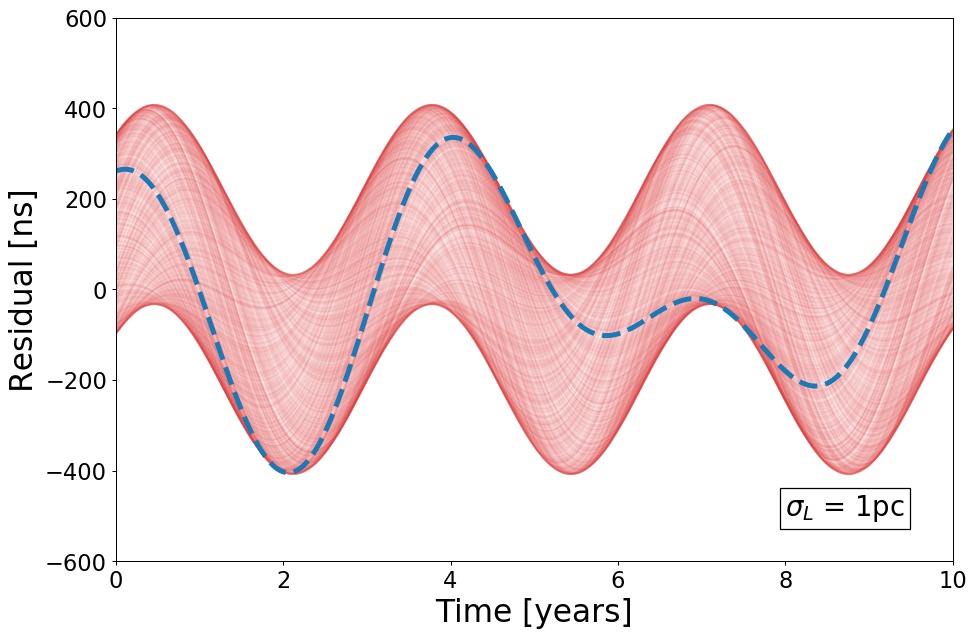}
    \caption{}
  \end{subfigure}
  \begin{subfigure}[t]{0.48\linewidth}
  \centering
    \includegraphics[width=0.75\linewidth]{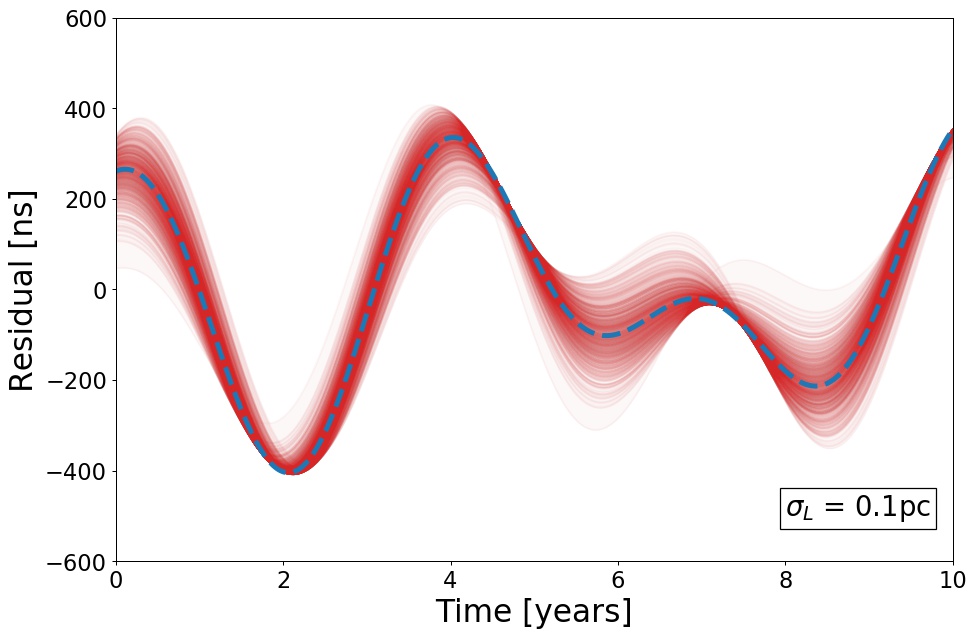}
    \caption{}
  \end{subfigure}
  \begin{subfigure}[t]{0.48\linewidth}
  \centering
    \includegraphics[width=0.75\linewidth]{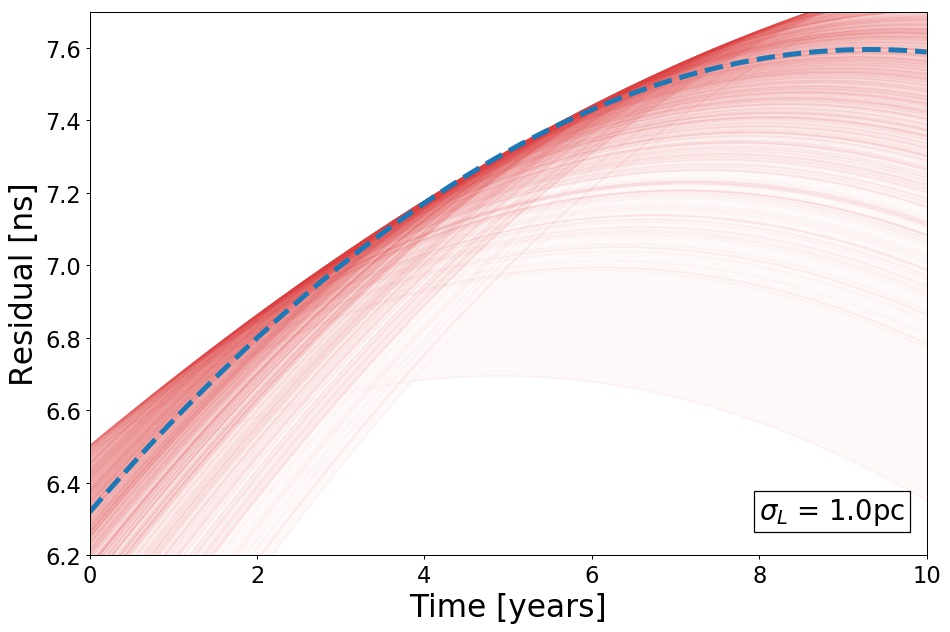}
    \caption{}
  \end{subfigure}
  \begin{subfigure}[t]{0.48\linewidth}
  \centering
    \includegraphics[width=0.75\linewidth]{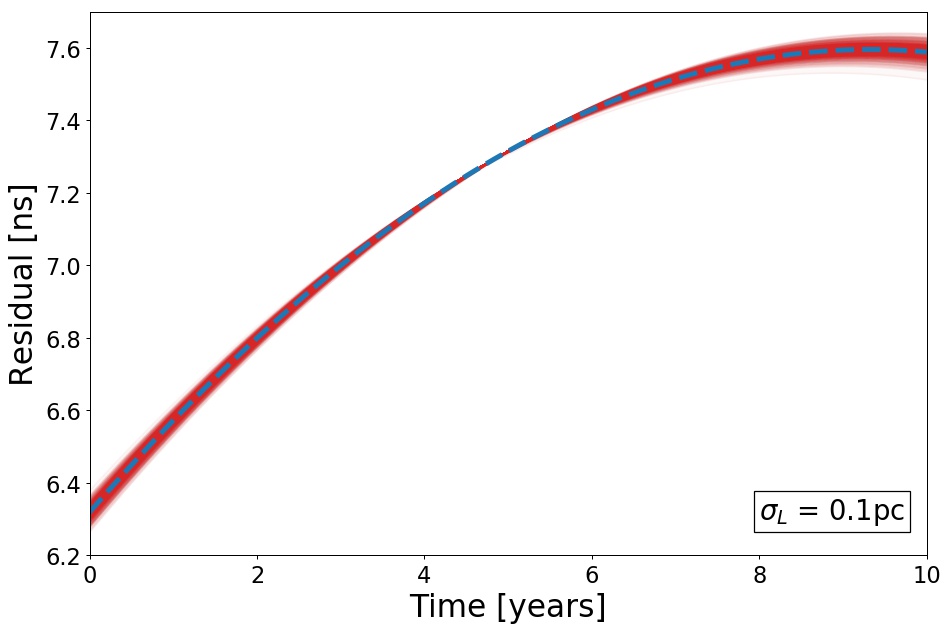}
    \caption{}
  \end{subfigure}
\caption[Pulsar Distance Uncertainties]{Here are four examples of timing residuals with simulated pulsar distance uncertainties, that help to visualize the pulsar distance wrapping problem.  The blue dashed lines represent the true timing residual at $L_\mathrm{true}$, and each realization plotted in red indicates the timing residual using the value of $L$ chosen from a Gaussian distribution centered at $L_\mathrm{true}$ with the $\sigma_L$ uncertainties indicated in each panel.  Each plot shows 1000 realizations, and they use pulsar 16 from Table~\ref{tab: pta reference}.  Panels \textbf{(a)} and \textbf{(b)} use source 1, and \textbf{(c)} and \textbf{(d)} use source 4 (Table~\ref{tab: source parameters reference}).  The key concept shown here is the relationship between $\Delta L_1$ (and $\lambda_\mathrm{gw}$), and the uncertainty $\sigma_L$. \\
\vspace{0.2cm}
For panels \textbf{(a)} and \textbf{(b)}, $\lambda_\mathrm{gw} = 1.017$ pc and $\Delta L_1 = 1.338$ pc.  We can see that in \textbf{(a)} the uncertainty is of the same order as the gravitational wavelength and $\Delta L_1$, and as a result the uncertainty envelope around the true timing residual is effectively saturated, with this pulsar alone parameter estimation would likely be impossible.  However, with the greater distance precision shown in \textbf{(b)} where the uncertainty is now well below the gravitational wavelength, the uncertainty envelope is restricted and thus we can now start to make an informed measurement of the true distance to the pulsar. \\
\vspace{0.2cm}
However in panels \textbf{(c)} and \textbf{(d)}, $\lambda_\mathrm{gw} = 30.522$ pc and $\Delta L_1 = 40.127$ pc.  In both of these cases the simulated uncertainty is well below the gravitational wavelength and $\Delta L_1$, therefore the uncertainty envelope in both is restricted.}
\label{fig:pulsar distance wrapping problem}
\end{figure}

In parameter estimation which includes the pulsar distances, this problem tends to be worse in the Fresnel formalism than in the plane-wave formalism because of the distance combination $L/R$.  Without any search boundaries on these two distances, arbitrarily large values of $L$ and $R$ can preserve their ratio, wrap around the phase interval, and still give the ``correct'' timing residual.  The high covariance between these parameters then introduces covariances with the other parameters in the model that then combine with $L$, which can produce a ``run-away'' effect in the parameter estimation where none of the parameters can be accurately measured.  However, if the search values of $L$ are set to lie within a finite region of parameter space, then this will help to allow for the parameter estimation to work.  So if our parameter estimation method is to recover the \textit{true} distance $L$ of a pulsar, we essentially need to restrict our parameter space to a search region on the order of the $\lambda_\mathrm{gw}$.  These concepts can be seen visually in Figure~\ref{fig:pulsar distance wrapping problem}.

In practice we found that this problem didn't pose a significant issue in our Fisher matrix analyses when using the plane-wave regime models IA and IB, but did pose a significant issue when using with the Fresnel regime models IIA and IIB.  In the case of the Fresnel regime models, if no restrictions were placed on the pulsar distance measurements at all then parameter estimation was not possible because the Fisher matrices could not be accurately inverted (their condition number exceeding $10^{14}$).  Therefore we addressed this problem in our simulations by adding an uncorrelated Gaussian prior to our knowledge of the pulsar distances in our Fisher matrix (\citeauthor{wittman_Fisher_for_beginners}).  This is accomplished by taking the pulsar distance sub-matrix in equation~\ref{eqn: Fisher block matrix} and adding to it $\bm{F}^{\bm{L'}} = \bm{F^L} + \left(\textbf{Cov}^{\bm{L}}\right)^{-1}$, where:
\begin{equation}
    \textbf{Cov}^{\bm{L}} = \text{diag}\left(\sigma_L^2\right) ,
\label{eqn:pulsar uncertainty}
\end{equation}
represents pulsar distance measurement constraints placed on our experiments, each pulsar with its own distance uncertainty $\sigma_L$.  In practice these uncertainties would likely come from electromagnetic observations of the pulsars.

In the case of the plane-wave regime models, we did find it possible to accurately invert the Fisher matrix and recover the source parameters without the need for this additional pulsar distance prior.  Therefore we conclude that the cause for this is the introduction of the distance parameter $R$ in the Fresnel regimes which is not in the plane-wave regimes.  As we discuss in detail in Chapter~\ref{ch:results - first investigation} the parameter $R$ is the most difficult parameter to measure from our models.  This parameter enters the Fresnel models IIA and IIB in the combination $L/R$, therefore it is highly sensitive to the pulsar distance wrapping problem and requires the additional pulsar distance prior in order to constrain the measurement.
\cleardoublepage
\chapter{Investigation of the Fresnel Models}\label{ch:results - first investigation}

In general we found from our Fisher matrix analyses of the Fresnel regimes (for both monochromatic and frequency evolving models), that \textit{none} of the model parameters can be recovered if there are no pulsar distance priors.  If the pulsar distances are completely free variables, then the introduction of the new parameter $R$ in the Fresnel regimes made parameter estimation in all of our simulations completely impossible, for the reasons explained in Chapter~\ref{ch:L-wrapping problem}.  This is unlike the plane-wave regimes, in which we found that it was possible to recover the source parameters in a Fisher matrix analysis (for both monochromatic and frequency evolving models) without any pulsar distance priors.

\begin{figure}
\centering
  \begin{subfigure}[t]{0.48\linewidth}
  \centering
    \includegraphics[width=\linewidth]{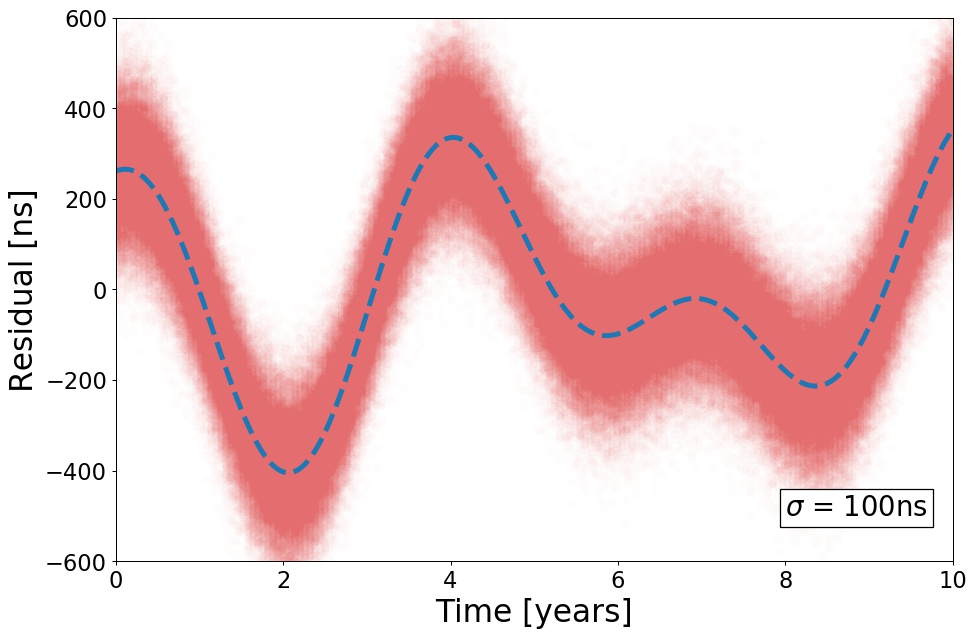}
    \caption{}
  \end{subfigure}
  \hfill
  \begin{subfigure}[t]{0.48\linewidth}
  \centering
    \includegraphics[width=\linewidth]{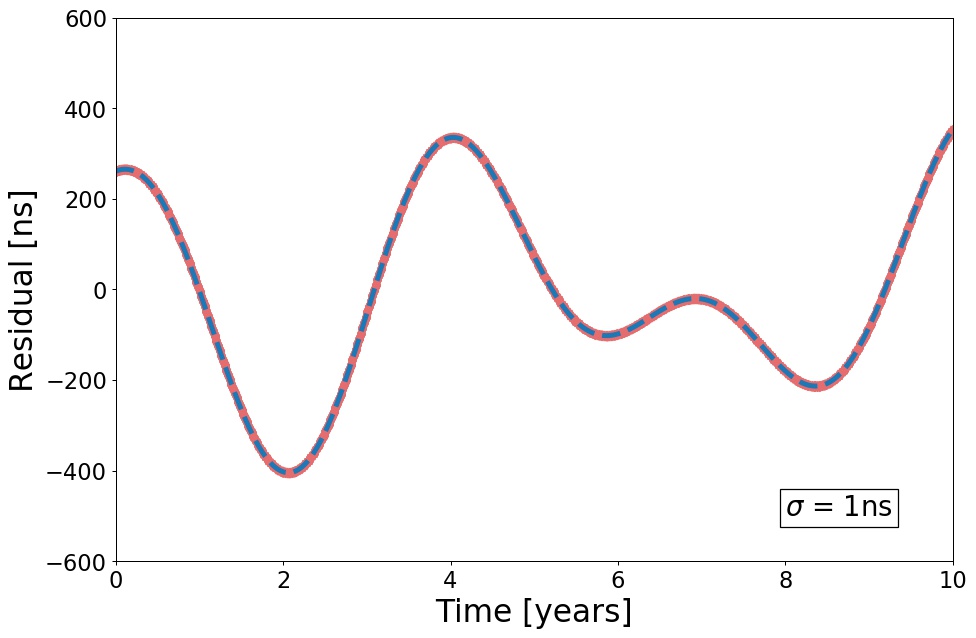}
    \caption{}
  \end{subfigure}
\caption[Pulsar Timing Uncertainty]{Here are two examples of timing residuals with simulated timing noise.  The blue dashed lines represent the true timing residual, and the red envelopes around them indicate the uncertainty region with Gaussian noise added on top the true timing residual (the timing uncertainties are indicated in each panel).  Each plot shows 1000 noise realizations, and they use source 1 and pulsar 16 from Tables~\ref{tab: source parameters reference} and~\ref{tab: pta reference}.}
\label{fig:timing uncertainty}
\end{figure}

For this study we used the PTA in Table~\ref{tab: pta reference}, an array of 40 of the pulsars from NANOGrav's PTA in~\cite{NG_11yr_data}.  A sky plot of this PTA is also shown for reference in Figure~\ref{fig: PTA pulsar sky plot locations}, as well as a visual overlay of the approximate locations on an image of the Milky Way Galaxy in Figure~\ref{fig: PTA MW map}.  In our simulations we timed each of these pulsars for an observation time of 10 years, with a timing cadence of 30 observations/year.  Each of these pulsars were timed at the exact same times, and all of the timings were evenly spaced over the observation period.  Additionally we simulate timing uncertainty $\sigma$ in the residuals (see equation~\ref{eqn: timing residual fisher matrix}) and pulsar distance uncertainty $\sigma_L$ (see equation~\ref{eqn:pulsar uncertainty}).  Figures~\ref{fig:pulsar distance wrapping problem} and~\ref{fig:timing uncertainty} show visual examples of these uncertainties.  In studies in this chapter we typically set to $\sigma = 100$ ns for the order of magnitude of present capabilities~\citep{Cordes_2010, Liu_2011, Arzoumanian_2014, NG_11yr_data}.  This assumption is meant to represent a best-case scenario, as real pulsar timing experiments model additional correlated red noise when estimating source parameters from their pulsar data sets~\citep{corrnoise_vanHaasteren2013, NG_11yr_data, NG_11yr_cw}.  Finally, in order to allow us to control the baseline distances to all of the pulsars in our PTA we introduced a PTA distance ``scale factor'' term.  Therefore a scale factor of 1 indicates that all of the pulsars in the PTA have their standard distances given by the $L$ column in Table~\ref{tab: pta reference}, and at most we increased the scale factor to 7 (because that would place our farthest pulsar \#~40 at roughly the distance of the Large Magellanic Cloud).

In a real pulsar timing experiment, there are other physical effects that create timing residuals which must also be included in a complete timing residual model~\citep{handbook_pulsar_astro, GWastro_Lee2011, NG_11yr_data}.  For simplicity we wanted to focus solely on the effects of the gravitational waves, without simultaneously modeling other timing residual sources, to help ensure that the gravitational wave parameters would not be degenerate with other model parameters.  The sources we consider in this work have high enough frequencies (with $\omega_0 > 10$ nHz) that we found their overall signal-to-noise ratio was not significantly impacted by subtracting a quadratic fit to the timing residual (which would roughly approximate of the impact of fitting a more complete timing model).  A simple analysis of this is provided in the next Section~\ref{sec: Simple SNR Analysis}.  This is in line with~\citet{toa_hazboun2019}, who provides a much more in-depth analysis of the general sensitivity of pulsar timing to gravitational waves.

In our work we use the coefficient of variation (CV) of a given model parameter $x$ as a proxy for that parameter's measureability:
\begin{equation}
    \mathrm{CV}_x \equiv \frac{\text{Standard Deviation of } x}{\text{Expectation Value of } x} = \begin{cases} \frac{\sigma}{\mu} ,  &\quad\text{Normal Distribution in $x$}  \\[8pt]
    \sqrt{e^{\sigma^2 \ln(10)^2} - 1} ,  &\quad\text{Log}_{10}\text{-Normal Distribution in $x$}
    \end{cases}
\label{eqn: CV}
\end{equation}
where $\mu$ and $\sigma$ are the distributions' parameters.  For a normally distributed $x$ parameter, $\mu$ and $\sigma$ are also the distribution's mean and standard deviation.  For a log-normally distributed $x$ parameter, the CV only depends on the log-normal $\sigma$ parameter.  Equation~\ref{eqn: CV} is effectively the fractional error of a given parameter $x$, or a measure of the dispersion of that parameter's distribution.  Once we have the covariance matrix of our parameters from our Fisher analysis, we take the $\sigma$ values from that matrix and compute the CVs based on how each respective parameter is distributed (in the case of the regular normal distribution, $\mu$ is the parameter value we inject into the simulation).  All parameters within a model are normally distributed in a Fisher matrix approximation (see again equation~\ref{eqn: Fisher/covariance posterior}), so we only need the second line of equation~\ref{eqn: CV} if we parametrize a model parameter $x$ as $\log_{10}(x)$.  Therefore a parameter with a small CV suggests to us that it would be measurable from the experimental set-up, while a CV of order unity or larger would suggest an unmeasurable parameter.  If we compute the Fisher matrix with a parametrization such that $\log_{10}(x)$ is \textit{normally} distributed, then the second line in equation~\ref{eqn: CV} gives us the CV of $x$ (which is log-normally distributed).

%---------------------------------------------------------------------------------
%---------------------------------------------------------------------------------
    \section{Simple Signal-to-Noise Ratio Analysis}\label{sec: Simple SNR Analysis}

    Here we outline a simple analysis to check that the gravitational wave signal-to-noise ratio (SNR) is not impacted by including other timing effects in the model for our considered sources.  This is important because if for example we considered a low frequency gravitational wave source whose timing residual over our simulated 10 year observation time didn't complete a full wave cycle and appeared parabolic (like a half wave cycle), then our parameter estimation results may be overly optimistic.  This is because a real timing experiment would simultaneously fit a quadratic polynomial and the gravitational wave signal to the timing data (as well as other models which we are not considering for simplicity here).  Therefore we would be less sensitive to a lower frequency source whose gravitational wave timing residual predicted by one of our models from Chapter~\ref{ch: The Continuous Wave Timing Residual} resembled a parabola.
    
\begin{figure}
        \centering
        \includegraphics[width=0.9\linewidth]{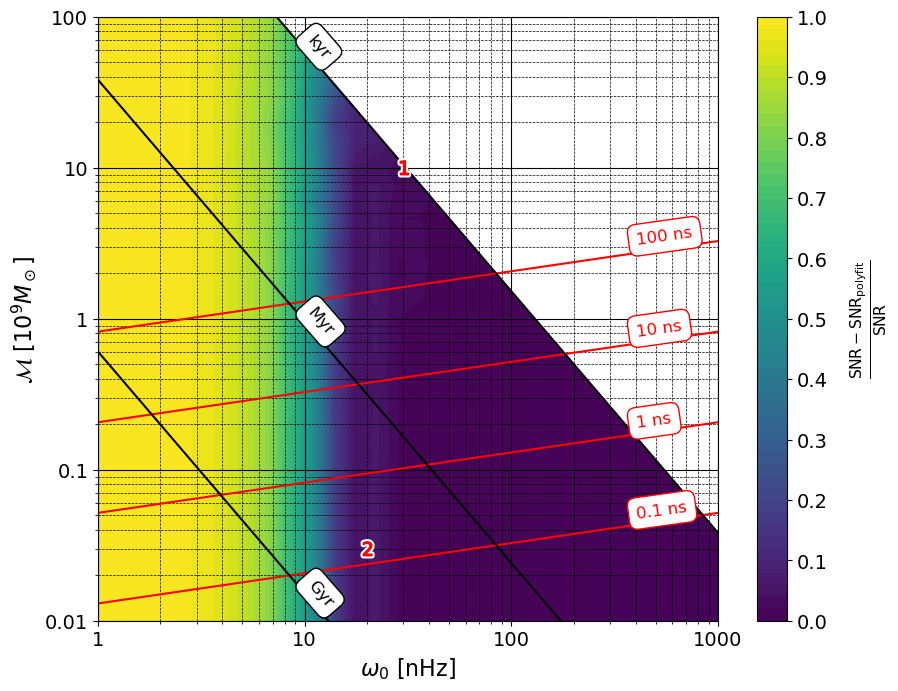}
        \caption[SNR Analysis: Source Orbital Frequency vs. Chirp Mass]{The fractional difference in the gravitational wave timing residual SNR defined in equation~\ref{eqn: SNR}, with vs. without a quadratic polynomial best-fit subtracted from the data, as a function of the source orbital frequency $\omega_0$ and chirp mass $\mathcal{M}$.  The fractional difference in SNR is independent of timing uncertainty $\sigma$, and we also found that it remains effectively unchanged for different values of $R$ (this specific result was computed for $R=100$ Mpc).  For this study we want to consider sources where the fractional difference in SNR is negligible, since we are only modelling gravitational wave effects in our Fisher matrix analyses.  Here we see a low fractional difference of less than $0.1$ at values beyond $\omega_0 > 10$ nHz.  Additionally contours of coalescence time $\Delta \tau_c$ (equation~\ref{eqn: time to coalescence}) are indicated, with a cut along $\Delta \tau_c = 1$ kyr due to assumption~\ref{as: t_obs << tau_c}.  Contours of $A_{E,\mathrm{res}}\left(\frac{R}{100 \ \mathrm{Mpc}}\right)$ are also indicated in red (see equation~\ref{eqn: earth term amplitude}), which serve as a useful proxy for how strong the timing residual signal is (especially when compared to the timing noise $\sigma$).  The fiducial sources from Table~\ref{tab: source parameters reference} are also indicated by their number 1 and 2 here, and have fractional SNR differences here of approximately $0.03$ and $0.06$, respectively.  This was computed using the IIB model (Section~\ref{subsec: Frequency Evolution CONJECTURE (Fresnel)}).}
        \label{fig: SNR}
    \end{figure}

    In order to help decide what source frequencies we should \textit{not} consider in our Fisher matrix analyses we first performed the following calculation.  Using following proxy for the SNR of our experiment:
    \begin{equation}
        \mathrm{SNR} = \sqrt{\sum_a \left(\frac{\mathrm{Res}_a}{\sigma} \right)^2} ,
    \label{eqn: SNR}
    \end{equation}
    we calculated the SNR of just the gravitational wave timing residuals for every pulsar at every time, using our IIB model (Section~\ref{subsec: Frequency Evolution CONJECTURE (Fresnel)}).  Then we recalculated the SNR, but first subtracted off a best-fit quadratic polynomial to the original gravitational wave timing residual:
    \begin{equation}
        \mathrm{Res}_\mathrm{polyfit} = \mathrm{Res} - \left( a + bt + ct^2\right) .
    \end{equation}
    This polynomial fitting procedure was done for every pulsar in the array, given its timing data, meaning in a PTA of 40 pulsars there were 120 polynomial best-fit parameters.  The result of this is given in Figure~\ref{fig: SNR}, which shows the fractional difference between these two SNR values.  If the fractional difference is low, then we can safely perform our Fisher matrix analysis on that source given that PTA without needing to model the additional parabolic timing residual.  Therefore we considered only sources in this study where the fractional difference in the SNR was small, that is with $\omega_0 > 10$ nHz.
    
    Further restrictions on the searchable region of parameter space come from consideration of the coalescence time (our fundamental assumption~\ref{as: t_obs << tau_c}), as well as the strength of the timing residual, which as a proxy can be estimated by the parameter $A_{E,\mathrm{res}}$ (equation~\ref{eqn: earth term amplitude}).  These are also indicated in Figure~\ref{fig: SNR}.

%---------------------------------------------------------------------------------
%---------------------------------------------------------------------------------
    \section{Measuring Source Distance}\label{sec: Measuring Source Distance}
    
    As a point of reference we used Source 1 in Table~\ref{tab: source parameters reference} as a fiducial reference.  This is a source worth considering because it has a low $\Delta \tau_c$ and is therefore strongly chirping, has a high $A_{E,\mathrm{res}}$ even when the source is at $R=100$ Mpc (which is good if our timing uncertainty $\sigma \sim 100$ ns), and a $\lambda_\mathrm{gw} \sim 1$ pc which sets the length scale of the typical wrapping cycle.

%---------------------------------------------------------------------------------
%---------------------------------------------------------------------------------
        \subsection{Figure-of-Merit (F.O.M)}\label{subsec: Figure-of-Merit}
        One of the primary goals of this work was to determine how well we could measure the source distance $R$ from the Fresnel corrections in our pulsar timing models.  As a starting question we asked, ``What is the most important factor in governing the recoverability of the source distance parameter?''  The main take-away we found is that fundamentally, in order to exploit the Fresnel corrections to our model to measure $R$, our precision is governed by the tightest we can constrain the pulsar distance uncertainties $\sigma_L$ on the \textit{farthest} pulsars in our PTA.
        
        To arrive at this conclusion we first simulated two PTA experiments, each with a selection of 39 ``near'' pulsars ($L < 1$ kpc) and a single ``distant'' pulsar (using the F.O.M. $L$ values indicated in Table~\ref{tab: pta reference}).  Source 1 was placed at a very nearby distance of $R=10$ Mpc.  In the first case we simulated a prior distance constraint only on the single distant pulsar to within its own wrapping cycle $\sigma_L = \Delta L_1$, while all near pulsars were unconstrained.  In the second case we did the opposite, placing no prior constraint on the position of the single distant pulsar and giving all nearby pulsars priors equal to their wrapping cycles.  For both scenarios we built up the size of the PTA by first simulating the timing array with only the single distant pulsar \#~40.  Then for every subsequent simulation we added in one additional nearby pulsar, starting at pulsar \#~1.  For every simulation we recorded the PTA's ability to measure the source distance parameter $R$ by calculating the $\mathrm{CV}_R$ value.
        \begin{figure}
            \centering
            \includegraphics[width=1\linewidth]{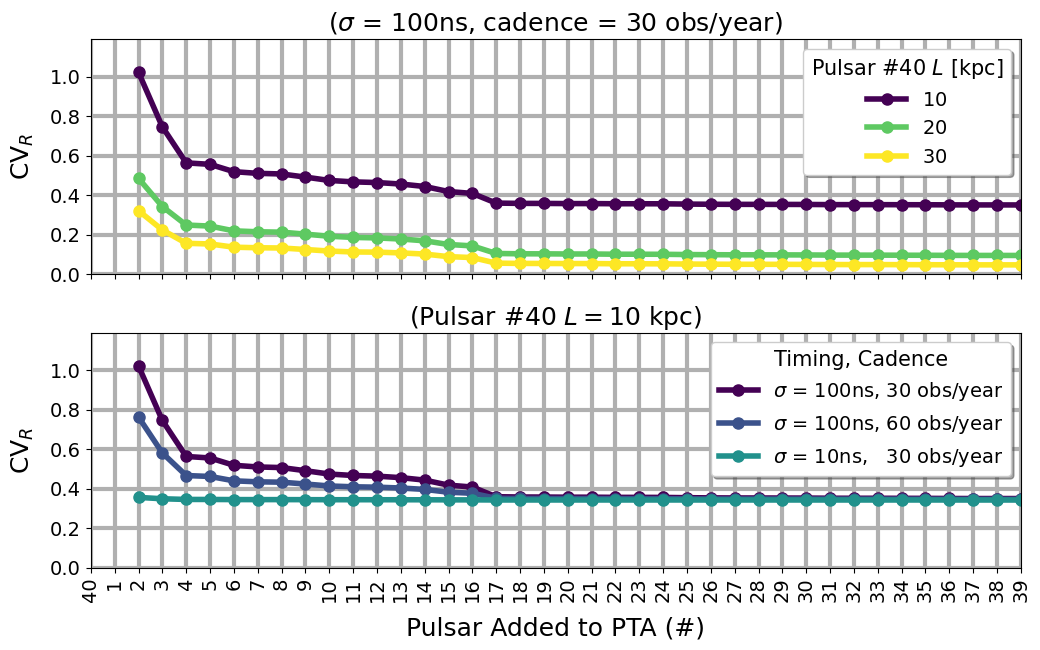}
        \caption[PTA Figure-of-Merit 1]{Two figure-of-merit studies investigating what is needed in a PTA to measure the source distance $R$ from Fresnel corrections.  See Table~\ref{tab: pta reference} for the pulsars and their F.O.M $L$ values which were used in these calculations.  Source 1 from Table~\ref{tab: source parameters reference} is simulated, at a distance $R=10$ Mpc (therefore $A_{E,\mathrm{res}}=21 \ \mu$s).  Note that because this is simply meant to be a figure-of-merit study to understand the general behavior of changing the pulsar and PTA parameters, we purposefully simulated a very near and loud source.  In both panels we first simulated the PTA with only the single ``distant'' pulsar (pulsar \#~40), placed at the distances indicated, with a distance uncertainty constraint of a wrapping cycle $\sigma_L = \Delta L_1$.  For every subsequent simulation we added in one additional ``nearby'' pulsar ($L < 1$ kpc) with no distance prior constraints, starting with pulsar \#~1 through pulsar \#~39.  In each simulation we calculated the $\mathrm{CV}_R$ value for that experiment.  In all cases, we found that we needed a minimum PTA size of 3 in order to accurately invert our Fisher matrix using SVD.  Interestingly beyond 18 pulsars in the PTA, adding additional pulsars didn't seem to improve the recoverability of $R$.  These results were computed using the IIB model (Section~\ref{subsec: Frequency Evolution CONJECTURE (Fresnel)}).}
            \label{fig: FOM - 39 near no-prior, 1 far prior}
        \end{figure}
        
        Only the first scenario succeeded in producing a reliable measurement of the source distance - the PTA with one well constrained distant pulsar, and many unconstrained nearby pulsars.  Results for this scenario are shown in the top panel of Figure~\ref{fig: FOM - 39 near no-prior, 1 far prior}.  This experimental set-up shows a marked improvement in the $\mathrm{CV}_R$ value as the distance to the furthest pulsar is increased (while still holding its $\sigma_L$ fixed).  However, in the opposite scenario when all nearby sources were known to within their wrapping cycles but the single distant pulsar was unconstrained, all measurements of $\mathrm{CV}_R$ were well above unity.  Therefore even when increasing the distance of the distant pulsar in the array, we found that prior knowledge only on the nearby pulsars was not enough to allow us to measure the value of the source distance from the Fresnel corrections.
        
        In summary, even if the distances of many nearby pulsars are known to a great deal of precision, we cannot recover $R$ unless we can strongly constrain the distance to the furthest pulsars in the array.  In fact that is \textit{all} we need in principle - we don't even need to have distance measurement constraints on the nearest pulsars in the array a priori.  Our conclusion for the reason that these two F.O.M. studies gave us these results is because nearby pulsars will likely have negligible Fresnel numbers, meaning that we can't probe their Fresnel corrections to measure $R$.  Even if these nearby pulsars have very well constrained distance measurements it is not enough since the actual magnitude of the Fresnel corrections are too small for these pulsars.  Therefore, in order to actually exploit knowledge coming from the Fresnel corrections in our models we need to have distant pulsars in the array, which have sufficiently high Fresnel numbers.  But due to the pulsar distance wrapping problem, if these distant pulsars are not well constrained then we once again lose the ability to measure the source distance $R$.
        
        We also explored the effects of increased observation cadence and improved timing uncertainty $\sigma$ on measuring $R$.  Using again a single constrained distant pulsar and 39 nearby pulsars, the bottom panel of Figure~\ref{fig: FOM - 39 near no-prior, 1 far prior} shows that if the PTA is small in size a higher timing cadence does help, but timing resolution makes the largest difference.  However, what is interesting is that regardless of the experimental timing uncertainty or cadence, all three schemes shown produce the same level of accuracy at recovering $R$ once our PTA is beyond about 17 pulsars in size.  This means a small PTA with very well timed pulsars is good, but if high timing precision can't be achieved then adding more pulsars to the PTA will help just as much.  Similarly cutting a PTA in half and doubling its observation cadence will produce roughly the same results.  
        
        More generally we observed during our studies that the $\mathrm{CV}$ of the model parameters tended to follow the power law $\mathrm{CV}\propto \left[\sigma / \sqrt{\mathrm{cadence}}\right]^p$.  The parameters $A$, $\iota$, $\psi$, and $\theta_0$ very strongly followed this law with $p=1$, and $\omega_0$ followed with $p\simeq 1$ in most cases.  For the remaining model parameters the $\mathrm{CV}$ still appeared to behave as a univariate function of $\sigma / \sqrt{\mathrm{cadence}}$, but the power law dependence was not as strong as with the previous parameters.  In most observed cases the power law could be fit with differing values of $p<1$ (based on the PTA set-up and source) for the parameters $\theta$, $\phi$, and $\mathcal{M}$, and still held valid over a large part of the $\sigma - \mathrm{cadence}$ parameter space.  However, we found that the $\mathrm{CV}_R$ was only a very weak function of this variable.
        
        As a final study, since we found that nearby pulsars don't help with the measurement of $R$ we therefore investigated the effect of adding additional constrained \textit{distant} pulsars into the PTA.  In Figure~\ref{fig: FOM - 30 near no-prior, 10 far prior} we started the initial PTA with pulsars \#~1-30 from Table~\ref{tab: pta reference} (again using their F.O.M. $L$ distances), and then consecutively added the final 10 (each of these with the distances indicated in Figure~\ref{fig: FOM - 30 near no-prior, 10 far prior}).  Again for this study, nearby pulsars were unconstrained, while each added distant pulsar was simulated with its own wrapping cycle prior $\sigma_L = \Delta L_1$.  As suspected, we can see that adding additional well constrained distant pulsars improves the measurement of $R$.  In a separate study we also found that if only the first of the distant pulsars was given a wrapping cycle prior constraint, then adding additional \textit{unconstrained} distant pulsars did not improve the measurement of $R$ notably.  So it appears that the only way to improve and actually probe the source distance from Fresnel corrections really is to have a PTA with numerous distance pulsars, whose distances are very well measured.
        \begin{figure}
            \centering
            \includegraphics[width=1\linewidth]{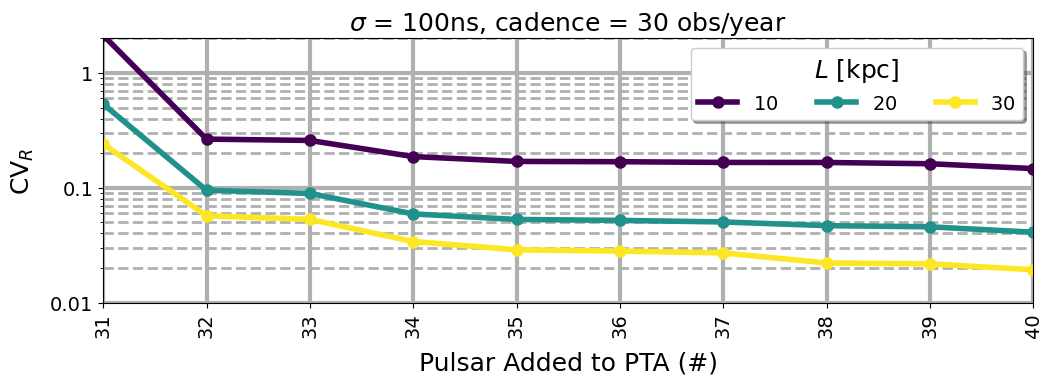}
        \caption[PTA Figure-of-Merit 2]{A figure-of-merit study that shows the improvement of measuring $R$ as additional constrained distant pulsars are added to a PTA.  Here the base PTA consists of nearby pulsars \#~1-30 from Table~\ref{tab: pta reference} using their F.O.M $L$ values, with no prior knowledge distance constraints.  Pulsars \#~31-40 are then added consecutively into the PTA, each with the $L$ distance indicated in the figure, and each with an uncertainty prior of its own wrapping cycle $\sigma_L = \Delta L_1$.  We used the same source and source distance $R$ as in Figure~\ref{fig: FOM - 39 near no-prior, 1 far prior}.  These results were computed using the IIB model (Section~\ref{subsec: Frequency Evolution CONJECTURE (Fresnel)}). }
            \label{fig: FOM - 30 near no-prior, 10 far prior}
        \end{figure}

%---------------------------------------------------------------------------------
%---------------------------------------------------------------------------------
        \subsection{Distant Sources}\label{subsec: distant sources}
        Based on our findings, a direct measurement of $R$ given current and near future standards would be difficult, unless perhaps the source happened to be very nearby, such as in the Virgo Cluster.  But for consideration of the direction of future technologies, distant sources on the order of $100$ Mpc to $1$ Gpc would require significant improvements in our PTA.  From our figure-of-merit studies we know that nearby pulsars don't contribute their Fresnel corrections towards improving the measurement of $R$ - for this we need many distant pulsars in the array with $\sigma_L$ uncertainties on the order of their wrapping distances.  Therefore we focused our investigation on what scenarios would give us good measurements of $R$, and how this might be useful in the future.
        
\begin{figure}
            \centering
              \begin{subfigure}[t]{0.49\linewidth}
              \centering
                \includegraphics[width=1\linewidth]{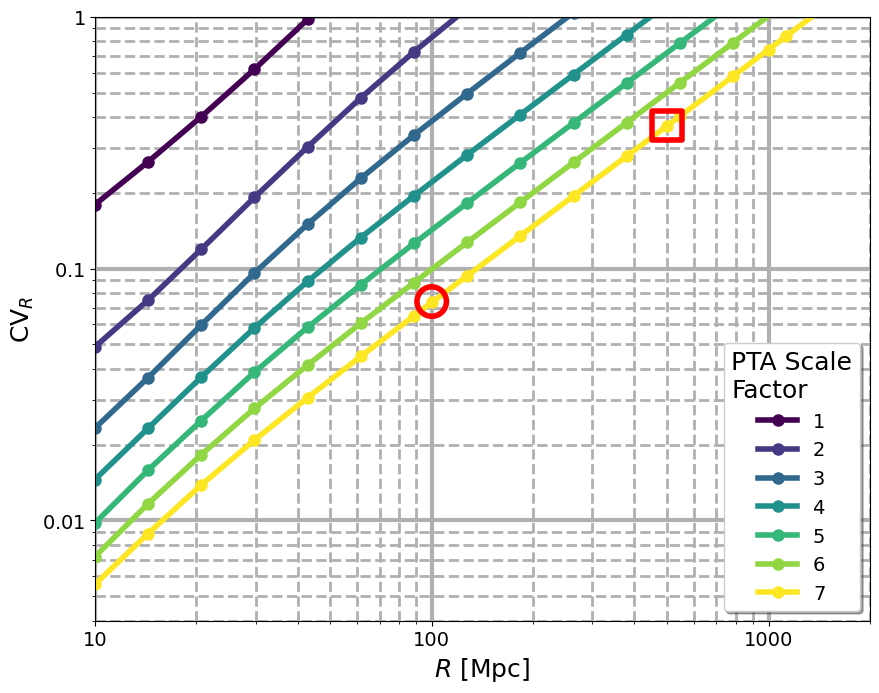}
              \end{subfigure}
              \hfill
              \begin{subfigure}[t]{0.49\linewidth}
              \centering
              \includegraphics[width=1\linewidth]{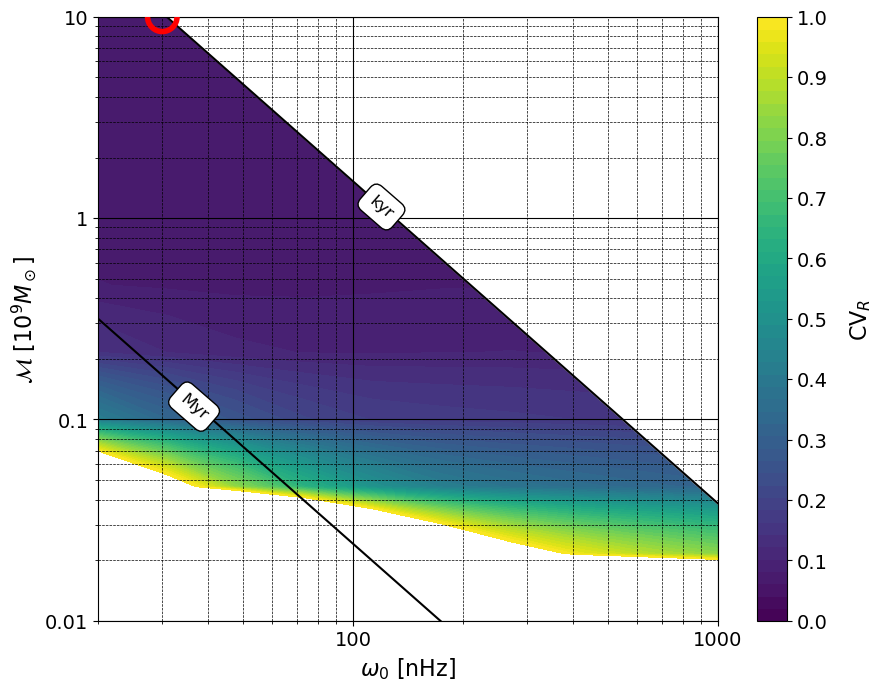}
              \end{subfigure}
        \caption[$\mathrm{CV}_R$ vs. $R$, $\omega_0$, and $\mathcal{M}$]{Source distance measurement as a function of distance and as a function of source intrinsic parameters $\mathcal{M}$ and $\omega_0$.  In both panels the timing uncertainty $\sigma=100$ ns, the pulsar distance uncertainty $\sigma_L = 1$ pc.  The red circle indicates corresponding points between the two graphs, and the red square corresponds to Figure~\ref{fig: s1 example triangle}.  Note, a scale factor of $7$ puts the furthest pulsar in our PTA at roughly the distance of the Large Magellanic Cloud.  These results were computed using the IIB model (Section~\ref{subsec: Frequency Evolution CONJECTURE (Fresnel)}).  \textbf{(Left)} Here we measure $R$ from Source 1 in Table~\ref{tab: source parameters reference}.  Note that $\sigma_L = 1 \ \mathrm{pc}\ \sim \lambda_\mathrm{gw}$ for this source.  \textbf{(Right)} These sources have the same angular parameters as our fiducial Source 1 (indicated in Table~\ref{tab: source parameters reference}) and are set at $R=100$ Mpc. The PTA has a scale factor of $7$.  Contours of coalescence time $\Delta \tau_c$ (equation~\ref{eqn: time to coalescence}) are indicated, with a cut along $\Delta \tau_c = 1$ kyr due to assumption~\ref{as: t_obs << tau_c}.  Bias leans towards lower frequencies, where using a fixed $\sigma_L$ value will benefit sources with larger wrapping cycles as compared to smaller ones (see Section~\ref{subsec: distant sources}).}
        \label{fig: s1 scaled PTA / M-omega0 space sf7 PTA}
        \end{figure}

        The left panel of Figure~\ref{fig: s1 scaled PTA / M-omega0 space sf7 PTA} shows how increasing the distances of all of the pulsars with the distance scale factor improves the measurement of $R$.  In effect we will need pulsars across the entire span of our Galaxy, and ideally out to the Large and Small Magellanic Clouds, with distance uncertainties on the order of $\Delta L_1$ or $\lambda_\mathrm{gw}$ in order to probe source distances beyond $100$ Mpc.  Since the Fresnel number scales as $L^2$, increasing the scale factor of our PTA yields a marked effect towards recovering $R$.  If, however, the pulsar distance uncertainties also scaled as $L^2$, which we would expect for measurements of pulsar distances made using parallax, then we found that these effects effectively cancelled each other out, resulting in no improvement in the recovery of $R$ even with larger PTAs.  Therefore it is crucial that as the pulsar distances in our PTA increases, that their uncertainties continue to remain of the order of the wrapping cycle (a quantity which is \textit{independent} of the pulsar's distance).
        
        Sources with certain physical properties will also be favored when measuring Fresnel corrections, as seen in the right panel of Figure~\ref{fig: s1 scaled PTA / M-omega0 space sf7 PTA}.  In terms of intrinsic parameters, strongly chirping sources with coalescence times $\Delta \tau_c \sim \mathrm{kyr} - \mathrm{Myr}$ will yield the best measurements of $R$ through Fresnel corrections.  The bias we see in this figure leans towards lower frequency sources, largely because these sources will have larger wrapping cycles and therefore require less pulsar distance precision than higher frequency sources (see equations~\ref{eqn:Delta L in monochromatic regimes} and~\ref{eqn: gw wavelength}).  Therefore keeping $\sigma_L = 1$ pc pinned means the lower frequency sources benefit from a higher degree of precision than the higher frequency sources.
        
\begin{figure}
            \centering
            \includegraphics[width=0.98\linewidth]{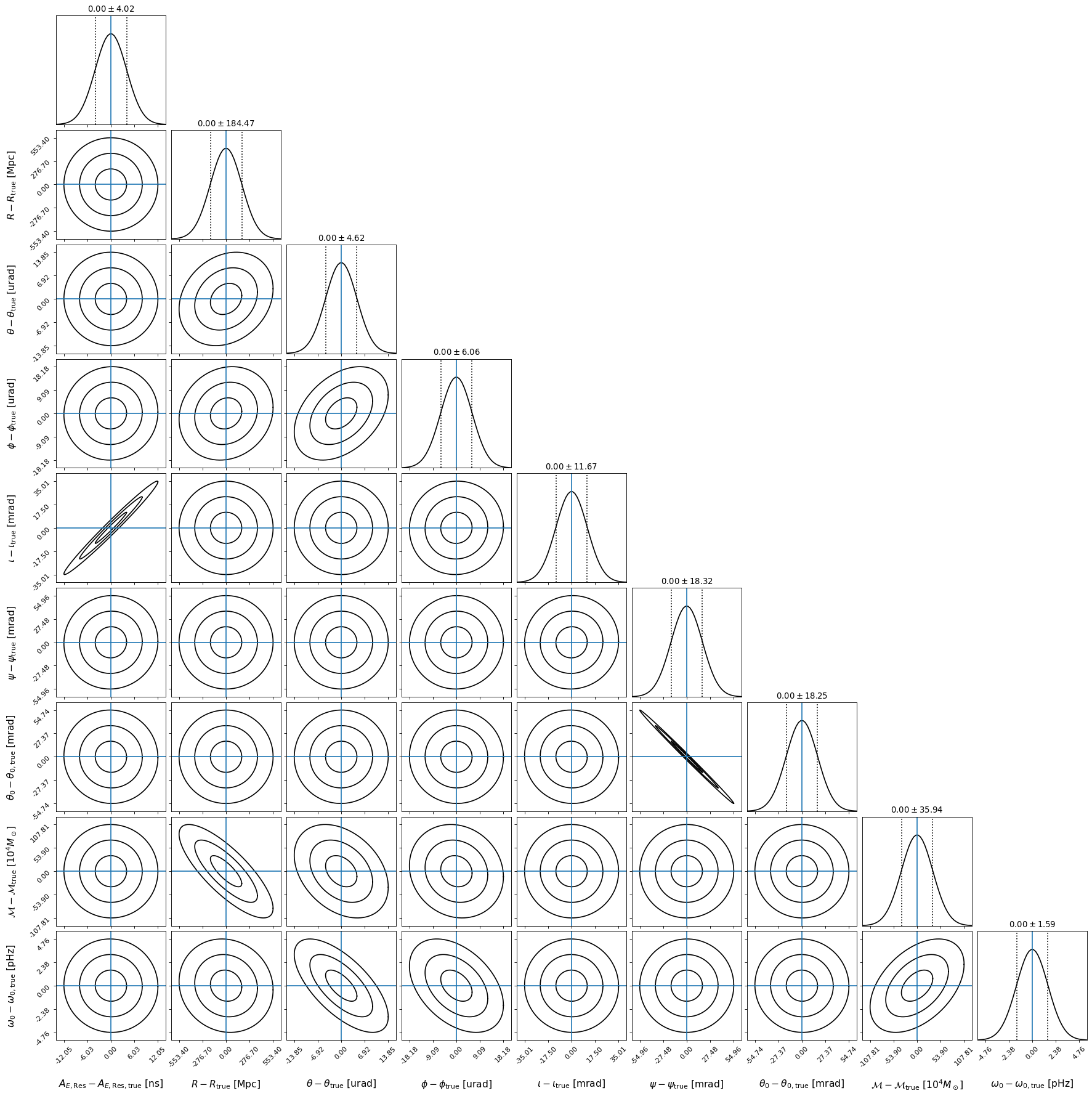}
        \caption[Fisher Triangle Plot Example - Source 1]{Fisher matrix measurements of the parameters for Source 1, which has been placed at $R=500$ Mpc (therefore $A_{E,\mathrm{res}}=414$ ns).  Here the PTA has a scale factor of $7$ and all pulsars have been given $\sigma_L = 1 \ \mathrm{pc}\ \sim \lambda_\mathrm{gw}$, with timing uncertainty $\sigma = 100$ ns.  This simulation and experimental set-up corresponds to the red square indicated in Figure~\ref{fig: s1 scaled PTA / M-omega0 space sf7 PTA}.  The Fisher matrix analysis predicts that this source can be recovered with:
        $\left\{ \mathrm{CV}_{A_{E,\mathrm{res}}}, \ \mathrm{CV}_R, \ \mathrm{CV}_\theta, \ \mathrm{CV}_\phi, \ \mathrm{CV}_\iota, \ \mathrm{CV}_\psi, \ \mathrm{CV}_{\theta_0}, \ \mathrm{CV}_\mathcal{M}, \ \mathrm{CV}_{\omega_0} \right\} = \left\{ 0.0097, \ 0.367, \ 3\times10^{-6}, \ 10^{-6}, \ 0.0148, \ 0.0175, \ 0.0182, \ 3.6\times10^{-5}, \ 5.3\times10^{-5} \right\}$.  These results were computed using the IIB model (Section~\ref{subsec: Frequency Evolution CONJECTURE (Fresnel)}).}
            \label{fig: s1 example triangle}
        \end{figure}

        While our focus so far has been on the recovery of the distance parameter, we also looked at the measurement of all source parameters and their covariances.  An example of this is shown in Figure~\ref{fig: s1 example triangle}, where Source 1 was placed at $R = 500$ Mpc.  Again this is a loud source and $\mathrm{CV}_{A_{E,\mathrm{res}}} = 0.0097$ for our simulated $\sigma = 100$ ns timing uncertainty.  In this case $\mathrm{CV}_R = 0.37$, while all remaining parameters had their respective values of $\mathrm{CV} \leq 0.018$.  Sky angles tend to be measured well for the reasons discussed in Section~\ref{sec: Source Localization}.  Plus as this is a highly chirping source, chirp mass and frequency are well measured.
         % Source ID:  20200916_20:43:42-54250  |   CV values:  0.009707 	0.368932 	0.000003 	0.000001 	0.014857 	0.017495 	0.018245 	0.000036 	0.000053
        
        For this particular example, we found that improving the timing uncertainty by an order of magnitude to $\sigma = 10$ ns improved the indicated $\mathrm{CV}$ values by one order of magnitude for the parameters $\left\{A_{E,\mathrm{res}}, \iota, \psi, \theta_0  \right\}$, while all remaining parameter $\mathrm{CV}$ values remained the same.  In general we found that these four parameters were most sensitive to timing accuracy, while mostly insensitive to pulsar distance accuracy.  While the measurement of $R$ is the opposite of this as we have already discussed, we found that the remaining four source parameters $\left\{\theta, \phi, \mathcal{M}, \omega_0 \right\}$ could benefit from both improvements to timing as well as pulsar distance uncertainties.

%---------------------------------------------------------------------------------
%---------------------------------------------------------------------------------
    \section{Source Localization}\label{sec: Source Localization}
    
    As we investigated in the previous section, high precision measurements of pulsar distances (to within $\sigma_L \sim \Delta L_1$) allow for the direct measurement of the source distance $R$ purely through Fresnel corrections.  In this section, we show that this level of precision can also improve localization of the source on the sky.  As we will explain, this improved sky localization mostly comes from phase and frequency parallax terms which can be exploited when the pulsar distance measurements are well constrained.  When we combine the sky localization with the recovered distance information, there exists potential for future PTAs to pinpoint the galaxy source of the gravitational waves, which could be of great benefit to astronomers seeking electromagnetic counterparts to these coalescing binary sources.
    
    To measure the solid angle on the sky that our uncertainty in the measurement of the source angles $\theta$ and $\phi$ sweeps out, we first compute the confidence ellipse area for those two parameters $\Delta A = \pi \chi^2 \sigma_{\theta^{'}} \sigma_{\phi^{'}}$ as outlined in~\citet{fisher_confidence_ellipses}\footnote{Note that in~\citet{fisher_confidence_ellipses}, equation 7 has a typo in that it is missing the factor of $\chi^2$ which we have corrected here in equation~\ref{eqn: sky localization}.}.  Here $\sigma_{\theta^{'}}$ and $\sigma_{\phi^{'}}$ are the measured uncertainties along the principle axes of the ellipse itself, and this quantity represents an area in the $\theta$-$\phi$ parameter space.  On the sky, we can define a small solid angle as $\Delta \Omega \approx \sin(\theta) \Delta\theta \Delta\phi$.  So to measure a small confidence ellipse on the sky we connect these two quantities by multiplying $\Delta A$ by $\sin(\theta)$ to get:
    \begin{equation}
        \Delta \Omega \approx \pi \chi^2 \sin(\theta) \sigma_{\theta^{'}} \sigma_{\phi^{'}} = \pi \chi^2 \sin(\theta) \sigma_\theta \sigma_\phi \sqrt{1 - \rho^2} .
    \label{eqn: sky localization}
    \end{equation}
    The uncertainties $\sigma_\theta$ and $\sigma_\phi$ and the correlation coefficient $\rho$ are are all measured from the inverse of the Fisher matrix.  In all of our results here we set $\chi^2 = 2.279$, which gives the approximate $68$\% likelihood area for our source.  In a similar calculation for small volume $\Delta V \approx \frac{4\pi}{3} R^2 \sin(\theta) \Delta R \Delta \theta \Delta \phi$ we compute the volume uncertainty region for our source as:
     \begin{equation}
         \Delta V \approx \frac{4\pi}{3}\chi^3 R^2 \sin(\theta) \sigma_{R^{'}} \sigma_{\theta^{'}} \sigma_{\phi^{'}},
     \label{eqn: volume localization}
     \end{equation}
     where again, primes on the uncertainty variables denote they are the uncertainties measured along the principle axes of the ellipsoid.  In 3D space, the approximate $68$\% likelihood volume of our source corresponds to a $\chi^2 = 3.5059$, which is what we used for all of our results shown here.  The values of the uncorrelated uncertainties along the principle axes $\sigma_{R^{'}}$, $\sigma_{\theta^{'}}$, and $\sigma_{\phi^{'}}$ were found by applying SVD to the 3D sub-matrix of our inverted Fisher matrix (for parameters $R$, $\theta$, and $\phi$).  This mathematically decomposes the matrix into three matrices, one of which is a rectangular diagonal matrix that contains the singular values, that is the uncorrelated uncertainties along the principle axes.
    
    In all of our timing regimes, part of the measurement of the sky angles $\theta$ and $\phi$ comes from the antenna patterns contained in the amplitudes of the Earth and pulsars terms of the timing residuals (see equations~\ref{eqn: antenna f} and~\ref{eqn: antenna F}).  Additionally, all timing regimes also have the ``phase parallax'' term $\left(1-\hat{r}\cdot\hat{p}\right) \frac{L}{c}$ which contains the sky angles, and appears in the phase $\Theta(t)$ of the pulsar terms of each model (see equations~\ref{eqn: Res(t) pw mono phase E and P}, \ref{eqn: Res(t) pw freq evo frequency phase E and P}, \ref{eqn: Res(t) fresnel mono heuristic phase E and P}, and~\ref{eqn: Res(t) fresnel freq evo phase E and P}).  In the frequency evolution regimes IB and IIB this becomes a ``phase-frequency parallax'' term, because not only does it appear in the pulsar term phase $\Theta(t)$ but also in the pulsar term frequency $\omega(t)$. Lastly, the Fresnel regimes add an additional small order correction $\frac{1}{2}\left(1-\left(\hat{r}\cdot\hat{p}\right)^2\right)\frac{L}{c}\frac{L}{R}$ to these phase/phase-frequency parallax terms.
    
    With this in mind we looked at examples of improvements to three PTA qualities which can improve sky localization:  timing precision $\sigma$, pulsar distance precision $\sigma_L$, and PTA distance scale factor.  Timing precision helps measure the sky angles from the overall amplitude of the timing residual.  Pulsar distance precision helps measure the sky angles from the phase/phase-frequency parallax terms due to the $\frac{L}{c}$ factor.  However this requires high precision measurements due to the pulsar distance wrapping problem.  It is partially for this reason that studies such as~\citet{NG_11yr_cw} choose to group these parallax terms into separate phase parameters for every individual pulsar in their models (which is a similar but alternative approach to introducing the pulsar \textit{distances} as free parameters in the model, as discussed in Section~\ref{sec: breaking the timing residual fisher matrix into submatrices} and Chapter~\ref{ch:L-wrapping problem}).  This approach sacrifices the additional sky localization information contained in these parallax terms, to help avoid the wrapping problem.  Finally, increasing the PTA scale factor can help improve the sky angle measurements by amplifying the chirping effects (due to the greater disparity between the Earth/pulsar frequencies), and boosting the size of the Fresnel corrections.
    
\begin{figure}
        \centering
        \includegraphics[width=1\linewidth]{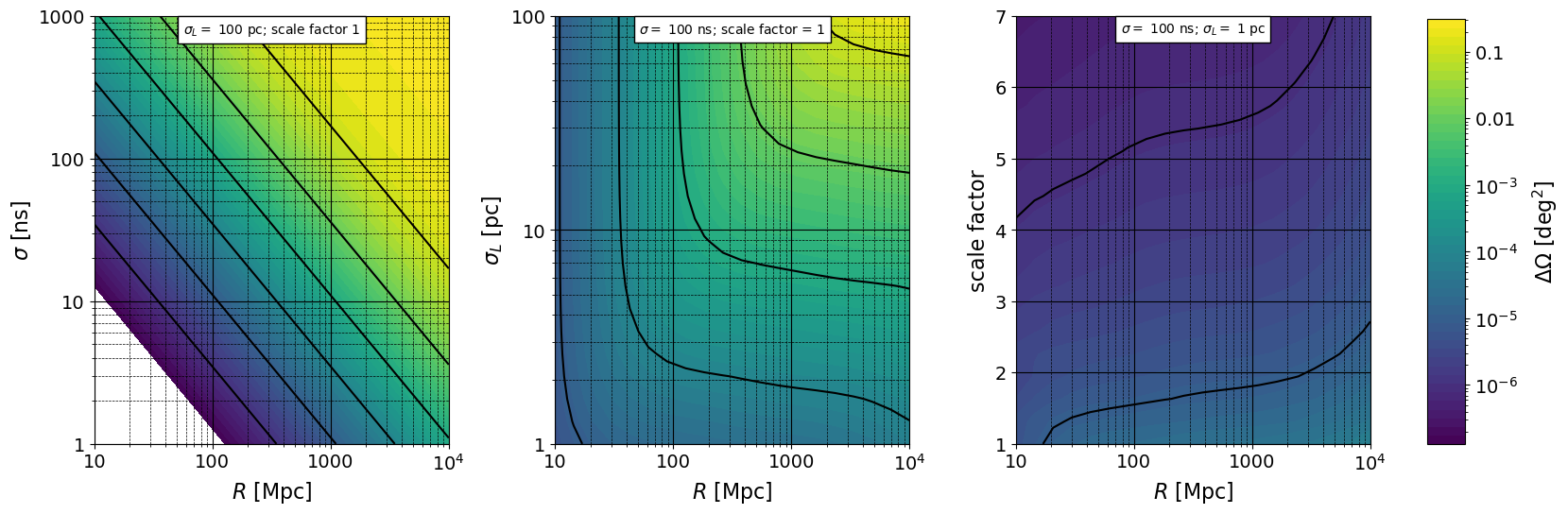}
    \caption[Source Localization 1 (PTA Parameters)]{Various effects of PTA improvements on source sky localization, shown here for the example of Source 1.  The bottom left corner of the first panel is excluded because our SVD matrix inversion failed our condition number requirement (see Section~\ref{sec: breaking the timing residual fisher matrix into submatrices}), therefore the Fisher matrix for these cases couldn't be stability inverted.  Nevertheless, we can clearly see the trend is that the overall sky accuracy improves as the source gets closer to us and the value of $\sigma$ decreases, as we would expect.  Compared to the other PTA parameters, timing precision improvements behave as a univariate function of $\sigma R$.  Furthermore, timing precision controls the largest range of magnitude differences in $\Delta \Omega$ as a function of source distance, followed by pulsar distance precision, then finally PTA scale size.  These results were computed using the IIB model (Section~\ref{subsec: Frequency Evolution CONJECTURE (Fresnel)}).}
        \label{fig: sky localization - comparison (contours)}
    \end{figure}

    Examples of improvements in these three PTA qualities as applied to Source 1 are shown in Figure~\ref{fig: sky localization - comparison (contours)}.  It is very difficult to disentangle the many different factors that are all competing to change the measurement of the source's sky localization (including the intrinsic source parameters which we discuss below).  Therefore we emphasize here that Figure~\ref{fig: sky localization - comparison (contours)} is only meant to serve as one specific illustration of the general comments that we have made so far, for the PTA set-ups indicated in each panel, and for one particular source.
    
    We find that unlike pulsar distance precision and PTA scale factor, timing precision behaves as a univariate function of $\sigma R$.  In the first panel of Figure~\ref{fig: sky localization - comparison (contours)} we expect that most of the knowledge of our sky angles is coming from the overall amplitude of the timing residual and not from the phase-frequency parallax or Fresnel corrections (due to the large $\sigma_L$ and standard scale factor).  The second and third panels show the additional benefits that can be gained when probing the information in the phase-frequency parallax and Fresnel corrections.  Recall that $\lambda_\mathrm{gw} \sim 1$ pc for Source 1.  In the middle panel the Fresnel corrections are not likely adding a great deal of information towards the sky angles themselves since $\sigma_L > \lambda_\mathrm{gw}$ for most of these cases.  However we still see that information can be gained through the phase-frequency parallax as the precision of the distance measurements to our pulsars improves.  As $\sigma_L$ decreases, we see for a given $\Delta \Omega$ resolution there are turnover points where even small improvements to the pulsar distance measurements can allow the same source to be localized with that precision at much greater distances.  Finally, with pulsar distance measurements on the order of the source's gravitational wavelength, the right panel shows that increasing the baseline distance between pulsars further improves our ability to measure the source's sky angles.  Overall we see that timing precision controls the largest range of magnitude differences in $\Delta \Omega$ as a function of source distance, followed by pulsar distance precision, then finally PTA scale size.
    
\begin{figure}
        \centering
        \includegraphics[width=1\linewidth]{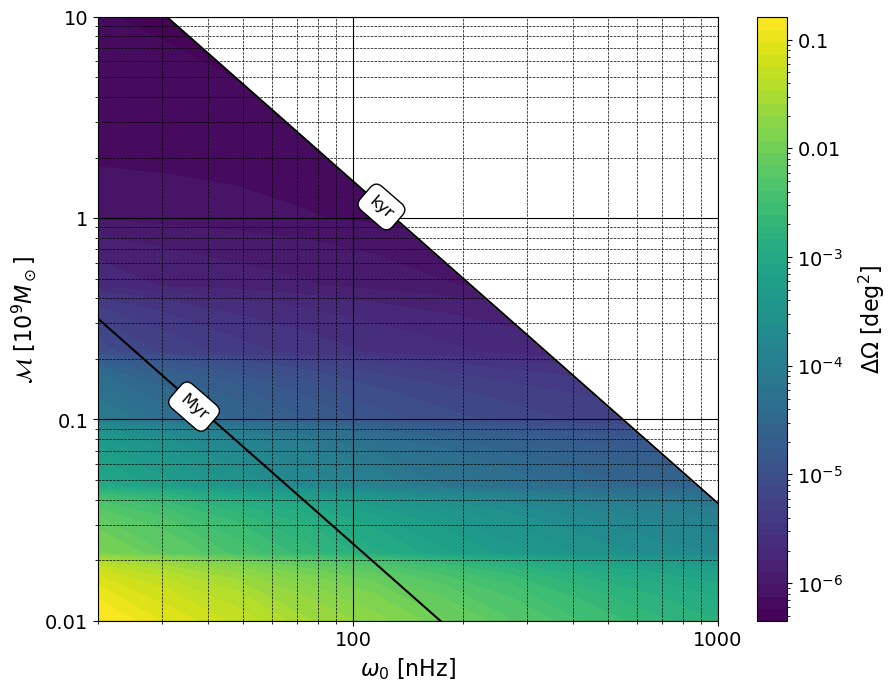}
        \caption[Source Localization 2 (Source Parameters)]{Source sky localization, as a function of source frequency and chirp mass.  These are for sources at $R=100$ Mpc, with the same angular parameters as our fiducial sources (indicated in Table~\ref{tab: source parameters reference}), timing uncertainty $\sigma = 100$ ns, pulsar distance uncertainty $\sigma_L = 1$ pc, and a PTA scale factor of $7$.  Contours of coalescence time $\Delta \tau_c$ (equation~\ref{eqn: time to coalescence}) are indicated, with a cut along $\Delta \tau_c = 1$ kyr due to assumption~\ref{as: t_obs << tau_c}.  Bias leans towards lower frequencies, where using a fixed $\sigma_L$ value will benefit sources with larger wrapping cycles as compared to smaller ones (see Section~\ref{subsec: distant sources}).  These results were computed using the IIB model (Section~\ref{subsec: Frequency Evolution CONJECTURE (Fresnel)}).  When we re-calculated this plot using the IB model (Section~\ref{subsec: Frequency Evolution (PW)}) and IA model (Section~\ref{subsec: Monochromatic Model (PW)}) we found that the relative difference in $\Delta \Omega$ between these two models and the original fully general IIB model was about $0.1$ and $0.7$, respectively.}
        \label{fig: sky localization - M-omega0}
    \end{figure}

    The source's intrinsic parameters $\mathcal{M}$ and $\omega_0$ also affect localization.  Figure~\ref{fig: sky localization - M-omega0} shows that for the same distance (here $R=100$ Mpc) and PTA set-up, different sources are localized to $\sim\mathcal{O}\left(0.1 \ \mathrm{deg}^2 \ - \ 10 \ \mathrm{arcsec}^2\right)$.  Notice that this PTA set-up matches that in the right panel of Figure~\ref{fig: s1 scaled PTA / M-omega0 space sf7 PTA} so we know that the Fresnel corrections are being probed for the sources where $\mathrm{CV}_R < 1$, since the distance parameter $R$ is measured entirely from these corrections in our model.

    In general we conclude that the Fresnel corrections themselves don't significantly improve sky localization of a source.  Figure~\ref{fig: sky localization - M-omega0} shows $\Delta \Omega$ predicted from the Fisher matrix analysis using the fully generalized IIB timing residual model (Section~\ref{subsec: Frequency Evolution CONJECTURE (Fresnel)}).  We repeated the same calculations of the sky angles for these sources using the IB model which removes the Fresnel corrections (Section~\ref{subsec: Frequency Evolution (PW)}), and the IA model which removes both the Fresnel corrections and the frequency evolution effects (Section~\ref{subsec: Monochromatic Model (PW)}), in order to compare the predictions between models.  It is important to note that this isn't directly comparable, because fundamentally these three models will compute different timing residuals for the same source (see again Figure~\ref{fig: model-regime comparison} - the level of difference will depend on what regime the source is actually in).  And the Fisher matrix analysis only asks how well can we recover the parameters \textit{from a given model}.  Nevertheless, comparing the results of these sets of calculations still gives us an idea of how different the sky localization is with vs. without the Fresnel corrections.  What we found was that the magnitude of the relative difference between $\Delta \Omega$ computed using the models IIB vs. IB and IIB vs. IA was at most about $0.1$ and $0.7$ for what is shown in Figure~\ref{fig: sky localization - M-omega0}, respectively.  This suggests that we lose some sky localization precision when we leave out Fresnel corrections, but we lose more precision when we don't account for frequency evolution.  However, even with these relative differences we still find that in the IB and IA models (which don't include Fresnel corrections) sky localization ranges from $\sim\mathcal{O}\left(0.1 \ \mathrm{deg}^2 \ - \ 10 \ \mathrm{arcsec}^2\right)$ depending on the intrinsic source parameters.  This along with the earlier observations and statements about Figure~\ref{fig: sky localization - comparison (contours)} is what leads us to believe sky localization itself is not greatly improved by the inclusion of Fresnel effects.    

    In their separate studies~\citet{CC_main_paper} reported sky localization could be measured with $\sim\mathcal{O}\left(1-10 \ \mathrm{deg}^2\right)$ while~\citet{DF_main_paper} reported sky localization could be measured with $\sim\mathcal{O}\left(100 \ \mathrm{arcsec}^2\right)$, which is a significant improvement.  The results and discussion here from our study help to explain why.  These studies use different physical models (with and without Fresnel corrections), different example sources and PTAs, and simulate very different pulsar distance uncertainties.  \citeauthor{CC_main_paper} used values of $\sigma_L \sim \mathcal{O}\left(1-100 \ \mathrm{pc}\right)$, while~\citeauthor{DF_main_paper} used $\sigma_L \sim \mathcal{O}\left(0.001 - 0.01 \ \mathrm{pc}\right)$.  Here we see in Figure~\ref{fig: sky localization - M-omega0} approximately five orders of magnitude difference in sky localization depending on the type of source, and in the middle panel of Figure~\ref{fig: sky localization - comparison (contours)} another five orders of magnitude difference between $\sigma_L = 1 \ - \ 100$ pc.  Therefore we easily found scenarios where $\Delta \Omega$ ranged from $\mathrm{arcsec}^2$ to $>10\ \mathrm{deg}^2$ precision, simply by accounting for these different effects, PTA qualities, and sources.
    
    While Fresnel corrections in our models may not add much precision to source sky localization, we can combine the distance measured from Fresnel corrections with this sky location, to pinpoint our source to within an uncertainty volume of space. An example for Source 1 is shown in Figure~\ref{fig: vol localization - scale factor / S1 localization - vol at 100 Mpc, sky at 1 Gpc}.  The results have only been given where all points have a corresponding $\mathrm{CV} < 1$ for the parameters $R$, $\theta$, and $\phi$.  Since the source distance $R$ is the hardest parameter to measure in general, our ability to localize the source to within some uncertainty volume in space is most directly determined by how well we can measure that parameter.  Some parts of the sky are less sensitive than others, and even if we lose the sensitivity within our PTA required to measure the volume localization $\Delta V$, we can still typically measure a sky location $\Delta \Omega$ with striking precision.  The right panel of Figure~\ref{fig: vol localization - scale factor / S1 localization - vol at 100 Mpc, sky at 1 Gpc} shows even at $1$ Gpc, Source 1 was measured here with sub-square arcminute precision no matter where it was located on the sky.
    \begin{figure}
        \centering
          \begin{subfigure}[t]{0.49\linewidth}
          \centering
            \includegraphics[width=1\linewidth]{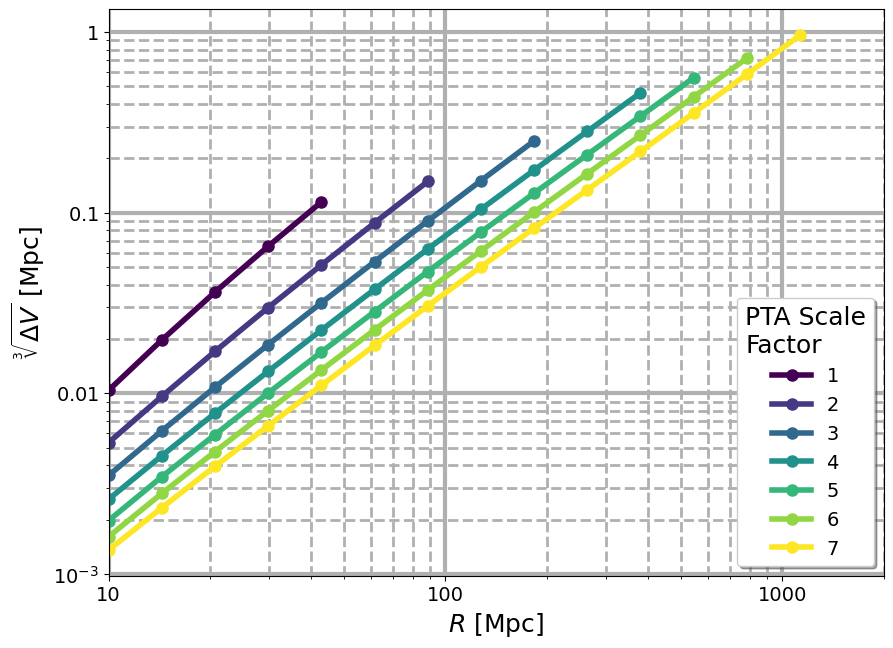}
          \end{subfigure}
          \hfill
          \begin{subfigure}[t]{0.49\linewidth}
          \centering
          \includegraphics[width=1\linewidth]{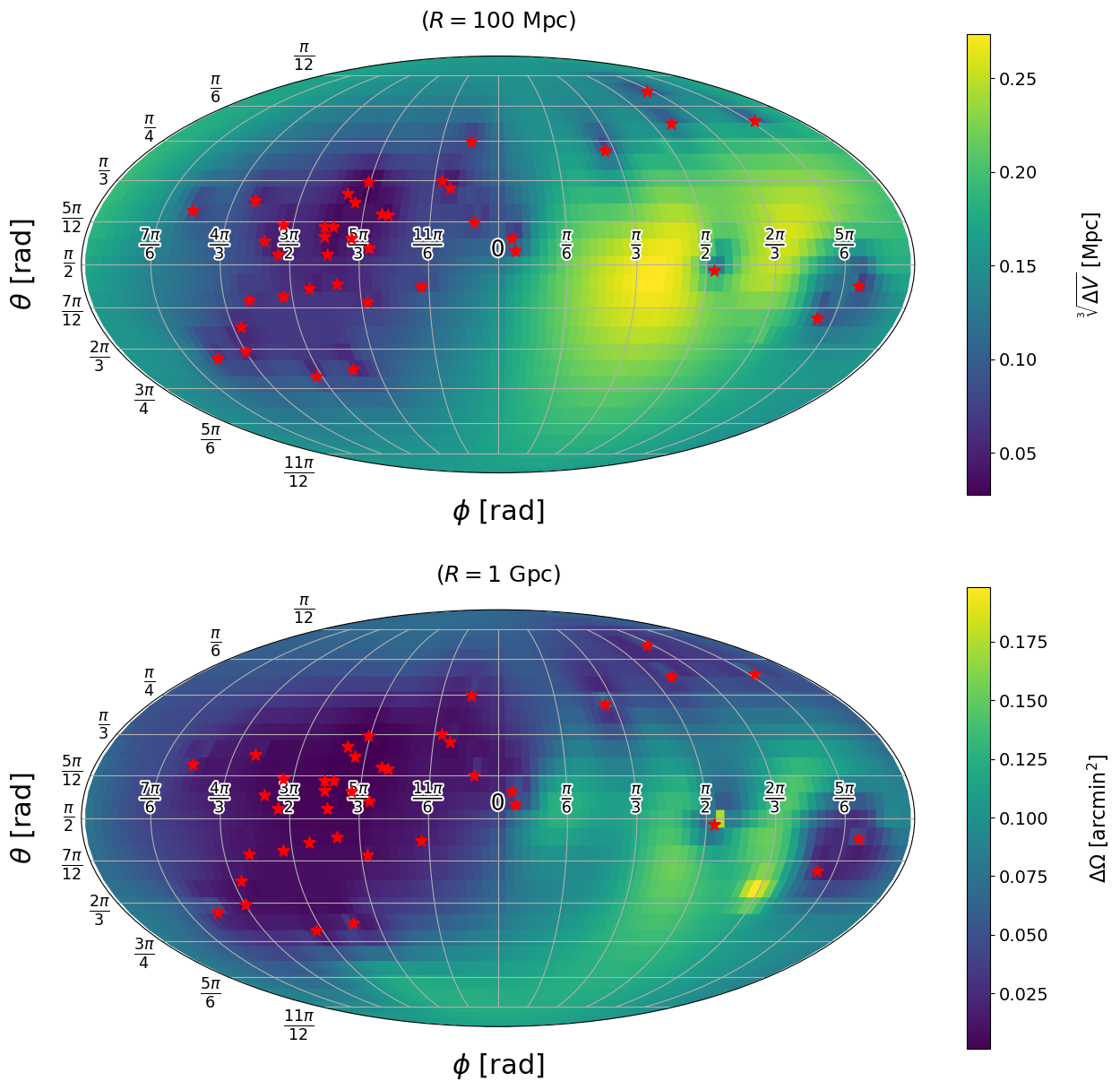}
          \end{subfigure}
    \caption[Source Localization 3 (Sky \& Volume Localization)]{Source sky and volume and localization calculated for Source 1, with timing uncertainty $\sigma = 100$ ns, and pulsar distance uncertainty $\sigma_L = 1 \ \mathrm{pc}\ \sim \lambda_\mathrm{gw}$.  The areas and volumes represent the $68$\% likelihood regions of our source.  These results were computed using the IIB model (Section~\ref{subsec: Frequency Evolution CONJECTURE (Fresnel)}).  \textbf{(Right)}~The top panel shows volume localization of the source at $R=100$ Mpc (therefore $A_{E,\mathrm{res}}=2 \ \mu$s).  In the bottom panel the source is at $R=1$ Gpc (therefore $A_{E,\mathrm{res}}=207$ ns), and although we can no longer measure its distance with this PTA (and hence calculate the volume localization), we can still measure the sky angles from the chirping in the signal to localize it on the sky.  In both the top and bottom panel the PTA has a scalefactor of $5$.}
    \label{fig: vol localization - scale factor / S1 localization - vol at 100 Mpc, sky at 1 Gpc}
    \end{figure}

%---------------------------------------------------------------------------------
%---------------------------------------------------------------------------------
    \section{Measuring Chirp Mass from a Monochromatic Source}\label{sec: Measuring Chirp Mass from a Monochromatic Source}
    
    Another important novelty of the Fresnel regime is that it allows a measurement of the source's chirp mass even if the source is producing monochromatic gravitational waves.  For a monochromatic source, Fresnel corrections break the $R$-$\mathcal{M}$ degeneracy in the IIA regime (see Section~\ref{subsec: Monochromatic Model (Fresnel)}).  As long as we can measure $A_{E,\mathrm{res}}$, $R$, and $\omega_0$, then from equation~\ref{eqn: earth term amplitude} we can infer the the chirp mass.  As pointed out in Section~\ref{sec: Simple SNR Analysis}, within the limitations of this study, low frequency sources below $\omega_0 < 10$ nHz require a greater analysis of the impact of other timing residual sources before confidently measuring the gravitational wave parameters and their uncertainties.  Therefore what we show here is simply meant to be a proof-of-concept.
    
    For this we use the fiducial Source 2 in Table~\ref{tab: source parameters reference}.  An ideal monochromatic source would have infinite $\Delta \tau_c$, so we chose a low chirp mass source on the edge of our SNR limitation.  At $R=100$ Mpc, the amplitude of such as source would be very small, so we also simulate a future timing experiment capable of $\sigma \sim 1$ ns timing uncertainties.  In a future study, it would be interesting to simulate the measurement of a higher chirp mass system with an orbital frequency $\omega_0 \sim 1$ nHz, as this type of source would produce a stronger amplitude.
    
\begin{figure}
        \centering
          \begin{subfigure}[t]{0.49\linewidth}
          \centering
          \includegraphics[width=1\linewidth]{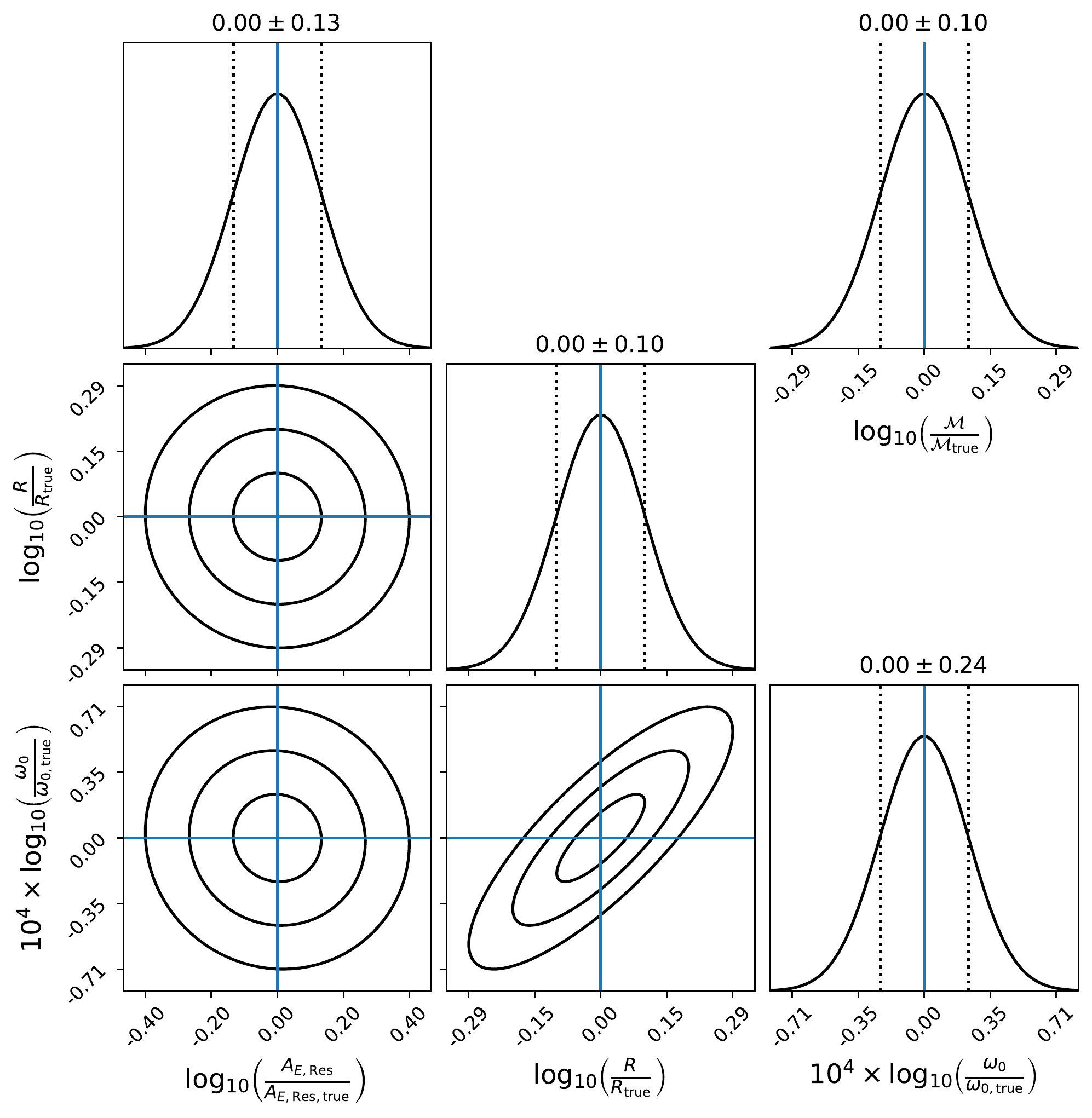}
          \end{subfigure}  % Source ID: 20200918_16:31:05-05345  |   CV values: 0.316293 	0.225834 	0.000007 	0.000004 	0.477479 	0.576421 	0.592171 	0.000054 --> CVM = 0.229168
          \hfill
          \begin{subfigure}[t]{0.49\linewidth}
          \centering
            \includegraphics[width=1\linewidth]{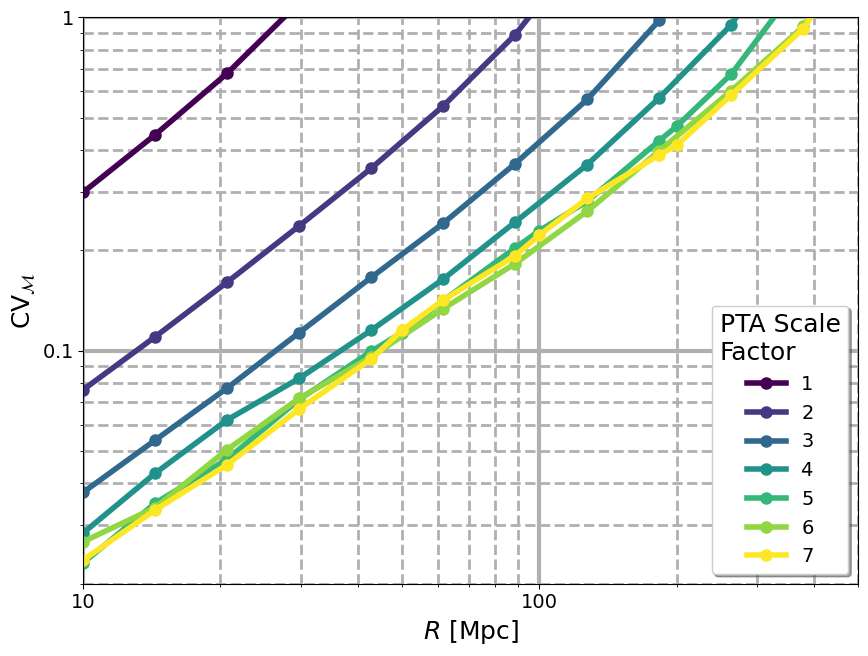}
          \end{subfigure}
    \caption[Regime IIA - Measuring $\mathcal{M}$]{Error propagation to the measurement of the source's chirp mass $\mathcal{M}$ for a monochromatic source recovered from the measurements of $A_{E,\mathrm{res}}$, $R$, and $\omega_0$, calculated here in both panels for Source 2.  Here all pulsars have been given timing uncertainty $\sigma = 1$ ns, and pulsar distance uncertainty $\sigma_L = 1.5 \ \mathrm{pc}\ \sim \lambda_\mathrm{gw}$.  Note that in this study we chose to parametrize the model in the log of the parameters $\left\{A_{E,\mathrm{res}}, \ R, \ \omega_0\right\}$.  These results were computed using the heuristic IIA model (Section~\ref{subsec: Monochromatic Model (Fresnel)}).  \textbf{(Left)} Here we show the recovery and correlations of the log of the parameters $\left\{ A_{E,\mathrm{res}}, \ R, \ \omega_0 \right\}$ in the triangle plot, and in the upper right corner we show the propagated uncertainty on the measurement of the log of $\mathcal{M}$.  Source 2 has been placed at $R=100$ Mpc (therefore $A_{E,\mathrm{res}}=0.15$ ns), and here the PTA has a scale factor of $5$.  In this particular example, we find that the uncertainty on the chirp mass is dominated by the pulsar distance uncertainty $\sigma_L$, which most strongly affects the recovery of $R$.  The Fisher matrix analysis predicts that this source can be recovered with:
    $\left\{ \mathrm{CV}_{A_{E,\mathrm{res}}}, \ \mathrm{CV}_{R}, \ \mathrm{CV}_\theta, \ \mathrm{CV}_\phi, \ \mathrm{CV}_\iota, \ \mathrm{CV}_\psi, \ \mathrm{CV}_{\theta_0}, \ \mathrm{CV}_{\omega_0} \right\} = \left\{ 0.32, \ 0.23, \ 7\times10^{-6}, \ 4\times10^{-6}, \ 0.48, \	0.58, \ 0.59, 5.4\times10^{-5} \right\}$.  Uncertainty propagation gives an inferred $\mathrm{CV}_{\mathcal{M}} = 0.23$ .  \textbf{(Right)} For small PTA scale factors, recovery of $\mathcal{M}$ is dominated by pulsar distance uncertainty $\sigma_L$.  As the scale factor increases, the larger Fresnel numbers make the measurement of $R$ more precise, and as a result for scale factors above about $5$ the measurements of $\mathcal{M}$ don't improve because they become dominated by timing uncertainty $\sigma$.  Note a scale factor of $7$ puts the furthest pulsar in our PTA at roughly the distance of the Large Magellanic Cloud.}
    \label{fig: source iia example - M recovery / source iia example - scalefactor plot}
    \end{figure}

    The left panel of Figure~\ref{fig: source iia example - M recovery / source iia example - scalefactor plot} shows an example of the uncertainty propagation to our measurement of the system's chirp mass.  We found that measuring $\mathcal{M}$ was most strongly correlated to measuring $A_{E,\mathrm{res}}$ and $R$.  Since measuring $A_{E,\mathrm{res}}$ depends on the timing uncertainty $\sigma$, and measuring $R$ depends on the pulsar distance uncertainty $\sigma_L$, this means different experimental set-ups may depend more on timing or parallax for the recovery of the system chirp mass.  As an example, this can be seen in the right panel of Figure~\ref{fig: source iia example - M recovery / source iia example - scalefactor plot}.  Initially, increasing the PTA scale factor notably improves the recovery of $\mathcal{M}$, since the higher Fresnel numbers and improved parallax measurements allow for better recovery of $R$ (similar to what was shown earlier in the left panel of Figure~\ref{fig: s1 scaled PTA / M-omega0 space sf7 PTA}).  However, around a scale factor of $5$ we reach the threshold where timing precision dominates over parallax in our ability to recover $\mathcal{M}$.  Moreover for this source we found that increasing $\sigma > 1$ ns more dramatically made the recovery of $\mathcal{M}$ dependent on the recovery of $A_{E,\mathrm{res}}$.

%---------------------------------------------------------------------------------
%---------------------------------------------------------------------------------
    \section{Conclusions}\label{sec: Conclusions - first investigation}
    
    The work in this and the preceding chapters constitutes the first major investigation of my PhD work.  To summarize everything thus far, in this work we motivate the importance of Fresnel corrections in the effects of gravitational waves on pulsar timing residual models, and develop what we call the ``Fresnel'' timing regime, separate from the previously established plane-wave regimes which are currently used in pulsar timing searches for continuous waves~\citep{PPTA_cw_2014,EPTA_cw_2016,NG_11yr_cw}.  We derive the Fresnel monochromatic model analytically, study it asymptotically, and provide a more simple and heuristically motivated model which is easier to implement in timing simulations and analyses.  With this we build a well motivated conjecture of the Fresnel frequency evolution timing residual model, which fully generalizes the previously studied plane-wave frequency evolution models with Fresnel order corrections to the phase and frequency of the timing residual.  Surrounding this we also offer a framework for understanding the relevant limits of each model, namely by considering the coalescence time of the source and the size of its Fresnel number with the pulsars in the timing array.

    Then we perform a Fisher matrix analysis for parameter estimation within these Fresnel regimes, using a representative NANOGrav-related PTA as the base for the experimental design within our simulations.  This was meant to build upon and bridge the gap between the studies by~\citet{CC_main_paper} and~\citet{DF_main_paper}.  The main challenge in measuring the source distance purely from Fresnel corrections is the pulsar distance wrapping problem.  We offer a new look at this problem and show how well measured the pulsar distances will need to be in order to accurately recover the source distance from our searches.  In general distance constraints are needed on the order of the wrapping cycle on the most distant pulsars in the PTA, but not on the nearby pulsars whose Fresnel corrections are negligible.
    
    One way to increase the Fresnel corrections in the timing residuals is by including pulsars at farther distances, since the Fresnel number scales as $F \propto L^2$ (see equation~\ref{eqn: Fresnel number}).  Intrinsic source properties $\mathcal{M}$ and $\omega_0$ also play an important role as they control both the amount of frequency evolution in the residuals and the characteristic wrapping cycle.  In our simulations we find that future PTAs with distant pulsars across the Milky Way Galaxy whose distances are constrained to the order of the wrapping cycle can measure the source distance with a $\mathrm{CV}_R \lesssim 0.1$ for sources at out to $\mathcal{O}(100 \ \text{Gpc})$ distances.
    
    High precision pulsar measurements not only allow us to probe the Fresnel corrections, but they also help localize the source through the combined phase-frequency parallax of all of the pulsars in the PTA.  The previous studies of source localization by \citeauthor{DF_main_paper} and \citeauthor{CC_main_paper} had shown many orders of magnitude difference in the sky angle ranging from $\mathcal{O}(100 \ \text{arcsec}^2 \ - \ 10 \text{deg}^2)$.  In this work we explain that this difference is due to many factors including the timing accuracy $\sigma$, the pulsar distance accuracy $\sigma_L$, the PTA scale factor, and the intrinsic source parameters.  We also show how knowledge of the source distance through Fresnel corrections can be combined with the sky angle measurements in order to localize the source to a volume of space.  Even for sources near $\sim 1$ Gpc in distance could be localized to within a volume $\Delta V < 1 \ \text{Mpc}^3$.  This would allow us to localize the galaxy source of the gravitational waves which could then be targeted with an in-depth multi-messenger follow-up, such as what was done recently in \citet{multimessenger_caitlin_2020}.
    
    A novelty of the Fresnel formalism is that it offers a way of measuring the chirp mass of a monochromatic source, which is something that cannot be done using the plane-wave formalism.  As long as the amplitude, distance, and frequency parameters can be measured, then the chirp mass can be calculated.  Furthermore, the Fresnel formalism opens a path for future cosmological studies using PTAs.  As discussed recently in~\citet{DOrazio_main_paper}, when the Fresnel models are generalized to a cosmologically expanding universe, then the distance parameter $R$ which comes from the Fresnel corrections becomes the comoving distance $D_c$ (assuming a spatially flat universe).  If the source is also significantly chirping, then the system's observed (redshifted) chirp mass can be measured.  Combining this with the measured observed frequency and amplitude parameters produces a measurement of the system's luminosity distance $D_L$, which is related to the comoving distance through $D_L = (1 + z)D_c$.  Therefore in the full Fresnel frequency evolution regime IIB, frequency evolution effects allow for the measurement of the source $D_L$, and Fresnel corrections allow for the measurement of the source $D_c$.  If both of these distances are recovered from the timing residual model then a measurement of $H_0$ can be obtained (as shown in \citeauthor{DOrazio_main_paper}).  Or alternatively if only one of these two distances can be measured accurately, but localization of the gravitational wave source and multimessenger counterparts identify the host galaxy, then the source host galaxy's redshift can be measured.  Combining the redshift with either of the measured source distances once again provides a measurement of $H_0$.
    
    While we acknowledge that many of our ``best-case'' scenarios required simulating idealized PTAs well beyond our current experimental capabilities, timing and pulsar measurements will continue to improve~\citep{Lam_2018}.  The Five hundred meter Aperture Spherical Telescope is now operational and offers exciting prospects for discovering many new pulsars~\citep{FAST_smits2009}.  The MeerKAT and MeerTime projects could provide high precision timing of pulsars with absolute timing errors on the order of 1 ns~\citep{MeerKAT_bailes2020}, and the future Square Kilometer Array is going to provide even greater sensitivity and will greatly increase the number of pulsars that we can time~\citep{SKA_janssen2015}.  Furthermore~\citet{GWastro_Lee2011} and~\citet{Ldist_smitts2011} discuss how pulsar distance measurements can be improved through timing parallax, which will improve with these more sensitive future experiments and could help push the pulsar distance precision closer to what is needed for this work.  In general this suggests the potential for future studies that could be made possible from the Fresnel timing regimes.  For example, seeing the direct impact that PTA size has on parameter estimation hopefully offers strong motivation to actively search for and include more distant pulsars with higher precision distance measurements in our current PTAs.  \citet{FAST_smits2009} even discusses the possibility of detecting extragalactic pulsars.  While it is unlikely such pulsars would have the precision constraints necessary for the measurements proposed in this work, it still points to new frontiers in the future of PTA science which could make this work possible.
    
    Finally, we reiterate here that in this work thus far we have explicitly assumed a flat static universe for the sake of simplicity and to help isolate the measurement of the distance parameter $R$ (see assumption~\ref{as:empty}).  More generally, for an expanding universe the fully generalized Fresnel frequency evolution model contains both a luminosity and a comoving distance parameter.  We will explore this generalized model in greater detail in Chapter~\ref{ch:results - H0 measurement}, and the intriguing cosmological implications that the Fresnel regime may offer to future PTAs.

\cleardoublepage
\chapter{Generalizing the Fresnel Models \& Measuring the Hubble Constant}\label{ch:results - H0 measurement}

\hspace{0.5cm} \begin{minipage}{0.9\linewidth}
\fbox {
    \parbox{\linewidth}{
    \underline{\textbf{Update}}

    At the time of finishing this dissertation, the work shown in this chapter was still a work-in-progress.  It was refined and published in~\cite{mcgrath2022H0}.  Some changes have been updated into this dissertation, but there may still be changes that were implemented in the final paper that were not yet realized here.
    }
}
\end{minipage}\\

\noindent Of notable interest is the ability of a gravitational wave based experiment to measure cosmological parameters such as the Hubble constant $H_0$.  These new gravitational wave based method measurements will become very important in helping to resolve the current tension between experimental measurements of $H_0$ \citep{gw170817_h0, Feeney_standardsirens}.  In this study we are motivated by the recent work of~\citet{DOrazio_main_paper} and myself~\citet{mcgrath2020fresnel} (the results of which were given in the previous Chapter~\ref{ch:results - first investigation}).  \citeauthor{DOrazio_main_paper} demonstrated how a measurement of the Hubble constant could be made from a purely gravitational wave-based method, by measuring the source's luminosity distance $D_L$ and it's comoving distance $D_c$.  In this approach $D_L$ is recovered from frequency evolution in the timing residual signal, and $D_c$ comes from probing the curvature of the wavefront across the Earth-pulsar baseline of the PTA experiment.  In this chapter we show that more generally, the distance measured from the curvature of the wavefront is the ``parallax distance'' $D_\mathrm{par}$, which is equivalent to $D_c$ in the flat universe case.

For comparison of other experimental methods that seek to measure the Hubble constant, the ``standard sirens'' approach \citep{Schutz_1986, Holz_2005} produces a measurement of $H_0$ through a hybrid gravitational wave/electromagnetic technique, from a source producing messengers in both of these channels.  Here the luminosity distance $D_L$ is measured from gravitational waves from a chirping source, while the redshift $z$ is measured from the host galaxy through electromagnetic observations.  Combining these measurements of $D_L$ and $z$ then allow an inference of $H_0$ to be made.  As an example of a purely gravitational wave-based measurement of $H_0$, \citet{messenger_gwH0} explained that gravitational wave signals from binary neutron star (BNS) mergers could be used to obtain both measurements of the source's luminosity distance $D_L$ and redshift $z$.  This would require well constrained knowledge of the neutron star equation of state from many BNS detections, which may be possible with future Einstein Telescope era gravitational wave observatories.  The approach presented in this paper is also purely gravitational wave-based, but here we are trading the redshift measurement for a \textit{second} distance measurement made from the Fresnel wavefront curvature effects.

In this study we take the models we have developed in the previous chapters and further generalize them to a cosmologically expanding universe.  So far our previous results were made under the fundamental assumption that the universe was flat and static.  Here we remove the assumption that the universe is static on cosmological scales and re-derive the formulae in a cosmologically expanding universe.  To list the important changes we make in this chapter: \\

\begin{minipage}{0.9\linewidth}
\fbox {
    \parbox{\linewidth}{
    \textbf{Assumptions:} 
    \begin{enumerate}
        \item The universe is described by the FLRW metric with comoving coordinates $(t, r, \theta, \phi)$ and curvature constant $k$.
    \end{enumerate}
    }
}
\end{minipage}

\begin{minipage}{0.9\linewidth}
\fbox {
    \parbox{\linewidth}{
    \begin{enumerate}
        \item[2.] The FLRW scale factor $a$ does not change appreciably over the time scale of the travel time of the photon from the pulsar to the Earth: $\dot{a}(t_0) \Delta t_{\mathrm{p}\rightarrow\mathrm{E}}$. \label{as: scale factor evolution}
        \item[3.] All ``local'' distances and times between the pulsar and the Earth are on a Minkowski background, $g_{\mu\nu} = \eta_{\mu\nu} = \text{diag}(-c^2, 1, 1, 1)$.  The transition from the global cosmological FLRW background to the local Minkowski background requires that all points of interest have comoving coordinate distance separations much smaller than the background curvature of spatial slices ($r^2 \ll \frac{1}{|k|}$), and that the observation time of our experiment is much smaller than the present day age of the universe ($\frac{t_\mathrm{obs}}{t_0} \ll 1$). \label{as: metric transition}
    \end{enumerate}
    }
}
\end{minipage}\\

In general we find that in order to recover the parallax distance parameter (and hence measure $H_0$), we require highly accurate measurements of the distances to the pulsars in our array.  The results presented in this work also demonstrate how the same methods used to measure $H_0$ also yield improved measurements of the pulsar distances, which is an additional important result in its own right.

%---------------------------------------------------------------------------------
%---------------------------------------------------------------------------------
    \section{Incorporating Cosmological Expansion \& the Hubble Constant}\label{sec: incorporating cosmological expansion and H0}

    In our cosmological framework we divide spacetime into two types of frames: a ``global cosmological frame'' where the background metric is Friedmann\textendash Lema\^{i}tre\textendash Robertson \textendash Walker (FLRW), and ``local frames'' where the background metric is Minkowski (see assumption~\ref{as: metric transition}).  The gravitational waves (as described by the metric perturbation below) are generated in the local source frame, propagate over the cosmological frame, and reach the Milky Way galaxy where we assume they are in the local observer frame.  Therefore the cosmological effects of an expanding universe only need to be considered during the gravitational waves propagation between the local source and observer frames.
    
    As a general starting point we begin by assuming that the FLRW metric describes our global spacetime background between the Earth and our gravitational wave source:
    \begin{align}
        ds^2 &= -c^2 dt^2 + a^2(t)\left[ \frac{dr^2}{1-kr^2} + r^2 d\Omega^2 \right], \nonumber \\
        &= a^2 \left[ -c^2 d\eta^2 + d\chi^2 + r(\chi)^2 d\Omega^2 \right] ,
    \label{eqn: FLRW metric}
    \end{align}
    where $d\Omega^2 \equiv d\theta^2 + \sin^2(\theta)d\phi^2$, $a(t)$ is the universe scale factor, and $k$ is the spacetime curvature constant. The coordinates $(t, r, \theta, \phi)$ are the comoving coordinates, while the version shown in the second line is made through the change to conformal time $d\eta \equiv \frac{dt}{a}$ (i.e. $\eta \equiv \int^t \frac{dt^{'}}{a}$) and the spatial coordinate transformation:
    \begin{equation}
        r(\chi) \equiv \begin{cases}
                    \frac{1}{\sqrt{k}}\sin\left(\sqrt{k}\chi\right) ,  & \quad k > 0 \\
                    \chi \ , & \quad k = 0 \\
                    \frac{1}{\sqrt{|k|}}\sinh\left(\sqrt{|k|}\chi\right) , & \quad k < 0 
                  \end{cases}
    \label{eqn: r(chi) definition}
    \end{equation}
    Here we are using the convention that the curvature parameter carries units of $\left[\mathrm{length}^{-2}\right]$, where $k>0$, $k=0$, or $k<0$ for closed, flat, and open spatial geometries, respectively.  This therefore implies that our time dependent scale factor $a$ is unitless.  We also choose to normalize $a(t_0) \equiv 1$ at the present day time $t_0$, but we will often write it explicitly in our derivations here.  Therefore, the Gaussian curvature of a spatial slice of our universe at the present time is $k$.
    
    It is well understood that solving the wave equation for the gravitational wave metric perturbation in the FLRW metric introduces the redshift parameter $z$ into the amplitude of the wave~\citep[see for example,][]{maggiore_2008}.  Without an independent measurement of the source's redshift (either from electromagnetic observations, or as we will explain, from effects of the curvature of the wavefront itself), this redshift parameter cannot be disentangled from the $R$, $\mathcal{M}$, and $\omega_0$ parameters.  The result is that we can only observe the luminosity distance $D_L = (1+z)R$, a redshifted system chirp mass $\mathcal{M}_\mathrm{obs} \equiv (1+z) \mathcal{M}$, and a redshifted orbital frequency parameter $\omega_{0,\mathrm{obs}} = \frac{\omega_0}{(1+z)}$.
    
    Next we need to calculate the expression for the retarded time of the wave~\citep[see for example,][]{greens_functions_1993}.  Crucially, if we make the assumption that the $a(\eta)$ does not evolve appreciably over the time it takes a photon to travel from our pulsar to the Earth (see assumption~\ref{as: scale factor evolution}), then $\eta_\mathrm{ret} - \eta \approx \frac{1}{a(t_0)} \left(t_\mathrm{ret} - t\right)$ and we can write:
    \begin{align}
        &\eta_\mathrm{ret} = \eta - \frac{\left|\vec{x}^{'} - \vec{x}\right|}{c} , \nonumber \\
        & \therefore \quad t_\mathrm{ret} \approx t - a(t_0)\frac{\left|\vec{x}^{'} - \vec{x}\right|}{c} .
    \label{eqn: tret - eta and t versions}
    \end{align}
    Next we must calculate $\left|\vec{x}^{'} - \vec{x}\right|$ in the three possible spatial geometries of our universe.  The details of this are shown in the box below, and the result is that regardless of the background curvature of the spatial slices of our cosmology, the generalized expression for the retarded time evaluated at the pulsar's position is (compared to equation~\ref{eqn: tret along Earth-pulsar baseline at x=L}):
    \begin{equation}
        t_\mathrm{ret}\left(t,L\hat{p}\right) \approx t \ \ - \underbrace{\frac{D_c}{c}}_{\mathrm{``Far} \ \mathrm{Field"}} \  + \ \ \underbrace{\left(\hat{r}\cdot\hat{p}\right)a(t_0)\frac{L}{c}}_{\mathrm{``Plane-Wave"}} \ \ - \ \ \underbrace{\frac{1}{2}\left(1-\left(\hat{r}\cdot\hat{p}\right)^2\right)a^2(t_0)\frac{L}{c}\frac{L}{D_\mathrm{par}}}_{\mathrm{``Fresnel"}} \ \ + \ \ \ldots 
    \label{eqn: tret - pulsar term}
    \end{equation}
    Now the source distance measurement $R$ in our Fresnel number proxy (equation~\ref{eqn: Fresnel number}) is replaced with the more general parallax distance $D_\mathrm{par}$, and the orbital frequency $\omega_0$ parameter is replaced with its observed $\omega_{0,\mathrm{obs}}$ quantity:
    \begin{align}
        F &\equiv \frac{L^2}{\lambda_\mathrm{gw,obs} D_\mathrm{par}} , \label{eqn: Fresnel number generalized} \\
        &\mathrm{and}\quad \lambda_\mathrm{gw,obs} \equiv \frac{\pi c}{\omega_{0,\mathrm{obs}}} \approx (30.5 \text{ pc}) \left(\frac{1\text{ nHz}}{\omega_{0,\mathrm{obs}}}\right) .
    \end{align}

    There are two notable and interesting results that come out of the expression in equation~\ref{eqn: tret - pulsar term}.  The first is that we find that \textit{two} cosmological distance measurements to the same source appear when working out to the Fresnel term in the expansion - the line-of-sight comoving distance $D_c$, and the parallax distance $D_\mathrm{par}$.  In practice, however, we can only measure $D_\mathrm{par}$ from our timing residual models.  This is because in addition to $\frac{D_c}{c}$ appearing as the first term in the expression for the retarded time measured at the pulsar, it also appears in the expression for the retarded time measured at the Earth (for the Earth, just set $L=0$ in equation~\ref{eqn: tret - pulsar term}).  Therefore, when the rest of our timing residual model is worked out (see for example the IIB model in equations~\ref{eqn: Res(t) fr freq evo - updated} and~\ref{eqn: Res(t) fresnel freq evo phase E and P - updated}), we can simply choose the fiducial time $t_0 = -\frac{D_c}{c}$ and all dependence on $D_c$ vanishes (this was a point of discussion in the derivations of the four timing models in Chapter~\ref{ch: The Continuous Wave Timing Residual}).
    
    The second is that in principle we could also choose to include the cosmological curvature constant $k$ as a parameter to attempt to directly measure it from the gravitational wave signal.  Equation~\ref{eqn: Dpar - R - DM - DC - DL relationship} provides the connection between $D_\mathrm{par}$ and $D_L$, and therefore would introduce both $z$ and $k$ as additional model parameters.  However, since the Fresnel corrections are already smaller order corrections and $k$ is currently understood to be very close to zero, we will restrict our attention in this work to a geometrically flat universe and assume $k=0$.

    Therefore, working under the assumption of a geometrically flat universe, the relationships given in Appendix~\ref{app: cosmological distances} provide us the connection to the Hubble constant:
    \begin{equation}
        k = 0 \quad \longrightarrow \quad \begin{cases}
            D_L = (1+z) D_\mathrm{par} , \\[4pt]
            H_0 = \frac{c}{R}\int^z_0 \frac{dz^{'}}{E(z^{'})} = \frac{c}{D_\mathrm{par}}\int^{D_L/D_\mathrm{par}-1}_0 \frac{dz^{'}}{E(z^{'})} .
        \end{cases}
    \label{eqn: Hubble constant}
    \end{equation}
    By procuring a measurement of both the distances $D_L$ and $D_\mathrm{par}$ from our pulsar timing model (see Section~\ref{sec: model IIB - revisited}) we can directly measure the source's redshift.  By then assuming values of the cosmological density parameters which appear in the Hubble function (see equation~\ref{eqn: expansion rate}), we can directly measure the Hubble constant $H_0$.  In practice in our work, we can also bypass equation~\ref{eqn: Hubble constant} when using the small redshift approximation, which doesn't require assumptions of the density parameters (see the discussion of equation~\ref{eqn: low redshift R approximation}). \\
    
    \begin{minipage}{0.9\linewidth}
    \fbox {
        \parbox{\linewidth}{
        \underline{\textbf{The Retarded Time Calculation}}

        The goal is to calculate $\left|\vec{x}^{'} - \vec{x}\right|$ in equation~\ref{eqn: tret - eta and t versions}, where $\vec{x}^{'}$ is the source location and $\vec{x}$ is the field point.  For our problem, we will primarily be interested in setting our field point at the pulsar.  With the center of our coordinate system at the Earth, let the source be at the coordinate $r=R$ (i.e. $\chi = \chi_S$) with unit vector $\hat{r}$, and let the pulsar be at the coordinate $r=L$ (i.e. $\chi=\chi_P$) with unit vector $\hat{p}$.\\
        
        Next we can use the law of cosines for Euclidean, spherical, and hyperbolic geometries:
        \begin{equation}
            \begin{cases}
                \begin{tabular}{r l r}
                    $\cos\left(\sqrt{k}\left|\vec{x}^{'} - \vec{x}\right|\right)$ &$= \cos\left(\sqrt{k}\chi_S\right)\cos\left(\sqrt{k}\chi_P\right) $ & \\
                    &$\qquad+ \sin\left(\sqrt{k}\chi_S\right)\sin\left(\sqrt{k}\chi_P\right)\left(\hat{r}\cdot\hat{p}\right) ,$ &$k > 0$ \\
                    $\left|\vec{x}^{'} - \vec{x}\right|$ &$= \chi_S^2 + \chi_P^2 - 2\chi_S\chi_P\left(\hat{r}\cdot\hat{p}\right) ,$ &$k = 0$ \\
                    $\cosh\left(\sqrt{|k|}\left|\vec{x}^{'} - \vec{x}\right|\right)$ &$= \cosh\left(\sqrt{|k|}\chi_S\right)\cosh\left(\sqrt{|k|}\chi_P\right) $ & \\
                    &$\qquad- \sinh\left(\sqrt{|k|}\chi_S\right)\sinh\left(\sqrt{|k|}\chi_P\right)\left(\hat{r}\cdot\hat{p}\right) .$ &$k < 0$
                \end{tabular}
            \end{cases}
        \end{equation}
        }
    }
    \end{minipage}   
        
    \begin{minipage}{0.9\linewidth}
    \fbox {
        \parbox{\linewidth}{
        Recall from Section~\ref{sec: incorporating cosmological expansion and H0} that at the present time the Gaussian curvature is $k$.  One approach to solving these equations is to Taylor expand in the small parameter $\sqrt{|k|}\chi_P$ in the closed and open universe cases, and in the small parameter $\frac{\chi_P}{\chi_S}$ the flat universe case.  The result is:
        \begin{equation}
            \left|\vec{x}^{'} - \vec{x}\right| \approx \chi_S - \left(\hat{r}\cdot\hat{p}\right)\chi_P + \frac{1}{2}\left( 1 - \left(\hat{r}\cdot\hat{p}\right)^2 \right) \begin{cases}
                \frac{\chi_P^2}{\frac{1}{\sqrt{k}}\tan\left(\sqrt{k}\chi_S\right)} \ , \quad & k>0\\
                \frac{\chi_P^2}{\chi_S} \ , \quad & k=0\\
                \frac{\chi_P^2}{{\frac{1}{\sqrt{|k|}}\tanh\left(\sqrt{|k|}\chi_S\right)}} \ . \quad & k<0\\
            \end{cases}
        \label{eqn: source-field separation}
        \end{equation}
        Additionally, using equation~\ref{eqn: r(chi) definition} we see that $\chi_P \approx L$ for each universe geometry case (Taylor expanding once again for the closed and open cases in $\sqrt{|k|}\chi_P$).    

        Consider for a moment a gravitational wave propagating directly from the source to the Earth.  The wave leaves it's position at redshift $z$ and time $t_\mathrm{ret}$, and arrives at Earth at redshift $z=0$ at time $t$ (recall the boundary conditions from Appendix~\ref{app: cosmological distances}).  In this case our field position $\vec{x} = 0$, so $\eta_\mathrm{ret} = \eta - \frac{\chi_S}{c}$ from equation~\ref{eqn: tret - eta and t versions}.  Therefore using equation~\ref{eqn: line-of-sight comoving distance Dc} we can write:
        \begin{align*}
            -\frac{\chi_S}{c} = \eta_\mathrm{ret} - \eta &= \int^{t_\mathrm{ret}}_t \frac{dt^{'}}{a} , \\
            &= -\frac{D_c}{a(t_0)c}.
        \end{align*}
        This gives us a connection between $\chi_S$ and the comoving coordinate distance $D_c$, namely that $\chi_S = \frac{D_c}{a(t_0)}$, which we can now substitute into equation~\ref{eqn: source-field separation}.  The result is that we can now see that the relevant distance which appears in the Fresnel term in our Taylor expansion is the parallax distance $D_\mathrm{par}$ (equation~\ref{eqn: Dpar - R - DM - DC - DL relationship}).
        }
    }
    \end{minipage}

    \begin{minipage}{0.9\linewidth}
    \fbox {
        \parbox{\linewidth}{
        Therefore the final form of equation~\ref{eqn: source-field separation} which is then used in equation~\ref{eqn: tret - pulsar term} is:
        \begin{equation}
            \left|\vec{x}^{'} - \vec{x}\right| \approx \frac{D_c}{a(t_0)} - \left(\hat{r}\cdot\hat{p}\right)L + \frac{1}{2}\left( 1 - \left(\hat{r}\cdot\hat{p}\right)^2 \right) a(t_0)\frac{L^2}{D_\mathrm{par}} .
        \label{eqn: source-field separation - final}
        \end{equation}
        Note that in the flat universe case $D_\mathrm{par} = D_c$, which is consistent with the result of \citet{DOrazio_main_paper} given that this was their working assumption.

        }
    }
    \end{minipage}\\ \\

%---------------------------------------------------------------------------------
%---------------------------------------------------------------------------------
    \section{Generalizing From a Static to Expanding Universe}\label{sec: generalizing the flat, static model}

    First we need to update our previous expressions equations~\ref{eqn: h plus and cross} and~\ref{eqn: amplitude of metric perturbation} to include the effects of cosmological expansion as described in Section~\ref{sec: incorporating cosmological expansion and H0}:
    \begin{equation}
        \begin{cases}
        h_{+}(t_\mathrm{obs}) &\equiv -h(t_\mathrm{obs}) \cos\big(2\Theta(t_\mathrm{obs})\big), \\
        h_{\times}(t_\mathrm{obs}) &\equiv -h(t_\mathrm{obs}) \sin\big(2\Theta(t_\mathrm{obs})\big), \\
        h(t_\mathrm{obs}) &\equiv \frac{4(G\mathcal{M}_\mathrm{obs})^{5/3}}{c^4 D_L}\omega(t_\mathrm{obs})^{2/3}, \\
        h_{0,\mathrm{obs}} &\equiv \frac{4(G\mathcal{M}_\mathrm{obs})^{5/3}}{c^4 D_L}\omega_{0,\mathrm{obs}}^{2/3} ,
      \end{cases} \label{eqn: h+x(t) & h(t) - updated}
    \end{equation}    
    and the angular phase and frequency functions $\Theta(t_\mathrm{obs})$ and $\omega(t_\mathrm{obs})$ from equations~\ref{eqn:monochrome freq}, \ref{eqn:monochrome phase}, \ref{eqn:freq evolution freq}, \ref{eqn:freq evolution phase}, \ref{eqn: time to coalescence}, and~\ref{eqn: thetac} to:
    \begin{align}
        \underset{(A)}{\text{Monochromatism}} \hspace{0.5cm} & \begin{cases}\begin{tabular}{l l}
            $\omega(t_\mathrm{obs})$ & $= \omega_{0,\mathrm{obs}}$ , \\ 
            $\Theta(t_\mathrm{obs})$ & $= \theta_0 + \omega_{0,\mathrm{obs}} (t_\mathrm{obs} - t_{0,\mathrm{obs}})$ , 
            \end{tabular}\end{cases} \\[4pt]
        \underset{(B)}{\text{Frequency Evolution}} \quad & \begin{cases} \begin{tabular}{l l l}
            $\omega(t_\mathrm{obs})$ & $= \omega_{0,\mathrm{obs}}\left[1 - \frac{t_\mathrm{obs}-t_{0,\mathrm{obs}}}{\Delta\tau_{c,\mathrm{obs}}} \right]^{-3/8}$ , & \\
            $\Theta(t_\mathrm{obs})$ & $= \theta_0 + \theta_c\left[ 1 - \left(\frac{\omega(t_\mathrm{obs})}{\omega_{0,\mathrm{obs}}}\right)^{-5/3} \right]$ , & \\
            $\Delta\tau_{c,\mathrm{obs}}$ & $\equiv \frac{5}{256}\left(\frac{c^3}{G\mathcal{M}_\mathrm{obs}}\right)^{5/3} \frac{1}{\omega_{0,\mathrm{obs}}^{8/3}}$ , \\
            &$= (1+z)\Delta\tau_c$ , \\
            $\theta_c$ & $=  \frac{1}{32}\left(\frac{c^3}{G\mathcal{M}_\mathrm{obs}\omega_{0,\mathrm{obs}}}\right)^{5/3}$ , \\
            &$\equiv \frac{8}{5}\Delta\tau_{c,\mathrm{obs}} \omega_{0,\mathrm{obs}} = \frac{8}{5}\Delta\tau_c \omega_0$ .
            \end{tabular}\end{cases} \label{eqn: frequency evolution (B) - updated}
    \end{align}
    Here $t_{0,\mathrm{obs}}$ here (and below) denotes the fiducial time for the model (note that $t_\mathrm{obs}=0$ would correspond to the present day start of our experiment on Earth).  The frequency evolution regime is governed by the physically significant quantities $\Delta \tau_{c,\mathrm{obs}}$ which is the ``observed coalescence time,'' and $\theta_c$ which is the ``coalescence angle'' (the total angle swept out before the system coalesces).  Notice that now in the FLRW universe the coalescence time is redshifted in our local observer frame but the angle is not, the number of orbital cycles the binary system will complete is independent of reference frame.
    
    Once in the local observer frame the gravitational waves will affect the timing of local pulsars.  The math and derivation at this point remains mostly unchanged to what was shown in Chapter~\ref{ch: The Continuous Wave Timing Residual}.  The only minor change is the notational distinction that we are now working with ``observed'' quantities (rather than the source frame quantities from a static universe).  Therefore (referring back to equations~\ref{eqn: Delta T / T} and~\ref{eqn:timing residual}) the gravitational wave induced fractional shift of the pulsar's period $T$ is now:
    \begin{equation}
        \frac{\Delta T}{T}(t_\mathrm{obs}) \approx \frac{1}{2}\hat{p}^i\hat{p}^j E^{\hat{r}\textsc{A}}_{ij} \int\limits^{t_\mathrm{obs}}_{t_\mathrm{obs} - \frac{L}{c}} \frac{\partial h_{\textsc{A}}\Big(t_\mathrm{obs,ret}\left(t^{'}_\mathrm{obs},\vec{x}\right)\Big)}{\partial t^{'}_\mathrm{obs}}\Bigg\rvert_{\vec{x}=\vec{x}_{0}\left(t^{'}_\mathrm{obs}\right)} dt^{'}_\mathrm{obs} ,
    \label{eqn: Delta T / T - updated}
    \end{equation}
    where $t_\mathrm{obs}$ is the time a pulsar's photon is observed arriving at Earth, $t_\mathrm{obs,ret}$ is the retarded time of the gravitational wave, and $\vec{x}_{0}(t_\mathrm{obs})$ is the spatial path of the photon between the pulsar and the Earth.  Finally the gravitational wave induced timing residual is now:
    \begin{equation}
        \mathrm{Res}(t_\mathrm{obs}) = \int\frac{\Delta T}{T}\left(t^{'}_\mathrm{obs}\right) dt^{'}_\mathrm{obs} = \int\frac{T_\mathrm{obs} \left(t^{'}_\mathrm{obs}\right) - T}{T} dt^{'}_\mathrm{obs}  = \ \mathrm{Obs}(t_\mathrm{obs}) - \mathrm{Exp}(t_\mathrm{obs}) . \label{eqn:timing residual - updated}
    \end{equation}
    The key ideas behind this work again stem from consideration of the retarded time $t_\mathrm{obs,ret}(t_\mathrm{obs},\vec{x})$ in equation~\ref{eqn: Delta T / T - updated}, which will now be replaced with the new result for equation~\ref{eqn: tret - eta and t versions} (for example, the pulsar term retarded time will be equation~\ref{eqn: tret - pulsar term}).

%---------------------------------------------------------------------------------
%---------------------------------------------------------------------------------
    \section{The Fresnel, Frequency Evolution Model - Revisited}\label{sec: model IIB - revisited}

    Chapter~\ref{ch: The Continuous Wave Timing Residual} details the four gravitational wave induced timing residual model regimes IA-IIB, characterized by frequency evolution and curvature effects.  Generalizing to a flat cosmologically expanding universe does not change the math of the derivation behind those four models, it only changes the interpretation of the model parameters as has already been discussed as well as adds one new parameter.  Crucially, the luminosity distance $D_L$ becomes the parameter measured from the amplitude of the metric perturbation (equation~\ref{eqn: h+x(t) & h(t) - updated}), while the parallax distance $D_\mathrm{par}$ (i.e. $D_M$ or $D_c$ in our flat universe) becomes the distance parameter measured in the retarded time.  Furthermore we now recognize that the chirp mass and orbital frequency parameters are measured in the observer frame of reference, and are no longer equivalent to their values in the source frame.  This means the most general IIB model now becomes:
    \begin{align}
        \overline{\mathrm{Res}}(t_\mathrm{obs}) &= \frac{F^\textsc{A}}{4} \left[ \frac{h_\textsc{A}\Big(\omega_{0E},\Theta_E-\frac{\pi}{4}\Big)}{\omega_{0E}} - \frac{h_\textsc{A}\Big(\overline{\omega}_{0P},\overline{\Theta}_P-\frac{\pi}{4}\Big)}{\overline{\omega}_{0P}} \right] , \label{eqn: Res(t) fr freq evo - updated} \\
        &\mathrm{for}\quad \textsc{A} \in [+, \times], \nonumber \\[4pt]
        &\underset{\left(t_{0,\mathrm{obs}} \ = \ -\frac{D_c}{c}\right)}{\mathrm{where}}\quad \begin{cases}
            \omega_{0E} &\equiv \omega_{0,\mathrm{obs}} , \\
            \overline{\omega}_{0P} &\equiv \omega_{0,\mathrm{obs}}\left[1 + \frac{\left(1-\hat{r}\cdot\hat{p}\right)\frac{L}{c} + \frac{1}{2}\left(1-\left(\hat{r}\cdot\hat{p}\right)^2\right)\frac{L}{c}\frac{L}{D_\mathrm{par}}}{\Delta\tau_{c,\mathrm{obs}}}  \right]^{-3/8} , \\
            \theta_{0E} &= \theta_0 , \\
            \overline{\theta}_{0P} &= \theta_0 + \theta_c\left(1-\left[1+ \frac{\left(1-\hat{r}\cdot\hat{p}\right)\frac{L}{c} + \frac{1}{2}\left(1-\left(\hat{r}\cdot\hat{p}\right)^2\right)\frac{L}{c}\frac{L}{D_\mathrm{par}}}{\Delta\tau_{c,\mathrm{obs}}}\right]^{5/8}\right) , \\
            \Theta_E & = \theta_{0E} + \omega_{0E}t_\mathrm{obs} , \\
            \overline{\Theta}_P & = \overline{\theta}_{0P} + \overline{\omega}_{0P}t_\mathrm{obs} ,
            \end{cases}\label{eqn: Res(t) fresnel freq evo phase E and P - updated} \\
            &\hspace{1.6cm}\begin{cases}
                \Delta\tau_{c,\mathrm{obs}} & \equiv \frac{5}{256}\left(\frac{c^3}{G\mathcal{M}_\mathrm{obs}}\right)^{5/3} \frac{1}{\omega_{0,\mathrm{obs}}^{8/3}} , \\
                \theta_c & =  \frac{1}{32}\left(\frac{c^3}{G\mathcal{M}_\mathrm{obs}\omega_{0,\mathrm{obs}}}\right)^{5/3} , \\
            \end{cases} \tag{\ref{eqn: frequency evolution (B) - updated} r} \\
           &\hspace{1.6cm}\begin{cases}
                h_{+} \equiv -h(\omega) \cos\big(2\Theta\big), \\
                h_{\times} \equiv -h(\omega) \sin\big(2\Theta\big),
            \end{cases} \tag{\ref{eqn: h+x(t) & h(t) - updated} r} \\
            &\hspace{1.6cm}h(\omega) \equiv \frac{4(G\mathcal{M})^{5/3}}{c^4 D_L}\omega^{2/3} , \tag{\ref{eqn: h+x(t) & h(t) - updated} r}\\
            &\hspace{1.6cm}\begin{cases}
                \begin{tabular}{l l l}
                    $f^{+}$ &$\equiv \frac{\hat{p}^i\hat{p}^j e^{\hat{r}+}_{ij}}{\left(1-\hat{r}\cdot\hat{p}\right)}$ &$= \frac{\left(\hat{p}\cdot\hat{\theta}\right)^2 - \left(\hat{p}\cdot\hat{\phi}\right)^2}{\left(1-\hat{r}\cdot\hat{p}\right)}$ , \\
                    $f^{\times}$ &$\equiv \frac{\hat{p}^i\hat{p}^j e^{\hat{r}\times}_{ij}}{\left(1-\hat{r}\cdot\hat{p}\right)}$ &$= \frac{2\left(\hat{p}\cdot\hat{\theta}\right)\left(\hat{p}\cdot\hat{\phi}\right)}{\left(1-\hat{r}\cdot\hat{p}\right)}$ ,
                \end{tabular}
            \end{cases} \tag{\ref{eqn: antenna f} r} \\
            &\hspace{1.6cm}\begin{cases}
                \begin{tabular}{l l l}
                    $F^+$ &$\equiv \frac{\hat{p}^i\hat{p}^j E^{\hat{r}+}_{ij}}{\left(1-\hat{r}\cdot\hat{p}\right)}$ &$= \frac{1}{2}\left(1+\cos^2(\iota)\right) \left[ \cos(2\psi) f^+ + \sin(2\psi) f^\times \right]$ , \\
                    $F^\times$ &$\equiv \frac{\hat{p}^i\hat{p}^j E^{\hat{r}\times}_{ij}}{\left(1-\hat{r}\cdot\hat{p}\right)}$ &$= \cos(\iota) \left[ -\sin(2\psi) f^+ + \cos(2\psi) f^\times \right]$ ,
                \end{tabular}
            \end{cases} \tag{\ref{eqn: antenna F} r} \\
            &\hspace{1.6cm}\begin{cases}
                \begin{tabular}{r l l r}
                    $\hat{r}$ &$= \big[\sin(\theta)\cos(\phi),$ &$\sin(\theta)\sin(\phi),$ &$\cos(\theta) \big]$, \\
                    $\hat{\theta}$ &$= \big[\cos(\theta)\cos(\phi),$ &$\cos(\theta)\sin(\phi),$ &$\sin(\theta) \big]$, \\
                    $\hat{\phi}$ &$= \big[-\sin(\phi),$ &$\cos(\phi),$ &$0 \big]$, \\
                    $\hat{p}$ &$= \big[\sin(\theta_p)\cos(\phi_p),$ &$\sin(\theta_p)\sin(\phi_p),$ &$\cos(\theta_p) \big]$.
                \end{tabular}
                \end{cases} \tag{\ref{eqn:source basis vectors}, \ref{eqn: p hat} r} 
    \end{align}
    
\begin{figure}
        \centering
        \includegraphics[width=0.9\linewidth]{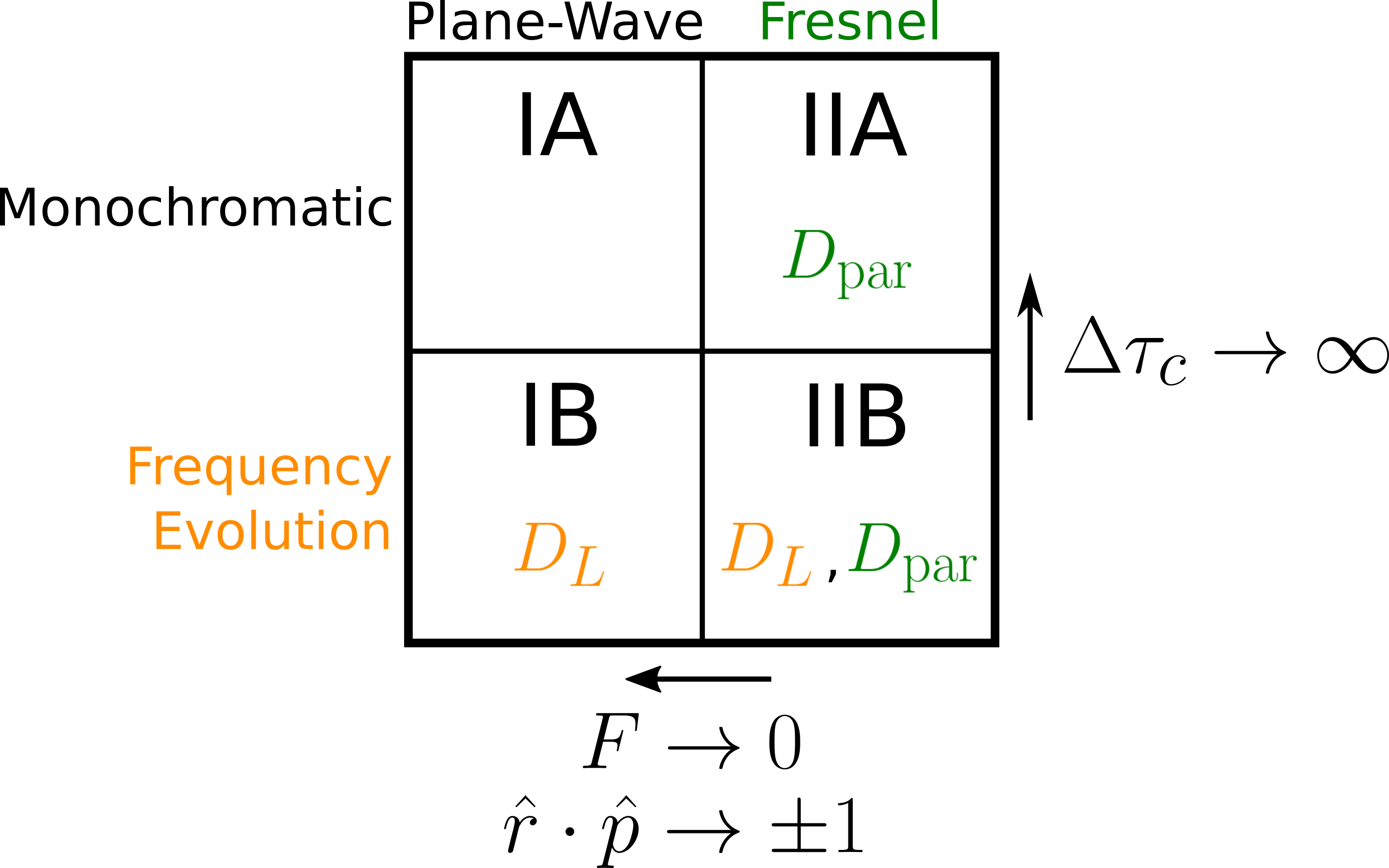}
    \caption[Gravitational Wave Timing Regime Classification - Revisited]{A revisited and revised version of Figure~\ref{fig: 4 regimes}.  Importantly, in a cosmologically expanding universe the frequency evolution regimes allow for the direct measurement of the source luminosity distance $D_L$, and the Fresnel regimes allow for the direct measurement of the source parallax distance $D_\mathrm{par}$.  In the fully general IIB model, both of these distances can be independently measured, and therefore through equation~\ref{eqn: Hubble constant} a value of $H_0$ can be inferred.}
        \label{fig: 4 model regimes - updated}
    \end{figure}

    The introduction of $D_L$ and $D_\mathrm{par}$ into the timing residual model is significant in how they break the degeneracy between other parameters within the four model regimes.  Following the discussion in Section~\ref{sec:regime degeneracies}, frequency evolution allows for the measurement of the source's chirp mass $\mathcal{M}_\mathrm{obs}$, while Fresnel corrections allow for measurement of the source's parallax distance $D_\mathrm{par}$.  This means that regime IB can measure the source's luminosity distance (through equation~\ref{eqn: observed earth term amplitude} from measurements of $A_{E,\mathrm{res}}$, $\mathcal{M}_\mathrm{obs}$, and $\omega_{0,\mathrm{obs}}$ parameters), but not the parallax distance.  Regime IIA can measure the source's parallax distance, but not the luminosity distance.  And regime IIB can measure both distances $D_L$ and $D_\mathrm{par}$.  This is summarized in Figure~\ref{fig: 4 model regimes - updated}.
    
    Therefore by using the IIB model, under the flat universe assumption if we can recover the source's two distance parameters $D_L$ and $D_\mathrm{par}$, equation~\ref{eqn: Hubble constant} lets us make a direct measurement of the source redshift as well as the Hubble constant.

%---------------------------------------------------------------------------------
%---------------------------------------------------------------------------------
    \section{Methods}\label{sec:methods}
    
    The main two analyses we used for the studies presented in this Chapter were Fisher matrix and Markov Chain Monte Carlo (MCMC) analyses.  As shown in Chapters~\ref{ch: Fisher matrix analysis} and~\ref{ch:results - first investigation}, a Fisher matrix is a useful tool for quick parameter estimation forecasting of a model under specified experimental constraints.  Computationally, this is efficient and can quickly allow us to test many different experiments.  But it does require that one assume a real search (such as with an MCMC analysis) could successfully identify the \textit{true} mode from any potential secondary modes.  This limitation is again due to the fact that the Fisher matrix is only meant to approximate the shape of the posterior near the maximum likelihood.  If for example the posterior is multi-modal (see Section~\ref{ch:L-wrapping problem}), then the Fisher matrix will not capture this behavior.  For a more in-depth (and computationally expensive) targeted study of the posterior, we use MCMC.
    
    As we did in Chapters~\ref{ch: Fisher matrix analysis} and~\ref{ch:results - first investigation}, for this work we assume that the timing residual data is the sum of the underlying gravitational wave induced residual plus some random noise, that is $\overrightarrow{\mathrm{Res}} = \overrightarrow{\mathrm{Res}}_\mathrm{true} + \vec{N}$.  For simplicity the noise we model is white noise, uncorrelated between observations/pulsars, and for each experiment the timing uncertainty $\sigma$ is uniform across all pulsars, that is $\bm{\Sigma} = \mathrm{diag}\left[\sigma^2\right]$.  Therefore the likelihood function we propose and use in this work is:
    \begin{align}
        \mathcal{L}\left(\overrightarrow{\mathrm{Res}}\mid\vec{\theta}\right) &= \frac{1}{\sqrt{(2\pi)^d \mathrm{det}(\bm{\Sigma})}}\exp{\left[-\frac{1}{2}\bigg( \overrightarrow{\mathrm{Res}} - \overrightarrow{\mathrm{Res}}\left(\vec{\theta}\right) \bigg)^T \bm{\Sigma}^{-1} \bigg( \overrightarrow{\mathrm{Res}} - \overrightarrow{\mathrm{Res}}\left(\vec{\theta}\right) \bigg) \right]} , \nonumber \\
        &= \frac{1}{\sqrt{(2\pi)^d}\sigma^d} \prod_c \exp{\left[-\frac{1}{2}\frac{1}{\sigma^2}\left( \mathrm{Res}^c - \mathrm{Res}^c\left(\vec{\theta}\right) \right)^2 \right]} .
    \tag{\ref{eqn: likelihood function} common $\sigma$}
    \end{align}
    With this the Fisher matrix is again:
    \begin{equation}
        \textsf{F}_{ij} = \sum_c \frac{1}{\sigma^2}\left(\frac{\partial \mathrm{Res}^c\left(\vec{\theta}_\mathrm{true}\right)}{\partial \theta_i}\right) \left(\frac{\partial \mathrm{Res}^c\left(\vec{\theta}_\mathrm{true}\right)}{\partial \theta_j}\right) .
    \tag{\ref{eqn: timing residual fisher matrix} common $\sigma$}
    \end{equation}
    We also continue to use the coefficient of variation (CV) as a useful metric for quantifying the measurability of a given parameter when performing a Fisher matrix analysis, as shown in equation~\ref{eqn: CV}.
    
    Model parametrization was chosen as follows.  In this cosmological framework, the gravitational wave timing residual in the full IIB regime is governed by the source parameters: $\vec{s} = \{D_L, D_\mathrm{par}, \theta, \phi, \iota, \psi, \theta_0, \mathcal{M}_\mathrm{obs}, \omega_{0,\mathrm{obs}} \}$.  Following the explanation given in Section~\ref{sec:regime degeneracies} we swap the $D_L$ parameter in our model with the ``observed Earth term timing residual amplitude'' parameter, defined as:
    \begin{equation}
        A_{E,\mathrm{obs}} \equiv \frac{h_{0,\mathrm{obs}}}{4 \omega_{0,\mathrm{obs}}} = \frac{\left(G\mathcal{M}_\mathrm{obs}\right)^{5/3}}{c^4 D_L \omega_{0,\mathrm{obs}}^{1/3}} \approx (140 \text{ ns}) \left(\frac{\mathcal{M}_\mathrm{obs}}{10^9 M_\odot}\right)^{5/3} \left(\frac{100\text{ Mpc}}{D_L}\right) \left(\frac{1\text{ nHz}}{\omega_{0,\mathrm{obs}}}\right)^{1/3} .
    \label{eqn: observed earth term amplitude}
    \end{equation}
    Log-parameters were also used for the non-angular parameters: $\left\{\log_{10}\left(A_{E,\mathrm{obs}}/A_{E,\mathrm{obs},\mathrm{true}}\right),\right.$ $\left.\log_{10}\left(D_\mathrm{par}/D_\mathrm{par,true}\right), \log_{10}\left(\mathcal{M}_\mathrm{obs}/\mathcal{M}_\mathrm{obs,\mathrm{true}}\right), \log_{10}\left(\omega_{0,\mathrm{obs}}/\omega_{0,\mathrm{obs},\mathrm{true}}\right) \right\}$.
    
    There are two approaches one can take to measure $H_0$ from the model.  The first is to use the described parametrization thus far, and make measurements of $A_{E,\mathrm{obs}}$, $\mathcal{M}_\mathrm{obs}$, $\omega_{0,\mathrm{obs}}$ (equation~\ref{eqn: observed earth term amplitude} then gives us $D_L$), and $D_\mathrm{par}$.  These measurements could be from either a Fisher and/or MCMC analysis.  Then we use equation~\ref{eqn: Hubble constant} to calculate $H_0$.  Note that this does require we make some assumptions for the density parameters that go into the Hubble function in that expression (see equation~\ref{eqn: expansion rate}).
    
    The second approach is to swap out the parallax distance parameter $D_\mathrm{par}$ and replace it directly with $H_0$ in our model, using the small redshift approximation.  For a flat $k=0$ universe and for $z \ll 1$, $z \approx \frac{H_0}{c} D_L$, so we can combine equations~\ref{eqn: Dpar - R - DM - DC - DL relationship} and~\ref{eqn: observed earth term amplitude} to write:
    \begin{equation}
        D_\mathrm{par} \approx \left[ \frac{c^4 A_{E,\mathrm{obs}} \omega_{0,\mathrm{obs}}^{1/3}}{\left(G\mathcal{M}_\mathrm{obs}\right)^{5/3}} + \frac{H_0}{c}\right]^{-1} . \qquad (\text{for } k = 0, \ z \ll 1) 
    \label{eqn: low redshift R approximation}
    \end{equation}
    This second approach is particularly useful when working with the two source problem described in Section~\ref{sec: 2 sources}, as equation~\ref{eqn: low redshift R approximation} will reduce the dimensionality of the model by one ($D_\mathrm{par,1}$ and $D_\mathrm{par,2}$ are both replaced by $H_0$), therefore giving us the direct \textit{joint} posterior of $H_0$ from the two sources.  With the first approach, one must make an estimate of the joint posterior on $H_0$ (for example, using a kernel density estimate) using the measured parameters from both sources.  The second approach builds into our model the additional knowledge that $H_0$ is a constant irrespective of the source we are detecting.  But again the approximation is truly only valid for $z \ll 1$.  In practice we find that the fractional difference between the value of $A_{E,\mathrm{obs}}$ using equation~\ref{eqn: observed earth term amplitude} vs. re-arranging equation~\ref{eqn: low redshift R approximation} is less than $0.1$ out until $D_\mathrm{par}\simeq 1.5$ Gpc.\footnote{Note that for this calculation we used the flat $\Lambda$CDM cosmology model in the PYTHON Astropy.Cosmology package to get $D_L$ from our injected value of $D_\mathrm{par}$ and $H_0$ (and assuming $\Omega_m = 0.3$) when solving equation~\ref{eqn: observed earth term amplitude}.}
    
    For our priors, we required all of the physical parameters be non-negative ($A_{E,\mathrm{obs}}$, $\mathcal{M}_\mathrm{obs}$, $\omega_{0,\mathrm{obs}}$, $D_L$, $D_\mathrm{par}$, $L$, $H_0$), and that $D_\mathrm{par} \leq D_L$ (which assumes $z > 0$ in equation~\ref{eqn: R - DM - DC - DL relationship}).  For the angular parameters we placed the general boundaries: $0 < \theta, \ \iota, \ \theta_0 < \pi$, and $0 < \phi, \ \psi < 2\pi$.
    
    For our MCMC simulations we used the PYTHON emcee package~\citep{emcee}.  For the jump proposal we found that a combination of differential evolution proposals and Gaussian Metropolis steps worked well.  The Metropolis jumps were made using a scaled version of the inverse Fisher matrix for the injection as the covariance matrix for the Gaussian distribution.  We found that if our covariance matrix was multiplied by some small scale-factor (for example, $0.25$), then these jumps helped improve the mean jump acceptance fraction, especially when the walkers were being initialized about the true mode.  However in general we found that the differential evolution jumps drastically improved the mixing of the walkers, which improved the overall efficiency of the MCMC simulation.
    
    To assess the convergence of the samples in our simulations, the integrated autocorrelation time of our parameters was periodically checked.  We typically ran our simulations for $100,000$ iterations, but if the chain reached longer than 100 times the estimated autocorrelation time of every model parameter, and if this estimate changed by less than 1\%, then we ended the simulation there.  The walkers were typically initialized in very small Gaussian ``balls'' about the true injectioned parameters using the inverse Fisher matrix prediction multiplied by some small scale-factor.  This helped them more quickly find and explore the local distributions around the true parameters.
    %  ----> https://emcee.readthedocs.io/en/stable/tutorials/monitor/

    Finally, in Chapter~\ref{ch:results - first investigation} we chose to simulate a NANOGrav-like PTA, simulating pulsars that are currently timed by the NANOGrav collaboration.  In this chapter, however, we decided to construct our own fiducial PTA ad hoc.  In order to make the PTA physically motivated, we used a simple Milky Way galaxy structure model to create a density distribution of pulsars $n_\mathrm{psr}$, in order to sample pulsars from the galactic disk~\citep{schneider_2015}:
    \begin{align}
        n_\mathrm{psr} \ \ \propto \ \ &\exp\left(-\frac{r}{h_R}\right)\Bigg[\exp\left(-\frac{|z|}{h_\mathrm{thin}}\right) + 0.02\exp\left(-\frac{|z|}{h_\mathrm{thick}}\right) \Bigg] \nonumber \\
        &\qquad \times \exp\left(-\frac{1}{2}\left[\left(\frac{x}{R}\right)^2 + \left(\frac{y}{R}\right)^2 + \left(\frac{z}{R}\right)^2\right]\right)H(r-r_\mathrm{min}) ,
    \label{eqn: fiducial PTA}
    \end{align}
    where the center of the coordinate system is at the Earth's position, the Galactic center is at 
    %$x_\mathrm{MW} \approx 25$ kpc,
    %
    $x_\mathrm{MW} \approx 8$ kpc, and $r \equiv \sqrt{(x-x_\mathrm{MW})^2 + y^2}$.  Here $h_R$ is the scale-length of the Galactic disk ($h_R \approx 3.5$ kpc), $h_\mathrm{thin}$ is the scale height of the thin disk ($h_\mathrm{thin} \approx 0.325$ kpc), and $h_\mathrm{thick}$ is the scale height of the thick disk ($h_\mathrm{thick} \sim 1.5$ kpc).  The first term in this expression creates the distribution of stars in the Milky Way, while the second term simply places preference on stars within a Gaussian ball centered on Earth.  The motivation for including this second term is simply the idea that we will likely be more sensitive to timing pulsars within some scale distance $R$ from the Earth.  We also placed a minimum distance $r_\mathrm{min}$ on this sphere (hence the Heaviside function).  For our fiducial PTA we chose $R=5$ kpc and $r_\mathrm{min} = 0.5$ kpc.  An example of 1000 pulsars generated from this distribution is shown in Figure~\ref{fig: my fiducial PTA}.

%---------------------------------------------------------------------------------
%---------------------------------------------------------------------------------
    \section{Pulsar Distance Wrapping Problem (Revisited) \& the Error Envelope}\label{sec: Pulsar Distance Wrapping Problem & Error Envelope}
    
    Perhaps the most challenging problem in this work, in terms of parameter estimation, is the pulsar distance wrapping problem, explained in detail in Chapter~\ref{ch:L-wrapping problem}.  Although in the frequency evolution regimes IB and IIB this ``wrapping cycle'' doesn't mathematically exist as the frequencies are now time-dependent (which technically breaks this $L$ timing residual degeneracy), we still find some support in parameter estimation at $L$ modulo these cycle distances in equation~\ref{eqn:Delta L in monochromatic regimes}.
    
    Consider the problem now in terms of parameter estimation using the likelihood function equation~\ref{eqn: likelihood function}.  If we work entirely in one of the monochromatic regimes IA or IIA, meaning we use one of these models to both calculate the timing residuals of our injected simulation source $\overrightarrow{\mathrm{Res}}$ \textit{and} recover the parameters $\overrightarrow{\mathrm{Res}}\left(\vec{\theta}\right)$, then the likelihood function is perfectly multimodal at every $L \pm \Delta L_n$ for every pulsar in the PTA.  Therefore in either of these regimes, we cannot identify the true pulsar distance with this likelihood since all modes will have equal probability.
    
\begin{figure}
        \centering
        \includegraphics[width=0.78\linewidth]{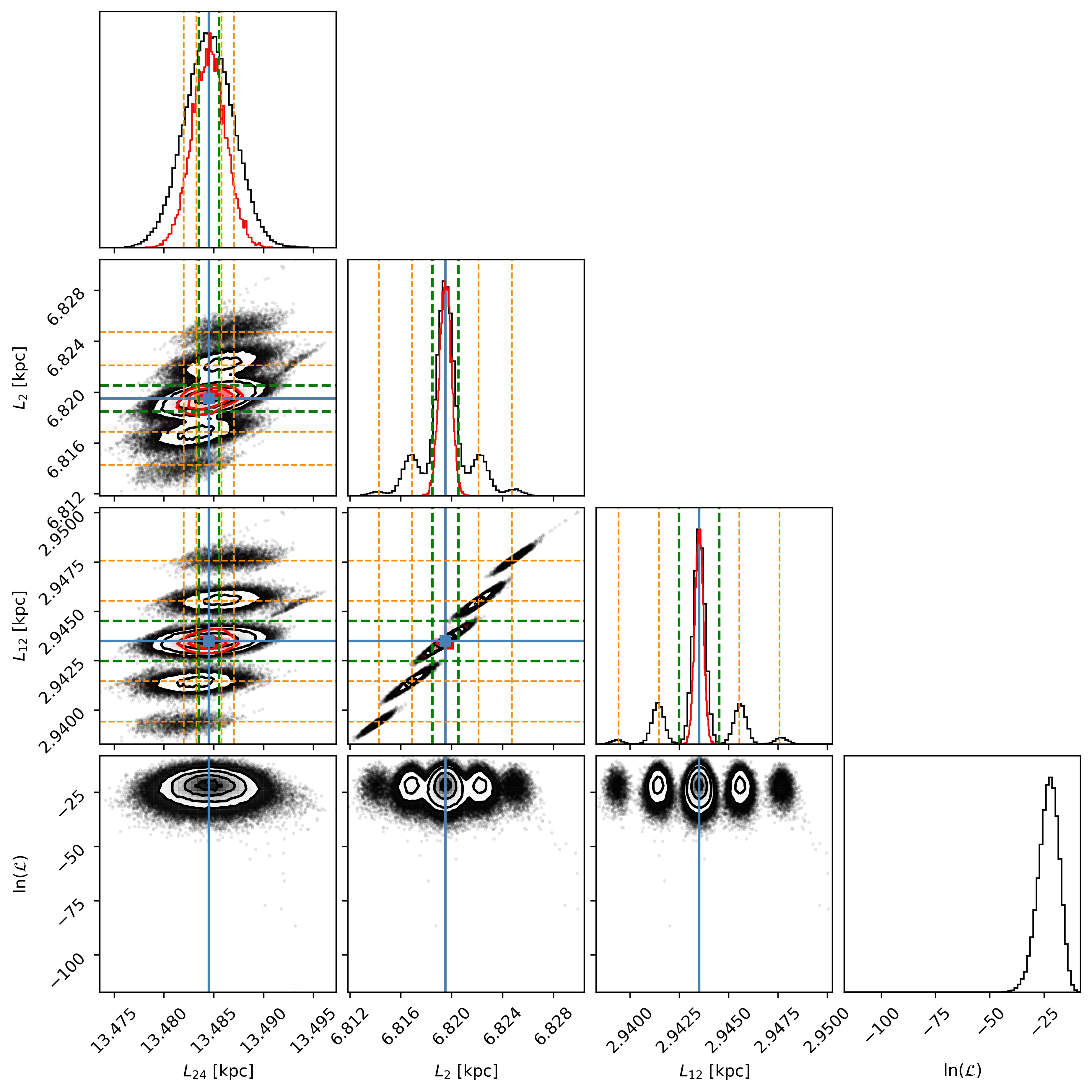}
    \caption[The Pulsar Distance ``Error Envelope'']{An example of the error envelope demonstrated from an MCMC simulation of a PTA (using the IIB model).  In the full simulation there were 40 pulsars (shown in Figure~\ref{fig: error envelope progression}) - this corner plot singles out three (namely the pulsars numbered 24, 2, and 12). To create a clear example, this simulation pinned the $D_\mathrm{par}$, $\theta$, and $\phi$ parameters at their true values, and set the timing uncertainty very low to $\sigma = 1$ ns across all pulsars.  The blue line shows the injected parameter value, the dashed green lines are $\pm \lambda_\mathrm{gw}$ references, the dashed orange lines are $\Delta L_n$ (for $n=\left\{-2,-1,1,2\right\}$), and the red histograms are the Fisher matrix predictions.  The bottom row shows the (unnormalized) log-likelihood values $\ln\left(\mathcal{L}\right)$.  Here we see that in both $L_2$ and $L_{12}$ the secondary modes are somewhat separated away from the primary mode, while the secondary modes in $L_{24}$ clearly blend to form an error envelope, which further widens the uncertainty.  We expect that there are likely additional modes for $L_{24}$ beyond just the first $\pm 2$ which the MCMC walkers simply had not yet begun to explore in the given simulation time.}
        \label{fig: IIB error envelope}
    \end{figure}

    If instead we work in one of the frequency evolution regimes IB or IIB, we find that this likelihood function is still multimodal for every pulsar for each pulsar's true distance modulo its wrapping cycle.  However, as moving out a wrapping cycle no longer returns the same exact timing residual as the true timing residual, these modes become less and less probable for every additional cycle away from the true distance.  \citet{CC_main_paper} identified this in their model as an ``error envelope'' which forms about the true pulsar distance.  The width of each of these secondary modes can blend with the primary mode, increasing the overall uncertainty that we would expect to recover on an individual pulsar distance parameter.  An example of this is shown in Figure~\ref{fig: IIB error envelope}.
    
\begin{figure}
        \centering
        \includegraphics[width=\linewidth]{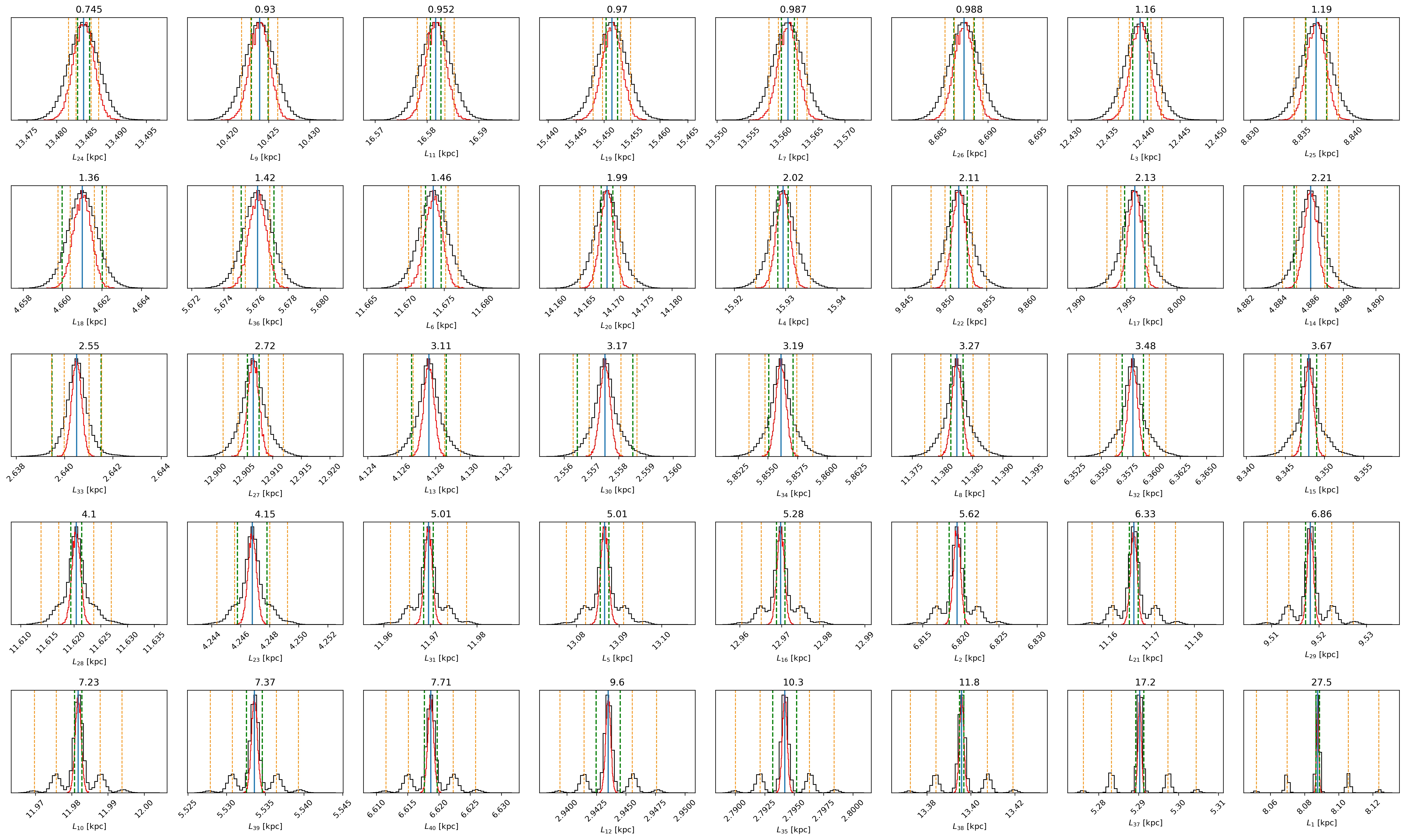}
    \caption[Classifying the Error Envelope]{The complete set of 1D posteriors for the same results shown in Figure~\ref{fig: IIB error envelope}.  Here we have sorted the pulsars in ascending value by their ratio of $\frac{\Delta L_1}{\sigma_{L,\mathrm{Fisher}}}$.  As we can see, this particular criteria successfully shows the progression of the error envelope, which motivated our classification in equation~\ref{eqn: error envelope criteria}.  For ratios less than about $4$, we see that uncertainties about each mode are both sufficiently large and sufficiently close to the primary mode that they blend together, and therefore create an otherwise wider error envelope about the true value.  However, for ratios between about $4-8$ these secondary modes begin to separate away from the primary mode far enough that they start to be distinguishable from the true mode.  Finally, for ratios greater than about $8$, we see that the secondary modes are completely distinguishable from the true mode.  In practice this therefore motivates us to search our results to see what the value of this ratio is for each of our pulsars.  If it is greater than $8$, then we would expect that the true distance to that pulsar would be identifiable.}
        \label{fig: error envelope progression}
    \end{figure}

    However, \citeauthor{CC_main_paper} didn't fully classify the properties of the error envelope.  Namely the geometric factors in the wrapping cycle equation~\ref{eqn:Delta L in monochromatic regimes} are of crucial importance in deciding if and how the error envelope will form for a given pulsar with a given source.  We find that the following are useful set of criteria for classifying the pulsar distance error envelope:
    \begin{equation}
        \frac{\Delta L_1}{\sigma_{L,\mathrm{Fisher}}} \sim \begin{cases}
        < 4 \ ,    &\text{Error Envelope}  \\
        4 - 8 \ ,  &\text{Error Envelope begins to separate}  \\
        > 8 \ ,    &\text{No Error Envelope} 
        \end{cases}
    \label{eqn: error envelope criteria}
    \end{equation}
    where $\sigma_{L,\mathrm{Fisher}}$ is the uncertainty of a given pulsar as predicted by the Fisher matrix analysis (that is, the standard deviation of a pulsar distance $L$ calculated from the inverse Fisher matrix, $\bm{C} = \bm{F}^{-1}$).  These criteria come from considering the uncertainty width about each mode and the separation of the modes from the true distance value (see for example Figure~\ref{fig: 2 Source Common Wrapping Cycle} for a visual).  In general the predictions on uncertainty coming from the Fisher matrix analysis are only valid around the true mode, and therefore cannot accurately forecast the multimodal behavior of the pulsar distances.  However in practice we have observed in our tests that the widths of the secondary modes at each of the pulsar distance wrapping cycles are approximately the same as the width of the primary mode.  Therefore, we use $\sigma_{L,\mathrm{Fisher}}$ as our proxy in defining equation~\ref{eqn: error envelope criteria}.  This effectively assumes that the behavior of the likelihood function near each wrapping cycle will look the same as the behavior of the likelihood function near the true mode.  As an example the pulsars shown in Figure~\ref{fig: error envelope progression} have been sorted explicitly by this criteria.

    The criteria in equation~\ref{eqn: error envelope criteria} is a rather useful tool, because it only requires a Fisher matrix calculation, it doesn't require a full MCMC simulation.  However, we see here that the criteria tracks the more precise results of an MCMC run very well.  Therefore this gives us a bridge between our Fisher and MCMC results.  It serves as a quick calculation tool that we can employ as a proxy to survey large parameter sets and PTA simulations.  For example, if we run a simulation and check the value of this ratio for all of our pulsars, we can quickly get a sense of whether or not we would likely be able to identify each pulsar's true position (if there is ``no error envelope'').  MCMC follow-ups can then be run to check the validity of these results.

%---------------------------------------------------------------------------------
%---------------------------------------------------------------------------------
    \section{The 1 Source Problem}\label{sec: 1 source}
    
\begin{figure}
        \centering
        \includegraphics[width=0.75\linewidth]{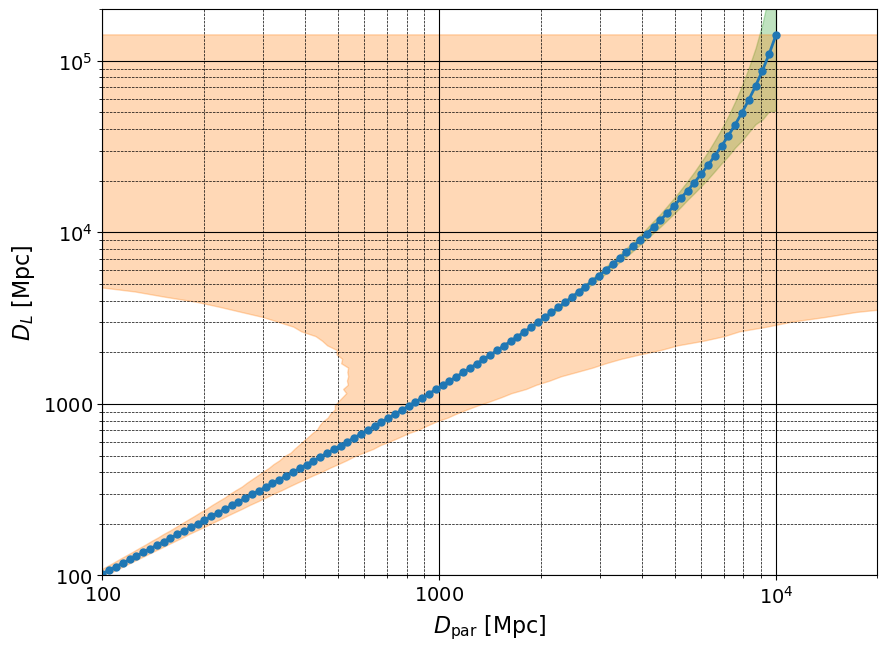}
    \caption[Uncertainties of $D_L$ vs. $D_\mathrm{par}$]{An example comparison of the Fisher predicted uncertainties of the two distance parameters $D_L$ and $D_\mathrm{par}$.  For this simulation, we set uniform uncertainties $\sigma = 100$ ns and $\sigma_L = 0.1 \lambda_\mathrm{gw}$, for a source with $\mathcal{M}_\mathrm{obs} = 10^9 \ \mathrm{M}_\odot$ and $\omega_{0,\mathrm{obs}} = 20$ nHz.  This was for a PTA of 100 pulsars (with $L_\mathrm{avg} = 8.3$ kpc and $L_\mathrm{max} = 19.9$ kpc).  The light orange region shows the uncertainty range on the parallax distance, while the light green region shows the uncertainty range on the luminosity distance.  Generally, the luminosity distance is easier to measure from the frequency chirping in our signal, as compared to the parallax distance from the Fresnel corrections.}
        \label{fig: 1 Source: R vs DL measurement}
    \end{figure}

    With only a single source generating a residual in the pulsar timing data, we find the biggest obstacle to measuring $H_0$ comes from the pulsar distance wrapping problem.  Our ability to recover $H_0$ accurately requires that we can measure the Fresnel corrections and the parallax distance $D_\mathrm{par}$.  However, this is the most challenging parameter to recover from the model.  An example showing the recovery of $D_L$ compared to $D_\mathrm{par}$ is shown in Figure~\ref{fig: 1 Source: R vs DL measurement}.

    Therefore like we found in Chapter~\ref{ch:results - first investigation}, with a single gravitational wave source we must know a priori our pulsar distances to sub-gravitational wavelength uncertainties in order to avoid the pulsar distance wrapping problem entirely.  Only then did we find that $D_\mathrm{par}$ could be recovered with sufficient accuracy that we could obtain a measurement of the Hubble constant through equation~\ref{eqn: Hubble constant}.  However, this level of precision is far beyond current capabilities, making this a very challenging measurement.  In conclusion, throughout our tests we did not find a very meaningful way of procuring the $H_0$ measurement in this way.
    
    What seems to us a perhaps more useful way of exploiting the Fresnel effects with just a single gravitational wave source is to instead \textit{reverse} the question.  If we already knew the \textit{source location} a priori, could we then use the Fresnel corrections in our model to better measure the distances to the pulsars in our PTA?  For this scenario, we imagine a case where our source has already been identified, perhaps by an electromagnetic counterpart~\citep{pta_multimessenger2021}.  To represent a best-case-scenario, we fix the distance and angular parameters ($D_L$, $D_\mathrm{par}$, $\theta$, $\phi$) in our model, and we \textit{do not} add any prior pulsar distance knowledge.  Now we look at how knowledge of our gravitational wave source can inform us about the distances to the pulsars in our PTA.
    
    An ideal measurement of a pulsar's distance would pinpoint its location to an uncertainty smaller than that pulsar's wrapping cycle for the given gravitational wave source.  Otherwise an error envelope will form.  So for a given known source, we ideally want to ask how many of the pulsars in our PTA will satisfy the ``no error envelope'' criteria in equation~\ref{eqn: error envelope criteria}?  In principle, if for a known source one of our pulsars satisfies this criteria, then a full Bayesian MCMC analysis may be able to distinguish the true distance from the secondary modes due to the wrapping cycle (see again Figure~\ref{fig: error envelope progression}), since these modes will be less probable.
    
\begin{figure}
        \centering
          \begin{subfigure}[b]{0.48\linewidth}
          \centering
            \includegraphics[width=1\linewidth]{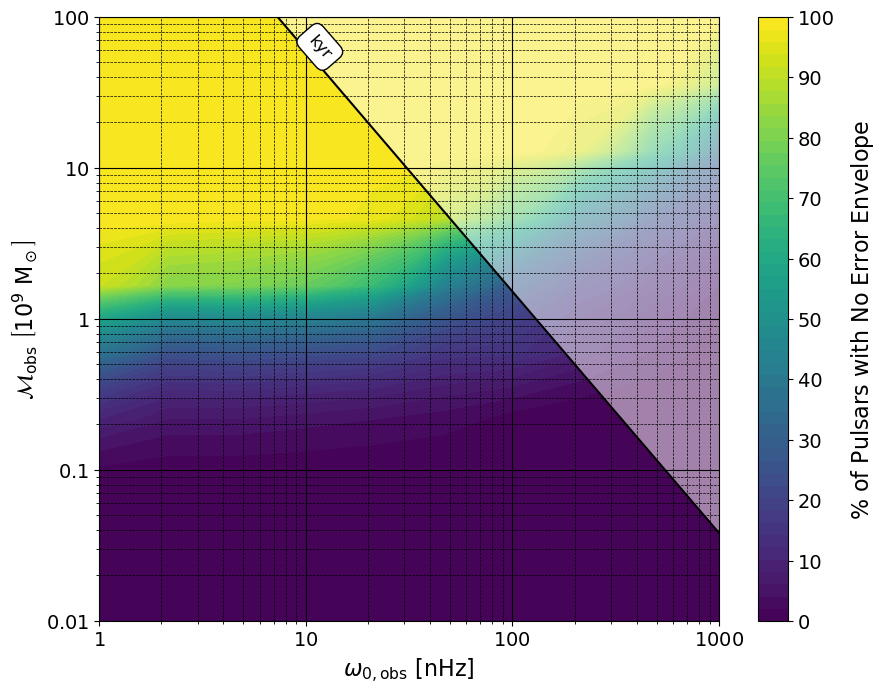}
          \end{subfigure}
          \hfill
          \begin{subfigure}[b]{0.48\linewidth}
          \centering
          \includegraphics[width=1\linewidth]{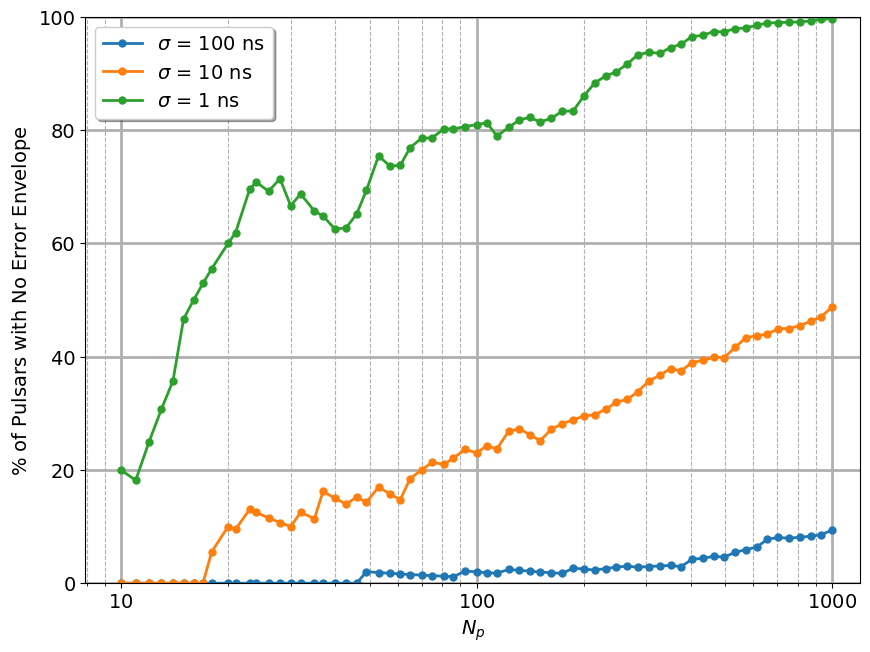}
          \end{subfigure}
        \caption[1 Known Source - Pulsar Distance Recovery]{An example of the total percentage of pulsars classified with ``no error envelope'' by equation~\ref{eqn: error envelope criteria}, for a single \textit{known} gravitational wave source (with pinned $D_\mathrm{par}$, $\theta$, and $\phi$ parameters).  For both figures, the source was simulated at $D_\mathrm{par} = 100$ Mpc.  These results were computed using the IIB model (Section~\ref{sec: model IIB - revisited}). \textbf{(Left)} In this simulation, $\sigma = 1$ ns, and there are 100 pulsars in the array (with $L_\mathrm{avg} = 8.3$ kpc and $L_\mathrm{max} = 19.9$ kpc).  As we did in Chapter~\ref{ch:results - first investigation}, the $\Delta \tau_{c,\mathrm{obs}} = 1$ kyr line is indicated for reference.  More massive sources are more likely to result in a higher number of pulsars in the array being identified with no error envelope.  \textbf{(Right)} Here the source was simulated with $\mathcal{M}_\mathrm{obs} = 3\times10^9 \ \mathrm{M}_\odot$ and $\omega_{0,\mathrm{obs}} = 30$ nHz.  Generally adding more pulsars in the array should help ``self-calibrate'' the PTA, and improve the measurements of \textit{all} of the pulsar distances in the array.}
    \label{fig: 1 Known Source - Pulsar Distance recovery}
    \end{figure}

    Interestingly, we find that the source's chirp mass and the number of pulsars in the array have a pivotal effect on deciding if the error envelope forms or not.  The left panel of Figure~\ref{fig: 1 Known Source - Pulsar Distance recovery} shows that around $\sim 10^9 \ \mathrm{M}_\odot$ there is a rapid transition from a low number of pulsars in the PTA satisfying the no error envelope criteria, to a complete $100$\% in the array satisfying this condition.  This is in spite of the fact that these pulsars all have different sky angles with respect to the source, which is one of the key factors in deciding the wrapping cycle distance (see equation~\ref{eqn:Delta L in monochromatic regimes}).  Furthermore, the right panel of Figure~\ref{fig: 1 Known Source - Pulsar Distance recovery} shows that increasing the number of pulsars in the array can also drive up the percentage of total pulsars which satisfies this criteria.  It seems that the network as a whole will ``self-calibrate'' as more pulsars are added, in the sense that with a known source the knowledge of the timing residual solution is shared across all pulsar distance parameters in the array, reducing the uncertainty around each pulsar's true mode.  Even with current timing standards of around $\sigma \sim 100$ ns, this suggests that future Square Kilometer Array-like PTAs with upwards of 1000 pulsars and a nearby known source $\mathcal{O}\left(100 \ \mathrm{Mpc}\right)$ could yield around $10$\% of those pulsars with distances known to $\mathcal{O}\left(\lambda_\mathrm{gw}\right)$.  With better timing, this could even reach as high as $100$\%.

%---------------------------------------------------------------------------------
%---------------------------------------------------------------------------------
    \section{The 2 Source Problem}\label{sec: 2 sources}
    
    The pulsar distance wrapping problem makes the measurement of $H_0$ from a single gravitational wave source very difficult, even with advanced future PTAs.  Seeing this as the primary challenging, we began to look for ways in which we could perhaps help break the wrapping cycle degeneracy to remove the secondary modes and the error envelope.
    
    So we asked the question, what if there were two (or more) gravitational wave sources in the collected timing residual data?  With a single source, every pulsar will have a specific wrapping cycle distance (equation~\ref{eqn:Delta L in monochromatic regimes}).  This distance primarily depends on the frequency of the source and the angular sky separation between the source and the pulsar.  And the degeneracy occurs because $L \pm \Delta L_n$ will produce another mode in the likelihood function.  But if there are two sources, then every pulsar will have \textit{two} wrapping cycle distances.  As long as the two sources differ in either frequency and/or angular sky position, then the key idea is that the two sets of wrapping cycle distances will not be the same. Therefore, the \textit{joint} likelihood for both sources will now only contain secondary modes in an individual pulsar's distance at common multiples of \textit{both} the wrapping cycle distances.
    
\begin{figure}
        \centering
        \includegraphics[width=1\linewidth]{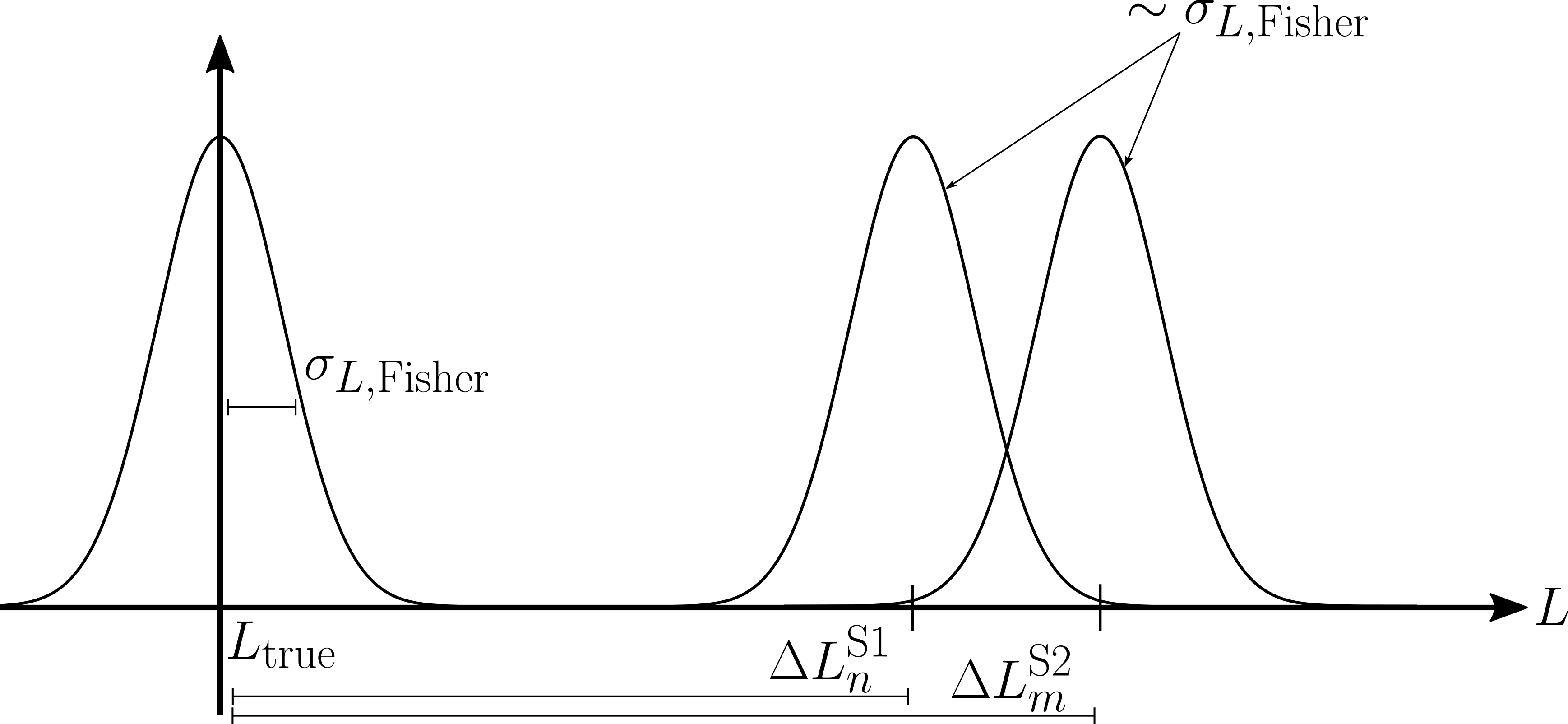}
    \caption[2 Sources - Common Wrapping Cycle]{Here we show a simplified visual representation of the 1D likelihood in any of our pulsar distance parameters, for two different sources ``S1'' and ``S2.''  As long as the sources are located on different parts of the sky and/or have different frequencies, they will have different sets of pulsar distance wrapping cycles (equation~\ref{eqn:Delta L in monochromatic regimes}).  In the \textit{individual} likelihoods, both sources will agree on the true pulsar distance $L_\mathrm{true}$ and will support that as the primary mode.  Secondary modes will then form at $\Delta L_n^\mathrm{S1}$ and $\Delta L_m^\mathrm{S2}$ for sources 1 and 2, respectively.  The \textit{joint} likelihood for both sources will find support at the distances where there is some common overlap between $\Delta L_n^\mathrm{S1}$ and $\Delta L_m^\mathrm{S2}$.  As a proxy, in this study we forecast that the width of these individual modes will be approximately the same width as predicted by the Fisher matrix analysis about the true mode, $\sigma_{L,\mathrm{Fisher}}$.  This then leads us to our criteria for predicting the strength of the new ``common'' secondary modes in the joint posterior (equation~\ref{eqn: 2 source mode condition}).  The location of these modes could be estimated as the average distance between the two individual modes, that is at $\left| \Delta L_n^\mathrm{S1} + \Delta L_m^\mathrm{S2} \right| / 2$.  Equation~\ref{eqn: 2 Source Common error envelope criteria} then provides a criteria for when that common mode will be close enough to the primary to blend the error envelope.}
        \label{fig: 2 Source Common Wrapping Cycle}
    \end{figure}

    Using the same logic as before to create the error envelope classification criteria in equation~\ref{eqn: error envelope criteria}, we can create two more criteria to try and predict \textit{where} we would get a double source common mode, and whether or not that new mode would contribute to the error envelope about the true pulsar distance.  Consider the illustration in Figure~\ref{fig: 2 Source Common Wrapping Cycle}.  The joint likelihood should find some support at distances where the uncertainties around the two separate wrapping cycle distances overlap.  The amount of overlap between these individual modes will decide how strongly the joint likelihood supports that distance, so as a rough proxy we could say:
    \begin{equation}
        \frac{\left|\Delta L_n^\mathrm{S1} - \Delta L_m^\mathrm{S2}\right|}{\sigma_{L,\mathrm{Fisher}}}  \sim \begin{cases}
        0 - 2 \ ,     &\text{``strong'' support}  \\
        2 - 4 \ ,   &\text{``weak'' support}  \\
        \end{cases}
    \label{eqn: 2 source mode condition}
    \end{equation}
    where $L_n^\mathrm{S1}$ and $L_n^\mathrm{S2}$ are the $n$th and $m$th wrapping cycles of source 1 and source 2, respectively.  As an example, if this quantity equalled $0$ then the $n$th and $m$th modes would directly overlap, hence there would be strong support for a mode here.  At a value of $2$ then center point of the $n$th and $m$th peaks would be $1-\sigma$ from each peak, and at a value of $4$ the center would be $2-\sigma$ from each peak.  For locations where there is strong support, we would then expect this new ``common'' wrapping cycle distance to be at approximately the average distance between these individual modes.
    
    Therefore we can then re-write our error envelope criteria equation~\ref{eqn: error envelope criteria} for the double source problem:
    \begin{equation}
        \frac{\left|\frac{\Delta L_n^\mathrm{S1} + \Delta L_m^\mathrm{S2}}{2}  \right|}{\sigma_{L,\mathrm{Fisher}}} \sim \begin{cases}
        < 4 \ ,    &\text{Error Envelope}  \\
        4 - 8 \ ,  &\text{Error Envelope begins to separate}  \\
        > 8 \ .    &\text{No Error Envelope} 
        \end{cases}
    \label{eqn: 2 Source Common error envelope criteria}
    \end{equation}
    This condition checks if the uncertainties around the common wrapping cycle between the $n$th and $m$th wrapping cycles of sources 1 and 2 will be close enough to the uncertainty around the true distance as to blend those uncertainties into a greater envelope.  Once again, in order to make this prediction we are assuming as a proxy that the width of the true mode predicted by the Fisher matrix for the joint source likelihood also approximates the secondary modes at the individual source wrapping cycles.

    We think that this has a great potential towards helping to remove the wrapping cycle problem for the pulsars in our array.  Considering any given pulsar, at minimum the new common wrapping cycle mode in the joint likelihood function will occur at the larger of the two wrapping cycle distances for the case where one wrapping cycle is already a multiple of the other.  But even better is when the two individual wrapping cycle distances are not multiples of one another - then the new common mode can be pushed further away from the true mode.  This should also increase the separation between modes in the joint likelihood in general, making it harder for an error envelope to form.
    
    Furthermore, from Section~\ref{sec: Pulsar Distance Wrapping Problem & Error Envelope} recall that in the regimes IB and IIB, frequency evolution itself nominally breaks the wrapping cycle degeneracy.  Therefore two sources with significant frequency evolution should even further constrain the pulsar distances.  In summary, with two sources the distance modes are further separated and with frequency evolution modes further from the true distance will find even less support in the likelihood.  All of this coupled with prior knowledge on our pulsar distances thanks to electromagnetic observations should now give us a total of three separate means of localizing the pulsars in our PTA.
    
    Mathematically, in our likelihood function equation~\ref{eqn: likelihood function} we now have $\overrightarrow{\mathrm{Res}}_\mathrm{true} = \overrightarrow{\mathrm{Res}}_\mathrm{true,1}$  $+ \overrightarrow{\mathrm{Res}}_\mathrm{true,2}$ for sources ``1'' and ``2.''  Fortunately adding a second source does not double our parameter space, since both sources will share the pulsar distance parameters $\vec{L}$.  We simply double the number of source parameters, so $\vec{\theta} = \vec{\theta}_1 \cup \vec{\theta}_2 = \left[\vec{s}_1, \vec{s}_2, \vec{L}\right]$. If we decompose our likelihood function, we can write it as the product of the likelihood function of just source 1, the likelihood of just source 2, and their cross terms.  That is the squared term in equation~\ref{eqn: likelihood function} becomes:
    \begin{align}
        &\left( \mathrm{Res}^c_1 + \mathrm{Res}^c_2 - \mathrm{Res}^c_1\left(\vec{\theta}\right) - \mathrm{Res}^c_2\left(\vec{\theta}\right) \right)^2  \nonumber \\
        &\hspace{2cm}\quad = \left( \mathrm{Res}^c_1 - \mathrm{Res}^c_1\left(\vec{\theta}\right) \right)^2 + \left( \mathrm{Res}^c_2 - \mathrm{Res}^c_2\left(\vec{\theta}\right) \right)^2  \nonumber \\
        &\hspace{2cm}\qquad \ + 2\bigg[ \mathrm{Res}^c_1\mathrm{Res}^c_2 - \mathrm{Res}^c_1\mathrm{Res}^c_1\left(\vec{\theta}\right) - \mathrm{Res}^c_2\mathrm{Res}^c_2\left(\vec{\theta}\right) \bigg. \nonumber \\
        &\hspace{2cm}\qquad \qquad \bigg. + \mathrm{Res}^c_1\left(\vec{\theta}\right)\mathrm{Res}^c_2\left(\vec{\theta}\right) \bigg]^2 .
    \end{align}
    
    Thanks to the two sets of wrapping cycles, the joint likelihood can help calibrate the PTA by more precisely identifying the \textit{true} pulsar distances.  At the same time, the more precisely recovered pulsar distances then better recover the parallax distance and hence the Hubble constant.  So the two source problem helps overcome the pulsar distance wrapping cycle obstacle, which has been the biggest challenge so far, and allows us to probe the Fresnel curvature effects in the model.  Critically, in all of the results that we now present here, \textit{we never specified any prior knowledge of our pulsar distances when performing any of our analyses}!
    
    Previously, even when performing a Fisher matrix analysis, the only way to accurately invert the Fisher matrix required that we first add in pulsar distance priors which would constrain the uncertainty on the pulsar distances down to the order of the wrapping cycle.  Otherwise, the Fisher matrix's condition number would be too large to allow us to accurately invert the matrix because of the strong covariances introduced from the pulsar distances and parallax distance parameter.  However, with two sources we found that this was no longer a required step for the Fisher-based results, further proving the benefit of combining the knowledge of two gravitational wave sources to parameter estimation!
    
    As mentioned in Section~\ref{sec:methods}, we can use the small redshift approximation (equation~\ref{eqn: low redshift R approximation}) to replace the parallax distance $D_\mathrm{par}$ parameter with $H_0$ directly.  With two sources this actually removes one of of the parameter dimensions (rather than having $D_\mathrm{par,1}$ and $D_\mathrm{par,2}$ parameters for the two sources, now we just have $H_0$) and gives us the joint posterior recovery of $H_0$.
    
\begin{figure}
        \centering
          \begin{subfigure}[b]{0.48\linewidth}
          \centering
            \includegraphics[width=1\linewidth]{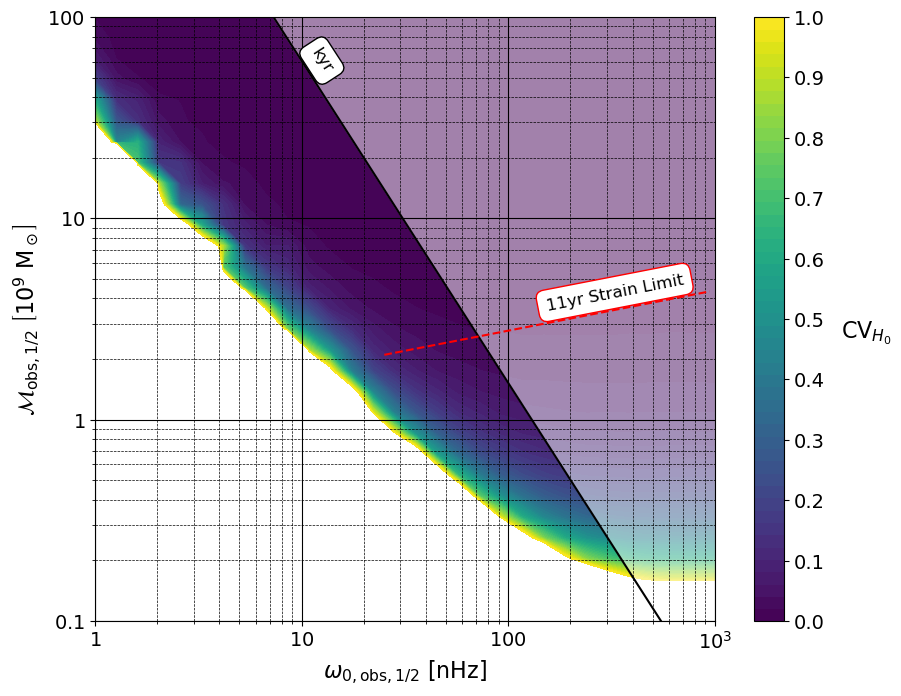}
          \end{subfigure}
          \hfill
          \begin{subfigure}[b]{0.48\linewidth}
          \centering
          \includegraphics[width=1\linewidth]{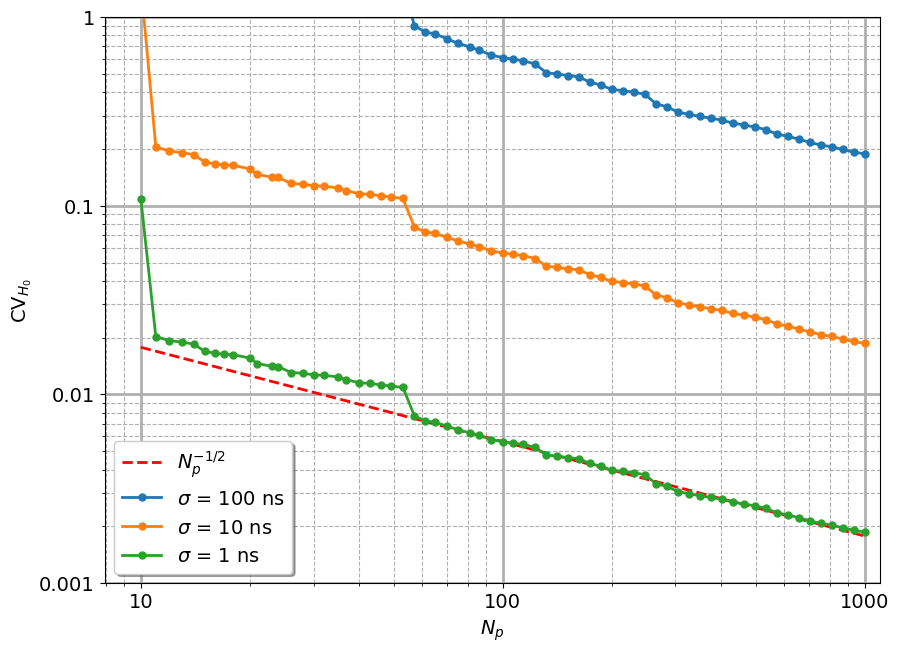}
          \end{subfigure}
        \caption[2 Sources - $H_0$ Recovery (Fisher Survey)]{$H_0$ recovery for the 2 source problem as a function of the observed source chirp mass and orbital frequency intrinsic parameters, and as a function PTA qualities (timing uncertainty and number of pulsars).  For simplicity, both sources are assumed to be completely identical (hence the ``$1/2$'' notation in the left panel for sources 1 and 2 in this figure) \textit{except} for being located on different parts of the sky (different $\theta$ and $\phi$ parameters).  \textbf{(Left)}  Here 100 pulsars ($L_\mathrm{avg} = 8.3$ kpc and $L_\mathrm{max} = 16.8$ kpc) are timed with $\sigma = 1$ ns uncertainty.  Both sources are placed at $D_\mathrm{par} =  100$ Mpc.  The results of \citet{NG_11yr_cw} are used to include a reference rough estimate of NANOGrav's 11yr continuous wave strain upper limit, $h_{0,\mathrm{11yr}} \approx 10^{-6} \frac{\omega_0}{\pi (1 \ \mathrm{Hz})}$.  \textbf{(Right)}  Here both sources are located at $D_\mathrm{par} = 100$ Mpc, with $\mathcal{M}_\mathrm{obs} = 10^9 \ \mathrm{M}_\odot$ and $\omega_{0,\mathrm{obs}} = 20$ nHz.  The PTA pulsar population used here has $L_\mathrm{avg} = 8.4$ kpc and $L_\mathrm{max} = 23.4$ kpc.  Improving the timing precision of our pulsars gives the largest improvement in the recovery of $H_0$.  But interestingly we also found that simply adding more pulsars to the array generally improves the measurement of $H_0$, scaling as roughly $\mathrm{CV}_{H_0} \sim \sigma/\sqrt{N_p}$.}
    \label{fig: 2 Sources - H0 recovery}
    \end{figure}

    Figure~\ref{fig: 2 Sources - H0 recovery} shows Fisher analysis surveys of the $H_0$ measurability in terms of intrinsic source chirp mass and frequency, as well as PTA characteristics like timing accuracy and number of pulsars.  Not surprisingly, we measure $H_0$ best from sources with high chirp mass and frequency, since these will have produce strong frequency evolution effects in the signal.  For reference, we include a rough estimate of the NANOGrav 11yr continuous wave strain upper limit, $h_{0,\mathrm{11yr}} \approx 10^{-6} \frac{\omega_0}{\pi (1 \ \mathrm{Hz})}$ \citep[see Figure 3 of][]{NG_11yr_cw}.
    
    As the right panel of Figure~\ref{fig: 2 Sources - H0 recovery} shows, timing precision makes a significant different in improving the measurement of $H_0$.  For an order of magnitude improvement in $\sigma$, we gain approximately an order of magnitude improvement in $\mathrm{CV}_{H_0}$.  But perhaps more interesting is the inclusion of additional pulsars in the PTA.  We find that even for a population of pulsars with fixed spatial extent, simply adding more pulsars improves the entire network's ability to recover $H_0$.  In summary, we observe that generally our ability to measure $H_0$ scales roughly as $\mathrm{CV}_{H_0} \sim \sigma/\sqrt{N_p}$.
    
\begin{figure}
        \centering
          \begin{subfigure}[b]{0.48\linewidth}
          \centering
            \includegraphics[width=1\linewidth]{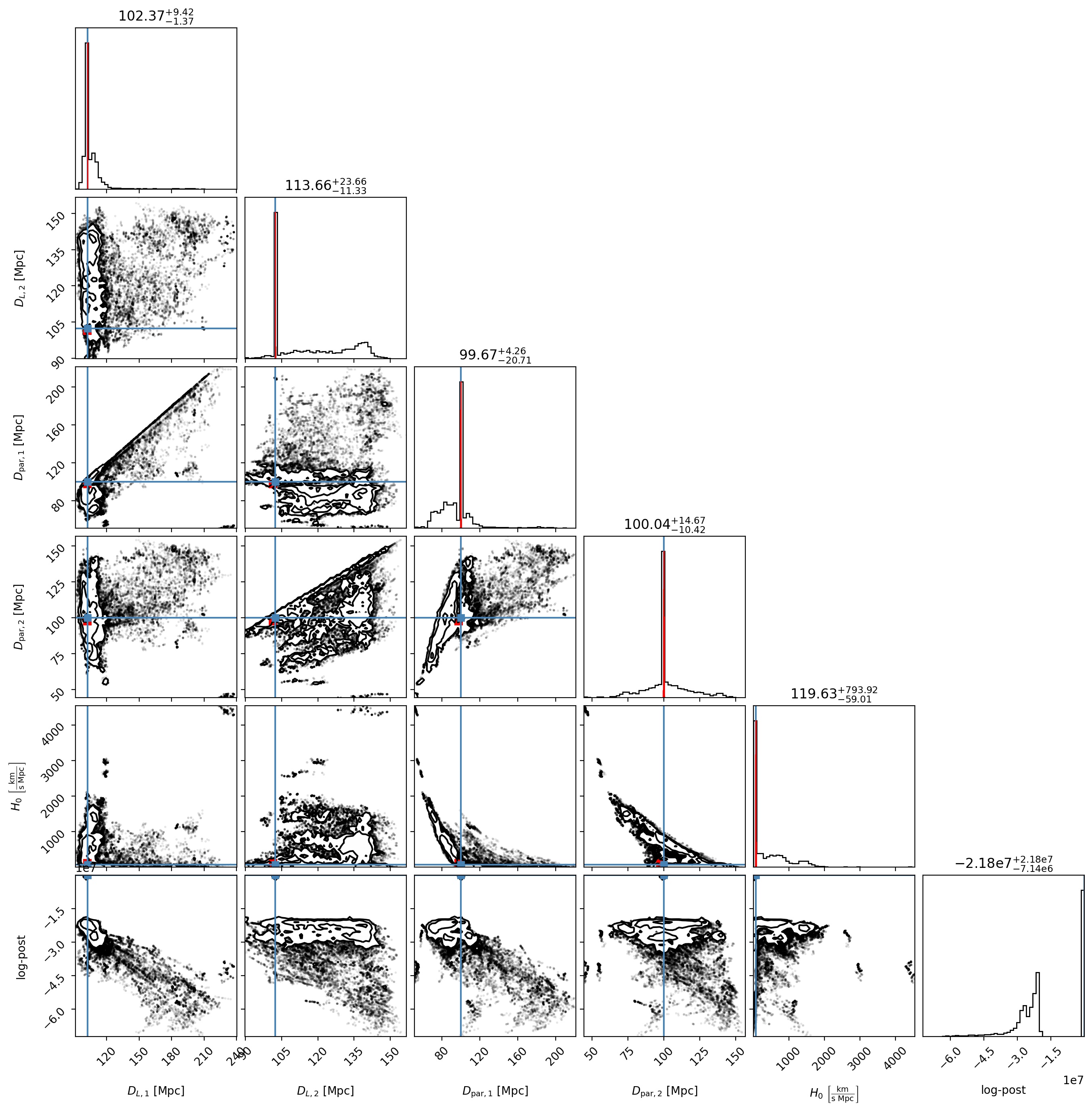}
          \end{subfigure}
          \hfill
          \begin{subfigure}[b]{0.48\linewidth}
          \centering
          \includegraphics[width=1\linewidth]{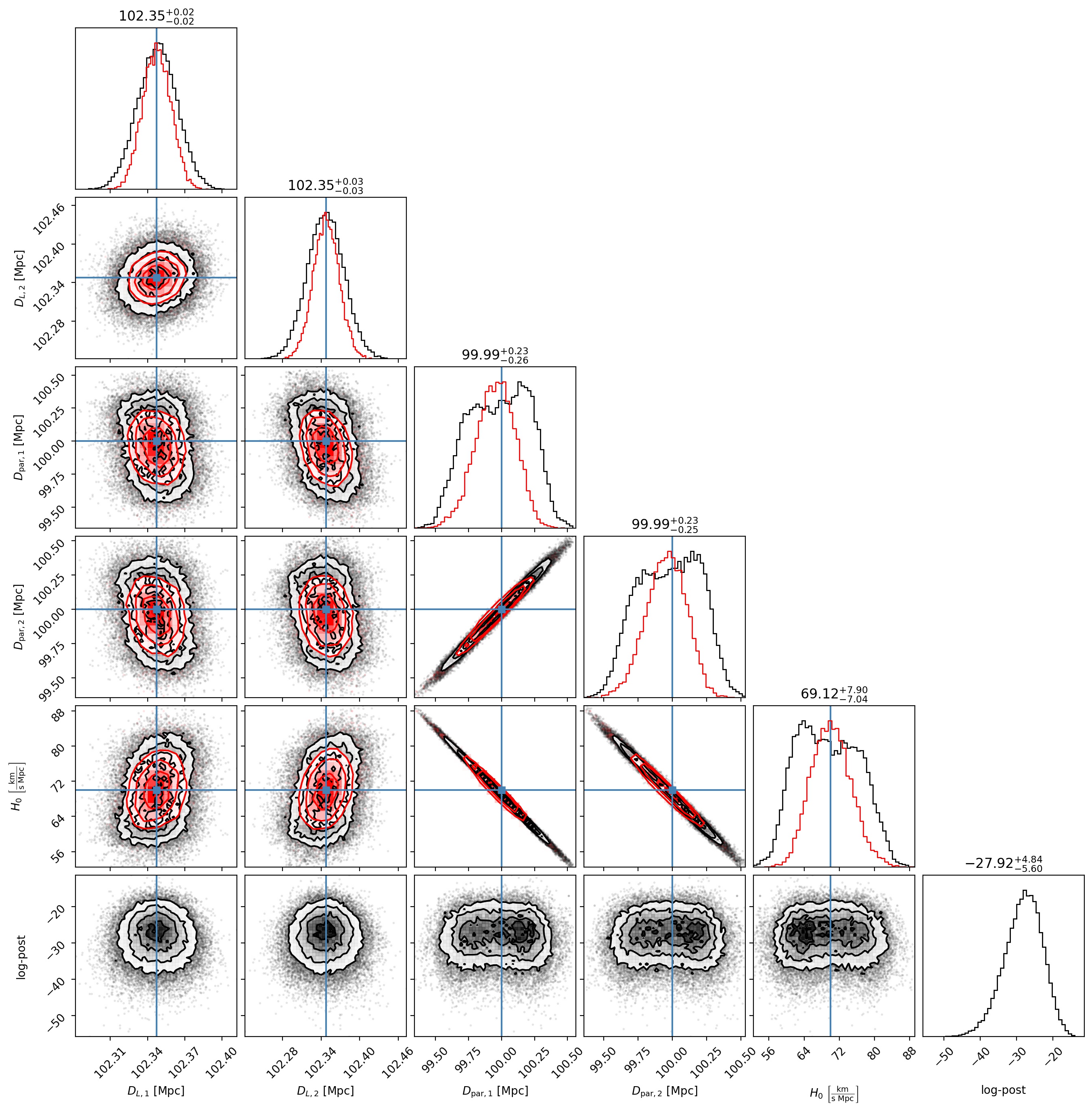}
          \end{subfigure}
        \caption[2 Sources - $H_0$ Recovery (MCMC Example)]{An MCMC targeted simulation of the recovery of $H_0$ for two sources (again, made identical for simplicity, except for having different sky angles $\theta$ and $\phi$).  Both panels are of the same simulation, showing just the recovery of the two sources' distances ($D_L$ and $D_\mathrm{par}$), and the joint recovery of $H_0$.  The bottom row of the corner plot shows the log-posterior values.  These are not normalized - what matters is simply the order of magnitude difference in values being spanned by the walkers.  The blue line indicates the injected true parameter, and the red histograms are the Fisher matrix based predictions.  In this example, $\mathrm{CV}_{H_0} = 0.065$.  \textbf{(Left)}  For the full simulation we initialized multiple ``balls'' of walkers - one about the true mode, and two additional balls around locations predicted by equation~\ref{eqn: 2 source mode condition}.  The bottom row shows many orders of magnitude difference between the walkers centered on the true mode and the walkers away from the true mode.  If we ran the MCMC simulation longer, we would expect that these low-probability walkers would eventually find and settle on the true mode.  \textbf{(Right)}  This is the same corner plot, but now with all of the walkers initialized away from the true mode removed from the data set.}
    \label{fig: 2 Sources - MCMC target}
    \end{figure}

    The results of a targeted MCMC search is shown in Figure~\ref{fig: 2 Sources - MCMC target}.  In order to help verify our previous discussion about how two sources help to break the pulsar distance wrapping cycle degeneracy, when initializing the walkers in our MCMC we used equation~\ref{eqn: 2 source mode condition} to predict where the nearest secondary modes would likely form in each pulsar distance parameter.  Then we initialized multiple ``balls'' of walkers - one about the true mode, and additional balls around the locations of these predicted secondary modes.  In the examples we tested, we found that the set of walkers near the true mode were able to explore the parameter space near that mode quite well, while the walkers placed at the secondary modes simply wandered around parameter space for the duration of our simulation without settling down.  This is actually what we expected to observe, because as we found in examples like Figure~\ref{fig: IIB error envelope} (with a single source) if the secondary modes did exist then our walkers could settle and explore those modes.  Given that our walkers did not find meaningful secondary modes in these simulations, this supports our conclusion about how two sources can help break the wrapping cycle degeneracy.  

%---------------------------------------------------------------------------------
%---------------------------------------------------------------------------------
    \section{Conclusions}\label{sec: conclusions - results2}
    
    In this work we have shown that future PTA experiments could make purely gravitational wave based measurements of the Hubble constant.  This is made possible largely by accounting for the Fresnel curvature effects in the wavefront across the Earth-pulsar baseline.  By using the fully general Fresnel frequency evolution timing residual model, we can obtain two separate distance measurements to the source: the luminosity distance $D_L$ (from the frequency evolution effects) and the parallax distance $D_\mathrm{par}$ (from the Fresnel effects).
    
    There do exist other gravitational wave-based techniques for measuring the Hubble constant.  These include the hybrid gravitational wave/electromagnetic standard sirens approach~\citep{Schutz_1986, Holz_2005, gw170817_h0}, and the pure gravitational wave-based binary neutron star merger approach~\citep{messenger_gwH0}.  Our proposed approach is also purely gravitational wave-based, and distinct in that it obtains two \textit{distance} measurements to the same source, rather than a distance and a redshift.  Therefore the systematic approach is different for this method, and could prove beneficial to be an important novel tool towards helping to resolve the Hubble tension.
    
    We find that the largest hurdle to overcome in making this measurement possible will likely be the pulsar distance wrapping problem.  Currently many pulsar distances are not known to a high degree of accuracy, therefore we include the pulsar distances as free parameters in our Bayesian model and parameter search.  However, due to the way the pulsar distance enters the timing residual model (described in Section~\ref{sec: Pulsar Distance Wrapping Problem & Error Envelope}) there exist degeneracies in the model at the pulsar distance wrapping cycles.  Because the parallax distance $D_\mathrm{par}$ is measured from Fresnel effects, it is highly sensitive to uncertainties in the pulsar distances (see equation~\ref{eqn: Fresnel number generalized}).  Therefore uncertainties in the distances to the pulsars in our array will propagate and potentially wash out our ability to accurately measure $H_0$.
    
    However, as we show in Section~\ref{sec: 1 source}, the Fresnel timing models could still be beneficial to measuring the \textit{pulsar distances}, if we are lucky enough to identify the source and its location through an electromagnetic counterpart.  With known distances and sky angles to our source, we can use the gravitational wave signal itself to help identify the true distances to the pulsars in our PTA, down to uncertainties of $\mathcal{O}\left(\lambda_\mathrm{gw}\right)$.  In fact we find that with this method the PTA will self-calibrate, in that as more and more pulsars are added to the array, the additional knowledge of the larger network of pulsars will work together to constrain each individual pulsar's location.
    
    While measuring $H_0$ from a single unknown gravitational wave source presents a very difficult challenge, we also investigated an alternative scenario - simultaneously searching over parameter space for \textit{two} sources.  The key insight that we show is that with two sources, the pulsar distance wrapping cycle degeneracy is further broken, thanks to the fact that two different sources will have different wrapping cycle distances for each pulsar.  The net effect of this is that two or more sources will actively calibrate the PTA by helping to identify the true pulsar distances from the degenerate secondary modes.  This when combined with the nominal wrapping cycle degeneracy break caused by a source's frequency evolution, and any prior electromagnetic knowledge of the pulsar distances, now give us three separate ways to constrain the distance to every pulsar in the PTA.
    
    For the two source problem we show that with sufficiently precise timing and a large enough array, we can find values of $\mathrm{CV}_{H_0} < 0.1$, suggesting measurements of the Hubble constant with relative uncertainties less than about $10$\%.  Adding more pulsars to the array scales as approximately $\mathrm{CV}_{H_0} \sim \sigma/\sqrt{N_p}$, so if future SKA-like PTAs could time hundreds of pulsars, this could also push down the uncertainty on our Hubble constant measurement.
    
    These results conclude the second major investigation of my PhD, generalizing the work of the previous chapters, and building a cosmological tool that could employ pulsar timing towards making novel measurements of our universe.  The results presented in this chapter are currently being prepared for publication.
    
    $\longrightarrow$ \textbf{Update:} This work was refined and published in~\cite{mcgrath2022H0}.

\cleardoublepage
\chapter{General Doppler Tracking}\label{ch: General doppler tracking}

Although the title of this dissertation acknowledges that the work presented here is for application in pulsar timing experiments, the background theory of all of this work generally falls under ``Doppler tracking'' \citep{doppler_estabrook, doppler_Tinto, doppler_armstrong}.  Thus for this chapter only, I present a ``side-project'' that I spent some time on during my PhD, working to generalize the results of the previous chapters to Doppler tracking experiments. 

In Chapter~\ref{ch: The Continuous Wave Timing Residual} we developed the theory behind the effect a gravitational wave will have on the observed period of a flashing pulsar.  The key idea was that in the presence of a gravitational wave, the observed pulsar period will differ periodically from what we would expect.  This difference, the timing residual, we can then measure experimentally, and from the difference then infer the parameters of the gravitational wave that is causing the fluctuation.

In that formalism we assumed a pulsar ``clock'' was sending a signal to the Earth, where the Earth was the center of our coordinate system.  This one-way transfer of information to the Earth is often referred to as a ``downlink.''  But we could also imagine that instead of a pulsar we launch a satellite with a clock which establishes this downlink and performs the experiment.  Building on this idea, if we are using a satellite to send us a signal, then this offers us the unique chance to also send a signal to the satellite.  If we use a clock on Earth to send the time to a satellite, it can similarly measure the fluctuation in the period of that Earth clock which it observes, and also attempt to measure the presence of a gravitational wave.  This one-way transfer of information from the Earth to a satellite is referred to as an ``uplink.''

If we use a master clock on Earth to first beam the time to a satellite, and if that satellite then re-transmits the signal directly back to Earth, we make use of both the uplink and downlink to extend the distance that the photons have to travel, and hence the amount of time that they will be affected by any passing gravitational wave.  We can then observe the final full ``round-trip'' signal as it returns to Earth and compare that back to our expectation of what we should see from the original master clock time, giving us once again an experimentally measured timing residual which can be compared to our theory.  In principle this original clock signal could be bounced back and forth any arbitrary number of times, and with each pass the photons would be altered by the presence of the gravitational wave.  The final observed times of our clock after whatever desired number of passes have been made would contain then the integrated effect of the gravitational wave with each pass.

This type of gravitational wave experiment is known as Doppler tracking, as it exploits the effect of Doppler shifted clock signals on satellites in order to measure the presence of gravitational waves.  And in all of this, we have been assuming that communication must always take place with Earth as one of the two ends, but one could imagine developing this same experiment but using two satellites which communicate with each other (and later send all of the data back to Earth).  Regardless of whether we use two satellites or one satellite and the Earth for the communication, there are a number of benefits that Doppler tracking has over strictly one-pass pulsar timing experiments:
\begin{enumerate}
    \item We have full control of the master clock.  Although the best pulsars rival the precision of our best atomic clocks, we cannot control their period or other physical effects which can sometimes slightly alter their rotation rates.
    \item Round-trip signals will look different from single-pass signals and will increase the overall amount of time the photons are traveling between relays, and hence the amount of time that they are affected by the presence of a gravitational wave.
    \item If using two satellites to perform the experiment, then all of the noise sources inherent to sending a light signal through the Earth's atmosphere to and from its surface are removed.
\end{enumerate}
The primary benefit of pulsar timing is size of those experiments.  The pulsars we tend to use are on the order of kiloparsecs away in distance, so although they only use the downlink part of the Doppler tracking method, that length scale is still far greater than what we can achieve by launching a satellite.

Nevertheless, motivation for this type of study may be taken from the following.  Atomic clocks are being developed to ever greater precision~\citep{atomicclock_record}, and because of this some scientists have already begun to consider future space-based Doppler tracking missions using atomic clocks to search for gravitational waves~\citep{loeb_gw_atomicclocks,GW_atomicclocks}.  Furthermore, given the success of past deep space missions such as the Voyager probes and the New Horizons spacecraft, scientists have also begun to develop plans for a future space-based probe which could reach 1000 AU~\citep{interstellar_probe_meeting}.  At a distance like this, the probe-Earth baseline may become large enough to be able to measure the Fresnel effects from gravitational wave sources within our own Galaxy, which then ties back into the primary focus of this dissertation.

Here we will build off of the previous work in Chapter~\ref{ch: The Continuous Wave Timing Residual} to derive the general round-trip Doppler tracking formula for an arbitrary number of uplinks/downlinks.  The solution process will be nearly identical to that chapter, except now we account for two-way motion of the photons (before we had only one-way motion).  Therefore most of our assumptions will carry over from before, but we name a few changes here:\\

\begin{minipage}{0.9\linewidth}
\fbox {
    \parbox{\linewidth}{
    \textbf{Assumptions:} 
    \begin{enumerate}
        \item The source we are looking for is still the same type of source we were considering before (i.e. a binary system), so all of the assumptions and ideas in Chapter~\ref{ch: GWs from a Binary Source} will still hold.
        \item Our coordinate system is attached to one of our two satellite endpoints in the Doppler tracking experiment.  If communication is between the Earth and a satellite, then we will consider the Earth to be the center of the coordinate system.
        \item The two satellites are located at fixed angular sky positions and distances from each other.
        \item Once the photons in our signal arrive at one of the satellite endpoints, they are immediately re-transmitted.  There are no time delays - the signal travels between satellites continuously.
        \item All additional assumptions from Chapter~\ref{ch: The Continuous Wave Timing Residual} still hold.
    \end{enumerate}
    }
}
\end{minipage}\\

%---------------------------------------------------------------------------------
%---------------------------------------------------------------------------------
    \section{The Observed Clock Period and Timing Residual}
    
    Again, we begin by asking what is the path that a photon traveling between satellites will take?  In order to carry over the notation we employed from Chapter~\ref{ch: The Continuous Wave Timing Residual}, our ``primary satellite'' will be the center of our coordinate system, and the ``secondary satellite'' will be a distance $L$ away with position vector given by equation~\ref{eqn: p hat}.  The primary satellite will send the master clock time to the secondary satellite (the uplink), and the time that the secondary clock observes will be immediately re-transmitted back to the primary satellite (downlink).  This will then be repeated an arbitrary $N$ number of times.
    
    We again use the metric in equation~\ref{eqn: spacetime interval} to set up the path that the photon will take (i.e. where $ds=0$) which gives us:
    \begin{align}
        \pm \ dr &= \frac{c dt}{\left[1+\hat{p}^i\hat{p}^j E^{\hat{r}\textsc{A}}_{ij} h_{\textsc{A}} \right]^{1/2}}, \tag{\ref{eqn: photon spacetime interval} r} \\
        &\approx \left[1 - \frac{1}{2}\hat{p}^i\hat{p}^j E^{\hat{r}\textsc{A}}_{ij} h_{\textsc{A}} \right] c dt . \nonumber
    \end{align}
    Radially outbound ``uplink'' photons take the ($+$) sign, and radially inbound ``downlink'' photons take the ($-$) sign. We attach the coordinate system to one of the two satellites so that the photons will travel along the radial path that links the two satellites, which means that $d\theta = d\phi = 0$.  If we use two satellites but reference them to the center of a coordinate system attached at the Earth's position, then $d\theta \neq d\phi \neq 0$ and the math is more complicated.
    
    Now consider a photon bouncing back and forth between the two endpoints.  The photon will leave at the emitted time $t_\mathrm{em}$, travel the distance $L$ along the uplink vector $\vec{x}_\uparrow$ and arrive at some time $t_1$, then return the distance $L$ along the downlink vector $\vec{x}_\downarrow$ arriving back at the first endpoint at some time $t_2$.  This will repeat some $N$ times and at the end of the full path it will be observed at time $t_\mathrm{obs}$.  Integrating equation~\ref{eqn: photon spacetime interval} gives us the photon's path:
    \begin{align*}
        +\int\limits^L_0 dr \ - \ \int\limits^0_L dr \ + \ \ldots \ + \ \int\limits^0_L dr \quad &\approx \quad \int\limits^{t_1}_{t_\mathrm{em}} \left[1-\frac{1}{2}\hat{p}^i\hat{p}^j E^{\hat{r}\textsc{A}}_{ij} h_{\textsc{A}}\left(\vec{x}_\uparrow\right) \right] cdt \\
        &\hspace{1cm}+ \ \int\limits^{t_2}_{t_1} \left[1-\frac{1}{2}\hat{p}^i\hat{p}^j E^{\hat{r}\textsc{A}}_{ij} h_{\textsc{A}}\left(\vec{x}_\downarrow\right) \right] cdt  \\
        &\hspace{1cm}+ \ldots \ + \ \int\limits^{t_\mathrm{obs}}_{t_N} \left[1-\frac{1}{2}\hat{p}^i\hat{p}^j E^{\hat{r}\textsc{A}}_{ij} h_{\textsc{A}}\left(\vec{x}_\downarrow\right) \right] cdt ,
    \end{align*}
    \begin{align}
        2 N L \quad \approx \quad c \left( t_\mathrm{obs} - t_\mathrm{em} \right) \ - \ \frac{1}{2}\hat{p}^i\hat{p}^j E^{\hat{r}\textsc{A}}_{ij}\left[ \int\limits^{t_1}_{t_\mathrm{em}}h_{\textsc{A}}\left(\vec{x}_\uparrow\right) cdt \ + \ \int\limits^{t_2}_{t_1}h_{\textsc{A}}\left(\vec{x}_\downarrow\right) cdt \right. \nonumber \\
        + \ldots \ + \ \left. \int\limits^{t_\mathrm{obs}}_{t_N}h_{\textsc{A}}\left(\vec{x}_\downarrow\right) cdt \right] .
    \end{align}
    Just like before we see that to ``zeroth order'' with no metric perturbation, the path the photon takes is $2 N L \approx c \left( t_\mathrm{obs} - t_\mathrm{em}\right)$, or $t_\mathrm{obs} \approx t_\mathrm{em} - 2NL$, which is what we would expect for flat unperturbed spacetime.  And just like before, since we are only
    interested in the solution to first order in the metric perturbation, we can use this to replace the limits of integration $t_1$, $t_2$, to $t_N$ in this expression with the approximate zeroth order time limits, and then evaluate the metric perturbation along the zeroth order path (the overall integrand is still first order).  To do this, we first note that in general, we can express the zeroth order path that the photon takes as follows:
    \begin{align}
        \vec{x}_0(t) \equiv \begin{cases} \vec{x}_{0,\uparrow}(t) \\
                                          \vec{x}_{0,\downarrow}(t)
                            \end{cases} 
            &\hspace{-0.3cm} \equiv \begin{cases}
                \Big[ct - \big(ct_\mathrm{em}+(2n-2)L\big)\Big]\hat{p} , \\
                \hspace{2cm} \rightarrow \ t_\mathrm{em}+(2n-2)\frac{L}{c} \leq t \leq t_\mathrm{em}+(2n-1)\frac{L}{c} ,\\
                \Big[(ct_\mathrm{em}+2nL)- ct\Big]\hat{p} , \\
                \hspace{2cm} \rightarrow \ t_\mathrm{em}+(2n-1)\frac{L}{c} \leq t \leq t_\mathrm{em}+2n\frac{L}{c} ,
            \end{cases} \nonumber \\[10pt]
            &\hspace{-0.3cm} \approx \begin{cases}
                \Big[ct - \big(ct_\mathrm{obs}+(2n-2-2N)L\big)\Big]\hat{p} , \\
                \hspace{2cm} \rightarrow \ t_\mathrm{obs}+(2n-2-2N)\frac{L}{c} \leq t \leq t_\mathrm{obs}+(2n-1-2N)\frac{L}{c} ,\\
                \Big[\big(ct_\mathrm{obs}+(2n-2N)L\big) - ct\Big]\hat{p} , \\
                \hspace{2cm} \rightarrow \ t_\mathrm{obs}+(2n-1-2N)\frac{L}{c} \leq t \leq t_\mathrm{obs}+(2n-2N)\frac{L}{c} ,
            \end{cases}
    \label{eqn:zeroth order photon path - N roundtrips}
    \end{align}
    where $n$ denotes the ``$n$th''-round-trip out of the total $N$ round-trips.  This describes the desired path - i.e. $\vec{x}_0 = 0$ when the photon begins along an uplink or arrives at the end of a downlink, and $\vec{x} = L\hat{p}$ when the photon arrives at the end up an uplink or begins along a downlink.  Now we can write:
    \begin{align}
        t_\mathrm{obs} \ \approx \ t_\mathrm{em} \ + \ 2N\frac{L}{c} \ + \ \frac{1}{2}\hat{p}^i\hat{p}^j E^{\hat{r}\textsc{A}}_{ij} \sum^N_{n=1} &\left[ \int\limits^{t_\mathrm{em}+(2n-1)\frac{L}{c}}_{t_\mathrm{em}+(2n-2)\frac{L}{c}} h_\textsc{A}\big(t,\vec{x}_{0,\uparrow}(t)\big) dt \right. \nonumber \\
        &\hspace{2cm}+  \left. \int\limits^{t_\mathrm{em}+2n\frac{L}{c}}_{t_\mathrm{em}+(2n-1)\frac{L}{c}} h_\textsc{A}\big(t,\vec{x}_{0,\downarrow}(t)\big) dt \right] .
    \label{eqn: tobs light pulse - N roundtrips}
    \end{align}
    
    As we did before, we have considered the path that a photon will travel between the satellites and now we have an expression equation~\ref{eqn: tobs light pulse - N roundtrips} which gives the difference in emitted and observed times of the photon at the end of $N$ round-trips.  Next we imagine that after the first clock time was emitted, one period later the second clock time is emitted, and we repeat the exact same steps above for a photon one period later.  Now the emitted time is $t_\mathrm{em}^{'} = t_\mathrm{em}+T$ and the observed time is $t_\mathrm{obs}^{'}$.  As we saw before:
    \begin{equation}
        \vec{x}^{'}_0(t) \equiv \vec{x}_0(t,t_\mathrm{em} \rightarrow t_\mathrm{em}^{'}) = \vec{x}_0(t-T) .
    \tag{\ref{eqn:zeroth order photon path prime} r}
    \end{equation}
    The expression one period later changes to:
    \begin{align}
        t_\mathrm{obs}^{'} &\approx t_\mathrm{em} + T + 2N\frac{L}{c} + \frac{1}{2}\hat{p}^i\hat{p}^j E^{\hat{r}\textsc{A}}_{ij} \sum^N_{n=1} \left[ \int\limits^{t_\mathrm{em}+T+(2n-1)\frac{L}{c}}_{t_\mathrm{em}+T+(2n-2)\frac{L}{c}} h_\textsc{A}\big(t,\vec{x}_{0,\uparrow}(t-T)\big) dt \right. \nonumber \\
        &\hspace{7cm}+  \left. \int\limits^{t_\mathrm{em}+T+2n\frac{L}{c}}_{t_\mathrm{em}+T+(2n-1)\frac{L}{c}} h_\textsc{A}\big(t,\vec{x}_{0,\downarrow}(t-T)\big) dt   \right], \nonumber \\
        & = t_\mathrm{em} + T + 2N\frac{L}{c} + \frac{1}{2}\hat{p}^i\hat{p}^j E^{\hat{r}\textsc{A}}_{ij} \sum^N_{n=1} \left[ \int\limits^{t_\mathrm{em}+(2n-1)\frac{L}{c}}_{t_\mathrm{em}+(2n-2)\frac{L}{c}} h_\textsc{A}\big(t+T,\vec{x}_{0,\uparrow}(t)\big) dt \right. \nonumber \\
        &\hspace{7cm}+  \left. \int\limits^{t_\mathrm{em}+2n\frac{L}{c}}_{t_\mathrm{em}+(2n-1)\frac{L}{c}} h_\textsc{A}\big(t+T,\vec{x}_{0,\downarrow}(t)\big) dt   \right],
    \label{eqn: tobs_prime light pulse - N roundtrips}
    \end{align}
    where just like before, in the second line we performed the substitution/coordinate-shift $\Tilde{t}=t-T$.  This shifts the limits of integration, and since the label of the coordinate of integration in an integral is arbitrary, we can let $\Tilde{t} \rightarrow t$.  The difference between equations~\ref{eqn: tobs light pulse - N roundtrips} and~\ref{eqn: tobs_prime light pulse - N roundtrips} is the observed clock period after $N$ round-trips between the satellites:
    \begin{align}
        T_\mathrm{obs} \equiv t^{'}_\mathrm{obs} - t_\mathrm{obs} &\approx T + \Delta T, \nonumber \\
        \mathrm{where} \quad \Delta T &= \frac{1}{2}\hat{p}^i\hat{p}^j E^{\hat{r}\textsc{A}}_{ij} \sum^N_{n=1} \left[ \int\limits^{t_\mathrm{em}+(2n-1)\frac{L}{c}}_{t_\mathrm{em}+(2n-2)\frac{L}{c}} \Big[h_\textsc{A}\big(t+T,\vec{x}_{0,\uparrow}(t)\big) - h_\textsc{A}\big(t,\vec{x}_{0,\uparrow}(t)\big)\Big] dt \right. \nonumber \\
        &\quad \quad \quad \quad \quad \quad + \left. \int\limits^{t_\mathrm{em}+2n\frac{L}{c}}_{t_\mathrm{em}+(2n-1)\frac{L}{c}} \Big[h_\textsc{A}\big(t+T,\vec{x}_{0,\downarrow}(t)\big) - h_\textsc{A}\big(t,\vec{x}_{0,\downarrow}(t)\big)\Big] dt   \right] ,\nonumber \\
        &\hspace{-1cm}= \frac{1}{2}\hat{p}^i\hat{p}^j E^{\hat{r}\textsc{A}}_{ij} \sum^N_{n=1} \left[ \int\limits^{t_\mathrm{em}+(2n-1)\frac{L}{c}}_{t_\mathrm{em}+(2n-2)\frac{L}{c}} \bigg[h_\textsc{A}\Big(t_{\mathrm{ret}}\big(t+T,\vec{x}_{0,\uparrow}(t)\big)\Big) - h_\textsc{A}\Big(t_{\mathrm{ret}}\big(t,\vec{x}_{0,\uparrow}(t)\big)\Big)\bigg] dt \right. \nonumber \\
        &\hspace{-1cm}\quad \quad \quad \quad \quad \quad + \left. \int\limits^{t_\mathrm{em}+2n\frac{L}{c}}_{t_\mathrm{em}+(2n-1)\frac{L}{c}} \bigg[h_\textsc{A}\Big(t_{\mathrm{ret}}\big(t+T,\vec{x}_{0,\downarrow}(t)\big)\Big) - h_\textsc{A}\Big(t_{\mathrm{ret}}\big(t,\vec{x}_{0,\downarrow}(t)\big)\Big)\bigg] dt   \right] .
    \label{eqn: observed pulsar period - N roundtrips}
    \end{align}
    Following the notation we introduced in equation~\ref{eqn: t^0_ret, w^0, Theta^0 notation definition}, we define:
    \begin{equation}
        \begin{cases}
            t^0_\mathrm{ret,\uparrow/\downarrow} &\equiv t_\mathrm{ret}\big(\vec{x}=\vec{x}_{0,\mathrm{\uparrow/\downarrow}}(t)\big) ,\\
            \omega^0_\mathrm{\uparrow/\downarrow} &\equiv \omega\left(t^0_\mathrm{ret,\uparrow/\downarrow}\right) ,\\
            \Theta^0_\mathrm{\uparrow/\downarrow} &\equiv \Theta\left(t^0_\mathrm{ret,\uparrow/\downarrow}\right) .
        \end{cases}
    \label{eqn: t^0_ret, w^0, Theta^0 UP/DOWN notation definition}
    \end{equation}

    As we did in equation~\ref{eqn: Delta T / T}, we can write the fractional change in the clock period with time due to the presence of a gravitational wave after $N$ round-trip passes in a number of useful ways:
    \begin{align}
      \frac{\Delta T}{T} &\approx \frac{1}{2}\hat{p}^i\hat{p}^j E^{\hat{r}\textsc{A}}_{ij} \sum^N_{n=1} \left[ \int\limits^{t_\mathrm{u,\uparrow}}_{t_\mathrm{l,\uparrow}} \frac{\partial h_{\textsc{A}}\left(t_\mathrm{ret}\left(t,\vec{x}\right)\right)}{\partial t}\Bigg\rvert_{\vec{x}=\vec{x}_{0,\uparrow}(t)} dt \ + \ \int\limits^{t_\mathrm{u,\downarrow}}_{t_\mathrm{l,\downarrow}} \frac{\partial h_{\textsc{A}}\left(t_\mathrm{ret}\left(t,\vec{x}\right)\right)}{\partial t}\Bigg\rvert_{\vec{x}=\vec{x}_{0,\downarrow}(t)} dt   \right], \nonumber \\
      &= \frac{1}{2}\hat{p}^i\hat{p}^j E^{\hat{r}\textsc{A}}_{ij} \sum^N_{n=1} \left[ \int\limits^{t_\mathrm{u,\uparrow}}_{t_\mathrm{l,\uparrow}} \frac{d h_{\textsc{A}}\left(t^0_\mathrm{ret,\uparrow}\right)}{d t^0_\mathrm{ret,\uparrow}} dt \ + \ \int\limits^{t_\mathrm{u,\downarrow}}_{t_\mathrm{l,\downarrow}} \frac{d h_{\textsc{A}}\left(t^0_\mathrm{ret,\downarrow}\right)}{d t^0_\mathrm{ret,\downarrow}} dt   \right], \nonumber \\
      &= \frac{1}{2}\hat{p}^i\hat{p}^j E^{\hat{r}\textsc{A}}_{ij} \sum^N_{n=1} \left[ \int\limits^{t_\mathrm{u,\uparrow}}_{t_\mathrm{l,\uparrow}}  \left[\frac{2}{3}h_\textsc{A}\left(t^0_\mathrm{ret,\uparrow}\right)\frac{\dot{\omega}\left(t^0_\mathrm{ret,\uparrow}\right)}{\omega^0_\uparrow} + 2h_{\textsc{A}}\left(\omega^0_\uparrow,\Theta^0_\uparrow+\frac{\pi}{4}\right)\omega^0_\uparrow\right]dt \right. \nonumber \\
      &\quad \quad \quad \quad \quad \quad + \left. \int\limits^{t_\mathrm{u,\downarrow}}_{t_\mathrm{l,\downarrow}} \left[\frac{2}{3}h_\textsc{A}\left(t^0_\mathrm{ret,\downarrow}\right)\frac{\dot{\omega}\left(t^0_\mathrm{ret,\downarrow}\right)}{\omega^0_\downarrow} + 2h_{\textsc{A}}\left(\omega^0_\downarrow,\Theta^0_\downarrow+\frac{\pi}{4}\right)\omega^0_\downarrow\right]dt \right] , \label{eqn: Delta T / T - N roundtrips} \\
      &\mathrm{where} \quad \begin{cases}
          \begin{tabular}{l l l}
              $t_\mathrm{u,\uparrow}$ &= $t_\mathrm{em}+(2n-1)\frac{L}{c}$ & $\approx t_\mathrm{obs}+(2n-1-2N)\frac{L}{c}$ , \\
              $t_\mathrm{l,\uparrow}$ &= $t_\mathrm{em}+(2n-2)\frac{L}{c}$ & $\approx t_\mathrm{obs}+(2n-2-2N)\frac{L}{c}$ , \\
              $t_\mathrm{u,\downarrow}$ &= $t_\mathrm{em}+2n\frac{L}{c}$ & $\approx t_\mathrm{obs}+(2n-2N)\frac{L}{c}$ , \\
              $t_\mathrm{l,\downarrow}$ &= $t_\mathrm{em}+(2n-1)\frac{L}{c}$ & $\approx t_\mathrm{obs}+(2n-1-2N)\frac{L}{c}$ . 
          \end{tabular}
      \end{cases} \nonumber
    \end{align}
    
    In equation~\ref{eqn: Delta T / T - N roundtrips} we need to evaluate the retarded time along the path of the photon as it travels the uplinks and downlinks between the satellites.  For a flat static universe, we found in Section~\ref{sec: Time Retardation} that:
    \begin{equation}
        t_\mathrm{ret}(t,x\hat{p}) = t  - \frac{R}{c}  +  \left(\hat{r}\cdot\hat{p}\right)\frac{x}{c}  -  \frac{1}{2}\left(1-\left(\hat{r}\cdot\hat{p}\right)^2\right)\frac{x}{c}\frac{x}{R}  +  \ldots ,
    \tag{\ref{eqn: tret along Earth-pulsar baseline} r}
    \end{equation}
    So explicitly evaluating the retarded time in equation~\ref{eqn: t^0_ret, w^0, Theta^0 UP/DOWN notation definition} along the photon path equation~\ref{eqn:zeroth order photon path - N roundtrips} gives us:
    \begin{align}
        t^0_\mathrm{ret,\uparrow} &= t - \frac{R}{c} + \left(\hat{r}\cdot\hat{p}\right)\frac{ct - ct_\mathrm{obs} - (2n-2-2N)L}{c} \nonumber \\
        &\hspace{2.5cm}- \frac{1}{2}\left(1-\left(\hat{r}\cdot\hat{p}\right)^2\right)\frac{\left(ct - ct_\mathrm{obs} - (2n-2-2N)L\right)^2}{cR} + \ldots, \label{eqn:tret 0 uplink} \\
        t^0_\mathrm{ret,\downarrow} &= t - \frac{R}{c} + \left(\hat{r}\cdot\hat{p}\right)\frac{ct_\mathrm{obs} + (2n-2N)L -ct}{c} \nonumber \\
        &\hspace{2.5cm}- \frac{1}{2}\left(1-\left(\hat{r}\cdot\hat{p}\right)^2\right)\frac{\left(ct_\mathrm{obs} + (2n-2N)L -ct\right)^2}{cR} + \ldots . \label{eqn:tret 0 downlink}
    \end{align}
    Now we can proceed to evaluate these integrals under the different formalisms we have discussed so far.

%---------------------------------------------------------------------------------
%---------------------------------------------------------------------------------
    \section{Plane-Wave Formalism (I)}\label{sec: Doppler PW Formalism}
    
    Keeping only terms in the plane-wave regime in equations~\ref{eqn:tret 0 uplink} and~\ref{eqn:tret 0 downlink} we have:
    \begin{align}
        t^0_\mathrm{ret,\uparrow} &\approx t - \frac{R}{c} + \left(\hat{r}\cdot\hat{p}\right)\frac{ct - ct_\mathrm{obs} - (2n-2-2N)L}{c} , \nonumber \\
        &= \left[ -\frac{R}{c} -\left(\hat{r}\cdot\hat{p}\right) \left(t_\mathrm{obs} + (2n-2-2N)\frac{L}{c}   \right)\right] + \left(1+\hat{r}\cdot\hat{p}\right)t , \\
        t^0_\mathrm{ret,\downarrow} &\approx t - \frac{R}{c} + \left(\hat{r}\cdot\hat{p}\right)\frac{ct_\mathrm{obs} + (2n-2N)L -ct}{c} , \nonumber \\
        &= \left[ -\frac{R}{c} +\left(\hat{r}\cdot\hat{p}\right) \left(t_\mathrm{obs} + (2n-2N)\frac{L}{c}   \right)\right] + \left(1-\hat{r}\cdot\hat{p}\right)t .
    \end{align}
    
    All of the math works out the same as it did in Section~\ref{sec: Plane-Wave Formalism} when solving the integral in equation~\ref{eqn: Delta T / T - N roundtrips}.  Once again, the crux of the plane-wave formalism is that the quantity $\frac{\partial t^0_\mathrm{ret,\uparrow/\downarrow}}{\partial t} = \left(1 \pm \hat{r}\cdot\hat{p}\right)$, which is time-independent and can therefore be directly pulled out of the integral.  This in turn means that the observed fractional change in the clock period doesn't directly depend on the functional form of the metric perturbation (e.g. whether it is monochromatic or frequency evolving), and only depends on the endpoints of the photon's motion (as it traverses the uplinks/downlinks).
    
    Notice here that $\frac{\partial t^0_\mathrm{ret,\uparrow/\downarrow}}{\partial t} = \left(1 \pm \hat{r}\cdot\hat{p}\right)$ will also result in two separate antenna patterns, one for the uplink and one for the downlink:
    \begin{align}
        \begin{cases}
            \begin{tabular}{l l l l}
                $f_{\uparrow/\downarrow}^{+}$ &$\equiv \frac{\hat{p}^i\hat{p}^j e^{\hat{r}+}_{ij}}{\left(1\pm\hat{r}\cdot\hat{p}\right)}$ &$= \frac{\hat{p}^i\hat{p}^j\hat{\theta}_i\hat{\theta}_j - \hat{p}^i\hat{p}^j\hat{\phi}_i\hat{\phi}_j}{\left(1\pm\hat{r}\cdot\hat{p}\right)}$ &$= \frac{\left(\hat{p}\cdot\hat{\theta}\right)^2 - \left(\hat{p}\cdot\hat{\phi}\right)^2}{\left(1\pm\hat{r}\cdot\hat{p}\right)}$ , \\
                $f_{\uparrow/\downarrow}^{\times}$ &$\equiv \frac{\hat{p}^i\hat{p}^j e^{\hat{r}\times}_{ij}}{\left(1\pm\hat{r}\cdot\hat{p}\right)}$ &$= \frac{\hat{p}^i\hat{p}^j\hat{\phi}_i\hat{\theta}_j + \hat{p}^i\hat{p}^j\hat{\theta}_i\hat{\phi}_j}{\left(1\pm\hat{r}\cdot\hat{p}\right)}$ &$= \frac{2\left(\hat{p}\cdot\hat{\theta}\right)\left(\hat{p}\cdot\hat{\phi}\right)}{\left(1\pm\hat{r}\cdot\hat{p}\right)}$ ,
            \end{tabular}
        \end{cases} \label{eqn: antenna f up/down}
    \end{align}
    \begin{align}
        \begin{cases}
            \begin{tabular}{l l l}
                $F_{\uparrow/\downarrow}^+$ &$\equiv \frac{\hat{p}^i\hat{p}^j E^{\hat{r}+}_{ij}}{\left(1\pm\hat{r}\cdot\hat{p}\right)}$ &$= \frac{1}{2}\left(1+\cos^2(\iota)\right) \left[ \cos(2\psi) f_{\uparrow/\downarrow}^+ + \sin(2\psi) f_{\uparrow/\downarrow}^\times \right]$ , \\
                $F_{\uparrow/\downarrow}^\times$ &$\equiv \frac{\hat{p}^i\hat{p}^j E^{\hat{r}\times}_{ij}}{\left(1\pm\hat{r}\cdot\hat{p}\right)}$ &$= \cos(\iota) \left[ -\sin(2\psi) f_{\uparrow/\downarrow}^+ + \cos(2\psi) f_{\uparrow/\downarrow}^\times \right]$ ,
            \end{tabular}
        \end{cases} \label{eqn: antenna F up/down}
    \end{align}
    where the $(+)$ sign corresponts to the uplink $(\uparrow)$ and the $(-)$ sign corresponds to the downlink $(\downarrow)$.  The downlink is the antenna pattern that we discussed before in Section~\ref{subsec:The Antenna Response}.  The uplink antenna pattern is the ``mirror'' of the downlink pattern, in the sense that the strong and weak parts of the antenna are reversed.  So considering as an example Figure~\ref{fig: antenna pattern example}, the respective uplink antenna pattern of that has the red and blue patterns in the same locations, but the strong and weak parts of that antenna pattern (corresponding to opposite parts of the sky) are flipped.
    
    Following the same steps as we did in Section~\ref{sec: Plane-Wave Formalism} we can write the expression for the fractional change in the clock's period as:
    \begin{align}
        \frac{\Delta T}{T}(t_\mathrm{obs}) &= \frac{1}{2} \sum^N_{n=1} \Bigg[ F_{\uparrow}^\textsc{A}\bigg( h_\textsc{A}\Big(\omega^0_P,\Theta^0_P\Big) -  h_\textsc{A}\Big(\omega^0_{E\mathrm{,\uparrow}},\Theta^0_{E\mathrm{,\uparrow}}\Big) \bigg) \nonumber \\
        &\hspace{2cm}+  F_{\downarrow}^\textsc{A}\bigg(h_\textsc{A}\Big(\omega^0_{E\mathrm{,\downarrow}},\Theta^0_{E\mathrm{,\downarrow}}\Big) -  h_\textsc{A}\Big(\omega^0_P,\Theta^0_P\Big)\bigg) \Bigg] , \label{eqn: Delta T / T pw - N roundtrips} \\
        &\hspace{-1cm}= \frac{1}{2}\hat{p}^i\hat{p}^j E^{\hat{r}\textsc{A}}_{ij} \sum^N_{n=1} \left[ \frac{h_\textsc{A}\Big(\omega^0_P,\Theta^0_P\Big) -  h_\textsc{A}\Big(\omega^0_{E\mathrm{,\uparrow}},\Theta^0_{E\mathrm{,\uparrow}}\Big)}{\left(1+\hat{r}\cdot\hat{p}\right)}  +  \frac{h_\textsc{A}\Big(\omega^0_{E\mathrm{,\downarrow}},\Theta^0_{E\mathrm{,\downarrow}}\Big) -  h_\textsc{A}\Big(\omega^0_P,\Theta^0_P\Big)}{\left(1-\hat{r}\cdot\hat{p}\right)} \right] , \nonumber \\
        &\hspace{-1cm}= \frac{1}{2} \hat{p}^i\hat{p}^j E^{\hat{r}\textsc{A}}_{ij} \sum^N_{n=1} \left[ -\frac{h_\textsc{A}\Big(\omega^0_{E\mathrm{,\uparrow}},\Theta^0_{E\mathrm{,\uparrow}}\Big)}{\left(1+\hat{r}\cdot\hat{p}\right)}  -2  \frac{\left(\hat{r}\cdot\hat{p}\right)h_\textsc{A}\Big(\omega^0_P,\Theta^0_P\Big)}{\left(1-\left(\hat{r}\cdot\hat{p}\right)^2\right)} + \frac{h_\textsc{A}\Big(\omega^0_{E\mathrm{,\downarrow}},\Theta^0_{E\mathrm{,\downarrow}}\Big)}{\left(1-\hat{r}\cdot\hat{p}\right)} \right] , \nonumber \\
        &\mathrm{where}\quad\begin{cases}
            \omega^0_{E\mathrm{,\uparrow}} &\equiv \omega\Big(t_\mathrm{obs}-\frac{R}{c} - (2N-2n+2)\frac{L}{c}\Big) , \\
            \omega^0_{E\mathrm{,\downarrow}} &\equiv \omega\Big(t_\mathrm{obs}-\frac{R}{c} - (2N-2n)\frac{L}{c}\Big) , \\
            \omega^0_{P} &\equiv \omega\Big(t_\mathrm{obs}-\frac{R}{c}-\left(2N-2n+1 - \hat{r}\cdot\hat{p}\right)\frac{L}{c}\Big) , \\
            \Theta^0_{E\mathrm{,\uparrow}} &\equiv \Theta\Big(t_\mathrm{obs}-\frac{R}{c} - (2N-2n+2)\frac{L}{c}\Big) , \\
            \Theta^0_{E\mathrm{,\downarrow}} &\equiv \Theta\Big(t_\mathrm{obs}-\frac{R}{c} - (2N-2n)\frac{L}{c}\Big) , \\
            \Theta^0_{P} &\equiv \Theta\Big(t_\mathrm{obs}-\frac{R}{c}-\left(2N-2n+1 - \hat{r}\cdot\hat{p}\right)\frac{L}{c}\Big) .
        \end{cases}\label{eqn: Delta T / T pw w^0_EP and Theta^0_EP - N roundtrips}
    \end{align}
    To get the timing residual we now integrate equation~\ref{eqn: Delta T / T pw - N roundtrips} over our observation time:
    \begin{equation}
        \mathrm{Res}(t) = \mathlarger{\int}\frac{\Delta T}{T}\left(t_\mathrm{obs}\right) dt_\mathrm{obs} .
    \tag{\ref{eqn:timing residual} r}
    \end{equation}

%---------------------------------------------------------------------------------
%---------------------------------------------------------------------------------
        \subsection{Plane-Wave, Monochromatic (IA)}\label{subsec: Doppler Monochromatic Model (PW)}
        
        For a monochromatic gravitational wave the frequency remains constant, so $\omega^0_{E\mathrm{,\uparrow/\downarrow}} = \omega^0_{P} = \omega_0$, and integration of equation~\ref{eqn: Delta T / T pw - N roundtrips} proceeds just as in Section~\ref{subsec: Monochromatic Model (PW)}.  The final result can be expressed compactly as:
        \begin{align}
            \mathrm{Res}(t) &= \frac{1}{4\omega_0} \sum^N_{n=1} \Bigg[ F_{\uparrow}^\textsc{A}\bigg(h_\textsc{A}\Big(\Theta_P-\frac{\pi}{4}\Big) -  h_\textsc{A}\Big(\Theta_{E\mathrm{,\uparrow}}-\frac{\pi}{4}\Big)\bigg)  \nonumber \\
            &\hspace{2cm}+  F_{\downarrow}^\textsc{A}\bigg(h_\textsc{A}\Big(\Theta_{E\mathrm{,\downarrow}}-\frac{\pi}{4}\Big) -  h_\textsc{A}\Big(\Theta_P-\frac{\pi}{4}\Big)\bigg) \Bigg] , \label{eqn: Res(t) pw mono - N roundtrips} \\
            &\mathrm{for}\quad \textsc{A} \in [+, \times], \nonumber \\
        \nonumber \\
            &\underset{\left(t_0 \ = \ -\frac{R}{c}\right)}{\mathrm{where}}\quad\quad \begin{cases}
                \Theta_{E\mathrm{,\uparrow}} &\equiv \Theta\Big(t-\frac{R}{c}-(2N-2n+2)\frac{L}{c}\Big) \\
                &\equiv \theta_0 + \omega_0\Big(t-(2N-2n+2)\frac{L}{c}\Big) , \\
                \Theta_{E\mathrm{,\downarrow}} &\equiv \Theta\Big(t-\frac{R}{c}-(2N-2n)\frac{L}{c}\Big) \\
                &\equiv \theta_0 + \omega_0\Big(t-(2N-2n)\frac{L}{c}\Big) , \\
                \Theta_{P} &\equiv \Theta\Big(t-\frac{R}{c}-\left(2N-2n+1-\hat{r}\cdot\hat{p}\right)\frac{L}{c}\Big) \\
                &\equiv \theta_0 + \omega_0\Big(t-\left(2N-2n+1-\hat{r}\cdot\hat{p}\right)\frac{L}{c}\Big) ,
                \end{cases}\label{eqn: Res(t) pw mono phase E and P - N roundtrips} \\
            &\hspace{2cm}\begin{cases}
                h_{+} \equiv -h_0 \cos\big(2\Theta\big), \\
                h_{\times} \equiv -h_0 \sin\big(2\Theta\big),
            \end{cases} \tag{\ref{eqn: h0 plus and cross} r} \\
            &\hspace{2cm}h_0 \equiv \frac{4(G\mathcal{M})^{5/3}}{c^4 R}\omega_0^{2/3} , \tag{\ref{eqn: amplitude 0 of metric perturbation} r}\\
            &\hspace{2cm}\begin{cases}
                \begin{tabular}{l l l}
                    $f_{\uparrow/\downarrow}^{+}$ &$\equiv \frac{\hat{p}^i\hat{p}^j e^{\hat{r}+}_{ij}}{\left(1\pm\hat{r}\cdot\hat{p}\right)}$ &$= \frac{\left(\hat{p}\cdot\hat{\theta}\right)^2 - \left(\hat{p}\cdot\hat{\phi}\right)^2}{\left(1\pm\hat{r}\cdot\hat{p}\right)}$ , \\
                    $f_{\uparrow/\downarrow}^{\times}$ &$\equiv \frac{\hat{p}^i\hat{p}^j e^{\hat{r}\times}_{ij}}{\left(1\pm\hat{r}\cdot\hat{p}\right)}$ &$= \frac{2\left(\hat{p}\cdot\hat{\theta}\right)\left(\hat{p}\cdot\hat{\phi}\right)}{\left(1\pm\hat{r}\cdot\hat{p}\right)}$ ,
                \end{tabular}
            \end{cases} \tag{\ref{eqn: antenna f up/down} r} \\
            &\hspace{2cm}\begin{cases}
                \begin{tabular}{l l l}
                    $F_{\uparrow/\downarrow}^+$ &$\equiv \frac{\hat{p}^i\hat{p}^j E^{\hat{r}+}_{ij}}{\left(1\pm\hat{r}\cdot\hat{p}\right)}$ &$= \frac{1}{2}\left(1+\cos^2(\iota)\right) \left[ \cos(2\psi) f_{\uparrow/\downarrow}^+ + \sin(2\psi) f_{\uparrow/\downarrow}^\times \right]$ , \\
                    $F_{\uparrow/\downarrow}^\times$ &$\equiv \frac{\hat{p}^i\hat{p}^j E^{\hat{r}\times}_{ij}}{\left(1\pm\hat{r}\cdot\hat{p}\right)}$ &$= \cos(\iota) \left[ -\sin(2\psi) f_{\uparrow/\downarrow}^+ + \cos(2\psi) f_{\uparrow/\downarrow}^\times \right]$ ,
                \end{tabular}
            \end{cases} \tag{\ref{eqn: antenna F up/down} r} \\
            &\hspace{2cm}\begin{cases}
                \begin{tabular}{r l l r}
                    $\hat{r}$ &$= \big[\sin(\theta)\cos(\phi),$ &$\sin(\theta)\sin(\phi),$ &$\cos(\theta) \big]$, \\
                    $\hat{\theta}$ &$= \big[\cos(\theta)\cos(\phi),$ &$\cos(\theta)\sin(\phi),$ &$\sin(\theta) \big]$, \\
                    $\hat{\phi}$ &$= \big[-\sin(\phi),$ &$\cos(\phi),$ &$0 \big]$, \\
                    $\hat{p}$ &$= \big[\sin(\theta_p)\cos(\phi_p),$ &$\sin(\theta_p)\sin(\phi_p),$ &$\cos(\theta_p) \big]$.
                \end{tabular}
                \end{cases} \tag{\ref{eqn:source basis vectors}, \ref{eqn: p hat} r} 
        \end{align}

%---------------------------------------------------------------------------------
%---------------------------------------------------------------------------------
        \subsection{Plane-Wave, Frequency Evolution (IB)}\label{subsec: Doppler Frequency Evolution (PW)}
        
        In the frequency evolution model the most important assumption that we continue to make is that the frequency of the gravitational waves are not changing appreciably on the observation time scale.  They do however evolve for every pass that the photon makes along the clock baseline.  Again we choose our fiducial time to be $t_0 =-\frac{R}{c}$.  Once more performing a Taylor expansion of equation~\ref{eqn: Delta T / T pw w^0_EP and Theta^0_EP - N roundtrips} in the small parameter $\frac{t_\mathrm{obs}}{\Delta \tau_c}$ gives us:
        \begin{equation}
            \underset{t_0 \ = \ -\frac{R}{c}}{\mathrm{choosing}}\quad\begin{cases}
                \begin{tabular}{ l l l }
                    $\omega^0_{E\mathrm{,\uparrow}}$ & $\approx \omega_0\left[1 + \frac{\left(2N-2n+2\right)\frac{L}{c}}{\Delta\tau_c}\right]^{-3/8}$ & $\equiv \omega_{0E\mathrm{,\uparrow}}$ , \\
                    $\omega^0_{E\mathrm{,\downarrow}}$ & $\approx \omega_0\left[1 + \frac{\left(2N-2n\right)\frac{L}{c}}{\Delta\tau_c}\right]^{-3/8}$ & $\equiv \omega_{0E\mathrm{,\downarrow}}$ , \\
                    $\omega^0_{P}$ & $\approx \omega_0\left[1 + \frac{\left(2N-2n+1-\hat{r}\cdot\hat{p}\right)\frac{L}{c}}{\Delta\tau_c}\right]^{-3/8}$ & $\equiv \omega_{0P}$ , \\[10pt]
                    $\Theta^0_{E\mathrm{,\uparrow}}$ & $\approx \left[\theta_0 + \theta_c\left(1-\left[1+ \frac{\left(2N-2n+2\right)\frac{L}{c}}{\Delta\tau_c}\right]^{5/8}\right)\right] + \omega_{0E\mathrm{,\uparrow}}t_\mathrm{obs}$ & \\
                    & $\equiv \theta_{0E\mathrm{,\uparrow}} + \omega_{0E\mathrm{,\uparrow}}t_\mathrm{obs}$ , \\
                    $\Theta^0_{E\mathrm{,\downarrow}}$ & $\approx \left[\theta_0 + \theta_c\left(1-\left[1+ \frac{\left(2N-2n\right)\frac{L}{c}}{\Delta\tau_c}\right]^{5/8}\right)\right] + \omega_{0E\mathrm{,\downarrow}}t_\mathrm{obs}$ & \\
                    & $\equiv \theta_{0E\mathrm{,\downarrow}} + \omega_{0E\mathrm{,\downarrow}}t_\mathrm{obs}$ , \\
                    $\Theta^0_{P}$ & $\approx \left[\theta_0 + \theta_c\left(1-\left[1+ \frac{\left(2N-2n+1-\hat{r}\cdot\hat{p}\right)\frac{L}{c}}{\Delta\tau_c}\right]^{5/8}\right)\right] + \omega_{0P}t_\mathrm{obs}$ & \\
                    & $\equiv \theta_{0P} + \omega_{0P}t_\mathrm{obs}$ ,
                \end{tabular}
            \end{cases} \label{eqn: Delta T / T pw w^0_EP and Theta^0_EP freq evo approx t0=-R/c - N roundtrips}
        \end{equation}
        
        Here we point out and emphasize and intriguing conceptual uniqueness that this Doppler tracking experiment has over our pulsar timing experiment.  In pulsar timing, the Earth-pulsar baseline distance is on the order of kiloparsecs, meaning light travel time for the photons are on the order of thousands of years.  The assumption with PTAs is that the gravitational wave frequency would not evolve over the observation time scales, but could potentially evolve over the Earth-pulsar ``baseline'' timescale, $\frac{L}{c}$.  So in PTA experiments, $t_\mathrm{obs} < \frac{L}{c}$ naturally.
        
        But with Doppler tracking experiments using human-made satellites, this baseline distance is much shorter, and we can easily imagine the observation time scale of our experiment lasting longer than this baseline timescale.  Thus potentially with Doppler tracking experiments, $\frac{L}{c} < t_\mathrm{obs}$.  While at first it may seem that this would fundamentally require us to evaluate this new problem differently than we did for PTAs, nothing actually changes mathematically (just conceptually).  The above Taylor approximations only required that $t_\mathrm{obs} \ll \tau_c$.  No requirements were made about the time scale of $\frac{L}{c}$ compared to $t_\mathrm{obs}$ or $\tau_c$.  But we can see than in the limit that $\frac{L}{c}\rightarrow 0$ we will just recover the monochromatic results as we would expect.
        
        We must now point out that in a Doppler tracking experiment, if $\frac{L}{c} < t_\mathrm{obs}$, then because $t_\mathrm{obs} \ll \tau_c$, necessarily $\frac{L}{c} \ll \tau_c$.  This case implies that every individual pass of the clock signal will experience an effectively monochrome gravitational wave, because with one pass we are also assuming now that not enough time is passing for the gravitational wave to evolve appreciably.  This also suggests that we may need a huge number of $N$ passes of the signal before we might start to suss out the frequency evolution effect within the timing residual.
        
        One final point of clarification to keep in mind in our discussion of Doppler tracking.  While it is possible that the observation time $t_\mathrm{obs} > \frac{L}{c}$, that does not also imply the observation time is greater than the total amount of time the signal is sent back and forth between our satellites, that is $t_\mathrm{obs} \ngtr N \frac{L}{c}$.  In fact, as we have just said, we will likely need to allow the clock signal to bounce back and forth between our satellites for a very long time, \textit{before} we then actually begin to ``observe'' the data.
        
        As a simple conceptual example, imagine two satellites separated by a distance of 1/2-light day, i.e. $\frac{L}{c} = 12$ hours (this would be a distance of about 86.6 A.U., or roughly 0.6 times the present distance to Voyager 1).  Let's say we generate a series of timestamps that span a single day, and bounce them continuously between our satellites for several years (such that $N\frac{L}{c} \sim \mathcal{O}\left(\mathrm{years}\right)$).  After that period, our actual ``observation'' time would only last a day.  Thus $\frac{L}{c} < t_\mathrm{obs} < N \frac{L}{c}$, but hopefully given the source and its rate of orbital evolution, over the timescale $N \frac{L}{c}$ the gravitational wave frequency would have had enough time to evolve appreciably within the signal that is observed over $t_\mathrm{obs}$.
        
        So in summary, if here in our Doppler tracking experiment $t_\mathrm{obs} < \frac{L}{c}$, then conceptually our time scales are the same as with the PTA time scales.  If $\frac{L}{c} < t_\mathrm{obs}$, that is still fine mathematically, but due to our assumption that $t_\mathrm{obs} \ll \tau_c$, we will likely need a very large number $N$ of passes, such that $N \frac{L}{c} \gg t_\mathrm{obs}$ in order to allow any frequency evolution to become measurable.

        Now returning to our approximations in equations~\ref{eqn: Delta T / T pw w^0_EP and Theta^0_EP freq evo approx t0=-R/c - N roundtrips}, the orbital frequencies are once again just constants within every integral under the assumption that $\frac{t_\mathrm{obs}}{\Delta \tau_c} \ll 1$, so we can continue with the integration of equation~\ref{eqn: Delta T / T pw - N roundtrips} just as we did in Section~\ref{subsec: Frequency Evolution (PW)}.  The final result is:
        
        \begin{align}
            \mathrm{Res}(t) &= \frac{1}{4} \sum^N_{n=1} \left[ F_{\uparrow}^\textsc{A} \left(\frac{h_\textsc{A}\Big(\omega_{0P},\Theta_P-\frac{\pi}{4}\Big)}{\omega_{0P}} -  \frac{h_\textsc{A}\Big(\omega_{0E\mathrm{,\uparrow}},\Theta_{E\mathrm{,\uparrow}}-\frac{\pi}{4}\Big)}{\omega_{0E\mathrm{,\uparrow}}}\right) \right. \nonumber \\
            &\hspace{1.85cm}\left. + F_{\downarrow}^\textsc{A} \left(\frac{h_\textsc{A}\Big(\omega_{0E\mathrm{,\downarrow}},\Theta_{E\mathrm{,\downarrow}}-\frac{\pi}{4}\Big)}{\omega_{0E\mathrm{,\downarrow}}} -  \frac{h_\textsc{A}\Big(\omega_{0P},\Theta_P-\frac{\pi}{4}\Big)}{\omega_{0P}}\right) \right] , \label{eqn: Res(t) pw freq evo - N roundtrips} \\
            &\mathrm{for}\quad \textsc{A} \in [+, \times], \nonumber \\
            \nonumber \\
            &\underset{\left(t_0 \ = \ -\frac{R}{c}\right)}{\mathrm{where}}\quad\begin{cases}
                \begin{tabular}{ l l }
                    $\omega_{0E\mathrm{,\uparrow}}$ & $\equiv \omega_0\left[1 + \frac{\left(2N-2n+2\right)\frac{L}{c}}{\Delta\tau_c}\right]^{-3/8}$ , \\
                    $\omega_{0E\mathrm{,\downarrow}}$ & $\equiv \omega_0\left[1 + \frac{\left(2N-2n\right)\frac{L}{c}}{\Delta\tau_c}\right]^{-3/8}$ , \\
                    $\omega_{0P}$ & $\equiv \omega_0\left[1 + \frac{\left(2N-2n+1-\hat{r}\cdot\hat{p}\right)\frac{L}{c}}{\Delta\tau_c}\right]^{-3/8}$ , \\
                    $\theta_{0E\mathrm{,\uparrow}}$ & $\equiv \theta_0 + \theta_c\left(1-\left[1+ \frac{\left(2N-2n+2\right)\frac{L}{c}}{\Delta\tau_c}\right]^{5/8}\right)$ , \\
                    $\theta_{0E\mathrm{,\downarrow}}$ & $\equiv \theta_0 + \theta_c\left(1-\left[1+ \frac{\left(2N-2n\right)\frac{L}{c}}{\Delta\tau_c}\right]^{5/8}\right)$ , \\
                    $\theta_{0P}$ & $\equiv \theta_0 + \theta_c\left(1-\left[1+ \frac{\left(2N-2n+1-\hat{r}\cdot\hat{p}\right)\frac{L}{c}}{\Delta\tau_c}\right]^{5/8}\right)$ , \\
                    $\Theta_{E\mathrm{,\uparrow}}$ & $\equiv \theta_{0E\mathrm{,\uparrow}} + \omega_{0E\mathrm{,\uparrow}}t_\mathrm{obs}$ , \\
                    $\Theta_{E\mathrm{,\downarrow}}$ & $\equiv \theta_{0E\mathrm{,\downarrow}} + \omega_{0E\mathrm{,\downarrow}}t_\mathrm{obs}$ , \\
                    $\Theta_{P}$ & $\equiv \theta_{0P} + \omega_{0P}t_\mathrm{obs}$ ,
                \end{tabular}
            \end{cases} \\
            &\hspace{1.6cm}\begin{cases}
                \Delta\tau_c \equiv \frac{5}{256}\left(\frac{c^3}{G\mathcal{M}}\right)^{5/3} \frac{1}{\omega_0^{8/3}} , \\
            \theta_c \equiv \frac{8}{5}\Delta\tau_c \omega_0 ,
            \end{cases} \tag{\ref{eqn: time to coalescence}, \ref{eqn: thetac} r} \\
            &\hspace{1.6cm}\begin{cases}
                h_{+} \equiv -h(\omega) \cos\big(2\Theta\big), \\
                h_{\times} \equiv -h(\omega) \sin\big(2\Theta\big),
            \end{cases} \tag{\ref{eqn: h plus and cross} r} \\
            &\hspace{1.6cm}h(\omega) \equiv \frac{4(G\mathcal{M})^{5/3}}{c^4 R}\omega^{2/3} , \tag{\ref{eqn: amplitude of metric perturbation} r}\\
            &\hspace{1.6cm}\begin{cases}
                \begin{tabular}{l l l}
                    $f_{\uparrow/\downarrow}^{+}$ &$\equiv \frac{\hat{p}^i\hat{p}^j e^{\hat{r}+}_{ij}}{\left(1\pm\hat{r}\cdot\hat{p}\right)}$ &$= \frac{\left(\hat{p}\cdot\hat{\theta}\right)^2 - \left(\hat{p}\cdot\hat{\phi}\right)^2}{\left(1\pm\hat{r}\cdot\hat{p}\right)}$ , \\
                    $f_{\uparrow/\downarrow}^{\times}$ &$\equiv \frac{\hat{p}^i\hat{p}^j e^{\hat{r}\times}_{ij}}{\left(1\pm\hat{r}\cdot\hat{p}\right)}$ &$= \frac{2\left(\hat{p}\cdot\hat{\theta}\right)\left(\hat{p}\cdot\hat{\phi}\right)}{\left(1\pm\hat{r}\cdot\hat{p}\right)}$ ,
                \end{tabular}
            \end{cases} \tag{\ref{eqn: antenna f up/down} r} \\
            &\hspace{1.6cm}\begin{cases}
                \begin{tabular}{l l l}
                    $F_{\uparrow/\downarrow}^+$ &$\equiv \frac{\hat{p}^i\hat{p}^j E^{\hat{r}+}_{ij}}{\left(1\pm\hat{r}\cdot\hat{p}\right)}$ &$= \frac{1}{2}\left(1+\cos^2(\iota)\right) \left[ \cos(2\psi) f_{\uparrow/\downarrow}^+ + \sin(2\psi) f_{\uparrow/\downarrow}^\times \right]$ , \\
                    $F_{\uparrow/\downarrow}^\times$ &$\equiv \frac{\hat{p}^i\hat{p}^j E^{\hat{r}\times}_{ij}}{\left(1\pm\hat{r}\cdot\hat{p}\right)}$ &$= \cos(\iota) \left[ -\sin(2\psi) f_{\uparrow/\downarrow}^+ + \cos(2\psi) f_{\uparrow/\downarrow}^\times \right]$ ,
                \end{tabular}
            \end{cases} \tag{\ref{eqn: antenna F up/down} r} \\
            &\hspace{1.6cm}\begin{cases}
                \begin{tabular}{r l l r}
                    $\hat{r}$ &$= \big[\sin(\theta)\cos(\phi),$ &$\sin(\theta)\sin(\phi),$ &$\cos(\theta) \big]$, \\
                    $\hat{\theta}$ &$= \big[\cos(\theta)\cos(\phi),$ &$\cos(\theta)\sin(\phi),$ &$\sin(\theta) \big]$, \\
                    $\hat{\phi}$ &$= \big[-\sin(\phi),$ &$\cos(\phi),$ &$0 \big]$, \\
                    $\hat{p}$ &$= \big[\sin(\theta_p)\cos(\phi_p),$ &$\sin(\theta_p)\sin(\phi_p),$ &$\cos(\theta_p) \big]$.
                \end{tabular}
                \end{cases} \tag{\ref{eqn:source basis vectors}, \ref{eqn: p hat} r} 
        \end{align}

%---------------------------------------------------------------------------------
%---------------------------------------------------------------------------------
    \section{Fresnel Formalism (II)}\label{sec: Doppler Fresnel Formalism}
    Note that the same assumptions in Section~\ref{sec: Fresnel Formalism} still apply here.  Keeping out to the Fresnel regime in equations~\ref{eqn:tret 0 uplink} and~\ref{eqn:tret 0 downlink} we can write (like we did in equation~\ref{eqn:tret 0 fr}):
    \begin{align}
        t^0_\mathrm{ret,\uparrow} &\approx \left(t_\mathrm{obs}-\frac{R}{c}+(2n-2-2N)\frac{L}{c}\right) + \frac{R}{c}\left(1+\hat{r}\cdot\hat{p}\right)u_\uparrow - \frac{1}{2}
        \frac{R}{c}\left(1-\left(\hat{r}\cdot\hat{p}\right)^2\right)u^2_\uparrow , \\
        t^0_\mathrm{ret,\downarrow} &\approx \left(t_\mathrm{obs}-\frac{R}{c}+(2n-2N)\frac{L}{c}\right) - \frac{R}{c}\left(1-\hat{r}\cdot\hat{p}\right)u_\downarrow - \frac{1}{2}
        \frac{R}{c}\left(1-\left(\hat{r}\cdot\hat{p}\right)^2\right)u^2_\downarrow , \\
        &\mathrm{where}\quad \begin{cases}
        u_\uparrow &\equiv \frac{ct-ct_\mathrm{obs}-(2n-2-2N)L}{R} , \\
        u_\downarrow &\equiv \frac{ct_\mathrm{obs}+(2n-2N)L-ct}{R} .
        \end{cases}
    \end{align}
    We can use these to perform the appropriate change of variables to $u_{\uparrow/\downarrow}$ as we did in Section~\ref{sec: Fresnel Formalism} to write:
    \begin{align}
          \frac{\Delta T}{T} &\approx \frac{1}{2}\hat{p}^i\hat{p}^j E^{\hat{r}\textsc{A}}_{ij}\sum^N_{n=1} \left[ \int\limits^{L/R}_{0} \left[\frac{2}{3}h_\textsc{A}\left(t^0_\mathrm{ret,\uparrow}\right)\frac{\dot{\omega}\left(t^0_\mathrm{ret,\uparrow}\right)}{\omega^0_\uparrow} + 2h_{\textsc{A}}\left(\omega^0_\uparrow,\Theta^0_\uparrow+\frac{\pi}{4}\right)\omega^0_\uparrow\right] \frac{R}{c}du_\uparrow \right. \nonumber \\
          &\hspace{1.6cm} \left. + \int\limits^{0}_{L/R} \left[\frac{2}{3}h_\textsc{A}\left(t^0_\mathrm{ret,\downarrow}\right)\frac{\dot{\omega}\left(t^0_\mathrm{ret,\downarrow}\right)}{\omega^0_\downarrow} + 2h_{\textsc{A}}\left(\omega^0_\downarrow,\Theta^0_\downarrow+\frac{\pi}{4}\right)\omega^0_\downarrow\right] \frac{-R}{c}du_\downarrow \right], \nonumber \\
          &= \frac{1}{2}\frac{R}{c}\hat{p}^i\hat{p}^j E^{\hat{r}\textsc{A}}_{ij}\sum^N_{n=1} \left[ \int\limits^{L/R}_{0} \left[\frac{2}{3}h_\textsc{A}\left(t^0_\mathrm{ret,\uparrow}\right)\frac{\dot{\omega}\left(t^0_\mathrm{ret,\uparrow}\right)}{\omega^0_\uparrow} + 2h_{\textsc{A}}\left(\omega^0_\uparrow,\Theta^0_\uparrow+\frac{\pi}{4}\right)\omega^0_\uparrow\right] du_\uparrow \right. \nonumber \\
          &\hspace{1.85cm}\left. + \int\limits^{L/R}_{0} \left[\frac{2}{3}h_\textsc{A}\left(t^0_\mathrm{ret,\downarrow}\right)\frac{\dot{\omega}\left(t^0_\mathrm{ret,\downarrow}\right)}{\omega^0_\downarrow} + 2h_{\textsc{A}}\left(\omega^0_\downarrow,\Theta^0_\downarrow+\frac{\pi}{4}\right)\omega^0_\downarrow\right] du_\downarrow \right] .
    \label{eqn: solving fr Delta T / T - N roundtrips}
    \end{align}

%---------------------------------------------------------------------------------
%---------------------------------------------------------------------------------
        \subsection{Fresnel, Monochromatic (IIA)}
        For a monochromatic gravitational wave, $\omega\left(t^0_\mathrm{ret,\uparrow/\downarrow}\right) = \omega_0$, $\dot{\omega} \equiv 0$, $h(t) = h_0$, and integration of equation~\ref{eqn: solving fr Delta T / T - N roundtrips} proceeds just as in Section~\ref{subsec: Monochromatic Model (Fresnel)}.  We begin by taking the same partial derivatives as we did in equation~\ref{eqn:partial h partial Theta at x0 in u} for both the uplinks and downlinks:
        \begin{align}
            h_\textsc{A}\left(\Theta^0_{\uparrow/\downarrow} + \frac{\pi}{4}\right) = &\left\{\begin{tabular}{r l r}
                $2 h \sin\left(2\Theta\left(t^0_\mathrm{ret,\uparrow}(u)\right)\right)$ & $= 2 h \sin\left(\textsf{A}_\uparrow + \textsf{B}_\uparrow u - \textsf{C}u^2\right) ,$ & $(+)$ \\
                $-2 h \cos\left(2\Theta\left(t^0_\mathrm{ret,\uparrow}(u)\right)\right)$ & $= -2 h \cos\left(\textsf{A}_\uparrow + \textsf{B}_\uparrow u - \textsf{C}u^2\right) ,$ &  $(\times)$ \\
                & & \\
                $2 h \sin\left(2\Theta\left(t^0_\mathrm{ret,\downarrow}(u)\right)\right)$ & $= 2 h \sin\left(\textsf{A}_\downarrow - \textsf{B}_\downarrow u - \textsf{C}u^2\right) ,$ & $(+)$ \\
                $-2 h \cos\left(2\Theta\left(t^0_\mathrm{ret,\downarrow}(u)\right)\right)$ & $= -2 h \cos\left(\textsf{A}_\downarrow - \textsf{B}_\downarrow u - \textsf{C}u^2\right) ,$ &  $(\times)$ 
            \end{tabular}\right. \label{eqn:partial h partial Theta at x0 in u - N roundtrips}\\
            \nonumber \\
            &\hspace{-2cm}\underset{\left(t_0 \ = \ -\frac{R}{c}\right)}{\mathrm{where}}\quad\quad \begin{cases}
            \textsf{A}_\uparrow &\equiv 2\theta_0 + 2\omega_0\left(t_\mathrm{obs} -(2N-2n+2)\frac{L}{c}\right) , \\
            \textsf{A}_\downarrow &\equiv 2\theta_0 + 2\omega_0\left(t_\mathrm{obs} -(2N-2n)\frac{L}{c}\right) , \\
            \textsf{B}_\uparrow &\equiv 2\omega_0\frac{R}{c}\left(1+\hat{r}\cdot\hat{p}\right) = \frac{2\textsf{C}}{\left(1-\hat{r}\cdot\hat{p}\right)} , \\
            \textsf{B}_\downarrow &\equiv 2\omega_0\frac{R}{c}\left(1-\hat{r}\cdot\hat{p}\right) = \frac{2\textsf{C}}{\left(1+\hat{r}\cdot\hat{p}\right)} , \\
            \textsf{C} &\equiv \omega_0\frac{R}{c}\left(1-\left(\hat{r}\cdot\hat{p}\right)^2\right) = \omega_0\frac{R}{c}\left(1-\hat{r}\cdot\hat{p}\right)\left(1+\hat{r}\cdot\hat{p}\right) .
            \end{cases}\label{eqn: f A B C expressions - up/down}
        \end{align}
        The downlink contributions are solved in exactly the same way as in Section~\ref{subsec: Monochromatic Model (Fresnel)}, the only change we have to make in the final answer is to use the slightly different terms $A \rightarrow A_\downarrow$ and $B \rightarrow B_\downarrow$.  However, for the uplinks there is an important difference.  Following what we had in equation~\ref{eqn: g integrals}, for the uplinks we need to solve:
        \begin{equation}
            \begin{cases}
                g_{+,\uparrow} \equiv \int\limits^{L/R}_0 \sin\left(\textsf{A}_\uparrow + \textsf{B}_\uparrow u - \textsf{C}u^2\right) du ,\\
                g_{\times,\uparrow} \equiv -\int\limits^{L/R}_0 \cos\left(\textsf{A}_\uparrow + \textsf{B}_\uparrow u - \textsf{C}u^2\right) du .
            \end{cases}\label{eqn: g integrals uplinks}
        \end{equation}
        We employ the same solution process as used in equation~\ref{eqn: exp^(i A B C)}, but this time noting that:
        \begin{equation*}
            \textsf{A}_\uparrow + \textsf{B}_\uparrow u - \textsf{C}u^2 = \left(\textsf{A}_\uparrow+\frac{\textsf{B}^2_\uparrow}{4\textsf{C}}\right) - \frac{\left(2\textsf{C}u-\textsf{B}_\uparrow\right)^2}{4\textsf{C}} \equiv \Phi_\uparrow - \frac{\pi}{2}x^2 ,
        \end{equation*}
        where for these uplink integrals, we make the substitution $\sqrt{\frac{\pi}{2}}x=\frac{2\textsf{C}u-\textsf{B}_\uparrow}{2\sqrt{\textsf{C}}}$.  The limits of integration under this transformation then become $\eta_{1,\uparrow} = \frac{-\textsf{B}_\uparrow}{\sqrt{2\pi\textsf{C}}}$ and $\eta_{2,\uparrow} = \eta_{1,\uparrow}\left[1 - \frac{2\textsf{C}}{\textsf{B}_\uparrow}\frac{L}{R}\right]$.  The end result is that for the uplinks, the definitions of $\Phi_\uparrow$, $\eta_{1,\uparrow}$, and $\eta_{2,\uparrow}$ are slightly different than what we had found originally.  However, the integral in equation~\ref{eqn: exp^(i A B C)} is still the same, which means that for the uplinks all of the subsequent math works out identically to the math for the downlinks (e.g. the solutions to the $g_{+,\uparrow}$ and $g_{\times,\uparrow}$ integrals are still of the form given in equation~\ref{eqn: g integrals solved}), we just use these new definitions in the final solution for the uplink contributions.
        
        Therefore, the final answer can be expressed as:
        \begin{align}
            \mathrm{Res}(t) &= \hat{p}^i\hat{p}^j E^{\hat{r}\textsc{A}}_{ij} \sqrt{\frac{\pi R/c}{8\omega_0\left(1-\left(\hat{r}\cdot\hat{p}\right)^2\right)}} \sum^N_{n=1} \Bigg[  \Big\{C\left(\eta_{2,\uparrow}\right)-C\left(\eta_{1,\uparrow}\right)\Big\} h_\textsc{A}\left(\Theta^{'}_\uparrow\right) \nonumber \\
            & \hspace{7cm} + \ \Big\{S\left(\eta_{2,\uparrow}\right)-S\left(\eta_{1,\uparrow}\right)\Big\} h_\textsc{A}\left(\Theta^{'}_\uparrow-\frac{\pi}{4}\right) \nonumber \\
            & \hspace{7cm} + \ \Big\{C\left(\eta_{2,\downarrow}\right)-C\left(\eta_{1,\downarrow}\right)\Big\} h_\textsc{A}\left(\Theta^{'}_\downarrow\right) \nonumber \\
            &\hspace{7cm} + \ \Big\{S\left(\eta_{2,\downarrow}\right)-S\left(\eta_{1,\downarrow}\right)\Big\} h_\textsc{A}\left(\Theta^{'}_\downarrow-\frac{\pi}{4}\right) \Bigg] , \label{eqn: Res(t) fresnel mono - N roundtrips} \\
            &\mathrm{for}\quad \textsc{A} \in [+, \times] , \nonumber \\
        \nonumber \\
            &\underset{\left(t_0 \ = \ -\frac{R}{c}\right)}{\mathrm{where}}\quad\quad \begin{cases}
                \Theta^{'}_\uparrow &\equiv \Theta\left(t+\frac{1}{2}\frac{R}{c}\frac{\left(1+\hat{r}\cdot\hat{p}\right)}{\left(1-\hat{r}\cdot\hat{p}\right)} - (2N-2n+2)\frac{L}{c}\right) \\
                &\equiv \theta_0 + \omega_0\left(t+\frac{1}{2}\frac{R}{c}\frac{\left(1+\hat{r}\cdot\hat{p}\right)}{\left(1-\hat{r}\cdot\hat{p}\right)} - (2N-2n+2)\frac{L}{c}\right) , \\[10pt]
                \Theta^{'}_\downarrow &\equiv \Theta\left(t+\frac{1}{2}\frac{R}{c}\frac{\left(1-\hat{r}\cdot\hat{p}\right)}{\left(1+\hat{r}\cdot\hat{p}\right)} - (2N-2n)\frac{L}{c}\right) \\
                &\equiv \theta_0 + \omega_0\left(t+\frac{1}{2}\frac{R}{c}\frac{\left(1-\hat{r}\cdot\hat{p}\right)}{\left(1+\hat{r}\cdot\hat{p}\right)} - (2N-2n)\frac{L}{c}\right) ,
                \end{cases}\label{eqn: Theta prime up/down} \\
            &\hspace{2cm}\begin{cases}
                \eta_{1,\uparrow} \equiv -\sqrt{\frac{2\omega_0 R/c \left(1+\hat{r}\cdot\hat{p}\right)}{\pi \left(1-\hat{r}\cdot\hat{p}\right)}} , \\
                \eta_{2,\uparrow} \equiv \eta_{1,\uparrow} \Big[1-\left(1-\hat{r}\cdot\hat{p}\right)\left(\frac{L}{R}\right)\Big]  , \\[10pt]
                \eta_{1,\downarrow} \equiv \sqrt{\frac{2\omega_0 R/c \left(1-\hat{r}\cdot\hat{p}\right)}{\pi \left(1+\hat{r}\cdot\hat{p}\right)}} , \\
                \eta_{2,\downarrow} \equiv \eta_{1,\downarrow} \Big[1+\left(1+\hat{r}\cdot\hat{p}\right)\left(\frac{L}{R}\right)\Big]  ,
            \end{cases} \label{eqn: vl,u up/down} \\
            &\hspace{2cm}\begin{cases}
                h_{+} \equiv -h_0 \cos\big(2\Theta\big), \\
                h_{\times} \equiv -h_0 \sin\big(2\Theta\big),
            \end{cases} \tag{\ref{eqn: h0 plus and cross} r} \\
            &\hspace{2cm}h_0 \equiv \frac{4(G\mathcal{M})^{5/3}}{c^4 R}\omega_0^{2/3}  , \tag{\ref{eqn: amplitude 0 of metric perturbation} r}\\
            &\hspace{2cm}\begin{cases}
                \begin{tabular}{l l l}
                    $f_{\uparrow/\downarrow}^{+}$ &$\equiv \frac{\hat{p}^i\hat{p}^j e^{\hat{r}+}_{ij}}{\left(1\pm\hat{r}\cdot\hat{p}\right)}$ &$= \frac{\left(\hat{p}\cdot\hat{\theta}\right)^2 - \left(\hat{p}\cdot\hat{\phi}\right)^2}{\left(1\pm\hat{r}\cdot\hat{p}\right)}$ , \\
                    $f_{\uparrow/\downarrow}^{\times}$ &$\equiv \frac{\hat{p}^i\hat{p}^j e^{\hat{r}\times}_{ij}}{\left(1\pm\hat{r}\cdot\hat{p}\right)}$ &$= \frac{2\left(\hat{p}\cdot\hat{\theta}\right)\left(\hat{p}\cdot\hat{\phi}\right)}{\left(1\pm\hat{r}\cdot\hat{p}\right)}$ ,
                \end{tabular}
            \end{cases} \tag{\ref{eqn: antenna f up/down} r} \\
            &\hspace{2cm}\begin{cases}
                \begin{tabular}{l l l}
                    $F_{\uparrow/\downarrow}^+$ &$\equiv \frac{\hat{p}^i\hat{p}^j E^{\hat{r}+}_{ij}}{\left(1\pm\hat{r}\cdot\hat{p}\right)}$ &$= \frac{1}{2}\left(1+\cos^2(\iota)\right) \left[ \cos(2\psi) f_{\uparrow/\downarrow}^+ + \sin(2\psi) f_{\uparrow/\downarrow}^\times \right]$ , \\
                    $F_{\uparrow/\downarrow}^\times$ &$\equiv \frac{\hat{p}^i\hat{p}^j E^{\hat{r}\times}_{ij}}{\left(1\pm\hat{r}\cdot\hat{p}\right)}$ &$= \cos(\iota) \left[ -\sin(2\psi) f_{\uparrow/\downarrow}^+ + \cos(2\psi) f_{\uparrow/\downarrow}^\times \right]$ ,
                \end{tabular}
            \end{cases} \tag{\ref{eqn: antenna F up/down} r} \\
            &\hspace{2cm}\begin{cases}
                \begin{tabular}{r l l r}
                    $\hat{r}$ &$= \big[\sin(\theta)\cos(\phi),$ &$\sin(\theta)\sin(\phi),$ &$\cos(\theta) \big]$, \\
                    $\hat{\theta}$ &$= \big[\cos(\theta)\cos(\phi),$ &$\cos(\theta)\sin(\phi),$ &$\sin(\theta) \big]$, \\
                    $\hat{\phi}$ &$= \big[-\sin(\phi),$ &$\cos(\phi),$ &$0 \big]$, \\
                    $\hat{p}$ &$= \big[\sin(\theta_p)\cos(\phi_p),$ &$\sin(\theta_p)\sin(\phi_p),$ &$\cos(\theta_p) \big]$.
                \end{tabular}
                \end{cases} \tag{\ref{eqn:source basis vectors}, \ref{eqn: p hat} r} 
        \end{align}
        (Make special note of the subtle differences of when $\left(1\pm\hat{r}\cdot\hat{p}\right)$ are used above in equations~\ref{eqn: Theta prime up/down} and~\ref{eqn: vl,u up/down}).

        \paragraph{Asymptotic Model}\mbox{}

        Once more we note the result when considering the asymptotic expansion of the Fresnel integrals.  Both $|\eta_{2,\uparrow/\downarrow}| \gg 1$ and $|\eta_{1,\uparrow/\downarrow}| \gg 1$, so long as we are not looking at a source that is nearly perfectly aligned or anti-aligned (where $\hat{r}\cdot\hat{p} \approx \pm 1$).  In this case, after replacing the Fresnel integrals with their asymptotic expansions given in equation~\ref{equation: Fresnel integrals} and dropping $\mathcal{O}\left(\frac{L}{R}\right)$ terms in the amplitude, the result simplifies to: 
        \begin{align}
            \mathrm{Res}(t) \approx \frac{1}{4\omega_0} \sum^N_{n=1}  &\left[-F_\uparrow^\textsc{A}\Bigg\{\frac{1}{1-\left(1-\hat{r}\cdot\hat{p}\right)\left(\frac{L}{R}\right)}\sin\bigg(\frac{\pi}{2}\eta_{2,\uparrow}^2\bigg)-\sin\bigg(\frac{\pi}{2}\eta_{1,\uparrow}^2\bigg)\Bigg\} h_\textsc{A}\left(\Theta^{'}\right) \right. \nonumber \\
            &\hspace{0.2cm}-  F_\uparrow^\textsc{A}\Bigg\{\frac{-1}{1-\left(1-\hat{r}\cdot\hat{p}\right)\left(\frac{L}{R}\right)}\cos\bigg(\frac{\pi}{2}\eta_{2,\uparrow}^2\bigg)+\cos\bigg(\frac{\pi}{2}\eta_{1,\uparrow}^2\bigg)\Bigg\} h_\textsc{A}\left(\Theta^{'}-\frac{\pi}{4}\right) \nonumber \\
            &\hspace{0.2cm}+  F_\downarrow^\textsc{A}\Bigg\{\frac{1}{1+\left(1+\hat{r}\cdot\hat{p}\right)\left(\frac{L}{R}\right)}\sin\bigg(\frac{\pi}{2}\eta_{2,\downarrow}^2\bigg)-\sin\bigg(\frac{\pi}{2}\eta_{1,\downarrow}^2\bigg)\Bigg\} h_\textsc{A}\left(\Theta^{'}\right) \nonumber \\
            &\hspace{0.2cm}+\left.  F_\downarrow^\textsc{A}\Bigg\{\frac{-1}{1+\left(1+\hat{r}\cdot\hat{p}\right)\left(\frac{L}{R}\right)}\cos\bigg(\frac{\pi}{2}\eta_{2,\downarrow}^2\bigg)+\cos\bigg(\frac{\pi}{2}\eta_{1,\downarrow}^2\bigg)\Bigg\} h_\textsc{A}\left(\Theta^{'}-\frac{\pi}{4}\right) \right] , \nonumber \\
            \approx \frac{1}{4\omega_0} \sum^N_{n=1}  &\left[-F_\uparrow^\textsc{A}\Bigg\{\sin\bigg(\frac{\pi}{2}\eta_{2,\uparrow}^2\bigg)-\sin\bigg(\frac{\pi}{2}\eta_{1,\uparrow}^2\bigg)\Bigg\} h_\textsc{A}\left(\Theta^{'}\right) \right. \nonumber \\
            & \hspace{0.2cm} -  F_\uparrow^\textsc{A}\Bigg\{-\cos\bigg(\frac{\pi}{2}\eta_{2,\uparrow}^2\bigg)+\cos\bigg(\frac{\pi}{2}\eta_{1,\uparrow}^2\bigg)\Bigg\} h_\textsc{A}\left(\Theta^{'}-\frac{\pi}{4}\right) \nonumber \\
            & \hspace{0.2cm} +  F_\downarrow^\textsc{A}\Bigg\{\sin\bigg(\frac{\pi}{2}\eta_{2,\downarrow}^2\bigg)-\sin\bigg(\frac{\pi}{2}\eta_{1,\downarrow}^2\bigg)\Bigg\} h_\textsc{A}\left(\Theta^{'}\right) \nonumber \\
            & \hspace{0.2cm} +\left.  F_\downarrow^\textsc{A}\Bigg\{-\cos\bigg(\frac{\pi}{2}\eta_{2,\downarrow}^2\bigg)+\cos\bigg(\frac{\pi}{2}\eta_{1,\downarrow}^2\bigg)\Bigg\} h_\textsc{A}\left(\Theta^{'}-\frac{\pi}{4}\right) \right] , \label{eqn: Res(t) fresnel mono asymptotic - N roundtrips}
        \end{align}
        and once again we see the Fresnel formalism solution more closely resembles that of the plane-wave solution, as they both share the same dominant amplitude term.

%---------------------------------------------------------------------------------
%---------------------------------------------------------------------------------
    \section{Future Directions}\label{sec: future directions}
    
    By the completion of my PhD and this dissertation, this was as far as I had taken the work on this side-project.  However, the next natural step would be to extend the Doppler tracking model to the fully general IIB regime.  Then we could carry out a similar investigation as was presented in Chapter~\ref{ch:results - first investigation}, in order to understand if the Fresnel regime offers an novel measurements to be made.
    
    Additionally, it would be interesting to alter the residuals calculated here for different types of gravitational wave sources, such as rotating deformed neutron stars~\citep{zimmermann1979,lasky_2015}.  In this chapter we simply carried over the mathematics for binary systems (namely for SMBHBs) producing gravitational waves, which is still certainly useful~\citep[see for example,][]{doppler_dorazio}.  In these binary systems, the coalescence of the source causes the frequency to ``spin-up'' and increase with time.  However, isolated deformed neutron stars would be expected to ``spin-down'' as energy is lost from the system from the gravitational radiation, causing their frequency to decrease with time~\citep{creighton_anderson_2011}.

\cleardoublepage

%%%%%%%%%%%%%%%%%%%%%%%%%%%%%%%%%%%%%%%%%%%%%%%%%%%%%%%%%%%%%%%%%%
% Quick references for some R packages used that I want in the bibliography
% \nocite{tikzDevice,plotly,reshape,Rcomputing,Florida2000}

%%%%%%%%%%%%%%%%%%%%%%%%%%%%%%%%%%%%%%%%%%%%%%%%%%%%%%%%%%%%%%%%%%%%
% The back matter

%%%%%%%%%%%%%%%%%%%%%%%%%%%%%%%%%%%%%%%%%%%%%%%%%%%%%%%%%%%%%%%%%%%%
%%%% LIST OF JOURNAL ABBREVIATIONS	
% If you don't write the journal names out in full in the bibliography then you need a list of journal abbreviations
% \include{chapters/Journals}

%%%%%%%%%%%%%%%%%%%%%%%%%%%%%%%%%%%%%%%%%%%%%%%%%%%%%%%%%%%%%%%%%%%%
%%%% 	BIBLIOGRAPHY	
% The bibliography itself can be single spaced with at least one extra space between items

% 1.8.1 Formatting the bibliography
% Include a complete bibliography at the end of the work. Arrange the bibliography alphabetically by the last name of the primary author. You may single-space citations, but leave one line of space between citations. If you use an article style format, where each chapter has its own separate bibliography, you must also include a cumulative bibliography at the end of the work.
% Verify any other requirements for formatting the bibliography at the end of the work. Certain disciplines/departments may require an alternate arrangement to the bibliography, for example, separating primary and secondary sources and then arranging each alphabetically by last name of author.

\newpage
\singlespace
\Urlmuskip=0mu plus 1mu\relax % Don't let the website links get all funky and break the page margins

\bibliographystyle{mnras} % <--- I changed the above bibliography style to the version MNRAS uses!  The problem with the above "apa-good" is that it lists ALL authors - and on those annoying LIGO collaboration papers with every member's name on it, the bibliography becomes unnecessarily long!  I couldn't figure out how to change it myself to suppress the long name lists and simply employ 'et al.', but MNRAS already does this, so I just downloaded their mnras.bst file and used it instead :)

\bibliography{references.bib} % keep this on, or you will get warnings about undefined citations

\begin{thebibliography}{}
\makeatletter
\relax
\def\mn@urlcharsother{\let\do\@makeother \do\$\do\&\do\#\do\^\do\_\do\%\do\~}
\def\mn@doi{\begingroup\mn@urlcharsother \@ifnextchar [ {\mn@doi@}
  {\mn@doi@[]}}
\def\mn@doi@[#1]#2{\def\@tempa{#1}\ifx\@tempa\@empty \href
  {http://dx.doi.org/#2} {doi:#2}\else \href {http://dx.doi.org/#2} {#1}\fi
  \endgroup}
\def\mn@eprint#1#2{\mn@eprint@#1:#2::\@nil}
\def\mn@eprint@arXiv#1{\href {http://arxiv.org/abs/#1} {{\tt arXiv:#1}}}
\def\mn@eprint@dblp#1{\href {http://dblp.uni-trier.de/rec/bibtex/#1.xml}
  {dblp:#1}}
\def\mn@eprint@#1:#2:#3:#4\@nil{\def\@tempa {#1}\def\@tempb {#2}\def\@tempc
  {#3}\ifx \@tempc \@empty \let \@tempc \@tempb \let \@tempb \@tempa \fi \ifx
  \@tempb \@empty \def\@tempb {arXiv}\fi \@ifundefined
  {mn@eprint@\@tempb}{\@tempb:\@tempc}{\expandafter \expandafter \csname
  mn@eprint@\@tempb\endcsname \expandafter{\@tempc}}}

\bibitem[\protect\citeauthoryear{{Abbott} et~al.,}{{Abbott}
  et~al.}{2016}]{gw150914}
{Abbott} B.~P.,  et~al., 2016, \mn@doi [Physical Review Letters]
  {10.1103/PhysRevLett.116.061102}, \href
  {https://ui.adsabs.harvard.edu/abs/2016PhRvL.116f1102A} {116, 061102}

\bibitem[\protect\citeauthoryear{{Abbott} et~al.,}{{Abbott}
  et~al.}{2017}]{gw170817_h0}
{Abbott} B.~P.,  et~al., 2017, \mn@doi [Nature] {10.1038/nature24471}, \href
  {https://ui.adsabs.harvard.edu/abs/2017Natur.551...85A} {551, 85}

\bibitem[\protect\citeauthoryear{{Acernese} et~al.,}{{Acernese}
  et~al.}{2015}]{virgo_ref}
{Acernese} F.,  et~al., 2015, \mn@doi [Classical and Quantum Gravity]
  {10.1088/0264-9381/32/2/024001}, \href
  {https://ui.adsabs.harvard.edu/abs/2015CQGra..32b4001A} {32, 024001}

\bibitem[\protect\citeauthoryear{Aggarwal et~al.,}{Aggarwal
  et~al.}{2019}]{NG_11yr_cw}
Aggarwal K.,  et~al., 2019, \mn@doi [The Astrophysical Journal]
  {10.3847/1538-4357/ab2236}, 880, 116

\bibitem[\protect\citeauthoryear{{Akutsu} et~al.,}{{Akutsu}
  et~al.}{2018}]{kagra_ref}
{Akutsu} T.,  et~al., 2018, \mn@doi [Progress of Theoretical and Experimental
  Physics] {10.1093/ptep/ptx180}, \href
  {https://ui.adsabs.harvard.edu/abs/2018PTEP.2018a3F01A} {2018, 013F01}

\bibitem[\protect\citeauthoryear{Albrecht et~al.,}{Albrecht
  et~al.}{2009}]{darkenergy_fisher}
Albrecht A.,  et~al., 2009, Findings of the Joint Dark Energy Mission Figure of
  Merit Science Working Group (\mn@eprint {arXiv} {0901.0721})

\bibitem[\protect\citeauthoryear{{Armstrong}}{{Armstrong}}{2006}]{doppler_armstrong}
{Armstrong} J.~W.,  2006, \mn@doi [Living Reviews in Relativity]
  {10.12942/lrr-2006-1}, \href
  {https://ui.adsabs.harvard.edu/abs/2006LRR.....9....1A} {9, 1}

\bibitem[\protect\citeauthoryear{Arzoumanian et~al.,}{Arzoumanian
  et~al.}{2014}]{Arzoumanian_2014}
Arzoumanian Z.,  et~al., 2014, \mn@doi [The Astrophysical Journal]
  {10.1088/0004-637x/794/2/141}, 794, 141

\bibitem[\protect\citeauthoryear{Arzoumanian et~al.,}{Arzoumanian
  et~al.}{2018}]{NG_11yr_data}
Arzoumanian Z.,  et~al., 2018, \mn@doi [The Astrophysical Journal Supplement
  Series] {10.3847/1538-4365/aab5b0}, 235, 37

\bibitem[\protect\citeauthoryear{{Arzoumanian} et~al.,}{{Arzoumanian}
  et~al.}{2020}]{multimessenger_caitlin_2020}
{Arzoumanian} Z.,  et~al., 2020, \mn@doi [The Astrophysical Journal]
  {10.3847/1538-4357/ababa1}, \href
  {https://ui.adsabs.harvard.edu/abs/2020ApJ...900..102A} {900, 102}

\bibitem[\protect\citeauthoryear{Babak et~al.,}{Babak
  et~al.}{2015}]{EPTA_cw_2016}
Babak S.,  et~al., 2015, \mn@doi [Monthly Notices of the Royal Astronomical
  Society] {10.1093/mnras/stv2092}, 455, 1665

\bibitem[\protect\citeauthoryear{{Bailes} et~al.,}{{Bailes}
  et~al.}{2020}]{MeerKAT_bailes2020}
{Bailes} M.,  et~al., 2020, \mn@doi [Publications of the Astronomical Society
  of Australia] {10.1017/pasa.2020.19}, \href
  {https://ui.adsabs.harvard.edu/abs/2020PASA...37...28B} {37, e028}

\bibitem[\protect\citeauthoryear{Belsley, Kuh  \& Welsch}{Belsley
  et~al.}{1980}]{condition_number_source}
Belsley D.~A.,  Kuh E.,   Welsch R.~E.,  1980, Regression Diagnostics:
  Identifying Influential Data and Sources of Collinearity.
John Wiley \& Sons, Ltd, \mn@doi{10.1002/0471725153}

\bibitem[\protect\citeauthoryear{{Caldwell}}{{Caldwell}}{1993}]{greens_functions_1993}
{Caldwell} R.~R.,  1993, \mn@doi [Physical Review D]
  {10.1103/PhysRevD.48.4688}, \href
  {https://ui.adsabs.harvard.edu/abs/1993PhRvD..48.4688C} {48, 4688}

\bibitem[\protect\citeauthoryear{Carroll}{Carroll}{2013}]{carroll_2013}
Carroll S.~M.,  2013, Spacetime and Geometry: An Introduction to General
  Relativity.
Pearson

\bibitem[\protect\citeauthoryear{{Chamberlin}, {Creighton}, {Siemens},
  {Demorest}, {Ellis}, {Price}  \& {Romano}}{{Chamberlin}
  et~al.}{2015}]{chamberlin2015}
{Chamberlin} S.~J.,  {Creighton} J. D.~E.,  {Siemens} X.,  {Demorest} P.,
  {Ellis} J.,  {Price} L.~R.,   {Romano} J.~D.,  2015, \mn@doi [Physical Review
  D] {10.1103/PhysRevD.91.044048}, \href
  {https://ui.adsabs.harvard.edu/abs/2015PhRvD..91d4048C} {91, 044048}

\bibitem[\protect\citeauthoryear{Coe}{Coe}{2009}]{fisher_confidence_ellipses}
Coe D.,  2009, Fisher Matrices and Confidence Ellipses: A Quick-Start Guide and
  Software (\mn@eprint {arXiv} {0906.4123})

\bibitem[\protect\citeauthoryear{{Corbin} \& {Cornish}}{{Corbin} \&
  {Cornish}}{2010}]{CC_main_paper}
{Corbin} V.,  {Cornish} N.~J.,  2010, arXiv e-prints, \href
  {https://ui.adsabs.harvard.edu/abs/2010arXiv1008.1782C} {p. arXiv:1008.1782}

\bibitem[\protect\citeauthoryear{{Cordes} \& {Shannon}}{{Cordes} \&
  {Shannon}}{2010}]{Cordes_2010}
{Cordes} J.~M.,  {Shannon} R.~M.,  2010, arXiv e-prints, \href
  {https://ui.adsabs.harvard.edu/abs/2010arXiv1010.3785C} {p. arXiv:1010.3785}

\bibitem[\protect\citeauthoryear{Creighton \& Anderson}{Creighton \&
  Anderson}{2011}]{creighton_anderson_2011}
Creighton J. D.~E.,  Anderson W.~G.,  2011, Gravitational-Wave Physics and
  Astronomy: An Introduction to Theory, Experiment and Data Analysis.
Wiley-VCH

\bibitem[\protect\citeauthoryear{{D'Orazio} \& {Loeb}}{{D'Orazio} \&
  {Loeb}}{2020}]{DOrazio_main_paper}
{D'Orazio} D.~J.,  {Loeb} A.,  2020, arXiv e-prints, \href
  {https://ui.adsabs.harvard.edu/abs/2020arXiv200906084D} {p. arXiv:2009.06084}

\bibitem[\protect\citeauthoryear{{Deller} et~al.,}{{Deller}
  et~al.}{2019}]{pulsar_parallax2019}
{Deller} A.~T.,  et~al., 2019, \mn@doi [The Astrophysical Journal]
  {10.3847/1538-4357/ab11c7}, \href
  {https://ui.adsabs.harvard.edu/abs/2019ApJ...875..100D} {875, 100}

\bibitem[\protect\citeauthoryear{{Deng} \& {Finn}}{{Deng} \&
  {Finn}}{2011}]{DF_main_paper}
{Deng} X.,  {Finn} L.~S.,  2011, \mn@doi [Monthly Notices of the Royal
  Astronomical Society] {10.1111/j.1365-2966.2010.17913.x}, \href
  {https://ui.adsabs.harvard.edu/abs/2011MNRAS.414...50D} {414, 50}

\bibitem[\protect\citeauthoryear{{Ellis}}{{Ellis}}{2013}]{Ellis_2013}
{Ellis} J.~A.,  2013, \mn@doi [Classical and Quantum Gravity]
  {10.1088/0264-9381/30/22/224004}, \href
  {https://ui.adsabs.harvard.edu/abs/2013CQGra..30v4004E} {30, 224004}

\bibitem[\protect\citeauthoryear{{Estabrook} \& {Wahlquist}}{{Estabrook} \&
  {Wahlquist}}{1975}]{doppler_estabrook}
{Estabrook} F.~B.,  {Wahlquist} H.~D.,  1975, \mn@doi [General Relativity and
  Gravitation] {10.1007/BF00762449}, \href
  {https://ui.adsabs.harvard.edu/abs/1975GReGr...6..439E} {6, 439}

\bibitem[\protect\citeauthoryear{{Feeney}, {Peiris}, {Williamson}, {Nissanke},
  {Mortlock}, {Alsing}  \& {Scolnic}}{{Feeney}
  et~al.}{2019}]{Feeney_standardsirens}
{Feeney} S.~M.,  {Peiris} H.~V.,  {Williamson} A.~R.,  {Nissanke} S.~M.,
  {Mortlock} D.~J.,  {Alsing} J.,   {Scolnic} D.,  2019, \mn@doi [Physical
  Review Letters] {10.1103/PhysRevLett.122.061105}, \href
  {https://ui.adsabs.harvard.edu/abs/2019PhRvL.122f1105F} {122, 061105}

\bibitem[\protect\citeauthoryear{{Foreman-Mackey}, {Hogg}, {Lang}  \&
  {Goodman}}{{Foreman-Mackey} et~al.}{2013}]{emcee}
{Foreman-Mackey} D.,  {Hogg} D.~W.,  {Lang} D.,   {Goodman} J.,  2013, \mn@doi
  [Publications of the Astronomical Society of the Pacific] {10.1086/670067},
  \href {https://ui.adsabs.harvard.edu/abs/2013PASP..125..306F} {125, 306}

\bibitem[\protect\citeauthoryear{Glen}{Glen}{2018}]{cromwell}
Glen S.,  2018, Cromwell's Rule: Simple Definition, \url
  {https://www.statisticshowto.datasciencecentral.com/cromwells-rule/}

\bibitem[\protect\citeauthoryear{Harrison}{Harrison}{2020}]{harrison_2020}
Harrison D.,  2020, The Multipole Expansion, \url
  {https://phys.libretexts.org/Bookshelves/Math_Methods_and_Pedagogy/Mathematical_Methods/The_Multipole_Expansion}

\bibitem[\protect\citeauthoryear{{Hazboun}, {Romano}  \& {Smith}}{{Hazboun}
  et~al.}{2019}]{toa_hazboun2019}
{Hazboun} J.~S.,  {Romano} J.~D.,   {Smith} T.~L.,  2019, \mn@doi [Physical
  Review D] {10.1103/PhysRevD.100.104028}, \href
  {https://ui.adsabs.harvard.edu/abs/2019PhRvD.100j4028H} {100, 104028}

\bibitem[\protect\citeauthoryear{{Hogg}}{{Hogg}}{1999}]{hogg_distances}
{Hogg} D.~W.,  1999, arXiv e-prints, \href
  {https://ui.adsabs.harvard.edu/abs/1999astro.ph..5116H} {pp
  astro--ph/9905116}

\bibitem[\protect\citeauthoryear{Holz \& Hughes}{Holz \&
  Hughes}{2005}]{Holz_2005}
Holz D.~E.,  Hughes S.~A.,  2005, \mn@doi [The Astrophysical Journal]
  {10.1086/431341}, 629, 15

\bibitem[\protect\citeauthoryear{{Janssen} et~al.,}{{Janssen}
  et~al.}{2015}]{SKA_janssen2015}
{Janssen} G.,  et~al., 2015, in Advancing Astrophysics with the Square
  Kilometre Array (AASKA14). p.~37 (\mn@eprint {arXiv} {1501.00127})

\bibitem[\protect\citeauthoryear{Kolkowitz, Pikovski, Langellier, Lukin,
  Walsworth  \& Ye}{Kolkowitz et~al.}{2016}]{GW_atomicclocks}
Kolkowitz S.,  Pikovski I.,  Langellier N.,  Lukin M.~D.,  Walsworth R.~L.,
  Ye J.,  2016, \mn@doi [Phys. Rev. D] {10.1103/PhysRevD.94.124043}, 94, 124043

\bibitem[\protect\citeauthoryear{{LIGO Scientific Collaboration} et~al.,}{{LIGO
  Scientific Collaboration} et~al.}{2015}]{ligo_ref}
{LIGO Scientific Collaboration} et~al., 2015, \mn@doi [Classical and Quantum
  Gravity] {10.1088/0264-9381/32/7/074001}, \href
  {https://ui.adsabs.harvard.edu/abs/2015CQGra..32g4001L} {32, 074001}

\bibitem[\protect\citeauthoryear{Lam}{Lam}{2020}]{michael_TOAvsPeriod}
Lam M.,  2020, Private communication

\bibitem[\protect\citeauthoryear{Lam, McLaughlin, Cordes, Chatterjee  \&
  Lazio}{Lam et~al.}{2018}]{Lam_2018}
Lam M.~T.,  McLaughlin M.~A.,  Cordes J.~M.,  Chatterjee S.,   Lazio T. J.~W.,
  2018, \mn@doi [The Astrophysical Journal] {10.3847/1538-4357/aac48d}, 861, 12

\bibitem[\protect\citeauthoryear{Lasky}{Lasky}{2015}]{lasky_2015}
Lasky P.~D.,  2015, \mn@doi [Publications of the Astronomical Society of
  Australia] {10.1017/pasa.2015.35}, 32, e034

\bibitem[\protect\citeauthoryear{{Lee}, {Wex}, {Kramer}, {Stappers}, {Bassa},
  {Janssen}, {Karuppusamy}  \& {Smits}}{{Lee} et~al.}{2011}]{GWastro_Lee2011}
{Lee} K.~J.,  {Wex} N.,  {Kramer} M.,  {Stappers} B.~W.,  {Bassa} C.~G.,
  {Janssen} G.~H.,  {Karuppusamy} R.,   {Smits} R.,  2011, \mn@doi [Monthly
  Notices of the Royal Astronomical Society]
  {10.1111/j.1365-2966.2011.18622.x}, \href
  {https://ui.adsabs.harvard.edu/abs/2011MNRAS.414.3251L} {414, 3251}

\bibitem[\protect\citeauthoryear{{Liu}, {Verbiest}, {Kramer}, {Stappers}, {van
  Straten}  \& {Cordes}}{{Liu} et~al.}{2011}]{Liu_2011}
{Liu} K.,  {Verbiest} J.~P.~W.,  {Kramer} M.,  {Stappers} B.~W.,  {van Straten}
  W.,   {Cordes} J.~M.,  2011, \mn@doi [Monthly Notices of the Royal
  Astronomical Society] {10.1111/j.1365-2966.2011.19452.x}, \href
  {https://ui.adsabs.harvard.edu/abs/2011MNRAS.417.2916L} {417, 2916}

\bibitem[\protect\citeauthoryear{Loeb \& Maoz}{Loeb \&
  Maoz}{2015}]{loeb_gw_atomicclocks}
Loeb A.,  Maoz D.,  2015, Using Atomic Clocks to Detect Gravitational Waves
  (\mn@eprint {arXiv} {1501.00996})

\bibitem[\protect\citeauthoryear{{Lorimer} \& {Kramer}}{{Lorimer} \&
  {Kramer}}{2004}]{handbook_pulsar_astro}
{Lorimer} D.~R.,  {Kramer} M.,  2004, {Handbook of Pulsar Astronomy}.
~ Vol. 4, Cambridge University Press

\bibitem[\protect\citeauthoryear{Maggiore}{Maggiore}{2008}]{maggiore_2008}
Maggiore M.,  2008, Gravitational Waves: Theory and Experiments.
~ Vol. 1, Oxford University Press

\bibitem[\protect\citeauthoryear{Maggiore}{Maggiore}{2018}]{maggiore_2018}
Maggiore M.,  2018, Gravitational Waves: Astrophysics and Cosmology.
~ Vol. 2, Oxford University Press

\bibitem[\protect\citeauthoryear{Manchester, Hobbs, Teoh  \& Hobbs}{Manchester
  et~al.}{2005}]{ATNF_catalogue}
Manchester R.~N.,  Hobbs G.~B.,  Teoh A.,   Hobbs M.,  2005, \mn@doi [The
  Astronomical Journal] {10.1086/428488}, 129, 1993

\bibitem[\protect\citeauthoryear{Marti, Hutson, Goban, Campbell, Poli  \&
  Ye}{Marti et~al.}{2018}]{atomicclock_record}
Marti G.~E.,  Hutson R.~B.,  Goban A.,  Campbell S.~L.,  Poli N.,   Ye J.,
  2018, \mn@doi [Phys. Rev. Lett.] {10.1103/PhysRevLett.120.103201}, 120,
  103201

\bibitem[\protect\citeauthoryear{{McGrath} \& {Creighton}}{{McGrath} \&
  {Creighton}}{2021}]{mcgrath2020fresnel}
{McGrath} C.,  {Creighton} J.,  2021, \mn@doi [Monthly Notices of the Royal
  Astronomical Society] {10.1093/mnras/stab1417}, \href
  {https://ui.adsabs.harvard.edu/abs/2021MNRAS.505.4531M} {505, 4531}

\bibitem[\protect\citeauthoryear{{McGrath}, {D'Orazio}  \&
  {Creighton}}{{McGrath} et~al.}{2022}]{mcgrath2022H0}
{McGrath} C.,  {D'Orazio} D.~J.,   {Creighton} J.,  2022, \mn@doi [Monthly
  Notices of the Royal Astronomical Society] {10.1093/mnras/stac2593}, \href
  {https://ui.adsabs.harvard.edu/abs/2022MNRAS.517.1242M} {517, 1242}

\bibitem[\protect\citeauthoryear{{Messenger} \& {Read}}{{Messenger} \&
  {Read}}{2012}]{messenger_gwH0}
{Messenger} C.,  {Read} J.,  2012, \mn@doi [Physical Review Letters]
  {10.1103/PhysRevLett.108.091101}, \href
  {https://ui.adsabs.harvard.edu/abs/2012PhRvL.108i1101M} {108, 091101}

\bibitem[\protect\citeauthoryear{Moore}{Moore}{2013}]{moore_2013}
Moore T.~A.,  2013, A General Relativity Workbook.
University Science Books

\bibitem[\protect\citeauthoryear{Nikulin}{Nikulin}{2011a}]{bernstein}
Nikulin M.,  2011a, Bernstein-von Mises theorem, \url
  {https://www.encyclopediaofmath.org/index.php/Bernstein-von_Mises_theorem}

\bibitem[\protect\citeauthoryear{Nikulin}{Nikulin}{2011b}]{cramer-rao}
Nikulin M.,  2011b, Rao-Cramér inequality, \url
  {https://www.encyclopediaofmath.org/index.php/Rao-Cramér_inequality}

\bibitem[\protect\citeauthoryear{Provornikova et~al.,}{Provornikova
  et~al.}{2021}]{interstellar_probe_meeting}
Provornikova E.,  et~al., 2021, {Unique heliophysics science opportunities
  along the Interstellar Probe journey up to 1000 AU from the Sun},
  \mn@doi{10.5194/egusphere-egu21-10504}, \url
  {https://doi.org/10.5194/egusphere-egu21-10504}

\bibitem[\protect\citeauthoryear{Schneider}{Schneider}{2015}]{schneider_2015}
Schneider P.,  2015, Extragalactic Astronomy and Cosmology: An Introduction,
  2nd edn.
Springer, p. 177–192, \mn@doi{10.1007/978-3-642-54083-7}

\bibitem[\protect\citeauthoryear{Schutz}{Schutz}{1986}]{Schutz_1986}
Schutz B.~F.,  1986, \mn@doi [Nature] {10.1038/323310a0}, 323, 310

\bibitem[\protect\citeauthoryear{{Smits}, {Lorimer}, {Kramer}, {Manchester},
  {Stappers}, {Jin}, {Nan}  \& {Li}}{{Smits} et~al.}{2009}]{FAST_smits2009}
{Smits} R.,  {Lorimer} D.~R.,  {Kramer} M.,  {Manchester} R.,  {Stappers} B.,
  {Jin} C.~J.,  {Nan} R.~D.,   {Li} D.,  2009, \mn@doi [Astronomy \&
  Astrophysics] {10.1051/0004-6361/200911939}, \href
  {https://ui.adsabs.harvard.edu/abs/2009A&A...505..919S} {505, 919}

\bibitem[\protect\citeauthoryear{{Smits}, {Tingay}, {Wex}, {Kramer}  \&
  {Stappers}}{{Smits} et~al.}{2011}]{Ldist_smitts2011}
{Smits} R.,  {Tingay} S.~J.,  {Wex} N.,  {Kramer} M.,   {Stappers} B.,  2011,
  \mn@doi [Astronomy \& Astrophysics] {10.1051/0004-6361/201016141}, \href
  {https://ui.adsabs.harvard.edu/abs/2011A&A...528A.108S} {528, A108}

\bibitem[\protect\citeauthoryear{{Soyuer}, {Zwick}, {D'Orazio}  \&
  {Saha}}{{Soyuer} et~al.}{2021}]{doppler_dorazio}
{Soyuer} D.,  {Zwick} L.,  {D'Orazio} D.~J.,   {Saha} P.,  2021, \mn@doi
  [Monthly Notices of the Royal Astronomical Society] {10.1093/mnrasl/slab025},
  \href {https://ui.adsabs.harvard.edu/abs/2021MNRAS.503L..73S} {503, L73}

\bibitem[\protect\citeauthoryear{{Taylor}, {van Haasteren}  \&
  {Sesana}}{{Taylor} et~al.}{2020}]{taylor2020}
{Taylor} S.~R.,  {van Haasteren} R.,   {Sesana} A.,  2020, \mn@doi [Physical
  Review D] {10.1103/PhysRevD.102.084039}, \href
  {https://ui.adsabs.harvard.edu/abs/2020PhRvD.102h4039T} {102, 084039}

\bibitem[\protect\citeauthoryear{Tinto}{Tinto}{1997}]{doppler_Tinto}
Tinto M.,  1997, Banach Center Publications, 41, 145

\bibitem[\protect\citeauthoryear{Vallisneri}{Vallisneri}{2008}]{michele_fisher}
Vallisneri M.,  2008, \mn@doi [Phys. Rev. D] {10.1103/PhysRevD.77.042001}, 77,
  042001

\bibitem[\protect\citeauthoryear{{Weinberg}}{{Weinberg}}{1972}]{weinberg1972}
{Weinberg} S.,  1972, Gravitation and Cosmology: Principles and Applications of
  the General Theory of Relativity.
New York: Wiley

\bibitem[\protect\citeauthoryear{Wittman}{Wittman}{otes}]{wittman_Fisher_for_beginners}
Wittman D.,  unpublished notes, Fisher Matrix for Beginners, UC Davis, \url
  {http://wittman.physics.ucdavis.edu/Fisher-matrix-guide.pdf}

\bibitem[\protect\citeauthoryear{{Xin}, {Mingarelli}  \& {Hazboun}}{{Xin}
  et~al.}{2021}]{pta_multimessenger2021}
{Xin} C.,  {Mingarelli} C. M.~F.,   {Hazboun} J.~S.,  2021, \mn@doi [The
  Astrophysical Journal] {10.3847/1538-4357/ac01c5}, \href
  {https://ui.adsabs.harvard.edu/abs/2021ApJ...915...97X} {915, 97}

\bibitem[\protect\citeauthoryear{Zee}{Zee}{2013}]{zee_2013}
Zee A.,  2013, Einstein Gravity in a Nutshell.
Princeton University Press

\bibitem[\protect\citeauthoryear{{Zhu} et~al.,}{{Zhu}
  et~al.}{2014}]{PPTA_cw_2014}
{Zhu} X.~J.,  et~al., 2014, \mn@doi [Monthly Notices of the Royal Astronomical
  Society] {10.1093/mnras/stu1717}, \href
  {https://ui.adsabs.harvard.edu/abs/2014MNRAS.444.3709Z} {444, 3709}

\bibitem[\protect\citeauthoryear{{Zimmermann} \& {Szedenits}}{{Zimmermann} \&
  {Szedenits}}{1979}]{zimmermann1979}
{Zimmermann} M.,  {Szedenits} E. J.,  1979, \mn@doi [Physical Review D]
  {10.1103/PhysRevD.20.351}, \href
  {https://ui.adsabs.harvard.edu/abs/1979PhRvD..20..351Z} {20, 351}

\bibitem[\protect\citeauthoryear{{van Haasteren} \& {Levin}}{{van Haasteren} \&
  {Levin}}{2013}]{corrnoise_vanHaasteren2013}
{van Haasteren} R.,  {Levin} Y.,  2013, \mn@doi [Monthly Notices of the Royal
  Astronomical Society] {10.1093/mnras/sts097}, \href
  {https://ui.adsabs.harvard.edu/abs/2013MNRAS.428.1147V} {428, 1147}

\makeatother
\end{thebibliography}

%%%%%%%%%%%%%%%%%%%%%%%%%%%%%%%%%%%%%%%%%%%%%%%%%%%%%%%%%%%%%%%%%%
%%%%% APPENDICES <--- UWM requested that they come *after* the bibliography!
% Put your appendices inside here, to maintain figure and table listings. Make sure to use \section{appendixA} to have some numbering for figures and tables.
\doublespace
\begin{appendices}
\chapter{Cosmological Distances}\label{app: cosmological distances}

For a gravitational wave propagating from our source to the Earth (``infalling''), we define the following ``boundary'' conditions:
\begin{align*}
    \textbf{Source:}  & \quad r = R, \quad z^{'} = z, \quad t^{'} = t_\mathrm{ret} , \\
    \textbf{Earth:} & \quad r = 0, \quad z^{'} = 0, \quad t^{'} = t , 
\end{align*}
where our reference frame is centered on the Earth.  From the Friedmann equations and the definition of the redshift in terms of the present day $t_0$ scale factor, $1+z = \frac{a(t_0)}{a(t)}$, we explicity list the following important quantities~\citep{weinberg1972,hogg_distances,schneider_2015}:
\begin{align}
    \textbf{Expansion Rate} \quad \left(\frac{\dot{a}(t)}{a(t)}\right)^2 &\equiv H^2 = H_0^2 \left[\frac{\Omega_r}{a^4(t)} + \frac{\Omega_m}{a^3(t)} + \Omega_\Lambda + \frac{\Omega_k}{a^2(t)}\right] ,  \nonumber \\
    &= H_0^2 \left[\frac{(1+z)^4}{a^4(t_0)}\Omega_r + \frac{(1+z)^3}{a^3(t_0)}\Omega_m + \Omega_\Lambda + \frac{(1+z)^2}{a^2(t_0)}\Omega_k\right] , \nonumber \\
    &\equiv H_0^2 E^2 ,  \label{eqn: expansion rate} \\
    \textbf{Curvature Density} \quad \Omega_k &\equiv a^2(t_0) \left[1-\frac{\Omega_r}{a^4(t_0)} - \frac{\Omega_m}{a^3(t_0)} - \Omega_\Lambda\right] = -k D_H^2 , \label{eqn: curvature density parameter} \\
    \textbf{Hubble Distance} \quad D_H &\equiv \frac{c}{H_0} , \label{eqn: Hubble distance} \\
    \substack{\textbf{Line-of-Sight} \\ \textbf{Comoving Distance}} \quad D_c &\equiv D_H \int^z_0 \frac{dz^{'}}{E} , \nonumber \\
    & = c a(t_0)\int^t_{t_\mathrm{ret}} \frac{dt^{'}}{a},  \label{eqn: line-of-sight comoving distance Dc} \\
    \textbf{Parallax Distance} \quad D_\mathrm{par} &\equiv a(t_0) \frac{R}{\sqrt{1-kR^2}} , \label{eqn: parallax distance Dpar}
\end{align}
where $H_0$ is the Hubble constant, $\Omega_r$, $\Omega_m$, and $\Omega_\Lambda$ are the radiation, matter, and vacuum density parameters, and $E$ is the ``Hubble function.''  For completeness here we will continue to write the present day scale factor explicitly, but typical convention is to normalize this to $a(t_0) \equiv 1$.  The two forms of the line-of-sight comoving distance in equation~\ref{eqn: line-of-sight comoving distance Dc} are made through the change of variables $t \rightarrow z$ using the definition of $z$ and equation~\ref{eqn: expansion rate}.

The luminosity distance of our source is defined by considering the energy flux $F_\mathrm{obs}$ (in this case, of our gravitational waves) as measured by the observer \citep{maggiore_2008}.  In a cosmologically expanding universe the observed energy $E_\mathrm{obs}$ is redshifted compared to the energy that was emitted in the source frame $E_\mathrm{s}$, such that $E_\mathrm{s} = (1+z) E_\mathrm{obs}$.  Additionally, time dilation means the time measured by the observer and source clocks are related by $t_\mathrm{s} = \frac{t_\mathrm{obs}}{(1+z)}$.  Finally, from the FLRW metric equation~\ref{eqn: FLRW metric}, the flux at time $t$ will be spread over a total area of $A(t) = 4\pi a^2(t) R^2$.  This means that we can write:
\begin{equation}
    F_\mathrm{obs}(t) \ \ \equiv \ \ \frac{L_\mathrm{obs}}{A(t)} \ \ = \ \ \frac{L_\mathrm{s}}{(1+z)^2 4\pi a^2(t) R^2} \ \ = \ \ \frac{L_\mathrm{s}}{4\pi \Big[(1+z)^2 a^2(t) R^2\Big]} \ \ \equiv \ \ \frac{L_\mathrm{s}}{4\pi D_L^2} ,
\label{eqn: luminosity distance flux relationship}
\end{equation}
where in the final equality we define our luminosity distance $D_L$ such that the observed flux matches our standard notion of flux, but as a function of the source frame luminosity $L_\mathrm{s}$.  Therefore at the present time, we have:
\begin{equation}
    D_L \equiv (1+z) a(t_0) R .
\label{eqn: DL - R relation}
\end{equation}

Next we draw the connection between the comoving coordinate distance $R$, the luminosity distance $D_L$, and the line-of-sight comoving distance $D_c$.  Integrating the metric equation~\ref{eqn: FLRW metric} along the radial ($d\theta=d\phi=0)$, null ($ds=0$), infalling path of the gravitational wave, and using equation~\ref{eqn: line-of-sight comoving distance Dc} gives us:
    \begin{align}
        R \ \ \equiv\ \  D_M \ \ &=\ \  \left.\begin{cases}
            \frac{1}{\sqrt{k}}\sin\left(\frac{\sqrt{k}}{a(t_0)}D_c\right), & \quad k > 0 \\[4pt]
            \frac{1}{a(t_0)}D_c, & \quad k = 0 \\[4pt]
            \frac{1}{\sqrt{|k|}}\sinh\left(\frac{\sqrt{|k|}}{a(t_0)}D_c\right), & \quad k < 0
        \end{cases}\right\} , \nonumber \\ 
        &=\ \  \left.\begin{cases}
            \frac{D_H}{\sqrt{|\Omega_k|}}\sin\left(\frac{\sqrt{|\Omega_k|}}{a(t_0)D_H}D_c\right), & \quad \Omega_k < 0 \\[4pt]
            \frac{1}{a(t_0)}D_c, & \quad \Omega_k = 0 \\[4pt]
            \frac{D_H}{\sqrt{\Omega_k}}\sinh\left(\frac{\sqrt{\Omega_k}}{a(t_0)D_H}D_c\right), & \quad \Omega_k > 0
        \end{cases}\right\} , \nonumber \\ 
        &=\ \  \frac{D_L}{a(t_0) (1+z)} ,
    \label{eqn: R - DM - DC - DL relationship}
    \end{align}
where $D_M$ is introduced as the ``transverse comoving distance'' following~\citeauthor{hogg_distances}, which is just a re-naming of the comoving coordinate distance.  Thus depending on the global universe geometry, the relationship between the various distances $R$, $D_L$, $D_c$, and $D_M$ are given by equation~\ref{eqn: R - DM - DC - DL relationship}, and notably in a flat universe (with $a(t_0)=1$ normalization) $R = D_M = D_c = D_L / (1+z)$.

One final and very important distance for our consideration is the ``parallax distance'' $D_\mathrm{par}$ in equation~\ref{eqn: parallax distance Dpar} \citep{weinberg1972}.  Substituting equation~\ref{eqn: R - DM - DC - DL relationship} into equation~\ref{eqn: parallax distance Dpar} lets us write:
\begin{align}
    D_\mathrm{par} \ \ \equiv\ \  \frac{a(t_0)R}{\sqrt{1-kR^2}} \ \ &=\ \ \frac{a(t_0)D_M}{\sqrt{1-kD_M^2}} , \nonumber \\ 
    &=\ \  \left.\begin{cases}
        \frac{a(t_0)}{\sqrt{k}}\tan\left(\frac{\sqrt{k}}{a(t_0)}D_c\right), & \quad k > 0 \\[4pt]
        D_c, & \quad k = 0 \\[4pt]
        \frac{a(t_0)}{\sqrt{|k|}}\tanh\left(\frac{\sqrt{|k|}}{a(t_0)}D_c\right), & \quad k < 0
    \end{cases}\right\} , \nonumber \\ 
    &=\ \  \frac{D_L}{(1+z)\sqrt{1-\frac{k D_L^2}{a^2(t_0)(1+z)^2}}} .
\label{eqn: Dpar - R - DM - DC - DL relationship}
\end{align}
As a final note, the closed universe case needs further special consideration, as the parallax distance not only diverges, but can take on \textit{negative} values, which is not immediately obvious from equation~\ref{eqn: Dpar - R - DM - DC - DL relationship}.  This is unique to just the closed geometry, and is discussed further in the box below. \\

\begin{minipage}{0.9\linewidth}
\fbox {
    \parbox{\linewidth}{
    \underline{\textbf{$D_\mathrm{par}$ in a Closed Universe ($k > 0$)}}\newline
    It should be noted that there are some interesting implications of the formulas as written here for the closed universe case.  For example, note that in equation~\ref{eqn: Dpar - R - DM - DC - DL relationship} the parallax distance becomes infinite at $R = R_\mathrm{max} = \frac{1}{\sqrt{k}}$.  Recalling equation~\ref{eqn: r(chi) definition}, this happens at $\chi = \frac{1}{\sqrt{k}}\frac{\pi}{2}$, but $\chi$ is defined on the interval $0\leq \chi \leq \frac{1}{\sqrt{k}}\pi$.  The coordinate $R$ decreases back to zero after it reaches its maximum value, but what happens to $D_\mathrm{par}$? \\
    
    %For a conceptual analogy, 
    %
    Conceptually, consider a point at the north pole of a globe generating some wave.  As the wave propagates away from the source, in the northern hemisphere two points on the wavefront (constant latitude) would produce a positive parallax measurement $D_\mathrm{par}$ back to the source (the wavefront curves \textit{toward} the source and \textit{away} from the direction of propagation).  At the equator, however, all points along the wavefront experience effectively a plane-wave.  Therefore there
    }
}
\end{minipage}  

\begin{minipage}{0.9\linewidth}
\fbox {
    \parbox{\linewidth}{
    is no well defined parallax distance at this location, the value of $D_\mathrm{par} \rightarrow \infty$, which is what we see at $\chi = \frac{1}{\sqrt{k}}\frac{\pi}{2}$ (i.e. $R = \frac{1}{\sqrt{k}}$). \\
    
    % Then interestingly, in the southern hemisphere two points on the wavefront (constant latitude) would now produce a \textit{negative} parallax measurement $D_\mathrm{par}$ back to the source (the wavefront curves \textit{away} the source and \textit{toward} the direction of propagation - toward the source antipode).  This is consistent when carefully considering the math.  Again using equation~\ref{eqn: r(chi) definition} we see that:
    % \begin{equation}
    %     D_\mathrm{par} = \frac{a(t_0)\sin\left(\sqrt{k}\chi\right)}{\sqrt{k}\sqrt{1-\sin^2\left(\sqrt{k}\chi\right)}} = \frac{a(t_0)\sin\left(\sqrt{k}\chi\right)}{\sqrt{k}\sqrt{\cos^2\left(\sqrt{k}\chi\right)}} = \frac{a(t_0)}{\sqrt{k}}\tan\left(\sqrt{k}\chi\right) , 
    % \end{equation}
    % but take special note that in the final equality here we are \textit{choosing} to take the \textit{positive} solution of the square root for values $0\leq \chi < \frac{1}{\sqrt{k}}\frac{\pi}{2}$ and the \textit{negative} solution of the square root for values $\frac{1}{\sqrt{k}}\frac{\pi}{2} < \chi \leq \frac{1}{\sqrt{k}}\pi$.  This makes the math consistent with our conceptual understanding of the parallax distance as just described, otherwise taking the positive solution of the square root would result in again positive parallax distance measurements to the source even when in the southern hemisphere.  This needs to be kept in mind when working with equation~\ref{eqn: Dpar - R - DM - DC - DL relationship} for the closed universe only, as the flat and open universe cases don't share this property.
    Then interestingly, in the southern hemisphere two points on the wavefront (constant latitude) would now produce a \textit{negative} parallax measurement $D_\mathrm{par}$ back to the source (the wavefront curves \textit{away} the source and \textit{toward} the direction of propagation - toward the source antipode).  This is consistent when carefully considering the math.  For closed geometry, the expression for $D_\mathrm{par}$ in equations~\ref{eqn: parallax distance Dpar} and~\ref{eqn: Dpar - R - DM - DC - DL relationship} really has a $\pm$ sign in front of the expression.  Again using equation~\ref{eqn: r(chi) definition} and including the $\pm$ sign we see that:
    \begin{equation*}
        D_\mathrm{par} = \pm\frac{a(t_0)\sin\left(\sqrt{k}\chi\right)}{\sqrt{k}\sqrt{1-\sin^2\left(\sqrt{k}\chi\right)}} = \pm\frac{a(t_0)\sin\left(\sqrt{k}\chi\right)}{\sqrt{k}\sqrt{\cos^2\left(\sqrt{k}\chi\right)}} = \frac{a(t_0)}{\sqrt{k}}\tan\left(\sqrt{k}\chi\right) . 
    \end{equation*}
    When written in terms of $\chi$, the sign ambiguity in equations~\ref{eqn: parallax distance Dpar} and~\ref{eqn: Dpar - R - DM - DC - DL relationship} is resolved.  In summary, for values $0\leq \chi < \frac{1}{\sqrt{k}}\frac{\pi}{2}$ (the northern hemisphere) the parallax distance is positive, for $\chi = \frac{1}{\sqrt{k}}\frac{\pi}{2}$ (the equator) the parallax distance diverges, and for values $\frac{1}{\sqrt{k}}\frac{\pi}{2} < \chi \leq \frac{1}{\sqrt{k}}\pi$ (the southern hemisphere) the parallax distance is negative.
    
    }
}
\end{minipage}

\chapter{Supporting Material}
\label{App: supporting material}

\section{Order of Magnitude Estimates}\label{app: Order of Magnitude Estimates}

The following are helpful order of magnitude estimates of important quantities that appear in this paper.
\begin{align}
    \Delta \tau_c &\approx (430 \text{ million years}) \left(\frac{10^9 M_\odot}{\mathcal{M}}\right)^{5/3} \left(\frac{1 \text{ nHz}}{\omega_0}\right)^{8/3} , \tag{\ref{eqn: time to coalescence} r} \\[8pt]
    \theta_c &\approx (3.5\times10^6 \text{ cycles}) \left(\frac{10^9 M_\odot}{\mathcal{M}}\right)^{5/3} \left(\frac{1 \text{ nHz}}{\omega_0}\right)^{5/3} , \tag{\ref{eqn: thetac} r} \\[8pt]
    F &\approx 0.0003 \left(\frac{\omega_0}{1\text{ nHz}}\right) \left(\frac{L}{1\text{ kpc}}\right)^2 \left(\frac{100\text{ Mpc}}{R}\right) , \tag{\ref{eqn: Fresnel number} r} \\[8pt]
    \lambda_\mathrm{gw} &\approx (30.5 \text{ pc}) \left(\frac{1\text{ nHz}}{\omega_0}\right) , \tag{\ref{eqn: gw wavelength} r} \\[8pt]
    h_0 &\approx 5.5\times10^{-16} \left(\frac{\mathcal{M}}{10^9 M_\odot}\right)^{5/3} \left(\frac{100\text{ Mpc}}{R}\right) \left(\frac{\omega_0}{1\text{ nHz}}\right)^{2/3} , \tag{\ref{eqn: amplitude 0 of metric perturbation} r} \\[8pt]
    A_\mathrm{E,res} &\approx (140 \text{ ns}) \left(\frac{\mathcal{M}}{10^9 M_\odot}\right)^{5/3} \left(\frac{100\text{ Mpc}}{R}\right) \left(\frac{1\text{ nHz}}{\omega_0}\right)^{1/3} , \tag{\ref{eqn: earth term amplitude} r} \\[8pt]
    \mathrm{Residual} &\sim A_\mathrm{E,res} . \label{eqn: residual order of magnitude}
\end{align}
The final equation~\ref{eqn: residual order of magnitude} of course ignores the many effects such as the antenna patterns and constructive/destructive interference effects of the Earth and pulsar terms, but as a rough proxy it is useful.

%---------------------------------------------------------------------------------
%---------------------------------------------------------------------------------
\newpage
\section{Data Reference Tables}\label{app: data reference}

\begin{table}[h!]
\centering
    \begin{tabular}{ c|c|c||c|c|c } 
             \hline
             & \begin{tabular}{c}$\mathcal{M}$ \\ $\left[10^9 \text{ M}_\odot\right]$\end{tabular} & \begin{tabular}{c}$\omega_0$ \\ $\left[\text{nHz}\right]$\end{tabular} & \begin{tabular}{c}$\Delta \tau_c$ \\ $\left[\text{kyr}\right]$\end{tabular} & \begin{tabular}{c} $F$ \\ $\left(L=1\text{ kpc, } \right.$ \\ $\left.R=100 \text{ Mpc}\right)$\end{tabular} & \begin{tabular}{c}$\lambda_\mathrm{gw}$ \\ $\left[\text{pc}\right]$\end{tabular} \\
             \hline
             Source 1 & 10   & 30 & 1.076             & 0.0098 & 1.017 \\
             Source 2 & 0.03 & 20 & $5.085\times10^4$ & 0.0066 & 1.526 \\
             \hline
    \end{tabular}
    \caption[Gravitational Wave Source Data (Chapter~\ref{ch:results - first investigation})]{These are the parameters of the example SMBHB sources we used in the study in Chapter~\ref{ch:results - first investigation}.  We fixed the angular parameters for each source at the same values: $\left\{ \theta, \ \phi, \ \iota, \ \psi, \ \theta_0 \right\} = \left\{ \frac{\pi}{2}, \ \frac{5\pi}{3}, \ \frac{\pi}{4}, \ \frac{\pi}{3}, \ 1 \right\}$ rad.  Source coordinate distance $R$ varied in each simulation and is indicated in the figures.  For quick reference we include the coalescence time of each source (computed from equation~\ref{eqn: time to coalescence}), a representative value of the Fresnel number (computed from equation~\ref{eqn: Fresnel number}) for that source placed at $R=100$ Mpc (e.g. the approximate distance of the Coma Cluster) with a pulsar at $L=1$ kpc, and the gravitational wavelength of each source (computed from equation~\ref{eqn: gw wavelength}).}
    \label{tab: source parameters reference}
\end{table}

\begin{table}
\centering
    \begin{tabular}{ l|c|c|c|c|c||c|c } 
            \hline
             & Name & \begin{tabular}{c}F.O.M. \\  $L \ \left[\text{kpc}\right]$\end{tabular} & \begin{tabular}{c}$L$ \\ $\left[\text{kpc}\right]$\end{tabular} & \begin{tabular}{c}$\theta_p$ \\ $\left[\text{rad}\right]$\end{tabular} & \begin{tabular}{c}$\phi_p$ \\ $\left[\text{rad}\right]$\end{tabular} &  \begin{tabular}{c} $F$ \\ $\left(\omega_0=10 \text{ nHz,}\right.$ \\ $\left.R=100\text{ Mpc} \right)$\end{tabular} &  \begin{tabular}{c} $\Delta L_1$ \\ $\left[\text{pc}\right]$ \\ $\left(\text{Source 1}\right)$ \end{tabular} \\  
             \hline
            1. & J0030+0451 & 0.53 & 0.36 & 1.49 & 0.13 & 0.0004 & 1.65 \\
            2. & J1744-1134 & 0.55 & 0.40 & 1.77 & 4.64 & 0.0005 & 5.39 \\
            3. & J2145-0750 & 0.57 & 0.53 & 1.71 & 5.70 & 0.0009 & 8.89 \\
            4. & J2214+3000 & 0.60 & 0.60 & 1.05 & 5.82 & 0.0012 & 3.68 \\
            5. & J1012+5307 & 0.63 & 0.70 & 0.64 & 2.67 & 0.0016 & 0.68 \\
            6. & J1614-2230 & 0.68 & 0.70 & 1.96 & 4.25 & 0.0016 & 2.08 \\
            7. & J1643-1224 & 0.68 & 0.74 & 1.79 & 4.38 & 0.0018 & 2.82 \\
            8. & J0613-0200 & 0.69 & 0.78 & 1.61 & 1.63 & 0.0020 & 0.54 \\
            9. & J0645+5158 & 0.70 & 0.80 & 0.66 & 1.77 & 0.0021 & 0.64 \\
            10. & J1832-0836 & 0.71 & 0.81 & 1.72 & 4.85 & 0.0021 & 12.13 \\
            11. & J2302+4442 & 0.72 & 0.86 & 0.79 & 6.03 & 0.0024 & 2.03 \\
            12. & J1918-0642 & 0.72 & 0.91 & 1.69 & 5.06 & 0.0027 & 45.35 \\
            13. & J0740+6620 & 0.72 & 0.93 & 0.41 & 2.01 & 0.0028 & 0.73 \\
            14. & J1455-3330 & 0.73 & 1.01 & 2.16 & 3.91 & 0.0033 & 1.27 \\
            15. & J1741+1351 & 0.73 & 1.08 & 1.33 & 4.63 & 0.0038 & 5.04 \\
            16. & J1909-3744 & 0.74 & 1.14 & 2.23 & 5.02 & 0.0043 & 4.46 \\
            17. & J2010-1323 & 0.75 & 1.16 & 1.80 & 5.28 & 0.0044 & 37.55 \\
            18. & J1713+0747 & 0.75 & 1.18 & 1.43 & 4.51 & 0.0046 & 3.92 \\
            19. & J1923+2515 & 0.76 & 1.20 & 1.13 & 5.08 & 0.0047 & 9.55 \\
            20. & B1855+09   & 0.77 & 1.20 & 1.40 & 4.96 & 0.0047 & 19.62 \\
            21. & J1024-0719 & 0.78 & 1.22 & 1.70 & 2.73 & 0.0049 & 0.57 \\
            22. & J0023+0923 & 0.78 & 1.25 & 1.41 & 0.10 & 0.0051 & 1.71 \\
            23. & J2043+1711 & 0.79 & 1.25 & 1.27 & 5.43 & 0.0051 & 16.20 \\
            24. & J1453+1902 & 0.80 & 1.27 & 1.24 & 3.90 & 0.0053 & 1.30 \\
            25. & J1853+1303 & 0.81 & 1.32 & 1.34 & 4.95 & 0.0057 & 15.40 \\
            26. & J1944+0907 & 0.82 & 1.36 & 1.41 & 5.17 & 0.0061 & 67.61 \\
            27. & J2017+0603 & 0.83 & 1.40 & 1.47 & 5.31 & 0.0064 & 130.45 \\
            28. & J2317+1439 & 0.84 & 1.43 & 1.31 & 6.10 & 0.0067 & 2.73 \\
            29. & J1738+0333 & 0.84 & 1.47 & 1.51 & 4.62 & 0.0071 & 5.49 \\
            30. & J1640+2224 & 0.84 & 1.50 & 1.18 & 4.36 & 0.0074 & 2.49 \\
            31. & J1910+1256 & 0.85 & 1.50 & 1.34 & 5.02 & 0.0074 & 20.71 \\
            32. & J0340+4130 & 0.85 & 1.60 & 0.85 & 0.96 & 0.0084 & 0.77 \\
            33. & J2033+1734 & 0.85 & 1.74 & 1.26 & 5.38 & 0.0099 & 17.61 \\
            34. & J1600-3053 & 0.85 & 1.80 & 2.11 & 4.19 & 0.0106 & 1.78 \\
            35. & J2229+2643 & 0.86 & 1.80 & 1.10 & 5.89 & 0.0106 & 3.48 \\
            36. & J1903+0327 & 0.87 & 1.86 & 1.51 & 4.99 & 0.0113 & 31.90 \\
            37. & B1937+21   & 0.88 & 3.50 & 1.19 & 5.15 & 0.0401 & 13.55 \\
            38. & J0931-1902 & 0.88 & 3.72 & 1.90 & 2.49 & 0.0453 & 0.54
    \end{tabular} 
\end{table}
\begin{table}
\centering
    \begin{tabular}{ l|c|c|c|c|c||c|c } 
            \hline
             & Name & \begin{tabular}{c}F.O.M.\\  $L \ \left[\text{kpc}\right]$\end{tabular} & \begin{tabular}{c}$L$ \\ $\left[\text{kpc}\right]$\end{tabular} & \begin{tabular}{c}$\theta_p$ \\ $\left[\text{rad}\right]$\end{tabular} & \begin{tabular}{c}$\phi_p$ \\ $\left[\text{rad}\right]$\end{tabular} &  \begin{tabular}{c} $F$ \\ $\left(\omega_0=10 \text{ nHz,}\right.$ \\ $\left.R=100\text{ Mpc} \right)$\end{tabular} &  \begin{tabular}{c} $\Delta L_1$ \\ $\left[\text{pc}\right]$ \\ $\left(\text{Source 1}\right)$ \end{tabular} \\  
             \hline
            39. & B1953+29   & 0.89 & 6.30 & 1.06 & 5.22 & 0.1300 & 7.96 \\
            40. & J1747-4036 & --- & 7.15 & 2.28 & 4.66 & 0.1675 & 2.80 \\
            \hline
    \end{tabular}
    \caption[PTA Data (Chapter~\ref{ch:results - first investigation})]{This is the pulsar timing array we used in the study in Chapter~\ref{ch:results - first investigation}, which consists of 40 pulsars from the NANOGrav PTA in~\citet{NG_11yr_data}.  The ``Figure-of-merit'' (F.O.M.) pulsar distance values were used only in the studies in Section~\ref{subsec: Figure-of-Merit}.  Note the final pulsar has no F.O.M. $L$ value because it was a variable and is indicated in the discussion and figures of that section.  The pulsar data in the $L$, $\theta_p$, and $\phi_p$ columns was taken from version 1.57 of the ATNF catalogue~\citep{ATNF_catalogue}.  For quick reference we include a representative value of the Fresnel number (computed from equation~\ref{eqn: Fresnel number}) for that pulsar with a source with an orbital frequency of $\omega_0=10$ nHz and a distance of $R=100$ Mpc (e.g. the approximate distance of the Coma Cluster), as well as the monochromatic plane-wave pulsar wrapping cycle distance computed from equation~\ref{eqn:Delta L in monochromatic regimes} for Source 1 (Table~\ref{tab: source parameters reference}).}
    \label{tab: pta reference}
\end{table}

\begin{figure}[hb]
    \centering
    \includegraphics[width=1\linewidth]{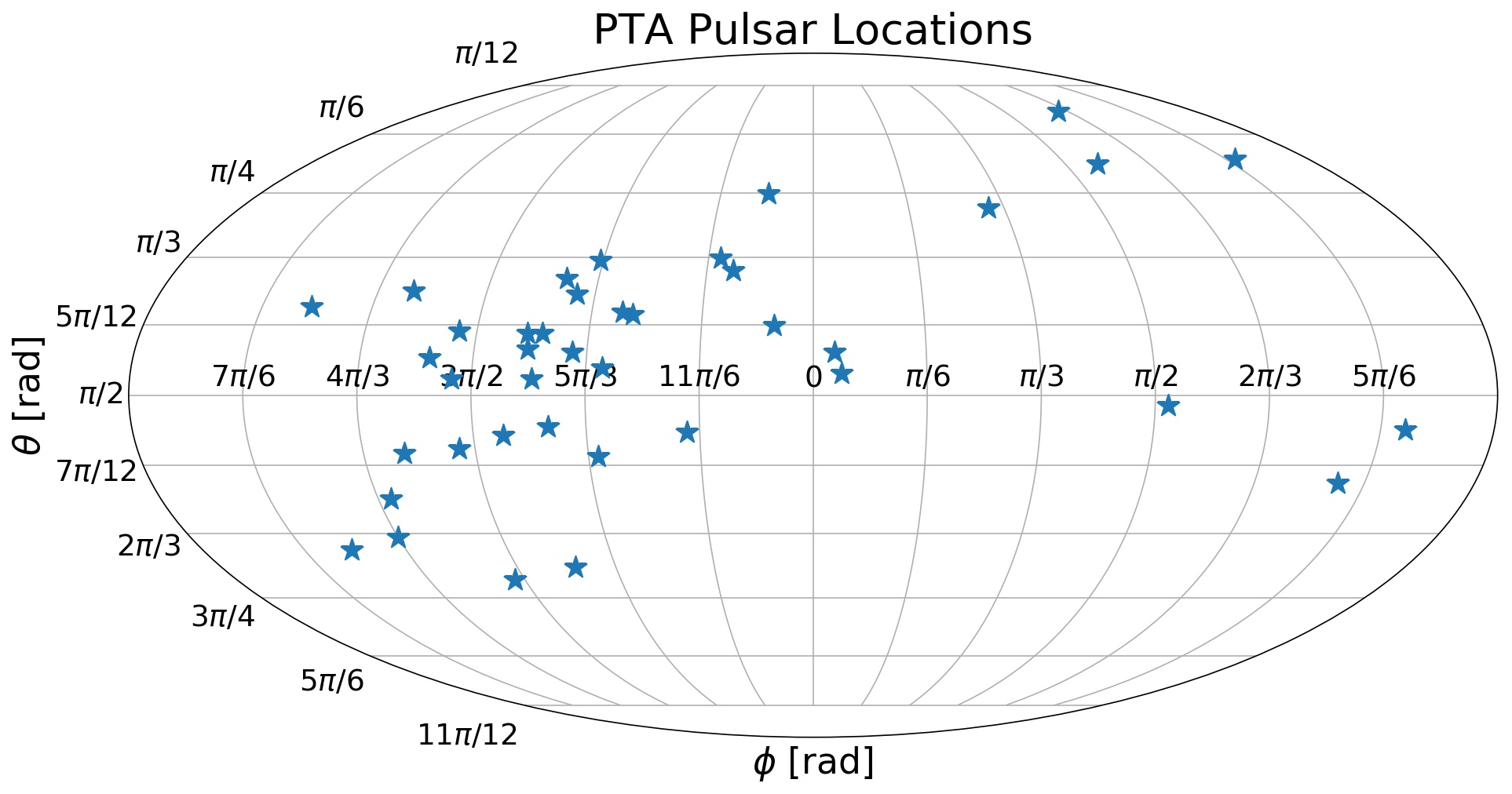}
    \caption[PTA Skymap (Chapter~\ref{ch:results - first investigation})]{A sky map of the pulsars in Table~\ref{tab: pta reference} that we used in our timing residual simulations in Chapter~\ref{ch:results - first investigation}.  See also Figure~\ref{fig: vol localization - scale factor / S1 localization - vol at 100 Mpc, sky at 1 Gpc}.}
    \label{fig: PTA pulsar sky plot locations}
\end{figure}

\begin{figure}[b]
\centering
  \begin{subfigure}[t]{0.55\linewidth}
  \centering
    \includegraphics[width=\linewidth]{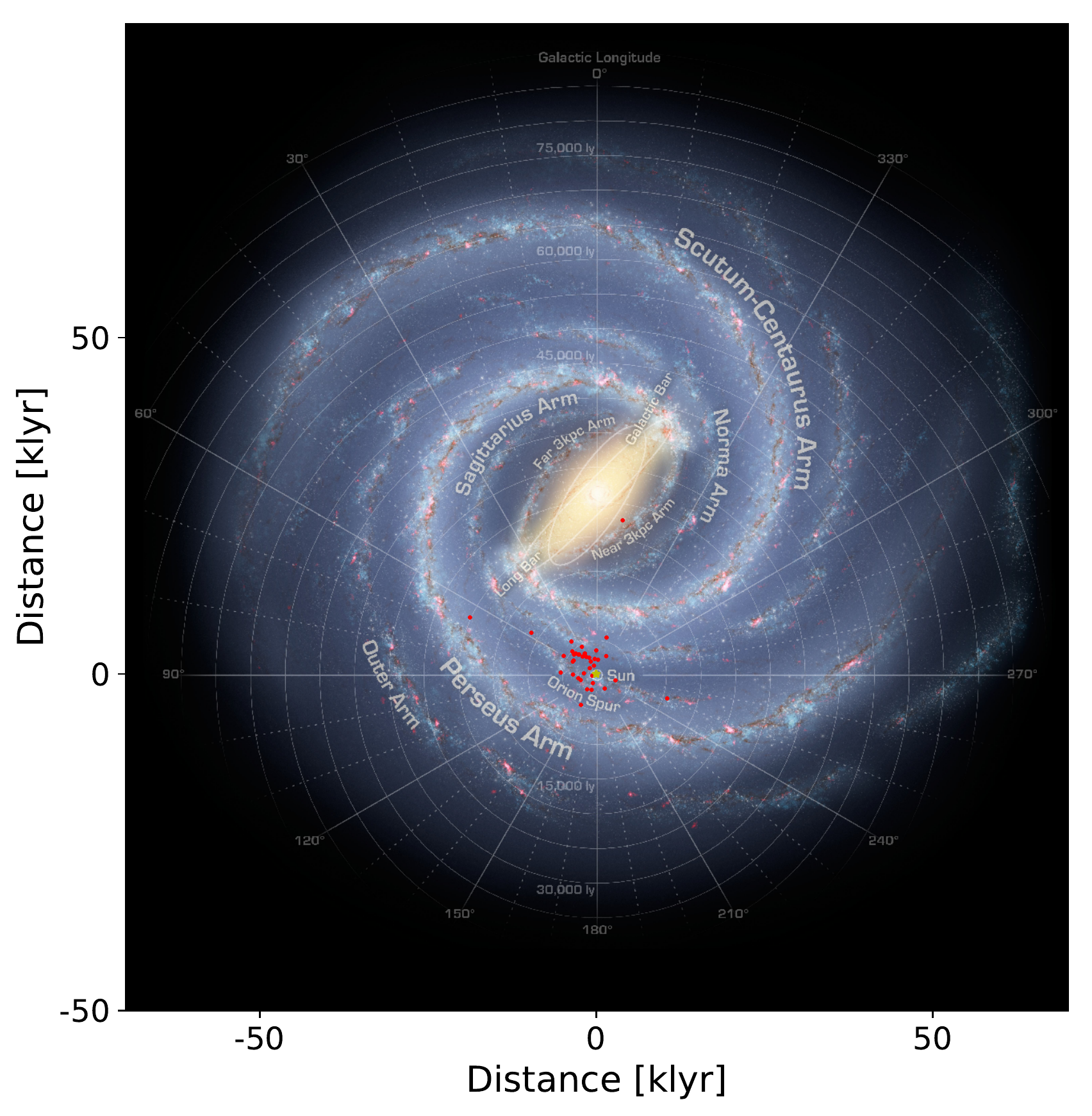}
  \end{subfigure}
  \hfill
  \begin{subfigure}[t]{0.55\linewidth}
  \centering
    \includegraphics[width=\linewidth]{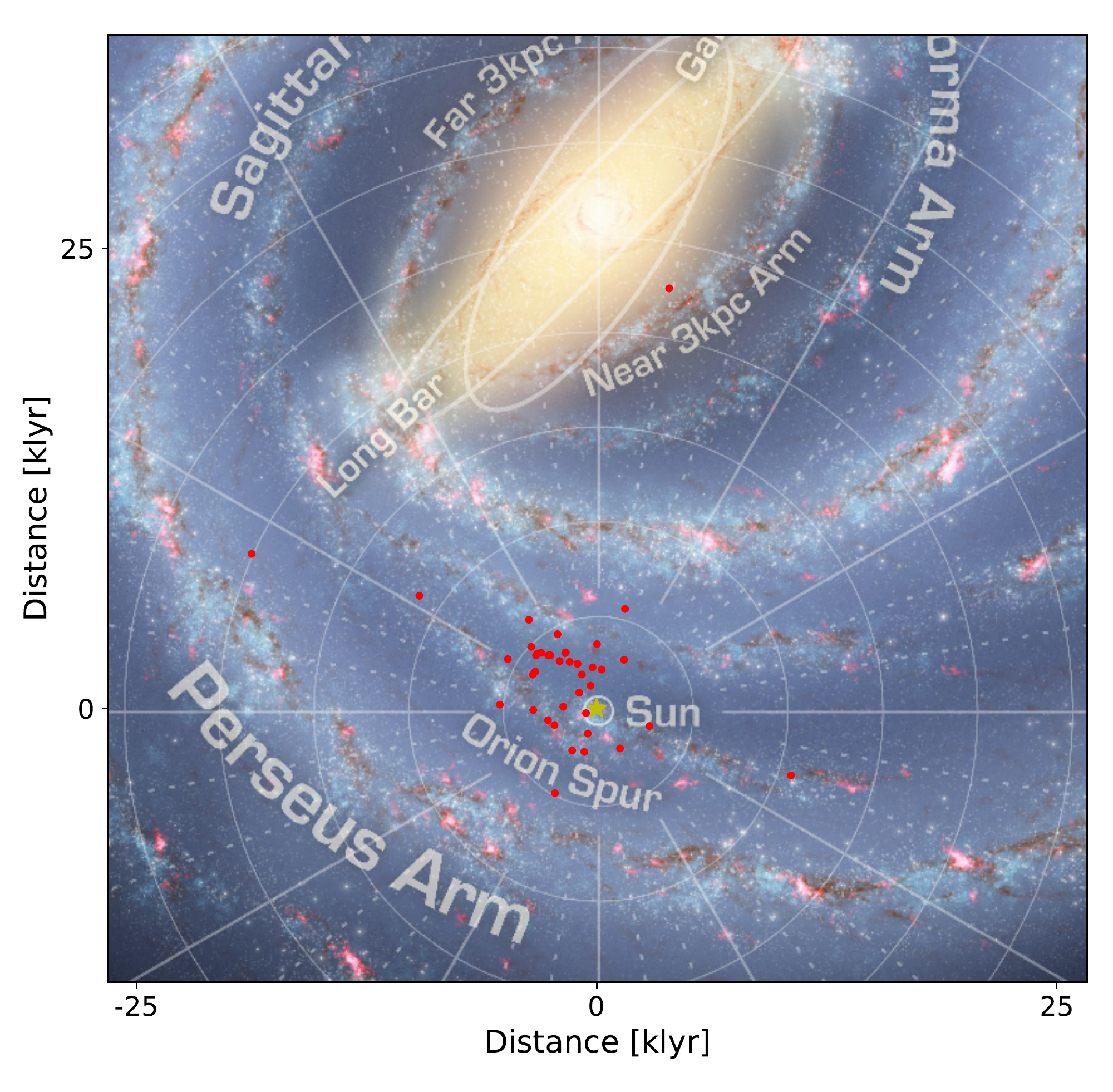}
  \end{subfigure}
\caption[PTA Pulsar Locations (Chapter~\ref{ch:results - first investigation}) in the Milky Way Galaxy]{The approximate locations of the pulsars in Table~\ref{tab: pta reference} (used in Chapter~\ref{ch:results - first investigation}) are overlaid here on a highly accurate artistic representation of the Milky Way Galaxy.  \textbf{Milky Way Image Credit}: NASA/JPL-Caltech/ESO/R. Hurt}
\label{fig: PTA MW map}
\end{figure}

\begin{figure}[b]
\centering
  \begin{subfigure}[t]{\linewidth}
  \centering
    \includegraphics[width=\linewidth]{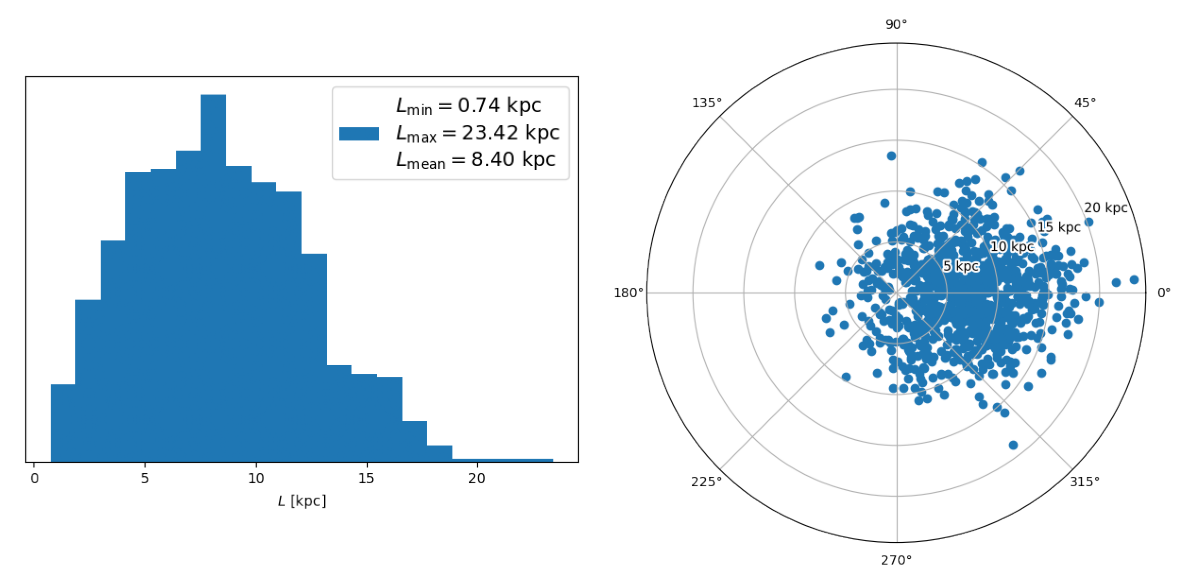}
  \end{subfigure}
  \\
  \begin{subfigure}[t]{\linewidth}
  \centering
    \includegraphics[width=\linewidth]{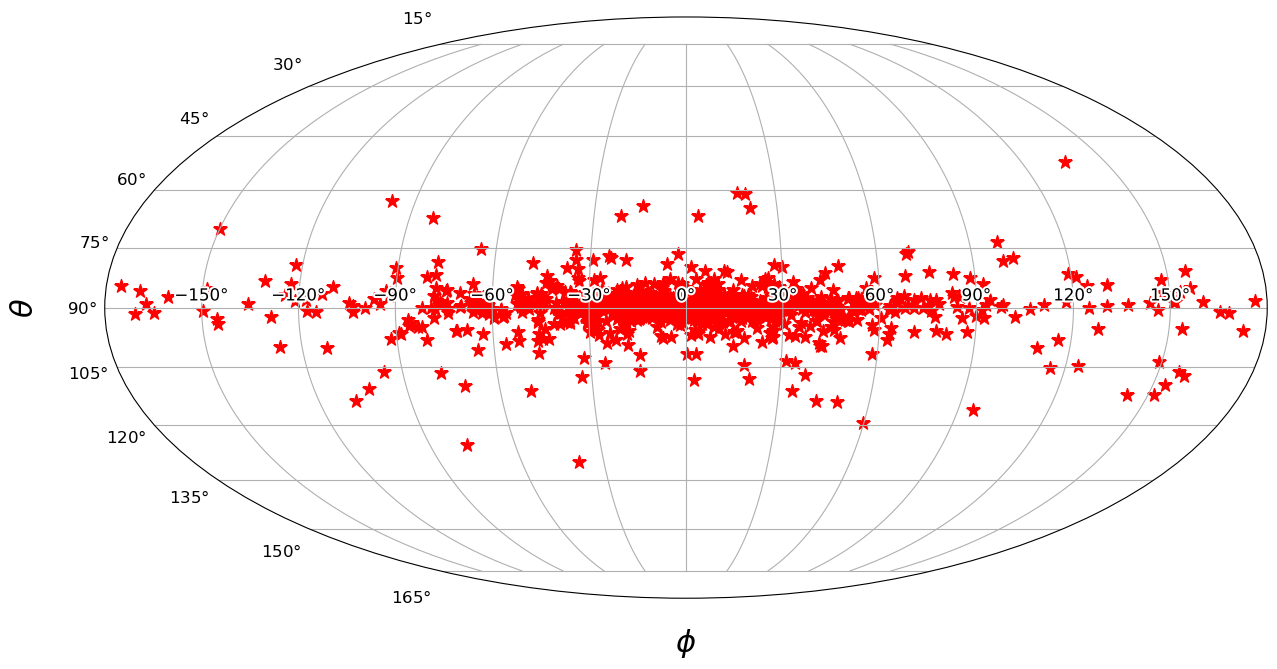}
  \end{subfigure}
    \caption[Fiducial PTA (Chapter~\ref{ch:results - H0 measurement})]{An example of the fiducial PTA we used for our studies in Chapter~\ref{ch:results - H0 measurement}, generated from the distribution given in equation~\ref{eqn: fiducial PTA}.  A population of 1000 pulsars are shown here.  The top left panel gives a histogram of their distances, the top right panel shows their distribution in the disk of the Milky Way (the center of the coordinate system here is at the Earth's position, with the Milky way at 
    %$x_\mathrm{MW} = 25$ kpc),
    %
    $x_\mathrm{MW} \approx 8$ kpc), and the bottom panel shows pulsar angular positions on the sky (Galactic center is at $\phi = 0^{\circ}$).}
\label{fig: my fiducial PTA}
\end{figure}

\end{appendices}

% %%%%%%%%%%%%%%%%%%%%%%%%%%%%%%%%%%%%%%%%%%%%%%%%%%%%%%%%%%%%%%%%%%%%
% %%%%% CURRICULUM VITAE 
% % Finally you must include your cv.  You can do that whatever way you like including by formatting it in a totally different program.

% % If you would like to grab it from some other source then be sure the page numbering is consecutive with the end of the bibliography and be sure it appears on the table of contents by adding a line such as: \addcontentsline{toc}{chapter}{Curriculum Vitae}

% % NOTE:  I had to add this combo of "\phantompage" and "\newpage" in order to get the TOC to point to the correct page.  Otherwise, without these the TOC seems to point back to the page that the Bibliography begins on.
% % REFERENCES:
% % ---> http://blog.bharatbhole.com/inserting-pages-from-an-external-pdf-document-within-a-latex-document/
% % ---> https://stackoverflow.com/questions/782187/latex-table-of-contents-links-to-wrong-section
% \phantomsection
% \newpage
% \phantomsection
% \addcontentsline{toc}{chapter}{Curriculum Vitae}
% \includepdf[pages=-]{chapters/CV}

\end{document}